\documentclass[aip, reprint]{revtex4-2}
\usepackage[utf8]{inputenc}
\usepackage{amsmath, amssymb, physics, hyperref, graphicx, multirow}
\usepackage{comment}
\newcommand{\angstrom}{\mbox{\normalfont\AA}}

\newcommand{\Sec}{Sec.}
\newcommand{\Subsec}{Sec.}
\newcommand{\Supplementary}{the Supplementary Material}

\draft 

\begin{document}

\title{Bayesian, frequentist, and information geometric approaches to parametric uncertainty quantification of classical empirical interatomic potentials}
\author{Yonatan Kurniawan}
\author{Cody L. Petrie}
\author{Kinamo J. Williams Jr.}
\author{Mark K. Transtrum}
\email{mktranstrum@byu.edu}
\affiliation{Department of Physics and Astronomy, Brigham Young University, Provo, Utah, 84604, USA}

\author{Ellad B. Tadmor}
\author{Ryan S. Elliott}
\author{Daniel S. Karls}
\affiliation{Department of Aerospace Engineering and Mechanics, University of Minnesota, Minneapolis, Minnesota, 55455, USA}

\author{Mingjian Wen}
\affiliation{Energy Technologies Area, Lawrence Berkeley National Laboratory, Berkeley, California, 94720, USA}

\date{\today}

\begin{abstract}
    In this paper, we consider the problem of quantifying parametric uncertainty in classical empirical interatomic potentials (IPs) using both Bayesian (Markov Chain Monte Carlo) and frequentist (profile likelihood) methods.
We interface these tools with the Open Knowledgebase of Interatomic Models and study three models based on the Lennard-Jones, Morse, and Stillinger--Weber potentials.
We confirm that IPs are typically sloppy, i.e., insensitive to coordinated changes in some parameter combinations.
Because the inverse problem in such models is ill-conditioned, parameters are unidentifiable.
This presents challenges for traditional statistical methods, as we demonstrate and interpret within both Bayesian and frequentist frameworks.
We use information geometry to illuminate the underlying cause of this phenomenon and show that IPs have global properties similar to those of sloppy models from fields such as systems biology, power systems, and critical phenomena.
IPs correspond to bounded manifolds with a hierarchy of widths, leading to low effective dimensionality in the model.
We show how information geometry can motivate new, natural parameterizations that improve the stability and interpretation of uncertainty quantification analysis and further suggest simplified, less-sloppy models.

\end{abstract}


\maketitle 

\section{INTRODUCTION}
\label{sec:introduction}
Interatomic potentials (IPs) are a foundational tool in computational materials science.\cite{Tadmor_Modeling_Materials}
They allow modelers to make efficient predictions of materials properties without reference to the complicated sub-atomic structure. 
Recently there has been considerable interest in applying methods of uncertainty quantification (UQ) to IPs.\cite{Wen_Shirodkar_Plechas_Kaxiras_Elliott_Tadmor_2017, Longbottom_Brommer_2019, Frederiksen_Jacobsen_Brown_Sethna_2004, wen2020dropout}
UQ assesses the reliability of materials predictions, leveraging tools from statistical inference.\cite{Chernatynskiy_Phillpot_LeSar_2013}
Statistical analysis of similar inverse problems in physics has motivated the study of \emph{sloppy models}.\cite{Brown_Sethna_2003}
Sloppy models lead to extremely ill-conditioned inverse problems and pose several challenges for standard statistical methods.\cite{Transtrum_Machta_Sethna_2010, Transtrum_Machta_Brown_Daniels_Myers_Sethna_2015}
This work considers the application of UQ to IPs in the context of sloppy models.
We focus on the structural uncertainties of the IPs as opposed to the random errors that come from the details of the simulation.\cite{Wan_Sinclair_Coveney_2021}
We find that many IPs are sloppy, which leads to challenges in interpreting UQ results, and use information geometry to mitigate some of these challenges.
(We will be using the terms ``IPs'' and ``models'' largely interchangeably in this paper, with ``models'' used in more generic settings, and ``IPs'' when specifically thinking of the training and predictions of interatomic models, i.e., potentials.)

Classical IPs have been widely used in materials science to circumvent the computational cost of quantum calculations, such as density functional theory (DFT), by approximating the interaction energy between atoms without considering the electrons.
However, a universal functional form that can describe all types of atomic bonding has not been discovered.
Thus, IPs are often designed for specific purposes, resulting in a plethora of models.\cite{Brenner_2000}
Efforts such as the Open Knowledgebase of Interatomic Models (OpenKIM)\cite{Tadmor_Elliott_Sethna_Miller_Becker_2011} aim to organize and standardize these IPs.

In the development of classical empirical IPs, the parameters are typically fit to match experimental or first principles data of some microscopic properties, such as the lattice parameters and elastic constants of single crystals, or the potential energy and atomic forces associated with various atomic configurations.\cite{Ercolessi_Adams_1994}
They are then used in conjunction with simulation codes to predict other properties that are not used in the fitting process.
Thus, UQ is relevant for assessing the reliability of these out-of-sample predictions.

\begin{figure*}[!ht]
    \centering
    \includegraphics[width=0.8\textwidth]{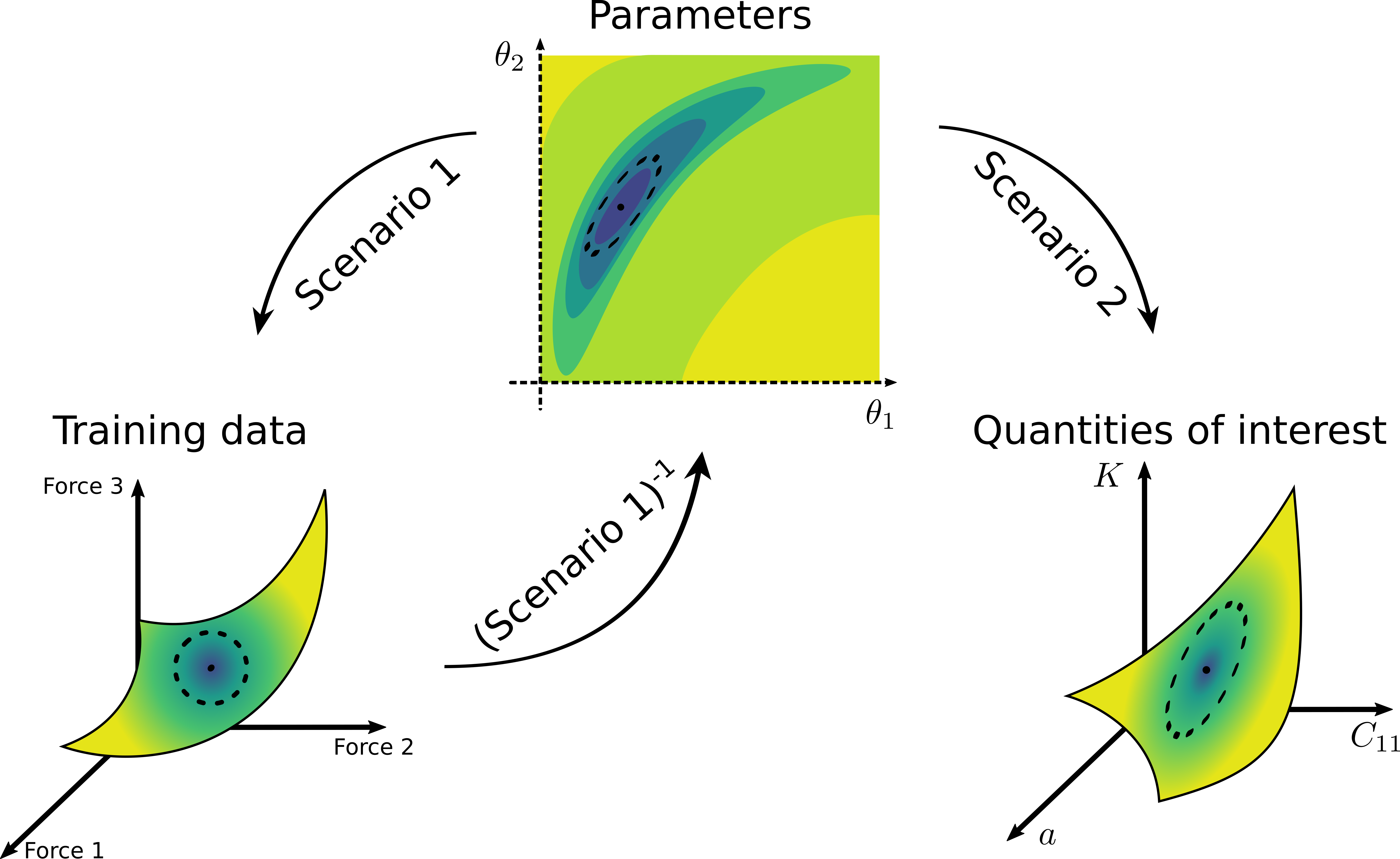}
    \caption[Uncertainty propagation in Information geometry]{
        Uncertainty propagation in information geometry.
        An IP is described by several parameters, collectively forming a parameter space (top center).
        By considering the predictions for all allowed parameter values for a given evaluation scenario, the model (IP) maps out a set of possible predictions in data space, known as the model manifold (bottom left, right).
        Uncertainty in training data (bottom left) is propagated through the inverse function to uncertainties in parameter space, represented here by contours of constant likelihood, i.e., cost, in the parameter space.
        Parametric uncertainty can then be propagated forward in another evaluation scenario to material quantities of interest, such as lattice constant, elastic constant, and bulk modulus (bottom right).
    }
    \label{fig:3_spaces}
\end{figure*}

In this paper, we are primarily interested in parametric uncertainty, i.e., uncertainty in the model's parameters,  which is quantified through, for example, a Bayesian posterior distribution or confidence regions on the parameter space.
In our formulation, the IP is used in two evaluation scenarios: one that makes predictions for the training data (e.g., energy and forces), and a second that makes predictions for other material quantities of interest.
By varying the parameters, the IP traces out a set of possible predictions for each scenario.
The set of predictions made by an IP for each evaluation scenario generates a corresponding \emph{model manifold},\cite{Transtrum_Machta_Sethna_2010} which can be studied using methods of information geometry. The IP parameters act as coordinates on the model manifold and distances on the manifold measure statistical identifiability.
The UQ process propagates uncertainties from training data to uncertainties in parameters via the (pseudo) inverse of the first evaluation scenario.
These uncertainties then propagate forward through the second scenario to give uncertainties in the predictions for the quantities of interest.
The inverse of an evaluation scenario is not given explicitly, but is only accessible through iterative evaluations of the scenario.
Consequently, the first uncertainty propagation process is the more challenging one and the focus of this study.
Given the parametric uncertainties, they can then be propagated to other quantities of interest, for example, as in Ref.~ \onlinecite{Hughes_Horstemeyer_Carino_Sukhija_Lawrimore_Kim_Baskes_2015}.
This entire process is illustrated in Fig.~\ref{fig:3_spaces}.

Many UQ methods have been developed to propagate uncertainties in data to uncertainties in parameters.\cite{Kennedy_OHagan_2001}
In materials science, Markov Chain Monte Carlo (MCMC) sampling of the Bayesian posterior is the most common approach.
Being a Bayesian method, this requires a prior distribution, and several prior distributions have been used, including uniform,\cite{Angelikopoulos_Papadimitriou_Koumoutsakos_2012, Rizzi_Jones_Debusschere_Knio_2013, Messerly_Shirts_Kazakov_2018, Vohra_Nobakht_Shin_Mahadevan_2018, Vohra_Mahadevan_2019, Dhaliwal_Nair_Singh_2019a, Dhaliwal_Nair_Singh_2019b} normal,\cite{De_Simon_Iglesias_Jones_Wood_2018} Jeffreys prior,\cite{Rizzi_Najm_Debusschere_Sargsyan_Salloum_Adalsteinsson_Knio_2012}, and maximum entropy.\cite{Cools-Ceuppens_Verstraelen_2017}
Other approaches to UQ include $F$-statistics estimations,\cite{Messerly_Knotts_Wilding_2017} ANOVA-based methods,\cite{Tschopp_Chris_Rinderspacher_Nouranian_Baskes_Gwaltney_Horstemeyer_2018} and multi-objective optimization.\cite{Mishra_Hong_Rajak_Sheng_Nomura_Kalia_Nakano_Vashishta_2018}
Other fields have used the profile likelihood method;\cite{Lu_Wang_Yan_Zhang_Xiao_2013, Chen_Li_Shin_Kim_2016, Wu_Xue_Liu_Ren_2019} to the best of our knowledge, ours is their first application to IPs.

The study of inverse problems in statistical physics has identified an important property of many multi-parameter models, known as \emph{sloppiness}.
Sloppy models are characterized by predictions that are insensitive to coordinated changes in combinations of parameters.\footnote{It is important to point out that models may manifest sloppiness in some evaluation scenarios, but not others. Thus ``sloppiness'' is a property of both the model and the specific set of predictions being considered.\cite{Machta_Chachra_Transtrum_Sethna_2013}}
Inverse problems for sloppy models are extremely ill-conditioned and, as we will show below, present obstacles for standard UQ methods, and there is broad interest in developing UQ methods that leverage the effective low-dimensionality of these problems to improve UQ performance.\cite{Constantine_Dow_Wang_2014, Cui_Martin_Marzouk_Solonen_Spantini_2014, Cui_Tong_2021}
Sloppiness was first systematically studied in 2003 by Brown and Sethna in the context of systems biology models.\cite{Brown_Sethna_2003}
The relevant object is the Fisher Information Matrix (FIM)\cite{brouwer2018underlying} that quantifies the information that data in a given evaluation scenario carry about parameters in a model.
Eigenvalues of the FIM provide a local measure of sloppiness.
For sloppy models, the FIM eigenvalues span many orders of magnitude and have many small eigenvalues.
These small eigenvalues correspond to sloppy combinations of parameters, i.e., those that are ill-constrained by data.\cite{White_Tolman_Thames_Withers_Mason_Transtrum_2016, Transtrum_Machta_Sethna_2010, Gutenkunst_Waterfall_Casey_Brown_Myers_Sethna_2007, Waterfall_Casey_Gutenkunst_Brown_Myers_Brouwer_Elser_Sethna_2006}

Although sloppy models are usually identified by their characteristic FIM spectrum, the theory of sloppy models is couched in the tools of information geometry, i.e., the application of differential geometry to statistics.\cite{Transtrum_Qiu_2014, Transtrum_Machta_Brown_Daniels_Myers_Sethna_2015}
The key object is the model manifold, which comprises the set of all predictions a model makes within a given evaluation scenario for all allowed parameter values.
As we show in Sec.~\ref{subsec:method_geodesic}, the model manifold is embedded in data space.
This space is useful because the distance corresponds to statistical identifiability.
In other words, points that are distant on the model manifold are statistically distinguishable, while nearby points are not.
Statistical identifiability therefore induces a Riemannian metric on the parameter space that is given by the FIM.
Using the FIM, one can, in principle, measure the ``information distance'' across the manifold.
We refer to such a distance as the manifold's width.
This width depends on the direction in which the distance is measured.
For sloppy models,  widths in some directions can be many orders of magnitude larger than in other directions.
This hierarchy of widths suggests that the model exhibits a low effective dimensionality.\cite{Transtrum_Machta_Sethna_2010, Transtrum_Machta_Sethna_2011,Machta_Chachra_Transtrum_Sethna_2013,quinn2019visualizing,quinn2019chebyshev}
When a model manifold is very thin, the parameters associated with the thin directions are unidentifiable from data.

Sloppy models are ubiquitous in many scientific fields including critical phenomena,\cite{Machta_Chachra_Transtrum_Sethna_2013} systems biology,\cite{Jeong_Zhuang_Transtrum_Zhou_Qiu_2018, White_Tolman_Thames_Withers_Mason_Transtrum_2016, Transtrum_Qiu_2016, Mannakee_Ragsdale_Transtrum_Gutenkunst_2016,  Transtrum_Machta_Brown_Daniels_Myers_Sethna_2015, Transtrum_Qiu_2012} power systems stability,\cite{transtrum2017information} particle accelerators,\cite{Gutenkunst_2007} and others.\cite{Waterfall_Casey_Gutenkunst_Brown_Myers_Brouwer_Elser_Sethna_2006}
In molecular modeling, it has been shown that IPs typically exhibit the same characteristic sloppy FIM spectrum;\cite{Frederiksen_Jacobsen_Brown_Sethna_2004, Wen_Li_Brommer_Elliott_Sethna_Tadmor_2016, Wen_Shirodkar_Plechas_Kaxiras_Elliott_Tadmor_2017, Longbottom_Brommer_2019} however, the techniques of information geometry have not yet found application to IPs.
As materials models increase in complexity, especially with the advent of machine learning models,\cite{zuo2020performance, deringer2019machine, wen2019hybrid, wen2020dropout} sloppiness will become increasingly relevant.
This underscores the importance of understanding the effects of sloppiness on IPs and other materials science models.

This work reports on several novel results applying sloppy model analysis to the UQ of IPs.
Since sloppy models are ill-conditioned, they pose unique challenges for standard UQ methods.
Results are often highly sensitive to details of the problem formulation and difficult to interpret.
We illustrate these challenges with both Bayesian and frequentist UQ methods, introducing the profile likelihood methods to the material modeling community.
A new contribution of this work is the application of information geometry techniques to bear on this problem and discuss how they can illuminate and mitigate these challenges.

This paper is organized as follows:
First, we precisely formulate the problem in \Subsec~\ref{subsec:method_cost} and discuss using the FIM as a local analysis of sloppiness in \Subsec~\ref{subsec:method_fim}.
Then, we discuss the Bayesian methods in \Subsec~\ref{subsec:method_mcmc}.
We follow Frederiksen et al.\cite{Frederiksen_Jacobsen_Brown_Sethna_2004} and introduce an effective sampling temperature to estimate the model accuracy.
This study is the first extensive study of the role of this sampling temperature in the identifiability of the model parameters.
We describe frequentist methods in section \ref{subsec:method_likelihood}; to the best of our knowledge, ours is the first application of profile likelihoods to IPs.
Finally, we describe information geometric tools, specifically geodesics, central to our sloppy model analysis in \Subsec~\ref{subsec:method_geodesic}.
This study is the first application of information geometry to interatomic potentials.
In particular, we demonstrate that the same global properties observed in other sloppy models (specifically manifold boundaries) are also present in IPs.
This opens up the possibility of using novel parameter reductions techniques, such as the manifold boundary approximation method (MBAM)\cite{Transtrum_Qiu_2014} for IP development.
\Subsec~\ref{subsec:interatomic_potentials_and_tests} presents the models used in this study, i.e., the IPs and quantities of interest.
We present the results for each method in \Sec~\ref{sec:results}.
Finally, we discuss the effects sloppiness has on UQ in \Sec~\ref{sec:discussion} and the prospect for accurate, efficient UQ in IPs generally.
From this cross-sectional survey, we make several recommendations that we hope will be beneficial for both IP developers and practitioners.



\section{METHODS}
\label{sec:methods}

In this section, we introduce the general methods that will be used later in this paper.
We pay particular attention to the mathematical assumptions that different methods require and the types of calculations that they enable.
Then, we introduce the IPs and data sets on which we conduct our study.

\subsection{Defining cost}
\label{subsec:method_cost}
The minimal elements for parametric UQ are (1) a collection of data, $\left\{y_m \right\}_{m = 1}^M$ (where $M$ is the number of data points), (2) a parameterized family of models that make predictions $\{f_m(\theta)\}_{m=1}^M$, and (3) a metric for comparing the model predictions to data, $\left\lVert \cdot \right\rVert$.
We assume that the model depends on $N$ parameters $\theta \in \mathcal{D} \subseteq \mathbb{R}^N$.
Here, $\mathcal{D}$ is the physically allowed domain; for example, it is common for some parameters to be restricted to positive values.
It is convenient to interpret both the data and model predictions as vectors in an $M$-dimensional \emph{data space}: $y_m \rightarrow \mathbf{y} \in \mathbb{R}^M$, $f_m(\theta) \rightarrow \mathbf{f}: \mathcal{D} \subseteq \mathbb{R}^N \rightarrow \mathbb{R}^M$.

The third requirement, a metric, defines a \emph{cost function} (also known as a loss function) that quantifies how well specific parameter values fit the available data, $C(\theta) = \left\lVert \mathbf{y} - \mathbf{f}(\theta) \right\rVert$.
The best fit parameters, denoted by $\theta^*$, minimize the cost.
By far the most common choice for a metric is (weighted) least squares
\begin{equation}
    \label{eq:cost_function}
    C(\theta) = \frac{1}{2} \sum_{m=1}^M r_m(\theta)^2,
\end{equation}
where we have introduced the \emph{residuals}
\begin{equation}
    \label{eq:residuals}
    r_m\left(\theta\right) = \frac{y_m - f_m\left(\theta \right)}{\sigma_m}
\end{equation}
that depends on the inverse weights $\sigma_m$ that act as error bars for each data point.

The cost function has a probabilistic interpretation as the negative log-likelihood
\begin{equation}
  \label{eq:probability}
  P(y | \theta) \sim \exp\left( -C(\theta) \right).
\end{equation}
Eq.~\eqref{eq:cost_function} corresponds to the case that residuals are independent and identically distributed Gaussian random variables: $r_m \sim \mathcal{N}(0,1)$, or equivalently $y_m \sim \mathcal{N}(f_m(\theta), \sigma_m^2)$.
Probability acts as a \emph{measure} on data space that induces a measure on parameters space that we use to quantify uncertainty.
For stochastic processes, the stochastic variation in the data is a natural measure, in which case the inverse weights, $\sigma_m$, are often taken to be the standard errors estimated from repeated observations.
Because they are often associated with experimental error bars, we refer to $\sigma_m$ as the ``error bars'' below, even though they may not be explicitly related to any experimental error.
When working with DFT data, one does not usually have an error bar that is exactly analogous to an experimental uncertainty.
However, there is a growing recognition of the need to estimate the errors in DFT calculations.\cite{Grossman_Schwegler_Draeger_Gygi_Galli_2004, Schunck_McDonnell_Sarich_Wild_Higdon_2015}
If such error estimates are available, they can be used as weights in the cost function.
Because predictions may be made for quantities that carry different physical units (e.g., energies vs forces), choosing $\sigma_m$ is minimally necessary for Eq.~\eqref{eq:cost_function} to be dimensionally consistent.
Deliberate selection of $\sigma_m$ is an important part of the UQ problem formulation since any eventual measure of the uncertainty in the model parameters will be derived from the choice of measure in data space.

A common choice for $\sigma$ that we advocate is to choose the weights to be some fractional value of the data,\cite{Fellinger_Park_Wilkins_2010} potentially including an additional cutoff to deal with near-zero data:\cite{Lenosky_Kress_Kwon_Voter_Edwards_Richards_Yang_Adams_1997}
\begin{equation}
  \label{eq:sigmam}
  \sigma_m = \sqrt{c_1^2 + c_2^2 \Vert y_m \Vert^2},
\end{equation}
where $c_1$ is the cutoff or padding term and $c_2$ is a constant that sets of the scale of the uncertainty from the data.
In general, choosing $\sigma_m$ as a fractional tolerance, e.g., 10\% of the data, is a reasonable choice and what we use in this study.
This choice corresponds to $c_1=0$ and $c_2=0.1$.
[The numerical values of our data are large enough that Eq.~\eqref{eq:sigmam} is insensitive to the choice of $c_1$, so we use zero.]

For IPs, errors in the data are often not the primary source of errors in the inferred model.
Because IPs are simplified functional forms that do not capture all of the nuances of quantum mechanics, there is an additional error due to missing physics.
We decompose the model error into contributions from the bias, $b$, representing missing physics, and errors in the data, $\epsilon$.
The data and the model are then related by
\begin{equation}
    \label{eq:model_bias_noise}
    y_m = f_m(\theta) + b_m + \epsilon_m.
\end{equation}
Conceptually the bias can be understood as accuracy of the model, and a fundamental problem is to estimate the bias of an IP for problem-specific applications.

To estimate model accuracy, Frederiksen et al.\cite{Frederiksen_Jacobsen_Brown_Sethna_2004} suggested tempering the cost by an effective temperature which modifies Eq.~\eqref{eq:probability} as
\begin{equation}
  \label{eq:probability_tempered}
  P(y | \theta) \sim \exp\left( -C(\theta)/T \right).
\end{equation}
This is to account for the model inadequacy, or bias, as discussed in Ref.~\onlinecite{Kennedy_OHagan_2001}.
Functionally, the temperature $T$ uniformly scales the weights in the residuals and can be adjusted as a coarse measure of the overall accuracy of the model.
In fact, the tempering technique is a common practice to improve UQ performance when the model does not match the data generating process.\cite{Miller_Dunson_2019}
A natural choice of temperature motivated by analogy with the equipartition theorem\cite{Frederiksen_Jacobsen_Brown_Sethna_2004} is
\begin{equation}
    \label{eq:natural_temperature}
    T_0 = \frac{2 C_0}{N},
\end{equation}
where $C_0$ is the minimum cost of the model and $N$ is the degrees of freedom of the model, which often is set to the number of parameters.
Since this choice of temperature includes the best fit cost, it incorporates information about the accuracy of the model with respect to the training data.
Using this natural temperature, we find that the bias far outweighs the errors in the data and our uncertainty estimates give a rough indication of the accuracy of an IP.

\begin{figure*}[!ht]
    \centering
    \includegraphics[width=0.9\textwidth]{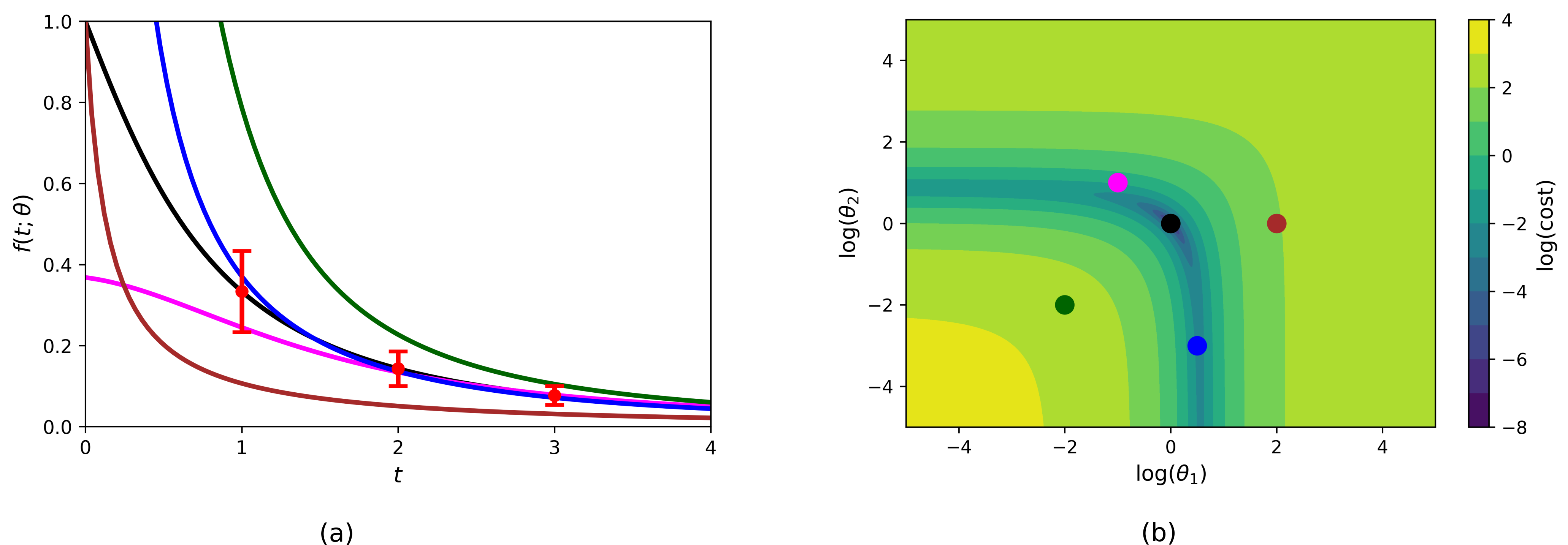}
    \caption[Time series and cost surface for the toy model]{
        Time series (a) and cost surface (b) for the toy model in Eq.~\eqref{eq:toy_model}.
        Data are given by the red points in (a), with the error bars set to 30\% of the data.
        Model predictions for different parameter values and their corresponding points on the cost surface are shown in matching colors.
        Contours in (b) represent the errors between the model predictions and data as calculated in Eq.~\eqref{eq:cost_function}.
    }
    \label{fig:ModelCostcontourCartoon}
\end{figure*}

To illustrate key ideas throughout this section, we use a two-parameter toy model of the form
\begin{equation}
    \label{eq:toy_model}
    f(t; \theta) = \frac{1}{t^2 + \theta_1 t + \theta_2}.
\end{equation}
We make predictions at times $t = 1.0, 2.0, 3.0$, i.e., $f_m(\theta) = f(t_m; \theta)$.
We restrict $\theta_i \geq 0$ (a common physical constraint on parameter values), which suggests working with the log-transformed parameter values.
We use $\mathbf{y} = [1/3, 1/7, 1/13]^T$ with tolerances $\sigma_m$ set to be $30\%$ of the data $\mathbf{y}$ for the purpose of visual clarity.
These data, along with model predictions for several values of the parameters, are shown in Fig.~\hyperref[fig:ModelCostcontourCartoon]{2(a)}.

It is useful to visually consider cost contours for this model, shown in Fig.~\hyperref[fig:ModelCostcontourCartoon]{2(b)}.
In general, we are interested in describing the regions in parameter space with low cost.
The model in Eq.~\eqref{eq:toy_model} is sloppy, as manifested by its insensitivity to coordinated variations in some parameter directions.
Because of this, the cost contours in the sloppy directions are elongated and the aspect ratio of the low-cost canyon around the minimum is very large.
Away from the best fit, many of the cost contours do not close; canyons stretch to the edges of parameter space and flatten into broad plateaus.
Extreme values of the parameters can have finite, and in many cases, very small cost.
These features are ubiquitous in sloppy models, and play a fundamental role in quantifying the parametric uncertainty.

For models with many parameters, direct visualization of the cost surface is not possible.
However, we have constructed this toy model to illustrate issues that are typical of high-dimensional parameter spaces.
In the next Secs. II B--II E, we describe several tools for analyzing the cost surface of multi-parameter models.


\subsection{Fisher information: Sloppy model analysis}
\label{subsec:method_fim}
To quantify the local geometry of the cost surface in a neighborhood of the best fit, we linearize the residuals about $\theta^*$,
\begin{equation}
  \label{eq:residuals_linear}
  r_m(\theta) \approx r_m(\theta^*) + \frac{\partial r_m}{\partial \theta} (\theta - \theta^*).
\end{equation}
To lowest order, the cost function becomes
\begin{equation}
  \label{eq:cost_quadratic}
  C(\theta) \approx C(\theta^*) + \frac{1}{2}(\theta - \theta^*)^T (J^T J) (\theta - \theta^*),
\end{equation}
where we have used the fact that $\nabla C = 0$ at $\theta^*$ and introduced the Jacobian of the residual function $J_{mn} = \partial r_m/\partial \theta_n = -(1/\sigma_m) \partial f_m/\partial \theta_n$ evaluated at $\theta^*$ .
The squared Jacobian appearing in Eq.~\eqref{eq:cost_quadratic} is the Fisher Information Matrix (FIM),
\begin{equation}
  \label{eq:FIM}
  \mathcal{I} = J^T J.
\end{equation}
The FIM is an important statistical quantity; its inverse is a lower bound on the covariance of an unbiased estimator of $\theta$, known as the Cram\'er--Rao bound.\cite{van_den_Bos_2007}

\begin{figure}[!ht]
    \centering
    \includegraphics[width=0.45\textwidth]{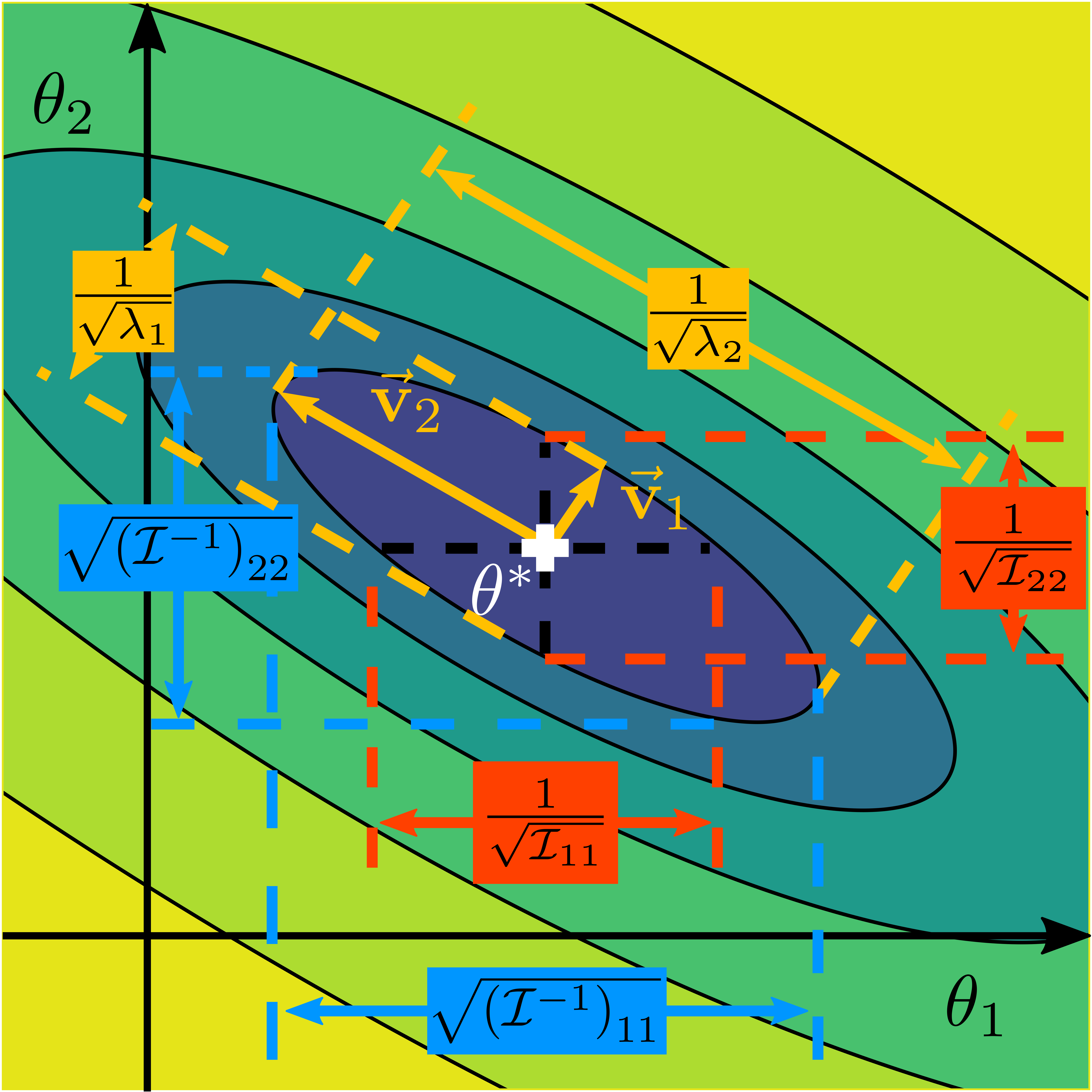}
    \caption[The FIM in parameter space]{
        The Fisher information describes the local geometry of the cost surface.
        Locally, cost contours are represented by ellipses where the axes are aligned with the eigenvectors of the FIM.
        The aspect ratio of the ellipse is given by the square root of the ratio of the eigenvalues.
    } 
    \label{fig:unit_ball}
\end{figure}

The local geometry of the cost surface around the best fit is described by the FIM, as we illustrate in Fig.~\ref{fig:unit_ball}.
Diagonals of the FIM describe the change in cost to each parameter individually, ignoring any potential correlations among parameters.
Cost contours form ellipses, aligned with the eigenvectors of the FIM, whose aspect ratio is given by the square root of the ratio of the eigenvalues.
Elongated directions are parallel to the eigenvectors with small eigenvalues, indicating that the data carry little information about those parameter combinations.
These parameter combinations are only weakly constrained by the data and have large uncertainties in their inferred values.
Projecting these ellipses onto the parameter axes estimates the uncertainty in each individual parameter, given by the diagonals of the inverse FIM.

\subsection{Bayesian analysis}
\label{subsec:method_mcmc}
The most common UQ methods in molecular modeling use a Bayesian framework.
In Bayesian statistics, the parametric uncertainty is described by a posterior distribution given by Bayes' theorem
\begin{equation}
    \label{eq:bayes}
	P\left(\theta \middle| \mathbf{y}\right) \propto L\left(\theta \middle|\mathbf{y}\right) \times \pi\left(\theta\right),
\end{equation}
where $L(\theta |\mathbf{y})$ and $\pi(\theta)$ are the likelihood and the prior distribution of the model's parameters, respectively.\cite{Press_Teukolsky_Vetterling_Flannery_2007, Gilks_Richardson_Spiegelhalter_1996}
The likelihood is functionally the same as the probability distribution of the observed data conditioned on the different values of the parameters, i.e., $L(\theta | \mathbf{y}) = P(\mathbf{y} | \theta)$.
As before, we temper the likelihood with a tunable ``sampling temperature.''
In a Bayesian context, this introduction is motivated by a formal analogy to the Boltzmann distribution.\cite{Brown_Sethna_2003, lamont2019correspondence, jaynes1957information}
The cost is analogous to the internal energy of a system, so low-temperature distributions are concentrated near the low-energy (i.e., low cost) region of parameter space.
Formally, the temperature uniformly scales the tolerances $\sigma_m$ in Eq.~\eqref{eq:residuals}.
In our case, the temperature scales the tolerances to be reflective of the model's accuracy and not just the error in the data itself.
Continuing the analogy, Eq.~\eqref{eq:bayes} becomes
\begin{equation}
    \label{eq:bayes_entropy}
    P(\theta \vert \mathbf{y}) \sim \exp\left( -\left( C(\theta) - S(\theta) T \right)/T \right),
\end{equation} 
in which the prior is analogous to entropy: $S = \log \pi$.

We sample from the posterior distribution using an MCMC algorithm.
There are several black-box libraries for MCMC sampling.
In this work, we used the \texttt{ptemcee} Python package, which utilizes an affine invariance property of the sampler.\cite{Foreman-Mackey_Hogg_Lang_Goodman_2013, Goodman_Weare_2010}
In addition, this method generates chains at different temperatures and mixes them with an appropriate acceptance probability.\cite{Vousden_Farr_Mandel_2016}
Traditionally, parallel tempering is utilized as a device to improve convergence rates by allowing walkers to skip over regions in parameter space with higher cost values and possibly find different minima (if they exist).\cite{Miller_Dunson_2019}
Our use of the sampling temperature to estimate the scale of the model accuracy [see Eq.~\eqref{eq:model_bias_noise}].
Previous results have shown that a natural sampling temperature given by $T_0 = 2C_0/N$ [see Eq.~\eqref{eq:natural_temperature}] gives a good estimate of the systematic errors in the model.\cite{Frederiksen_Jacobsen_Brown_Sethna_2004}
Thus, raising the temperature from $T = 1$ to $T = T_0$ transitions the sampling from the target accuracy to a more realistic estimate of the actual systematic errors.

For the toy model in Eq.~\eqref{eq:toy_model} and Fig.~\ref{fig:ModelCostcontourCartoon}, data are generated from the model itself, i.e., there is no bias.
In this case, varying the sampling temperature effectively changes the size of the error bars $\sigma_m$ in the model.
Of particular interest is how the samplers interact with the plateaus and canyons of the cost surface as the temperature is varied.

To assess the convergence, we simulate multiple chains and use the multivariate potential scale reduction factor (PSRF), denoted by $\hat{R}^p$\cite{Gelman_Rubin_1992, Brooks_Gelman_1998, Vats_Knudson_2020}.
The value of $\hat{R}^p$ is related to the ratio of the covariance between and within the independent chains, given by
\begin{equation}
    \label{eq:psrf}
    \hat{R}^p = \frac{n-1}{n} + \frac{m+1}{m} \lambda_\text{max} (W^{-1} B/n),
\end{equation}
where $n$ and $m$ are the numbers of iterations and chains, respectively, and $\lambda_\text{max}(A)$ denotes the largest eigenvalue of matrix $A$.
$B/n$ and $W$ are the variance between and within the independent chains, $\psi_j$,
\begin{equation}
    \begin{aligned}
        \frac{B}{n} &= \frac{1}{m-1} \sum_{j=1}^m \left( \bar{\psi}_j - \bar{\psi} \right) \left( \bar{\psi}_j - \bar{\psi} \right)^T \\
        W &= \frac{1}{m(n - 1)} \sum_{j=1}^m \sum_{t=1}^n \left( \psi_{jt} - \bar{\psi}_j \right) \left( \psi_{jt} - \bar{\psi}_j \right)^T.
    \end{aligned}
\end{equation}
Note that as $n \to \infty$, the value of $\hat{R}^p$ declines to one.\cite{Gelman_Rubin_1992, Brooks_Gelman_1998}
Thus, as the MCMC samples converge to a stationary distribution, the value of $\hat{R}^p$ approaches one; however, the converse is not necessarily true.
Common thresholds of $\hat{R}^p$ are in the range of 1.1 to 1.05.\cite{Brooks_Gelman_1998}
In this work, we have used the more stringent requirement ($\hat{R}^p < 1.05$).

\begin{figure*}[!ht]
    \centering
    \includegraphics[width=0.6\textwidth]{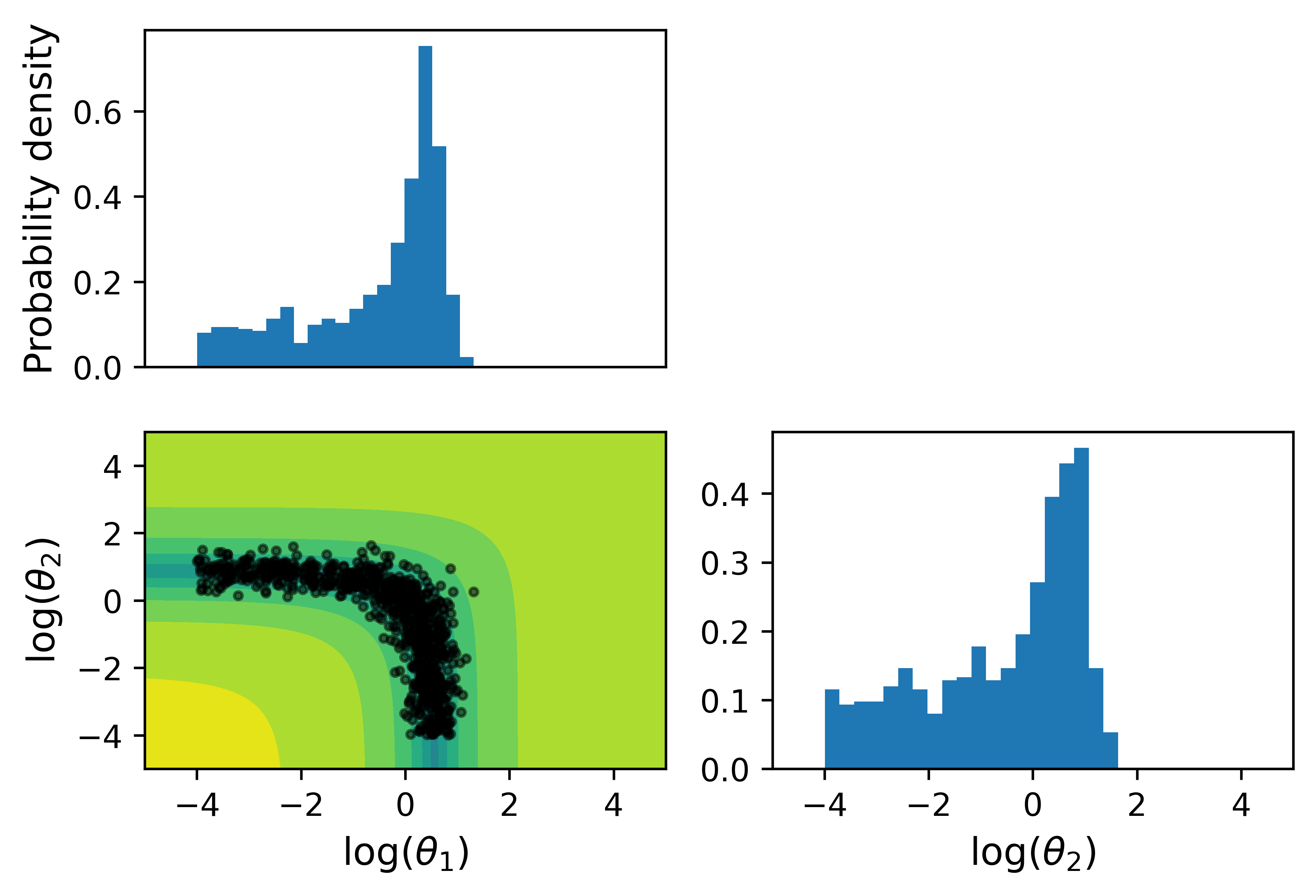}
    \caption[MCMC samples of the toy model]{
        MCMC samples of the model defined by Eq.~\eqref{eq:toy_model}.
        In the lower left frame, the samples are plotted on top of the cost contour.
        The univariate marginal distributions of the samples are given on the diagonal.
        We pick the sampling temperature $T=0.5$ to  highlight the effect of sloppiness to the Bayesian sampling, i.e., parameter evaporation.
    }
    \label{fig:mcmc_cartoon}
\end{figure*}

To illustrate, we sample the posterior of the model in Eq.~\eqref{eq:toy_model} with a uniform prior distribution that is non-zero over $(-4, 4)$.
Fig.~\ref{fig:mcmc_cartoon} shows the result of the sampling as an array of plots that summarize the sample.
Along the main diagonal, we plot the univariate marginal distributions, i.e., projection of the samples' distribution onto a parameter axis.
In the lower left frame of the array, we show samples in two-dimensional parameter space.
In higher dimensions, these plots are two-dimensional marginal distributions.
In this example, we have superimposed the samples on top of the cost contours [compare this to Fig.~ \hyperref[fig:ModelCostcontourCartoon]{2(b)}].

In Fig.~\ref{fig:mcmc_cartoon}, note that the MCMC samples are located along the canyon.
This is expected as the posterior is large in regions of high likelihood, i.e., low cost.
Thus, the samples accurately conform to the cost contours and quantify the uncertainty in the parameter estimates.
However, by inspecting Fig.~\ref{fig:mcmc_cartoon}, we can anticipate a potential problem when the cost contours have flat, elongated canyons that extend to extreme parameter values.
In this scenario, it will be common to ``evaporate'' parameters, i.e., have samples that extend over the full range of model parameters.
Parameter evaporation was first observed when sampling posterior distributions for sloppy models in systems biology;\cite{Gutenkunst_2007} however, the phenomenon occurs in molecular models as well, as we document below.
The term parameter evaporation continues the analogy to statistical mechanics.
In the language of statistical inference, one would say that the posterior does not concentrate with respect to the prior and is \emph{unidentifiable}.

Closely related to the parameter evaporation is the question of choosing the prior distribution, $\pi(\theta)$, on parameter space.
Here, we have used a uniform prior, a common choice for an uninformative prior.
However, we find the prior can strongly influence the posterior distribution in subtle ways.
Even the apparently innocuous uniform prior can introduce strong biases.
Note that in Fig.~\ref{fig:mcmc_cartoon}, the samples stop evaporating due to the boundaries of the prior distribution.
Consequently, the marginal distribution of the samples has a hard cutoff at this boundary.
With a broader prior, the posterior will be even wider, and samples may no longer be concentrated near the best fit.
The long canyons and broad plateaus of the cost surface would dominate the samples.
We intuitively explain this effect in terms of a trade-off between energy and entropy in the sampling process.
While the single most probable parameter value is the best fit (i.e., ground state), there are many more sub-optimal parameter values along the canyon.
In other words, the broad prior introduces a large entropy in some regions of the parameter space.
For broad priors, the entropic contribution dominates the sampling.

The effect is clearly demonstrated with a simple example.
Consider a one-dimensional cost function given by
\begin{align}
  \label{eq:SimpleCost}
  C(\theta) = \begin{cases}
    C_0 \quad & \theta < l \\
    C_0 + \Delta \quad & \theta > l
    \end{cases}
\end{align}
where $\theta$ is a non-negative parameter and $C_0$, $\Delta$ and $l$ are non-negative constants.
We take a prior $\pi(\theta) = \mathcal{U}(0,L)$, i.e., a uniform, flat prior from zero to a positive value of $L$.
After calculating the posterior distribution, we find the average cost for this scenario to be
\begin{align}
  \label{eq:AvgCost}
  \langle C \rangle = \begin{cases}
    C_0 \quad & L < l \\
    C_0 + \Delta \left(  \frac{L - l}{L + l \left( e^{\Delta/T} - 1\right)} \right) \quad & L > l.
    \end{cases}
\end{align}
Note that for very large $L$ (i.e., very broad prior), $\langle C \rangle \rightarrow C_0 + \Delta$.
That is to say, for a sufficiently broad prior, the posterior distribution is dominated by bad fits (large cost) because of their large entropic contribution.
This result holds for any non-zero sampling temperature and regardless of how bad the fit is (i.e., the size of $\Delta$).

As we will see in \Sec~\ref{sec:results}, the trade-off between entropy and energy is even more nuanced for ``sloppy'' cost landscapes with many dimensions.
Rarely is there an objectively ``correct'' prior, and any choice is almost certain to introduce artifacts into the statistics of the posterior.
In these cases, it is unclear to what extent the posterior accurately reflects the target uncertainty.
One solution is to sample with multiple priors and temperatures, a computationally expensive task, and try to assess the effect of prior and sampling temperature on the results.
Because this practice generates multiple posteriors, it potentially undermines the Bayesian paradigm in which a single posterior summarizes all the information one has about the parameters of a model.
However, we believe that these extra steps are an important intermediate analysis in understanding the effect of the prior on the posterior and necessary for constructing a reliable posterior.
Alternatively, one could use a formalism that does not require an a priori measure on parameter space.
This is the domain of frequentist statistics, which we discuss next.


\subsection{Frequentist analysis}
\label{subsec:method_likelihood}
Although much has been said about the philosophical differences between Bayesians and frequentists,\cite{Bayarri_Berger_2004, Wagenmakers_Lee_Lodewyckx_Iverson_2008} here we use a functional distinction.
The choice of prior in Sec.~\ref{subsec:method_mcmc} was a central question.
The prior acts as a measure on parameter space, i.e., a weight function whose integral generalizes the concept of lengths and volumes.\cite{athreya2006measure}
In the frequentist approach no such measure exists.
Without a measure on parameter space, we lose the machinery of a posterior distribution, but we also need fewer mathematical assumptions.
With no prior, the goal is to describe the set of parameter values that have small cost (i.e., below some statistically defined threshold) without attaching any (probabilistic) weight to regions of parameter space.
While there are many frequentist tools available, we use the profile likelihood.\cite{Cole_Chu_Greenland_2014}
To the best of our knowledge, profile likelihoods have not been previously applied to the study of interatomic potentials.
Importantly, profile likelihoods generate paths through the parameter space that we will compare with the geodesic curves generated by the information geometric analysis in section~\ref{subsec:method_geodesic}.

\begin{figure*}[!ht]
    \centering
    \includegraphics[width=0.6\textwidth]{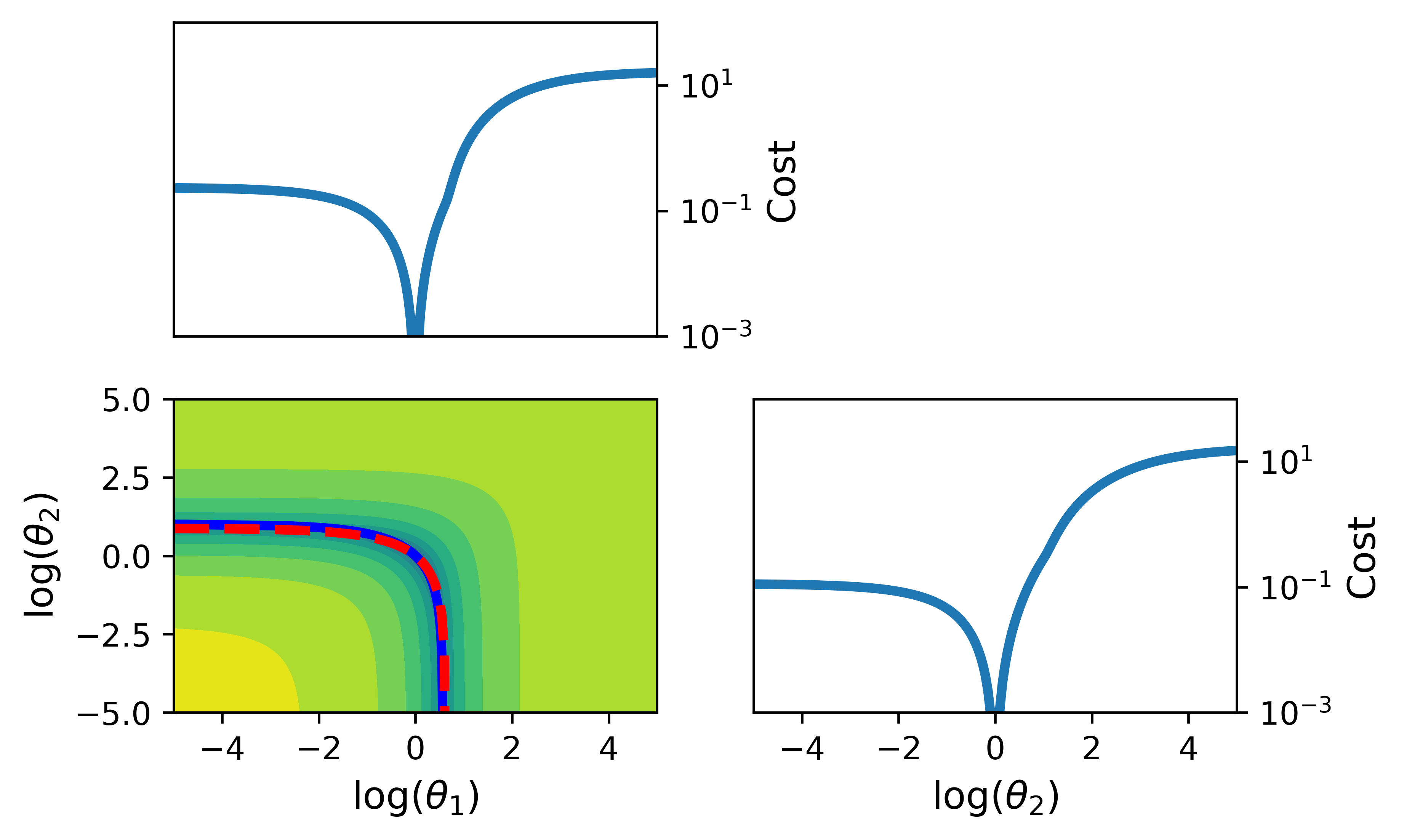}
    \caption[Profile likelihood of the model]{
        Profile likelihood for the model in Eq.~\eqref{eq:toy_model}.
        In the lower left frame, the red and blue curves show the paths traced from the profile likelihood computation for the parameter on the horizontal and vertical axes, respectively.
        The cost profiles, i.e., the cost along these paths, are given on the diagonal.}
    \label{fig:likelihood_cartoon}
\end{figure*}

The basic idea is to select one parameter, fix it to a constant value, and globally optimize the likelihood function (i.e., minimize the cost) over the remaining $N-1$ model parameters.\cite{Cole_Chu_Greenland_2014, Rolke_Lopez_Conrad_2005}
By varying the value to which the parameter is fixed, we trace out a ``profile'' of how the cost depends on this parameter in the context of the rest of the model.
The procedure is best understood through example, as we now demonstrate on the toy model from Eq.~\eqref{eq:toy_model} in Fig.~\ref{fig:likelihood_cartoon}.
As before, we summarize results with a two-by-two array of figures.
Consider the cost contours in the lower left frame of the plot array.
The red curve is the set of points obtained by fixing $\theta_1$ to a constant value and optimizing the cost over $\theta_2$.
The optimization searched over vertical slices of the parameter space for each value of $\theta_1$.
Similarly, the blue curve is the set of points found by fixing $\theta_2$ and optimizing over horizontal slices, i.e., over $\theta_1$.
Along the main diagonal, we plot the cost along each of these profile likelihood paths.

\begin{figure*}[!ht]
    \centering
    \includegraphics[width=0.8\textwidth]{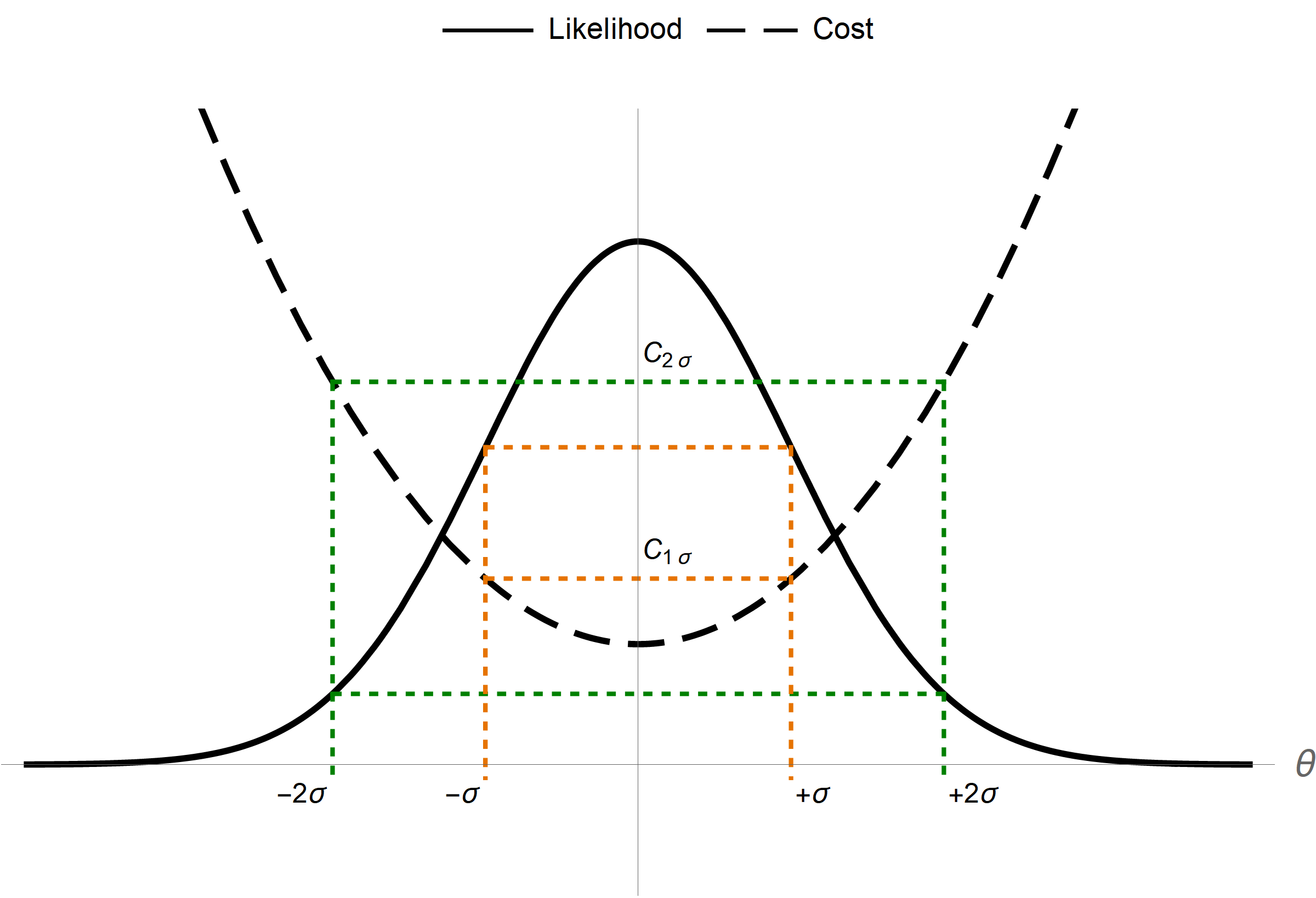}
    \caption[Confidence intervals of a Gaussian likelihood]{
        Confidence intervals for a Gaussian likelihood at several confidence levels.
        $C_{1\sigma}$ and $C_{2\sigma}$ correspond to the cost threshold at 68\% and 95\% confidence levels, respectively.
        The confidence interval of $\theta$ given 68\% confidence level spans from $-\sigma$ to $+\sigma$.
    }
    \label{fig:confidence_level}
\end{figure*}

By construction, the profile likelihood paths trace out the canyon on the cost contour.
By comparing the paths and the cost along the paths, we extract information about how variation of the parameters affects the variation in the predictions.
The paths also tell us how the parameters correlate with each other.
Statistical confidence levels correspond to an allowed error or cost threshold.
For a given confidence level, the uncertainty of an individual parameter is given by the width of the profile likelihood that has cost values lower than this cost threshold.
This idea is illustrated in Fig.~\ref{fig:confidence_level} for a simple Gaussian likelihood function.

To calculate profile likelihoods, we developed a Python package, \texttt{profile\_likelihood},\cite{Kurniawan_2021} that additionally interfaces with IPs from the OpenKIM repository at \url{https://openkim.org}.
We use the Levenberg--Marquardt algorithm with geodesic acceleration in the optimization process.\cite{Transtrum_Sethna_2012}

The profile likelihood analysis method has its own challenges and limitations.
Optimizing multi-dimensional cost functions can be challenging;\cite{Transtrum_Machta_Sethna_2010} however, by using the result of the previous optimization as the starting point of each iteration, convergence is relatively fast and stable.
Additionally, the profiling process effectively projects parameter curves onto the parameter axes.
As we will see in \Sec~\ref{sec:results}, if the cost canyon curves or bends over on itself, it will be missed by the profile likelihood.
More broadly, the construction is not invariant to reparameterization.
Just as priors may introduce artifacts in the Bayesian framework, the parameterization can introduce artifacts into the profile likelihood.
To avoid these issues, we next use information geometry to study the uncertainty in a parameterization-independent way.


\subsection{Information geometry}
\label{subsec:method_geodesic}
Information geometry is an approach to statistics in which a multi-parameter model is interpreted as a high-dimensional manifold.
We study this manifold using computational differential geometry that allows us to extract the key geometric and topological features of the model manifold.
These features shed light on issues related to UQ.

\begin{figure*}
    \centering
    \includegraphics[width=0.9\textwidth]{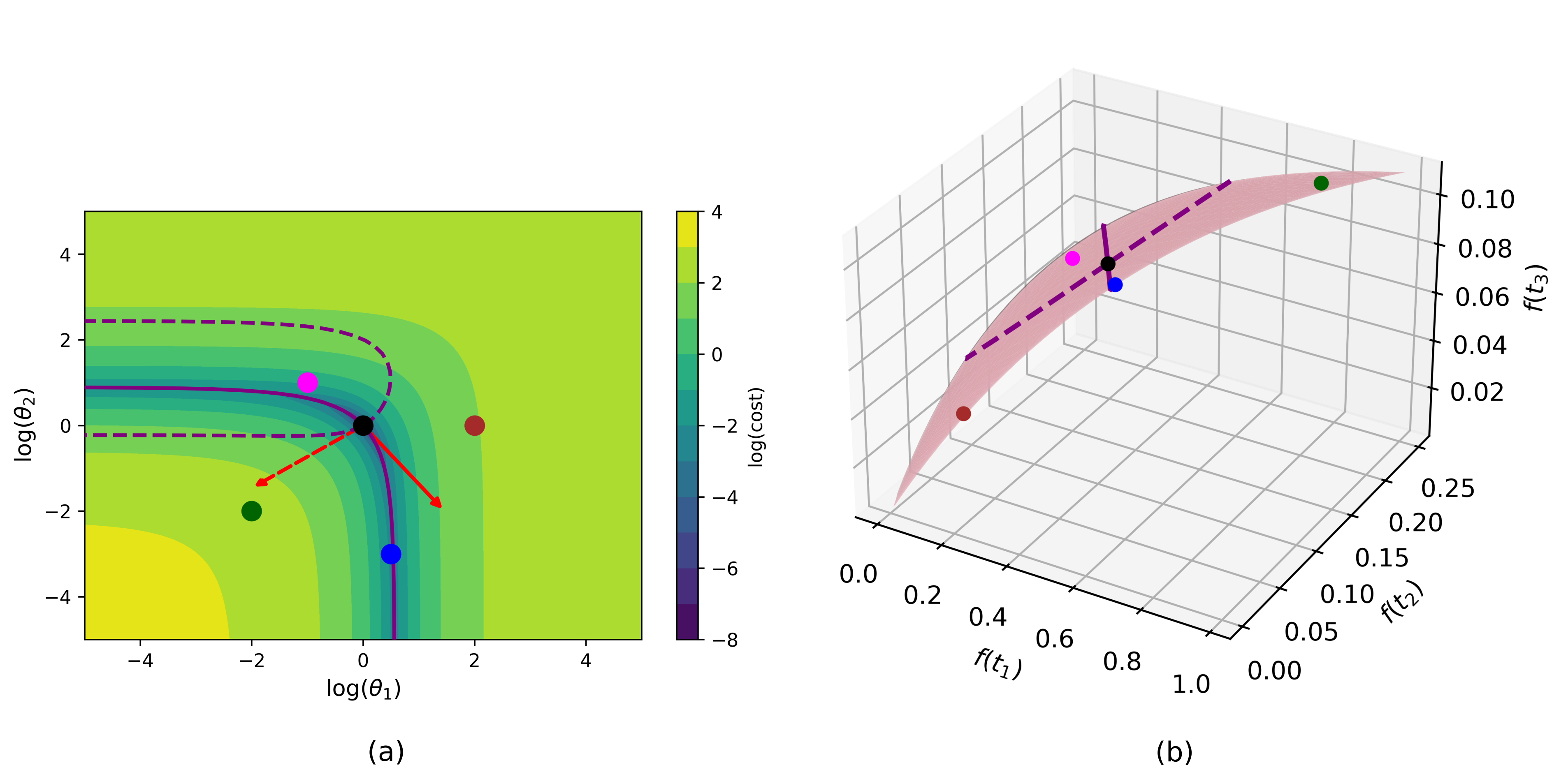}
    \caption[Model manifold of the toy model]{
        Model manifold of the toy model in Eq.~\eqref{eq:toy_model}.
        Each point in parameter space (a) corresponds to a set of predictions in data space (b).
        The set of all possible predictions trace out the model manifold.
        For reference, colored points are the same as those in Fig.~\ref{fig:ModelCostcontourCartoon}.
        The red arrows in parameter space shows the eigenvectors of the FIM.
        Geodesics (solid and dashed curves) relate manifold structures, such as manifold boundaries, to parameter space.
        }
    \label{fig:infogeo_toy}
\end{figure*}

As we have seen in \Subsec~\ref{subsec:method_cost}, a multi-parameter model makes a set of predictions for a given evaluation scenario, $f_m(\theta)$, that we interpret as a vector in data space.
That is to say, the model is a mapping between parameter space and data space,
\begin{equation}
  \label{eq:modelmapping}
  f: \mathcal{D} \subseteq \mathbb{R}^N \rightarrow \mathbb{R}^M.
\end{equation}
Conceptually, the model manifold is constructed by mapping all possible parameter values to their corresponding predictions in data space, i.e., the model manifold is the image of parameter space under the model map, illustrated for the toy model [Eq.~\eqref{eq:toy_model}] in Fig.~\ref{fig:infogeo_toy}.

Critically, note that the manifold is bounded by two one-dimensional segments.
We focus on the boundaries of the model manifold as they are the geometric feature most relevant to parameter uncertainty.
In parameter space, we have seen that there can be large or infinite uncertainties when cost contours do not close, i.e., confidence regions extend to the limits of the parameter domain.
Infinite, high-entropy regions of parameter space are mapped to finite regions near the boundary of the model manifold, and thus, these contours, i.e., the contours that do not close, are generic when the model manifold is bounded.
Since the cost corresponds to the distance in data space, the images of cost contours are approximately concentric circles on the model manifold.
Each segment of the boundary is a manifold of co-dimension one.
Associated with each boundary segment is a parameter or combination of parameters that are \emph{practically unidentifiable} at some level of statistical confidence.

We use geodesics, i.e., distance minimizing curves, on the model manifold to find the unidentifiable parameter combination associated with each boundary segment.\cite{Transtrum_Qiu_2014}
Geodesics are a natural extension of the local FIM analysis.
As such, the boundaries and parameter combinations they find identify the global sloppiness of the model.
Geodesics traverse the low cost regions in parameter space to find unidentifiable parameter combinations -- those whose uncertainty will be large -- which can be removed to reduce the complexity and sloppiness of the model.

We approximate geodesics curves along the model manifold by numerically solving the geodesic equation,
\begin{equation}
    \frac{\partial^2 \theta^i}{\partial \tau^2} = -\sum\limits_{j,k} \Gamma^i_{jk} \frac{\partial \theta^j}{\partial \tau} \frac{\partial \theta^k}{\partial \tau},
\end{equation}
where
\begin{equation}
    \Gamma^i_{jk} = \sum\limits_{l,m} \left(\mathcal{I}^{-1}\right)^{il} \frac{\partial y_m}{\partial \theta^l} \frac{\partial^2 y_m}{\partial \theta^j \partial \theta^k}
\end{equation}
are the so-called Christoffel symbols, $\mathcal{I}$ is the FIM, $\theta$ are the parameters, and $\tau$ is the arc length of the geodesic along the model manifold.
We numerically solve for the path of the geodesic by treating this equation as an initial value problem where the initial position is given by the nominal parameter values.
The  initial direction is given by a few of the sloppiest eigenvectors.
We provide a simple example script for calculating geodesics on GitHub.\cite{MBAMgithub}
To illustrate this, we again turn to the toy model in Eq.~\eqref{eq:toy_model}.
Figure~\ref{fig:infogeo_toy} shows two geodesics that pass through the best fit point on the model manifold, the solid and dashed curves in both parameter space (a) and on the model manifold (b).
Note how the geodesics rotate in parameter space to naturally follow the cost contours and align with the unidentifiable parameters.
As the geodesic curves approach the edge of the model manifold, we see that either $\theta_1 \rightarrow 0$ or $\theta_2 \rightarrow 0$.
From the correspondence between the parameter space picture and the data space picture, we deduce that the upper boundary segment on the model manifold corresponds to $\theta_1 \rightarrow 0$ while the lower segment corresponds to $\theta_2 \rightarrow 0$.
These limiting values indicate which combinations of parameters are unidentifiable and have unbounded uncertainties.
We often calculate several geodesics emanating from the nominal parameter values in several of the sloppiest eigenvectors of the FIM.
These geodesics will help us identify the least identifiable parameter (combinations) in the model.

In this simple example, the two boundary segments are already aligned with the parameters of the model.
In more realistic models, boundary segments often correspond to the coordinated combinations of bare parameters.
In these cases, we will use geodesics to identify the correlation and find a more natural, identifiable reparameterization.


\subsection{Interatomic potentials and tests}
\label{subsec:interatomic_potentials_and_tests}
In this study, we apply the methods described in \Subsec~\ref{subsec:method_fim}~-~\ref{subsec:method_geodesic} to empirical IPs taken from the OpenKIM repository.\cite{Tadmor_Elliott_Sethna_Miller_Becker_2011, elliott:tadmor:2011}
The OpenKIM framework has a standardized collection of models, data, and tests for computing materials properties that make the UQ process reproducible and transferable.
For this study, we chose the Lennard-Jones (LJ) and Morse potentials parameterized for silicon and nickel, respectively, to validate methods and demonstrate general principles on low-dimensional models.\footnote{We note that the LJ and Morse IPs do not provide good representations for silicon or nickel. They are simply chosen for their simplicity to help understand the UQ methodology.}
We then extend the investigation to the molybdenum disulfide (MoS$_2$) system using the more complex Stillinger--Weber (SW) potential.\cite{Wen_Shirodkar_Plechas_Kaxiras_Elliott_Tadmor_2017}

These potentials are categorized as cluster potentials.
Given a system with $N$ atoms, the total potential energy, $\mathcal{V}$, is
\begin{equation}
    \label{eq:total_energy}
    \mathcal{V} = \sum_{\substack{i, j = 1 \\ i < j}}^N \phi_2(\mathbf{r}_i, \mathbf{r}_j) +
        \sum_{\substack{i, j, k = 1 \\ i < j < k}}^N \phi_3(\mathbf{r}_i, \mathbf{r}_j, \mathbf{r}_k) +
        \dots,
\end{equation}
where $\phi_n$ denotes the $n$-body potential function and $\mathbf{r}_i$ is the position of atom $i$.

The Lennard-Jones (LJ) potential is a pair potential, i.e., Eq.~\eqref{eq:total_energy} only consists of the two-body (pair-wise) interaction term, and the higher order potential functions are set to zero.
The pair-wise interaction has two parameters, given by
\begin{equation}
    \label{eq:Lennard-Jones}
    \begin{aligned}
        \phi_{\text{LJ}}(r_{ij}) &= 4\epsilon \left( \left( \frac{\sigma}{r_{ij}} \right)^{12} -  \left( \frac{\sigma}{r_{ij}} \right)^6 \right) + \Delta, \\
        \Delta &= -4\epsilon \left( \left( \frac{\sigma}{r_\text{cut}} \right)^{12} -  \left( \frac{\sigma}{r_\text{cut}} \right)^6 \right),
    \end{aligned}
\end{equation}
where $r_{ij}=\|\mathbf{r}_i-\mathbf{r}_j\|$ is the distance between atoms $i$ and $j$.
The potential is only non-zero when $r_{ij} < r_\text{cut} = 7.91118~\angstrom$.
The parameter $\epsilon$ is an energy scaling factor in the potential, while $\sigma$ is related to the equilibrium distance of the pair interaction.
The shifting factor, $\Delta$, is chosen so that the potential is continuous at $r_\text{cut}$.\cite{OpenKIM_LJ_driver, Lennard-Jones1, Lennard-Jones2, Lennard-Jones3}

The Morse potential is also a pair potential, similar to the LJ potential.
The pair-wise interaction with three parameters is given by
\begin{equation}
    \label{eq:Morse}
    \begin{aligned}
        \phi_{\text{M}}(r_{ij}) &= \epsilon \left( -e^{-2C(r_{ij}-r_0)} + 2e^{-C(r_{ij}-r_0)} \right) + \Delta, \\
        \Delta &= -\epsilon \left( -e^{-2C(r_\text{cut}-r_0)} + 2e^{-C(r_\text{cut}-r_0)} \right),
    \end{aligned}
\end{equation}
where $\epsilon$ is an energy scaling factor, $r_0$ is the equilibrium distance, and $C$ controls the width of the potential well.
Again, the potential is only non-zero when $r_{ij} < r_\text{cut} = 9.75476~\angstrom$ and $\Delta$ is chosen such that the potential is continuous at $r_\text{cut}$.\cite{OpenKIM_Morse_driver, Morse_shifted_potential}

We use these pair potentials to predict the unrelaxed energy and forces of silicon (LJ potential) and nickel (Morse potential) atoms in an evaluation scenario comprised of a randomly perturbed body-centered triclinic configuration with periodic boundary conditions.
The lattice parameters are given as follows:\cite{OpenKIM_TPEAF_2, OpenKIM_TPEAF_driver}
\begin{align*}
    a &= 3.1287525~\angstrom & \alpha = 87.25318054968444 ^\circ \\
    b &= 3.15146~\angstrom & \beta = 93.34074777413502 ^\circ \\
    c &= 3.13121044~\angstrom & \gamma = 91.23134462011188 ^\circ
\end{align*}
We also use a random triclinic configuration with 64 silicon atoms\cite{OpenKIM_TPEAF_64} with several other cluster potentials in a broader survey of the Fisher information.
We generate artificial data for both of these models using the default parameter values reported in OpenKIM.
We calculate the energy and forces evaluated at the default parameters and add 10\% Gaussian noise (for forces, we use 10\% of the magnitude of the force vector).
The introduction of the noise in the data results in new sets of best fit parameters, which for the LJ potential are $\epsilon = 3.34545922$ eV and $\sigma = 1.98171508~\angstrom$, while for the Morse potential are $\epsilon = -0.24366083$ eV, $C = 0.65508552~\angstrom^{-1}$, and $r_0 = 4.48704315~\angstrom$.
Note that the default parameter values were found by fitting to experimental data.\cite{OpenKIM_LJ, OpenKIM_Morse}
Because these original training data were not generated by the model, the parameter values reflect some model inadequacy.
We will later explore how the parameter uncertainty responds to the scale of model inadequacy by tempering this likelihood for a range of sampling temperatures.

Note that these structures are not the ground states for silicon or nickel, but they are tests available in OpenKIM that are convenient for validating methods on low-dimensional models.
We will see that they clearly illustrate the problems that the sloppiness of the model brings to the standard UQ methods and the phenomena that we will discuss later are generic, at some confidence level, to any atomic configuration used.

We extend the analysis to the Stillinger--Weber (SW) potential for monolayer MoS$_2$,\cite{Wen_Shirodkar_Plechas_Kaxiras_Elliott_Tadmor_2017, OpenKIM_SW_MoS2, OpenKIM_SW_MoS2_driver} which contains both two-body and three-body interactions.
The two-body interaction takes the form
\begin{equation}
    \label{eq:SW_2body}
    \begin{aligned}
        \phi_2^{IJ}\left(r_{ij}\right) =&~
            A_{IJ} \left( B_{IJ} \left(\frac{\sigma_{IJ}}{r_{ij}}\right)^{p_{IJ}}
                - \left(\frac{\sigma_{IJ}}{r_{ij}}\right)^{q_{IJ}}\right) \\
           &\times \exp \left( \frac{\sigma_{IJ}}{r_{ij} - r_{IJ}^{\text{cut}}} \right),
    \end{aligned}
\end{equation}
where uppercase subscripts denote the types of atoms, e.g., $A_{IJ}$ is the parameter $A$ corresponding to interaction between atoms of type $I$ and type $J$.
The three-body term is given by
\begin{equation}
    \label{eq:SW_3body}
    \begin{aligned}
        \phi_3^{IJK} \left( r_{ij}, r_{ik}, \beta_{jik} \right) =&~
            \lambda_{JIK} \left( \cos \beta_{jik} - \cos \beta_{JIK}^0 \right)^2 \\
            &\times \exp \left( \frac{\gamma_{IJ}}{r_{ij} - r_{IJ}^\text{cut}}
                + \frac{\gamma_{IK}}{r_{ik} - r_{IK}^\text{cut}} \right),
    \end{aligned}
\end{equation}
with $\beta_{jik}$ being the angle between the $i$--$j$ and $i$--$k$ bonds.

We calibrate this potential to DFT data of the atomic forces in configurations near the equilibrium state at 750 K, as described in Ref.~\onlinecite{Wen_Shirodkar_Plechas_Kaxiras_Elliott_Tadmor_2017}.
Since the data are not generated from the IP, they contain bias and we will temper the likelihood over a range of temperatures to account for it.
Our formulation follows closely that of the original paper (e.g., we set $q_{IJ} = 0$, fix $\gamma$ to be the same for all types of interaction, and use the same training set); however, we make a few changes.
First, we allow parameters $p_{IJ}$ to take any positive real value and remove the relation between $\sigma_{IJ}$ and the equilibrium lattice constants of the system.
We also do not require $d\phi_2/dr|_{r=d} = 0$ at the equilibrium bond length $d$, which removes the constraint on $B_{IJ}$. 
The remaining free parameters are $A_{IJ}$, $B_{IJ}$, $p_{IJ}$, and $\sigma_{IJ}$ for each type of pair-wise interaction (Mo--Mo, Mo--S and S--S interactions), $\lambda_{IJK}$ for S--Mo--S and Mo--S--Mo interactions, and $\gamma$.

We again choose error tolerances to be 10\% of predicted values.
Note that this leads to non-uniform weighting factors in our cost function, unlike Ref.~\onlinecite{Wen_Shirodkar_Plechas_Kaxiras_Elliott_Tadmor_2017}.
Fitting this model leads to a new set of optimal parameter values listed in Table~\ref{tab:SW_parameters_2body} for the two-body interaction term and Table~\ref{tab:SW_parameters_3body} for the three-body interaction term.\cite{our_SW_model}
Other parameters that are not listed in these tables take the same values as listed in Ref.~\onlinecite{Wen_Shirodkar_Plechas_Kaxiras_Elliott_Tadmor_2017}, such as the cutoff radii and the reference bond angle.
The cost at the best fit is $1.390 \times 10^6$.
Because the fitting data are forces near equilibrium, the error bars are very small (leading to a large cost) with larger weight on configurations near equilibrium.
However, in our Bayesian analysis, we sample the posterior at many temperatures, effectively scaling these small error bars up to something more reasonable.
The process provides a systematic study of the role of error bars in quantifying parametric uncertainty in sloppy, molecular models.

\begin{table}[!ht]
    \centering
    \begin{ruledtabular}
        \begin{tabular}{c c c c}
             & \multicolumn{3}{c}{Interaction} \\
            \cline{2-4}
            Parameter & Mo--Mo & Mo--S & S--S \\
            \hline
            $A$ (eV) & 18.4310060 & 8.83861305 & 0.37463396 \\
            $B$ & 0.00641786 & 1.04793603 & 561.429270 \\
            $p$ & 4.73717813 & 8.26621744 & 2.66196913 \\
            $\sigma (\angstrom)$ & 6.16940454 & 1.92967991 & 0.41904814 \\
        \end{tabular}
    \end{ruledtabular}
    \caption[Fitted parameters of the two-body term in the SW potential for MoS$_2$]{
        Fitted parameters of the two-body term in the SW potential for MoS$_2$.
    }
    \label{tab:SW_parameters_2body}
\end{table}

\begin{table}[!ht]
    \centering
    \begin{ruledtabular}
        \begin{tabular}{c c}
          Parameter & value \\
          \hline
          $\lambda_{\text{S--Mo--S}}$ (eV) &  4.28784076 \\
          $\lambda_{\text{Mo--S--Mo}}$ (eV) & 14.4285026 \\
          $\gamma$ ($\angstrom$) & 1.53800500 \\
        \end{tabular}
    \end{ruledtabular}
    \caption[Fitted parameters of the three-body term in the SW potential for MoS$_2$]{
        Fitted parameters of the three-body term in the SW potential for MoS$_2$.
    }
    \label{tab:SW_parameters_3body}
\end{table}

After the calibration process, we propagate the parametric uncertainty of this potential in a second evaluation scenario to predict the uncertainty of the change in energy as a response to the lattice stretching and compression.
This calculation is done by creating MoS$_2$ unit cells with various in-plane lattice constants $a$ and then relaxing the atoms in the perpendicular, out-of-plane, direction.
We probe the calculation in the range $(a - a_0) \in [-0.5, 0.5] ~\angstrom$, where $a_0$ is the equilibrium lattice constant.
Then, we compare the uncertainty to the result in Ref.~\onlinecite{Wen_Shirodkar_Plechas_Kaxiras_Elliott_Tadmor_2017} qualitatively.



\section{RESULTS}
\label{sec:results}
\begin{figure}[!ht]
    \centering
    \includegraphics[width=0.45\textwidth]{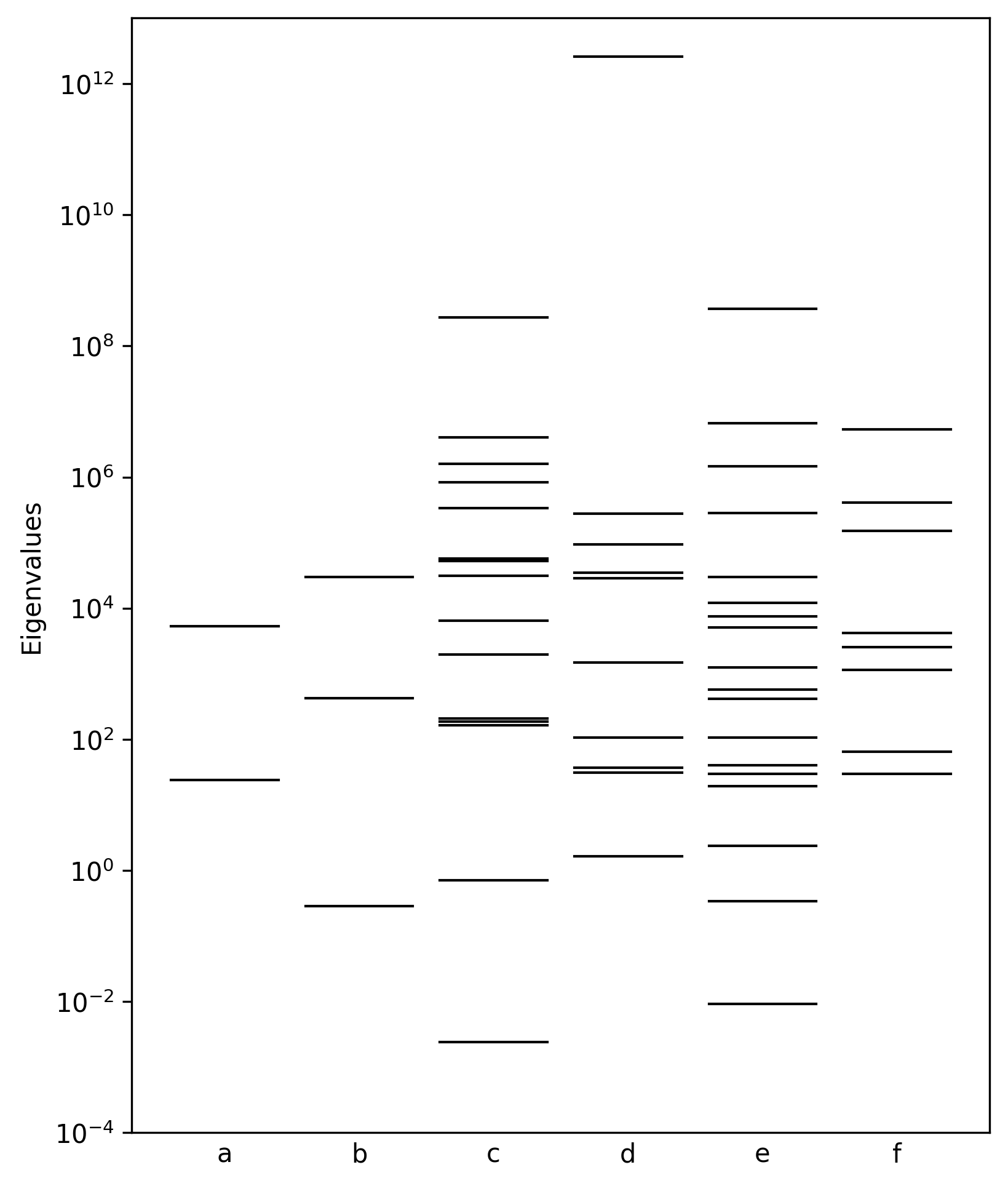}
    \caption[Eigenvalues of the FIM for various interatomic models]{
        Eigenvalues of the FIM for various IPs: (a) LJ for Si, (b) Morse for Ni, (c) SW for MoS$_2$, and (d) Khor--Das Sarma, (e) EDIP, and (f) SW, each for Si.
        For each model, the larger (smaller) eigenvalues represent stiff (sloppy) parameter combinations in the direction of their respective eigenvectors.
    }
    \label{fig:fim_eigenvalues}
\end{figure}

\begin{figure}[!ht]
    \centering
    \includegraphics[width=0.45\textwidth]{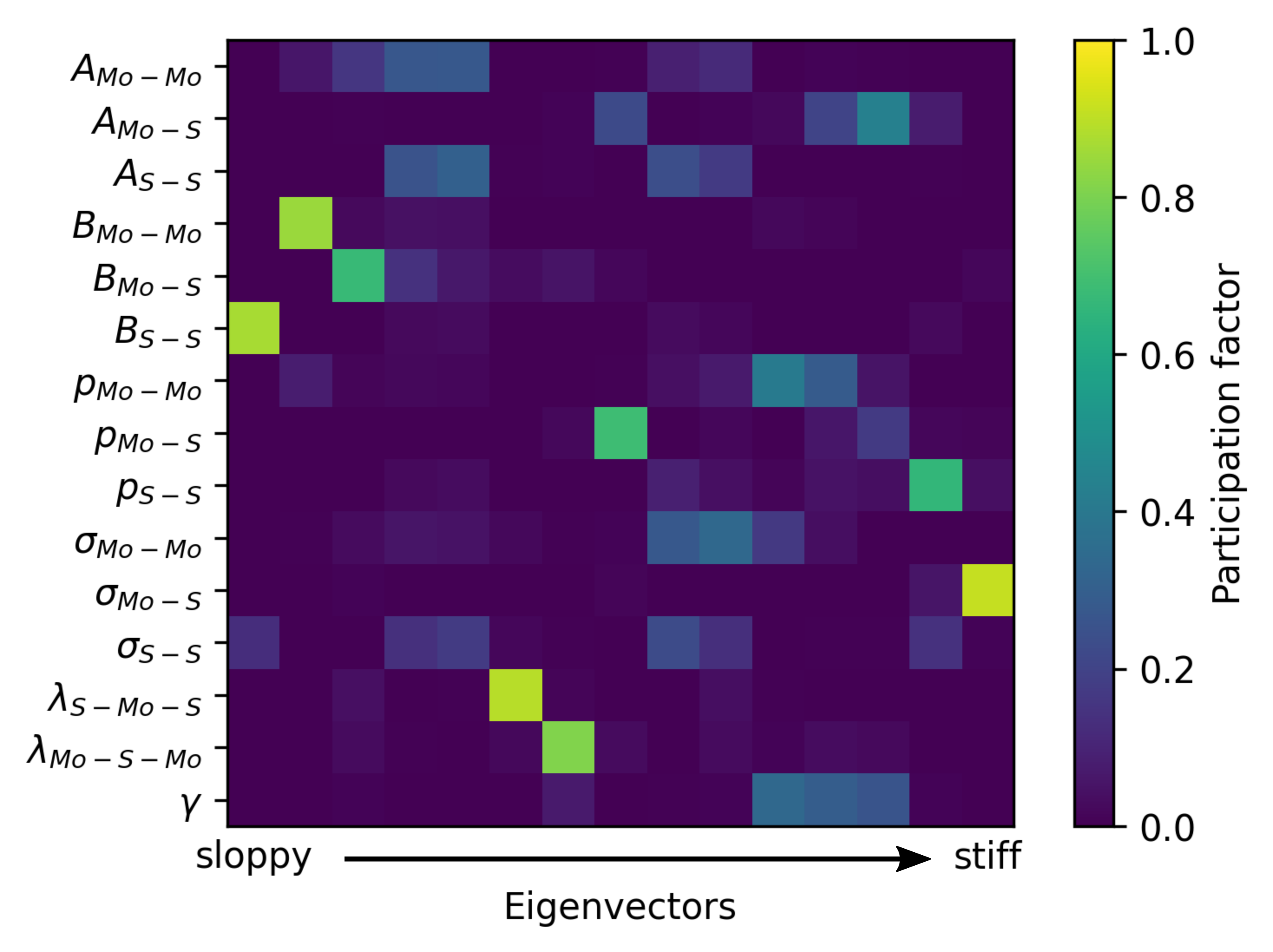}
    \caption[Participation factor of the SW potential for MoS$_2$]{
        Participation factor of the SW potential for MoS$_2$, calculated as the element-wise square of the eigenvectors of the FIM.
        Each column corresponds to an eigenvector, increasing in stiffness from left to right.
        The participation factor shows how much each parameter contributes to each eigenvector.
        The parameter direction is given by the logarithm of the labels on the vertical axis.
        The sloppiest eigenvector (the left most column) is mostly in the $\log(B_{\text{S--S}})$ direction.
    }
    \label{fig:fim_SW_eigenvectors}
\end{figure}

Figure~\ref{fig:fim_eigenvalues} shows the eigenvalues of the FIM for the models in \Subsec~\ref{subsec:interatomic_potentials_and_tests}, evaluated at the nominal values of the parameters.
We use these models to perform energy and forces calculations, as the first evaluation scenario, as explained in \Subsec~\ref{subsec:interatomic_potentials_and_tests}.
Note that they are sloppy; the eigenvalues cover many orders of magnitude, indicating that many parameters are unidentifiable from the data.
To illustrate that sloppiness is a general property of IPs, we also include the eigenvalues of the FIM, evaluated at the fitted parameters, for the Khor--Das Sarma potential\cite{OpenKIM_Threebody_driver, Threebody} (three-body potential), Environment Dependent Interatomic Potential (EDIP)\cite{OpenKIM_EDIP_driver, EDIP_a, EDIP_b} (bond-order potential), and the original SW potential for silicon\cite{OpenKIM_SW_driver, Stillinger-Weber, SW_Balamane_a, SW_Balamane_b} in predicting the energy and forces of the atoms in a random triclinic silicon configuration\cite{OpenKIM_TPEAF_driver, OpenKIM_TPEAF_64} (the original parameters for these potentials can be found in OpenKIM\cite{OpenKIM_Threebody, OpenKIM_EDIP, OpenKIM_SW}).
Figure~\ref{fig:fim_SW_eigenvectors} shows the participation factor\cite{Perez-arriaga_Verghese_Schweppe_1982, Garofalo_Iannelli_Vasca_2002} for the SW MoS$_2$ model, i.e., how much each parameter contributes to each eigenvector.
Participation factors are calculated as the element-wise square of the eigenvectors of the FIM.
We conclude that the sloppiest direction, indicated by the eigenvector with the smallest eigenvalue, is dominated by the parameter $B_{\text{S--S}}$.
Similarly, we can read off the participation factors of each parameter in the other eigendirections.

The FIM is a local calculation and computationally inexpensive compared to other methods discussed here.
As a result we recommend using the FIM as an initial step to UQ.
We will revisit the results from the FIM when we extend the analysis and compare the results to more global methods.

\begin{figure*}[!ht]
    \includegraphics[width=0.9\textwidth]{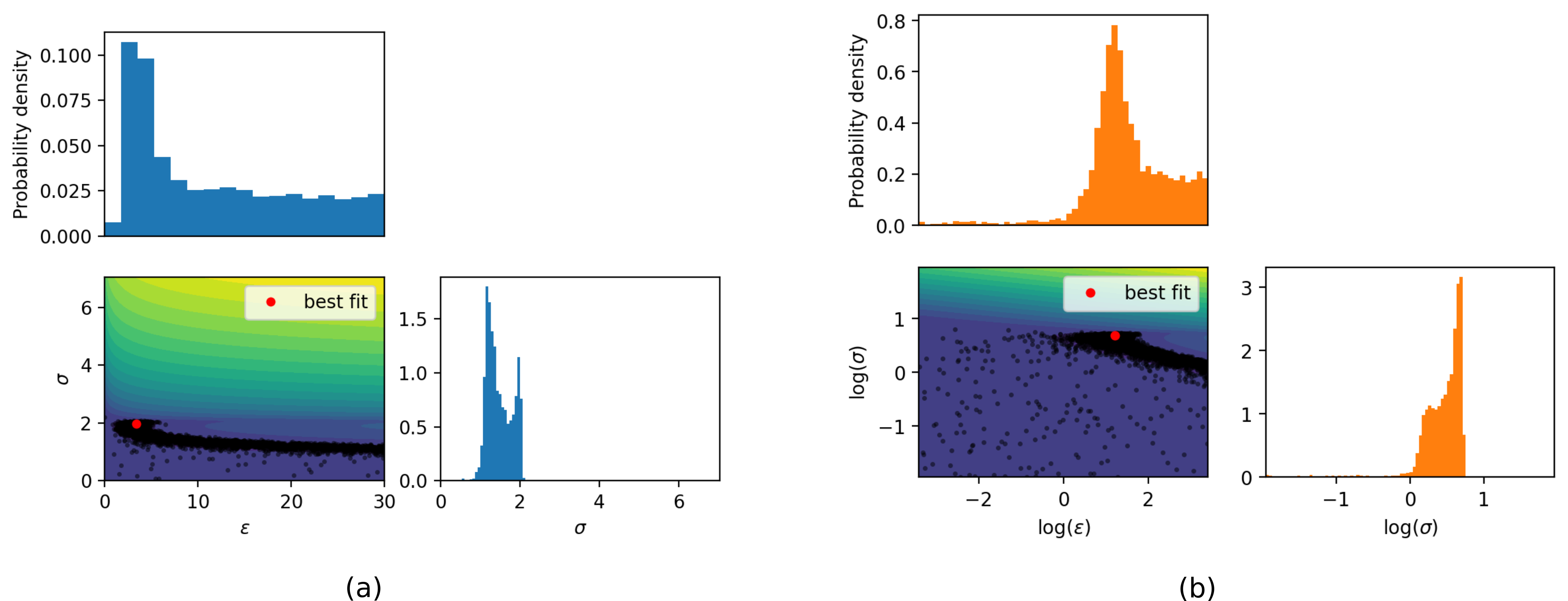}
    \caption[MCMC samples of the LJ potential]
        {MCMC samples for the LJ potential, sampled in (a) linear and (b) log parameter scales, at sampling temperature $T=21.5$.
        The original parameterization of the potential is given by the red dot.
        The samples are plotted against the cost contour on the lower left frame on each figure, with samples condensed around the low cost canyon.
        The marginal distributions are shown on the diagonal for each figure.
    }
    \label{fig:resultsLJ}
\end{figure*}

\begin{figure*}[!ht]
    \centering
    \includegraphics[width=\textwidth]{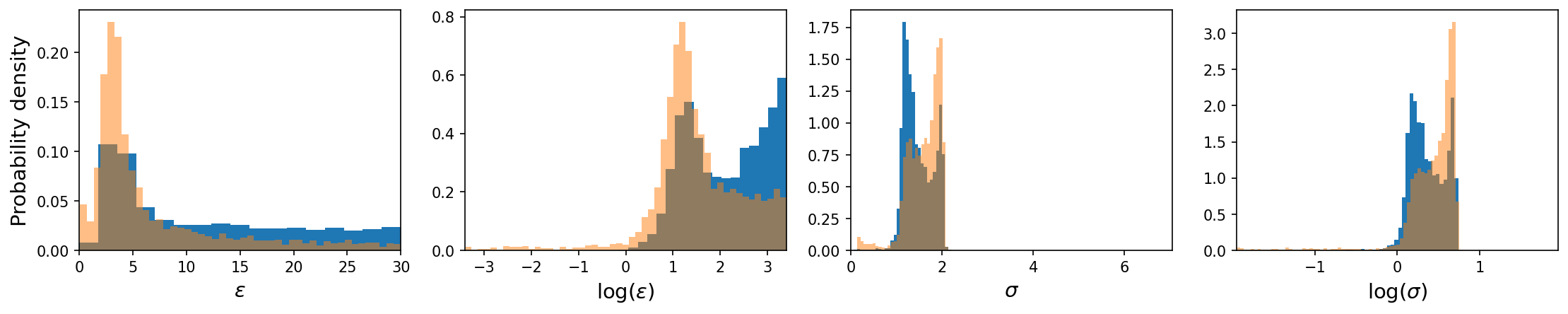}
    \caption[Comparison of the marginal distributions for the LJ potential]{
        Comparison of the marginal distributions for the MCMC simulations sampled in linear (blue) and log (orange) parameter scales.
        A uniform prior was used in both cases.
        Differences in the posterior distributions reflect the role of parameter scaling and choice of prior.
    }
    \label{fig:results_LJ_marginal_compare}
\end{figure*}

We now consider the results of the Bayesian analysis for the LJ potential.
Bayesian analysis requires a prior and a common choice by molecular modelers is the uniform prior.
Figure~\ref{fig:resultsLJ} shows these results sampled for uniform on both the linear [Fig.~\hyperref[fig:resultsLJ]{10(a)}] and log scales [Fig.~\hyperref[fig:resultsLJ]{10(b)}].
In both cases, we use a uniform prior in their respective parameter space, bounded by a rectangular region defined by $0<\epsilon<30$ and $0<\sigma<2^{-1/6} r_\text{cut}$ in linear parameter space, and $\left|\log(\epsilon)\right|<\log(30)$ and $\left|\log(\sigma)\right|<\log(2^{-1/6} r_\text{cut})$ in log parameter space.
The upper bound of $\sigma$ is chosen so that the pair-wise equilibrium length is less than the cutoff distance.
Although both priors are uniform and correspond to approximately the same range, they distribute prior probability weight differently and lead to different posterior distributions.
At first glance, the sampling is in good agreement with what is expected from the cost surface.
Samples dominate the regions of low cost and give a visual validation that the samples are converged.
Figure~\ref{fig:results_LJ_marginal_compare} compares the marginal distributions for each parameter on both linear and log scales.
Note how the parameter scaling and, by extension, the choice of prior, can have a strong impact on the posterior distribution.
On a log scale, there is a broad, flat plateau for large, negative values of $\log(\sigma)$ and $\log(\epsilon)$.
These choices affect how the uncertainties are interpreted and eventually propagated to new predictions.
This example illustrates the nuanced relationship between parameterization and prior selection.
These effects can be much more severe and harder to tease out in high-dimensional parameter spaces.

In more general terms, recall that the prior defines how one measures volume in parameter space.
Regions of the parameter space with large volume may dominate samples, in analogy to statistical mechanics in which high-entropy configurations can dominate an ensemble.
It is also related to (though not exactly the same as) the phenomenon known as Lindley's paradox in which Bayesian and frequentist approaches can give different results in a hypothesis test when a broad prior is used.\cite{Robert_2013}
This issue can become especially subtle for sloppy models in high dimensions.
These models are insensitive to coordinated changes in many parameters, indicating that there are large regions of parameter space with nearly identical fits, i.e., fits with high-entropy contributions to the posterior.
In these cases, large entropic contributions may dominate their relative frequency in the posterior.
The high dimensionality makes it difficult to quantify the role of the energy vs entropy in the final sample and, by extension, justify the choice of prior.

\begin{figure*}[!ht]
    \centering
    \includegraphics[width=0.9\textwidth]{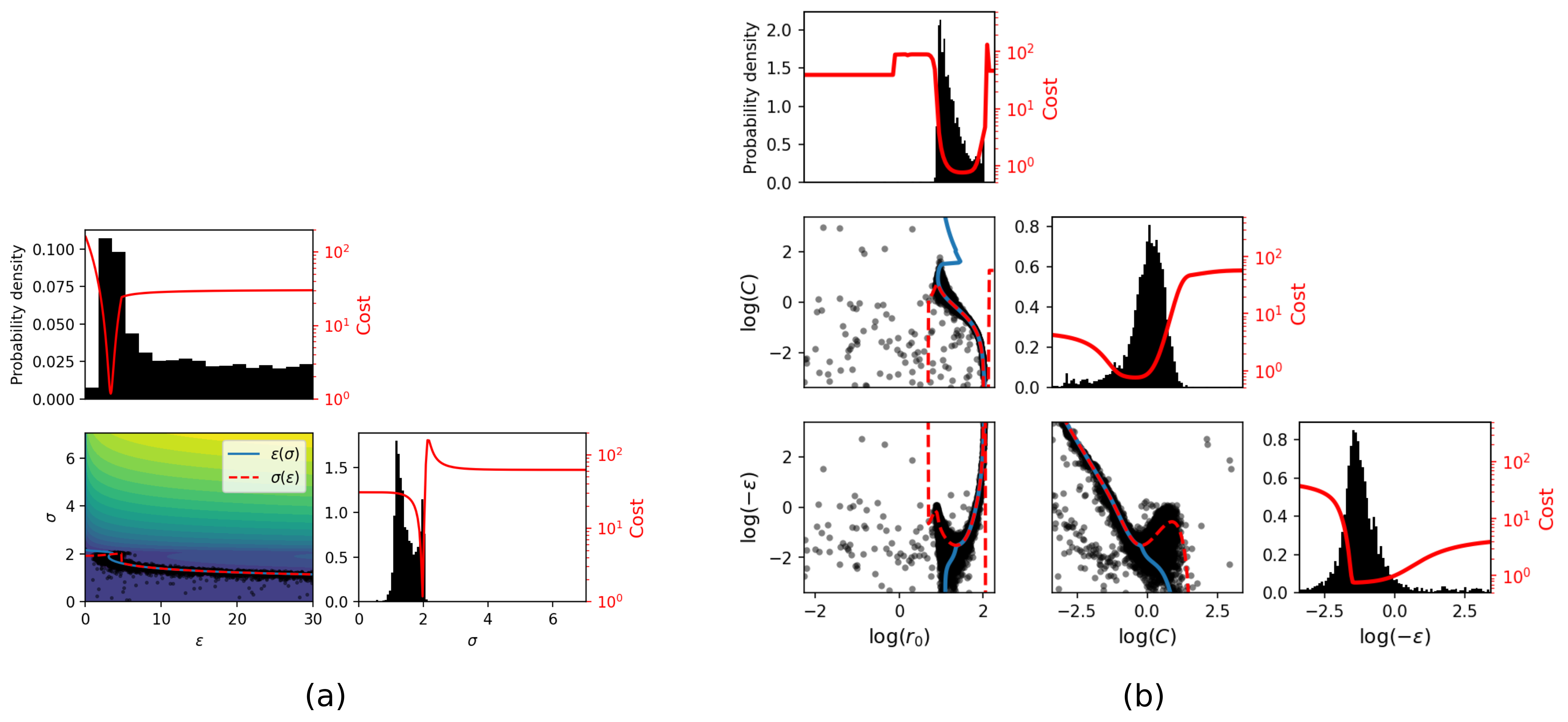}
    \caption[MCMC samples and profile likelihood for the LJ and Morse potentials]{
        MCMC samples and profile likelihood for the (a) LJ and (b) Morse potentials at sampling temperatures $T=21.5$ and $T=10.0$, respectively.
        We plot the cost surface for LJ because it only contains two parameters; in general, it is not possible to plot the cost surface, e.g., for Morse.
        On the lower triangle frames, the MCMC samples are plotted as the black points while the red and blue curves show the profile likelihood paths for the parameters on the horizontal and vertical axes, respectively.
        On the diagonal, we superimpose the cost profiles (red curves) on top of the marginal distribution of the MCMC samples.
        These plots show qualitative agreement between the two methods for low-dimensional models.
        MCMC samples are concentrated around the profile likelihood paths, indicating that the sampling is energy-dominated and there are no significant artifacts from the choice of prior.
        However, there are signs of large entropy regions that could dominate the sampling at higher temperatures, e.g., evaporation at large negative values of $\log(r_0)$.}
    \label{fig:resultsLJMorse}
\end{figure*}

For high-dimensional sloppy models, it is instructive to compare the results of the Bayesian and frequentist techniques, as done in Fig.~\ref{fig:resultsLJMorse} for the LJ and Morse potentials.
For the Morse potential, we also use a uniform prior, bounded by $\left|\log(r_0)\right| < \log(r_\text{cut})$, $\left|\log(C)\right| < \log(30)$, and $\left|\log(-\epsilon)\right| < \log(30)$.
These cases illustrate how the two methods agree in low-dimensional examples that are well-understood.
Note that the samples are very dense in regions around the paths of the profile likelihood, indicating that the posterior is energy-dominated and that there are not significant artifacts from the prior.
Furthermore, the marginal distributions of each parameter are congruous with the profile likelihoods (main diagonal).
However, there are hints of large-entropy regions that could become significant at higher sampling temperatures, for example, samples evaporating on the sub-optimal region at large negative values of $\log(r_0)$.

\begin{figure*}[!ht]
    \centering
    \includegraphics[width=0.9\textwidth]{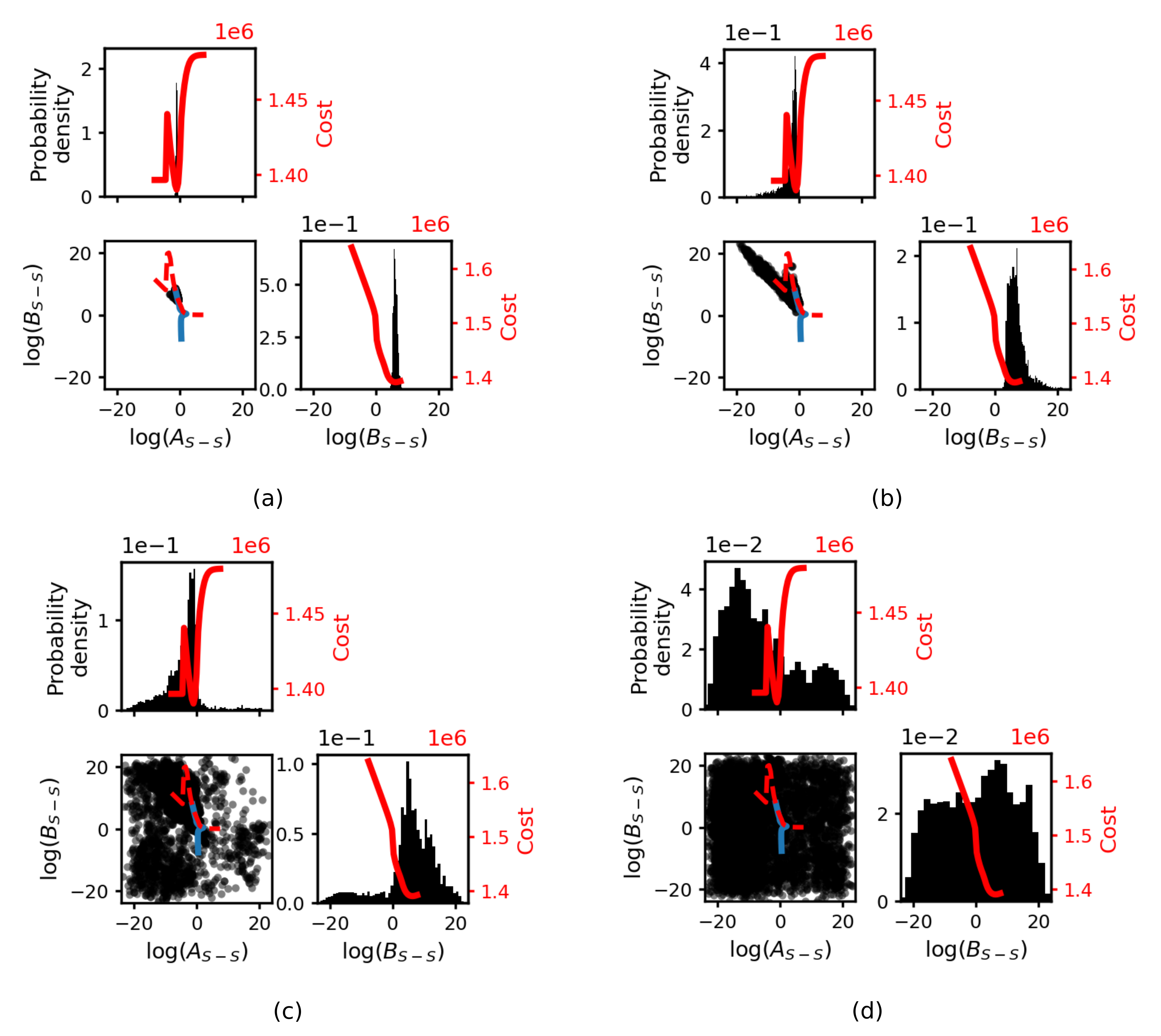}
    \caption[MCMC samples for the SW MoS$_2$ potential at several different temperatures]{
        MCMC samples and profile likelihood for parameters $A$ and $B$ of the S--S interaction in the SW MoS$_2$ potential with a uniform prior in log parameter space at sampling temperatures: (a) $T = 5.40\times10^{-3}~T_0$, (b) $T = 5.40\times10^{-2}~T_0$, (c) $T = 5.40\times10^{-1}~T_0$, and (d) $T = 5.40~T_0$, where $T_0$ is a natural temperature.      
        Note that different parameter combinations evaporate at different temperatures, e.g., $A_{\text{S--S}}$ and $B_{\text{S--S}}$ evaporate at lower temperature in a coordinated way (b), while they evaporate in all directions at higher temperature (c) and (d).
    }
    \label{fig:results_SW_compare}
\end{figure*}

We now turn to the SW model in Fig.~\ref{fig:results_SW_compare}.
We set the prior distribution to be uniform in a rectangular region, defined by $\left|\log(\theta_i)\right| < 24$, where $\theta_i$ are the parameters in this potential.
Figure~\ref{fig:results_SW_compare} summarizes a Bayesian sampling at four temperatures for parameters $A_\text{S--S}$ and $B_\text{S--S}$ ($T = 5.40 \times 10^{-3}~T_0, 5.40 \times 10^{-2}~T_0, 5.40 \times 10^{-1}~T_0, 5.40~T_0$, where $T_0 \approx 1.85 \times 10^5$ is the natural temperature).
The sampling results at other sampling temperatures for the other parameters can be found in \Supplementary.

At low temperatures, the profile likelihoods again agree with the Bayesian sampling.
Next, we increase the sampling temperature.
Recall that the temperature uniformly scales the error bars in Eq.~\eqref{eq:residuals}.
As the temperature rises, the uncertainty estimates in the parameters also increase; however, it does not increase uniformly in each of the parameters.
At some critical temperatures, the uncertainty in a particular parameter abruptly transitions to infinity.
For example, note that, from the spread of the samples, the uncertainties in the parameters $A_{\text{S--S}}$ and $B_{\text{S--S}}$ are relatively small at $T = 5.40 \times 10^{-3}~T_0$, but becomes effectively infinite at $T = 5.40 \times 10^{-1}~T_0$.
The posterior has transitioned from a distribution of low-temperature, energy-dominated samples to high-temperature, entropy-dominated samples.
The higher sampling temperature has ``evaporated'' the parameter.
We discuss this further in \Sec~\ref{sec:discussion}.

\begin{figure}
    \centering
    \includegraphics[width=0.45\textwidth]{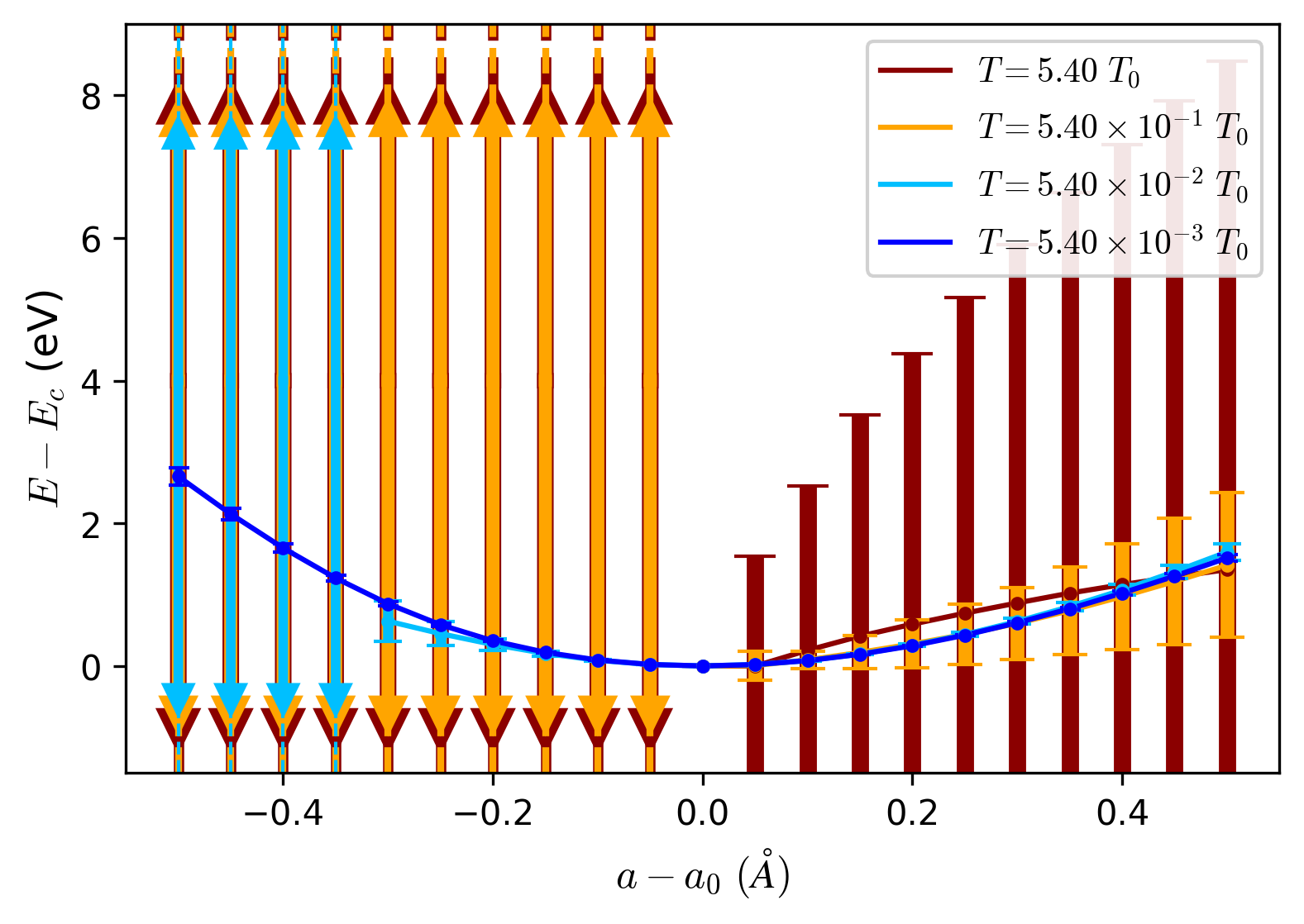}
    \caption[Propagated uncertainty of energy difference as a function of lattice stretching and compression]{
        Propagated uncertainty of changes in energy as a function of lattice stretching and compression.
        The parameters $a_0$ and $E_c$ are the equilibrium lattice constant and cohesive energy, respectively.
        The uncertainty of this quantity is calculated at several different temperatures from the ensembles in Fig.~\ref{fig:results_SW_compare}.
        Note that the uncertainty of the energy at higher temperature diverges to infinity.
        This is a results of parameter evaporation, where the evaporated parameters predict infinite energy.
    }
    \label{fig:results_SW_energylatconst}
\end{figure}

For the next step of UQ for this model, we propagate the parametric uncertainty and calculate the uncertainty of the change in energy as a response to lattice stretching and compression.\footnote{The MoS${_2}$ layer is constrained to remain flat, so that out-of-plane wrinkling is not possible.}
Figure~\ref{fig:results_SW_energylatconst} shows the uncertainty of this quantity of interest, calculated at several different sampling temperatures using the ensembles in Fig.~\ref{fig:results_SW_compare}.
Note that at lower sampling temperatures, such as at $T = 5.40 \times 10^{-3}~T_0$ (blue), the uncertainties of the predicted quantities are finite.
Moreover, the uncertainty in the tension domain ($a > a_0$) matches the distribution of predicted quantities from various models in Ref.~\onlinecite{Wen_Shirodkar_Plechas_Kaxiras_Elliott_Tadmor_2017}.
However, at higher temperatures, the uncertainties diverge as the MoS$_2$ sheet is compressed.

At higher temperatures, some of the MCMC walkers sample regions with extreme values of parameters, near the edge of the support of the prior.
These evaporated samples represent interactions with a very strong repulsive force in the compression domain.
The magnitude of the energy grows very fast as the lattice is compressed.
Consequently, the uncertainty of the energy in this domain diverges.
Furthermore, we are unable to propagate the uncertainty from the $T=5.40~T_0$ samples.
The ensemble at this temperature contains many samples representing extreme potentials, e.g., a semi-infinite square-well potential.

The phenomenon of parameter evaporation illustrated in Fig.~\ref{fig:results_SW_compare} has been observed previously.\cite{Gutenkunst_2007}
When a parameter evaporates, its marginal posterior distribution approaches its prior.
Evaporated parameters do not participate in the statistics of the model; they do not encode any information in the data and do not constrain future predictions.
In other words, the effective dimensionality of the model is reduced by the number of evaporated parameters.
However, evaporated parameters affect statistical methods, slowing down convergence in both MCMC sampling and profile likelihood optimization.
They also obscure interpretation since the evaporated parameters are often combinations of the bare parameters.
Performing UQ with these ``nuisance'' parameters is challenging.

\begin{figure}
    \centering
    \includegraphics[width=0.45\textwidth]{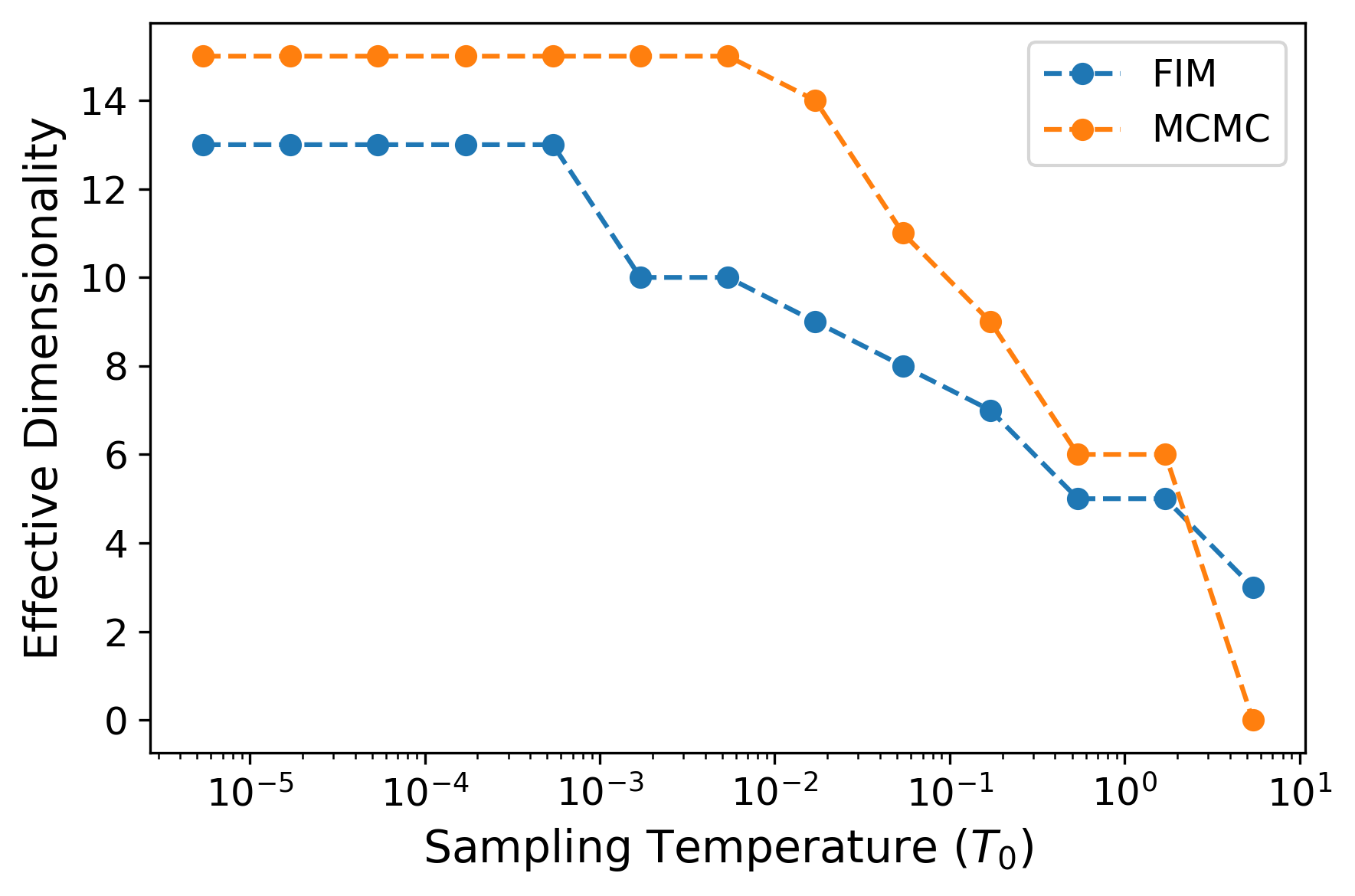}
    \caption[Comparison between the results from MCMC samples and the FIM]{
        Relation between sampling temperature and the effective dimensionality, obtained from the local (FIM) and global (MCMC) estimates.
        The effective dimensionality from the MCMC samples is the number of non-evaporated parameters at a given temperature.
        The local estimate is the number of eigenvalues of the Fisher information larger than a given temperature.
    }
    \label{fig:eigval_mcmc_compare}
\end{figure}

Parameter evaporation is a global manifestation of the ``sloppiness'' phenomenon.
Sloppiness was first recognized as the exponential distribution of FIM eigenvalues, as in Fig.~\ref{fig:fim_eigenvalues}, as a local property.
However, it was later shown using information geometry that sloppiness is a global property of the model and evaluation scenario.
For sloppy models, the entire model manifold is systematically compressed into an object of low effective dimensionality, and in many practical cases, the eigenvalues of the FIM (local property) are a good estimate for the widths of the model manifold (global property).\cite{Transtrum_Machta_Sethna_2011, Transtrum_Machta_Sethna_2010, Machta_Chachra_Transtrum_Sethna_2013, quinn2019visualizing, quinn2019chebyshev}
We check this correspondence for the case of the SW IP by comparing the eigenvalues in Fig.~\ref{fig:fim_eigenvalues} with the number of effective (non-evaporated) parameters in the model at each sampling temperature.
We consider a parameter ``evaporated" if the samples approach a boundary of the prior corresponding to this parameter.
Although, in general, there is a subtle difference between the evaporated and the non-evaporated parameters, the temperature ladder we use is sparse enough that there is a clear distinction between the two.
We show the comparison between the local estimate and the result deduced from MCMC in Fig.~\ref{fig:eigval_mcmc_compare}, and we find good agreement.

Sloppiness in high dimensions leads to cost contours that do not close and complicates the question of prior selection and the role of parameter-space entropy in the Bayesian posteriors.
We now use information geometry to better understand this phenomenon, first using the LJ model as a motivating example.

\begin{figure}[!ht]
    \centering
    \includegraphics[width=0.45\textwidth]{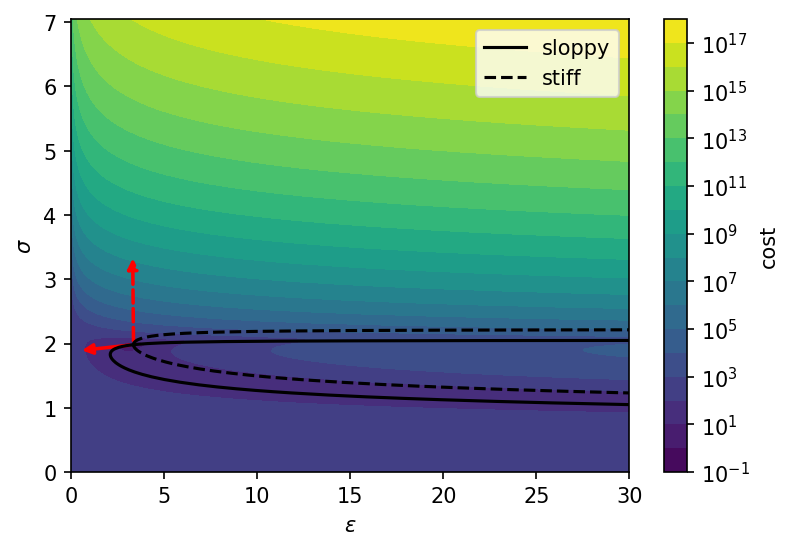}
    \caption[Geodesics for the LJ potential]{
        Geodesics for the LJ potential in the sloppy and stiff directions.
        Each geodesic starts at the best fit parameters and moves in the forward and backward directions of the sloppy and stiff eigenvectors of the FIM, shown by the corresponding arrows.
        By considering the difference in the scaling of the parameter axes, the eigenvectors are orthogonal to each other.
    }
    \label{fig:lj_geo_cost}
\end{figure}

Figure~\ref{fig:lj_geo_cost} shows four geodesics paths in parameter space radiating from the best fit.
Although difficult to discern from this figure, the geodesics that found boundaries took the same asymptotic form: $\log(\epsilon) \rightarrow \infty$, $\log(\sigma) \rightarrow -\infty$, with $\log(\epsilon)$ diverging six times faster than $\log(\sigma)$.
The significance of this result is more apparent when expressed in the so-called $AB$ form with $A = 4 \epsilon \sigma^{12}$ and $B = 4 \epsilon \sigma^6$.
As the original parameters, $\epsilon$ and $\sigma$, approach the boundary, they are correlated such that $A \rightarrow 0$ while $B$ remains finite.
This suggests that the $AB$ parameterization is a more natural parameterization, which from Eq.~\eqref{eq:Lennard-Jones} gives,
\begin{equation}
    \begin{aligned}
        \phi_\text{LJ}(r_{ij}) &= 4\epsilon \left( \sigma^{12} \left( \frac{1}{r_{ij}^{12}} - \frac{1}{r_\text{cut}^{12}} \right)
             - \sigma^{6} \left( \frac{1}{r_{ij}^{6}} - \frac{1}{r_\text{cut}^{6}} \right) \right)\\
          &= A \left( \frac{1}{r_{ij}^{12}} - \frac{1}{r_\text{cut}^{12}} \right) -
             B \left( \frac{1}{r_{ij}^{6}} - \frac{1}{r_\text{cut}^{6}} \right) \\
          &\rightarrow -B \left( \frac{1}{r_{ij}^{6}} - \frac{1}{r_\text{cut}^{6}} \right).
    \end{aligned}
    \label{eq:reduced_LJ}
\end{equation}

\begin{figure*}[!ht]
    \centering
    \includegraphics[width=0.9\textwidth]{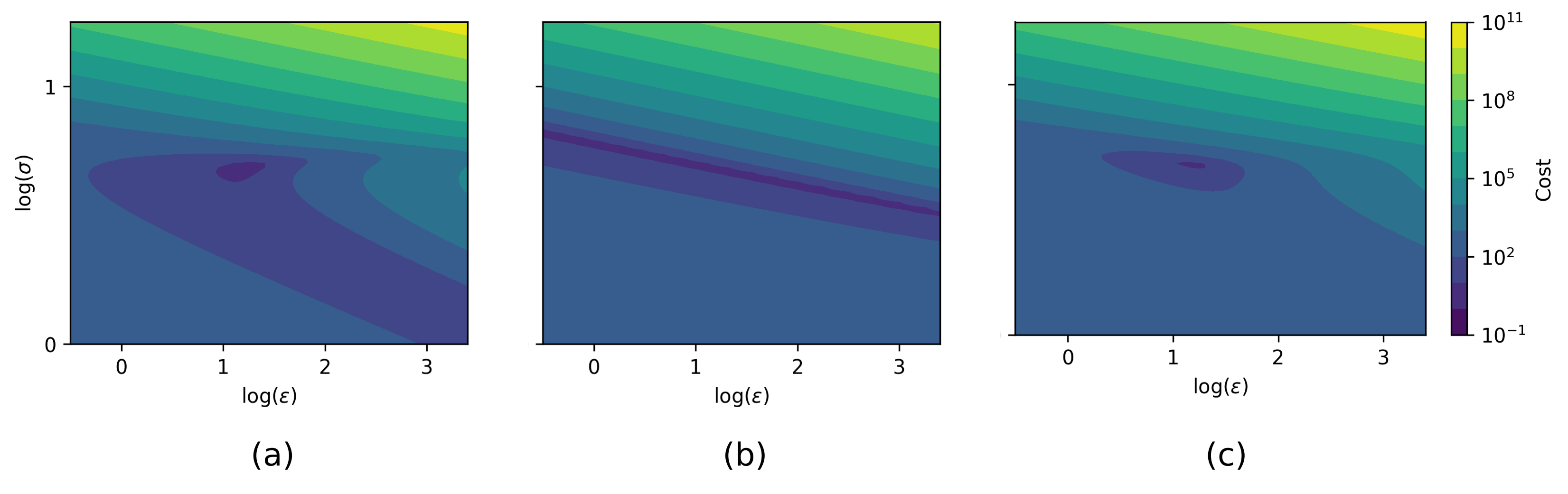}
    \caption{Cost surface of the LJ potentials with data that probe the (a) attractive, (b) repulsive, and (c) both regimes.
    When only the data in the attractive (repulsive) regime are used, parameter $A$ ($B$), which corresponds to repulsive (attractive) interaction in the LJ potential, is unidentifiable.
    This is shown by the direction of the low cost canyon in (a) and (b).
    When the data probe both regimes, these canyons are raised and parameters $A$ and $B$ become more identifiable.
    However, note that the plateau extending to $\{\epsilon, \sigma\} \to 0$ direction exists in all cases, which shows that there is a fundamental limitation in the LJ potential in predicting energy and forces.
    }
    \label{fig:cost_surface_LJ}
\end{figure*}

For this fitting problem, $B$ is the identifiable parameter combination.
There is a natural limit that removes the unidentifiable parameter, $A \rightarrow 0$, that leads to a physically interpretable reduced model, i.e., a purely attractive potential.
Although this reduced model loses the physics of the repulsive part of the potential, the data to which the model was fit included atomic configurations that only probed the attractive regime.
In the case when the data only probe the repulsive regime, $A$ becomes the identifiable parameter, and removing the unidentifiable parameter ($B \to 0$) will lead to a purely repulsive potential, instead.
This behavior is illustrated in Fig.~\ref{fig:cost_surface_LJ}.
Thus, the geometry (1) reflects the information content of the data, (2) explains the correlations among the inferred parameters, (3) isolates unidentifiable combinations of parameters, and (4) suggests reduced models for simplifying the statistics.

\begin{figure*}[!ht]
    \centering
    \includegraphics[width=0.6\textwidth]{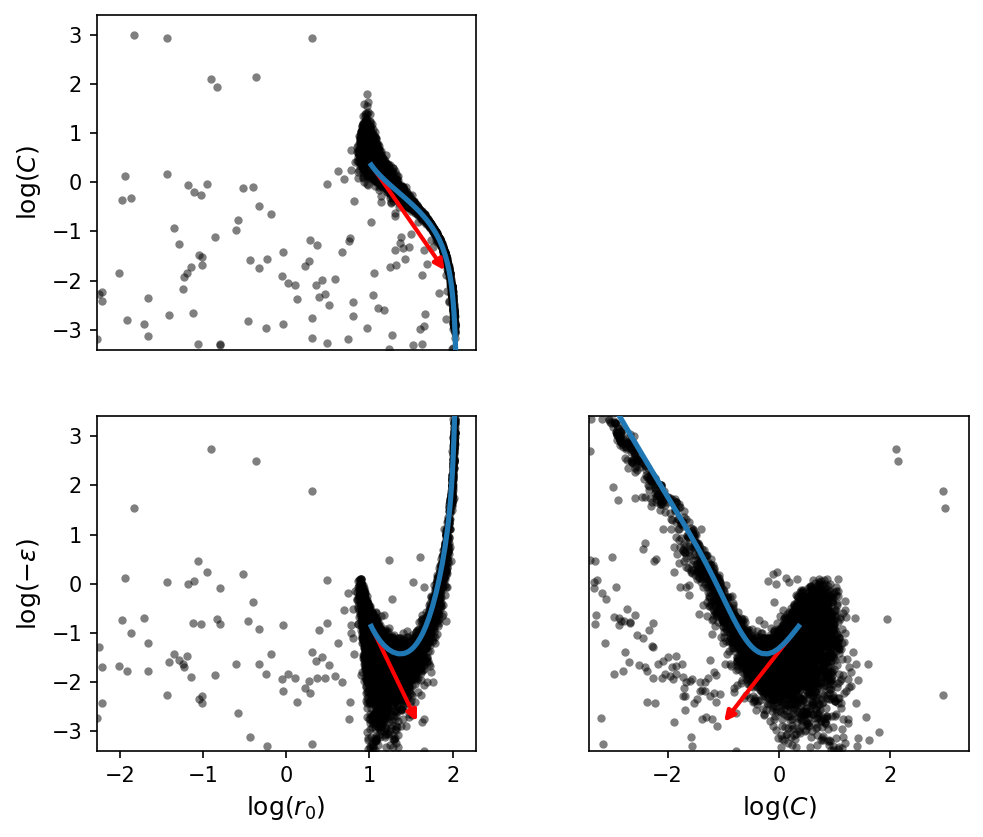}
    \caption[Geodesic for the Morse potential]{
        Geodesic in the sloppiest direction for the Morse potential.
        Geodesics (blue curves) are shown with the MCMC samples (black points) to illustrate that the geodesics follow the same low cost canyons as MCMC.
        These geodesics reveal specific parameter limits leading to boundaries of the model manifold, e.g., $C\rightarrow 0$ as $\epsilon \rightarrow -\infty$.}
    \label{fig:geo_mcmc_Morse}
\end{figure*}

An analogous calculation on the Morse potential reveals many of the same themes.
We find the geodesic initially aligned with the sloppiest eigenvector of the FIM, eventually approaches a boundary in which $C \rightarrow 0$ and $\epsilon \rightarrow -\infty$, as seen in Fig.~\ref{fig:geo_mcmc_Morse}.
Note that this geodesic aligns with the MCMC results, i.e., the low-cost canyon.
As with LJ, this geodesic suggests a natural reparameterization of the model,
\begin{align}
  k &= -2\epsilon C^2.
\end{align}
As $\epsilon$ and $C$ approach extreme values, the specific combination $k$ remains finite.
In this parameterization, $k$ and $r_0$ are the identifiable parameter combinations, while $C$ is unidentifiable since the geodesic approaches the boundary given by $C \rightarrow 0$.
Evaluating the limit $C \rightarrow 0$ at constant $k$ and $r_0$ leads to the simplified model,
\begin{align}
  \widetilde{\phi}_\text{M}(r_{ij}) &= \frac{1}{2} k (r_{ij} - r_0)^2 - \frac{1}{2} k (r_\text{cut} - r_0)^2.
\end{align}
In this limit, the model is a simple harmonic potential parameterized by an equilibrium position, $r_0$, and a stiffness, $k$.
This indicates that the configurations do not carry enough information about the IP's anharmonicity to constrain those parameter combinations.
Once again, the geometry reflects the information content of the data, explains observed correlations, isolates the unidentifiable combinations, and suggests alternative parameterizations and simplified models.

\begin{figure*}[!ht]
    \centering
    \includegraphics[width=0.6\textwidth]{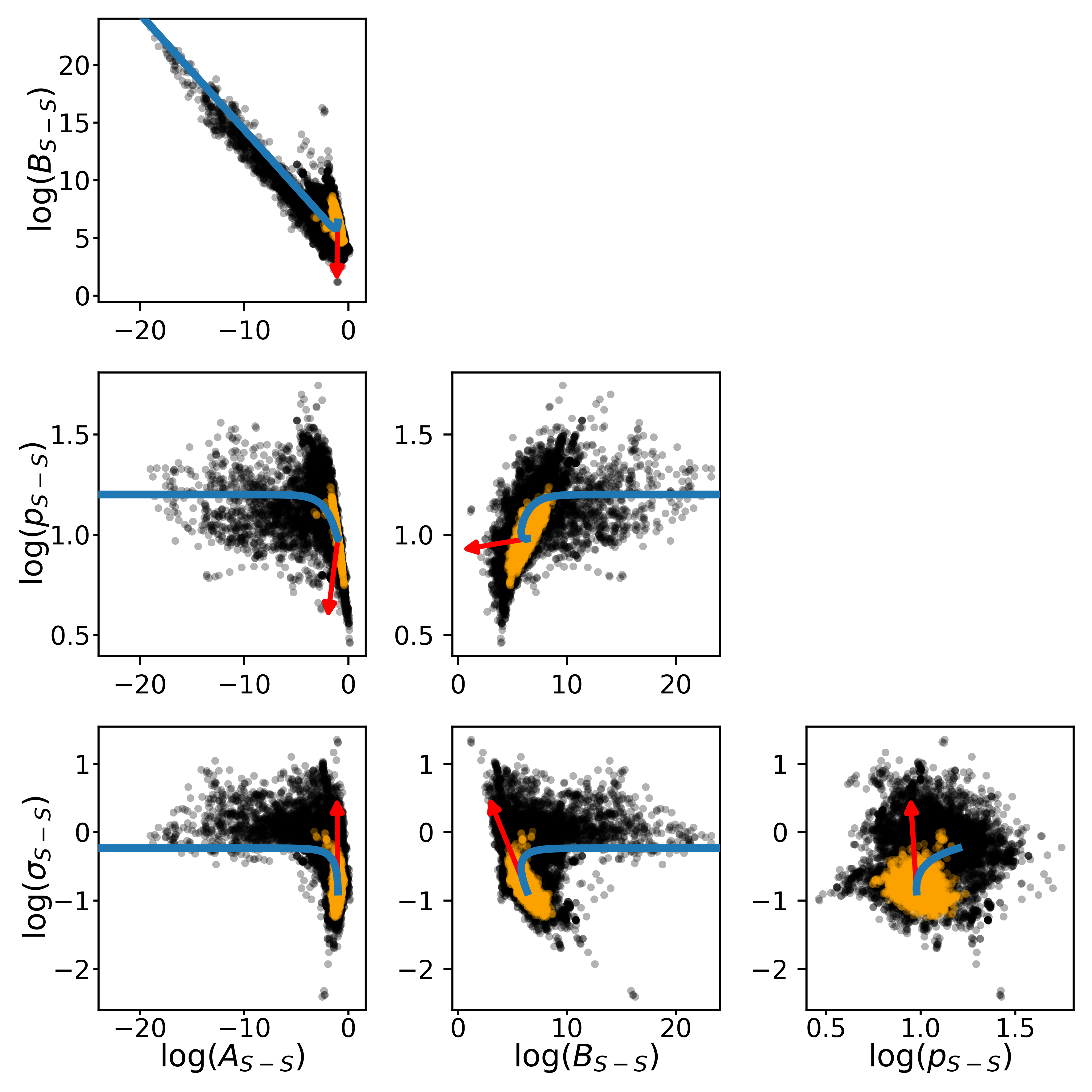}
    \caption[Geodesic for the SW Mo$S_2$ potential]{
        Geodesics in the sloppiest direction for the SW model.
        Geodesics (blue curves) are plotted with MCMC results at sampling temperatures of $5.40 \times 10^{-3}~T_0$ (orange points) and $5.40 \times 10^{-2}~T_0$ (black points).
        Axis scaling is set to show detail, not to reflect the boundaries of the uniform prior.
        The geodesic started in the local sloppy direction defined by the FIM, $B\rightarrow 0$ as $\sigma_\text{S--S} \rightarrow \infty$.
        Eventually, the geodesic turned to find the boundary given by the limit $A_\text{S--S}\rightarrow 0$ as $B_\text{S--S}\rightarrow \infty$.
        }
    \label{fig:geo_mcmc_SW_evaporation}
\end{figure*}

We now consider the SW potential.
As before, we calculate a geodesic in the sloppiest direction and find that it encounters the boundary $A_{\text{S--S}} \rightarrow 0$ and $B_{\text{S--S}} \rightarrow \infty$, as shown in Fig.~\ref{fig:geo_mcmc_SW_evaporation}.
Again, we reparameterize the model,
\begin{align}
    \theta = A_{\text{S--S}} B_{\text{S--S}}.
\end{align}
Both $A_{\text{S--S}}$ and $B_{\text{S--S}}$ are unidentifiable parameters, but there is an identifiable combination given by $\theta = A_{\text{S--S}} B_{\text{S--S}}$.
Note that as $A_{\text{S--S}} \rightarrow 0$ at constant $\theta$, $B_{\text{S--S}} \rightarrow \infty$, consistent with the evaporation in Fig.~\ref{fig:geo_mcmc_SW_evaporation}.
Furthermore, considering $A_{\text{S--S}}, 1/B_{\text{S--S}} \rightarrow 0$ at constant $\theta$ leads to the reduced form,
\begin{align}
  \widetilde{\phi}_2^{\text{S--S}}(r_{ij}) = \theta\left(\frac{\sigma}{r_{ij}}\right)^p \exp\left(\frac{\sigma}{r_{ij}-r^\text{cut}}\right).
\end{align}

\begin{figure}[!ht]
    \centering
    \includegraphics[width=0.45\textwidth]{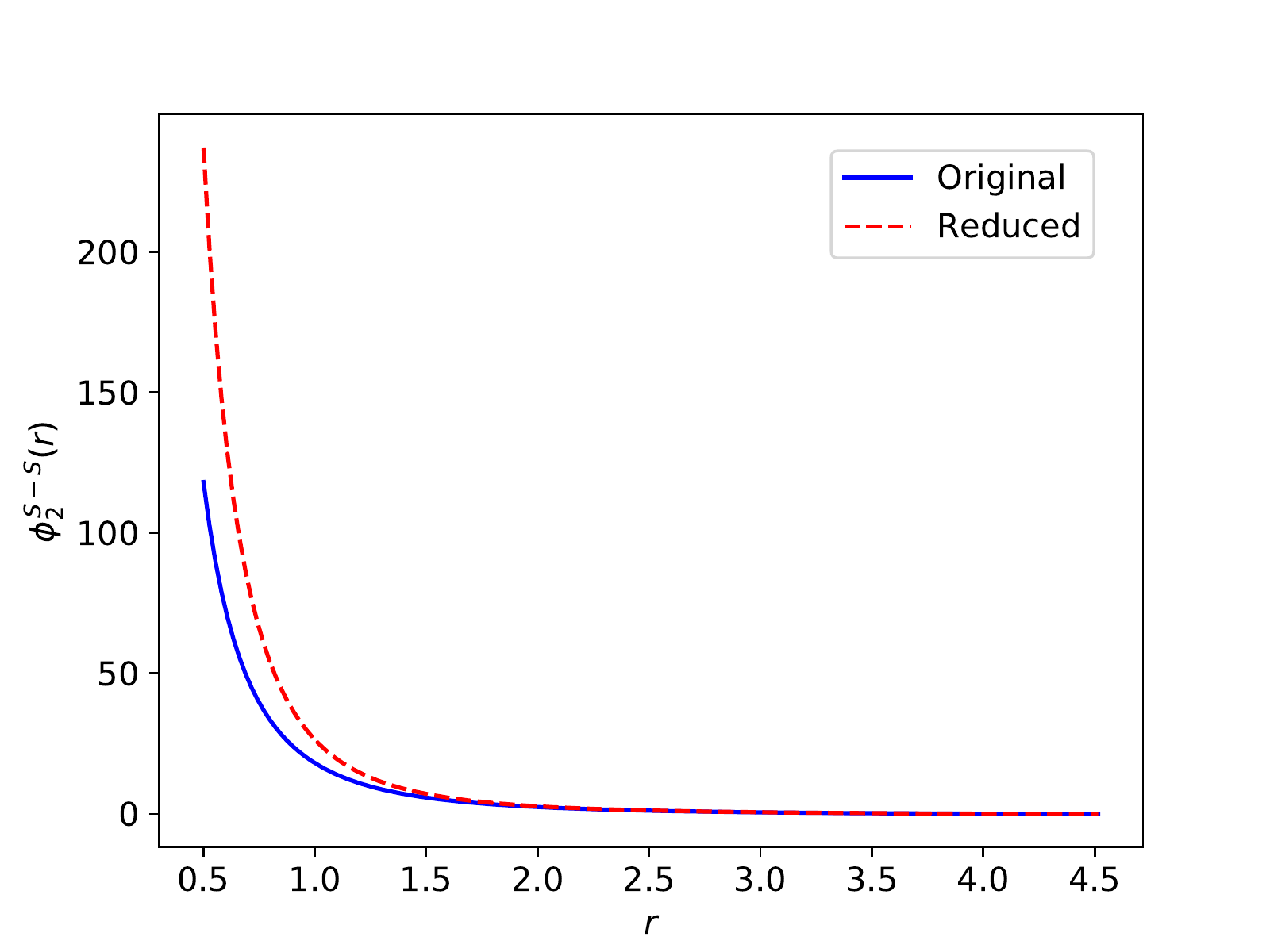}
    \caption[Reduced SW model for S--S interaction]{
        Reduced models on the boundaries have fewer identifiable parameters and abstract away irrelevant aspects of the physics.
        Geodesics identified the boundary defined by the coordinated limit $A_{\text{S--S}} \rightarrow 0$, $B_{\text{S--S}} \rightarrow \infty$.
        This figure compares the forms of the original and reduced two-body SW potentials for the S--S interaction.
        This reduction abstracts away details about the repulsive core of the potential.}
    \label{fig:SW_potential_comparison}
\end{figure}

Figure~\ref{fig:SW_potential_comparison} shows the plots of the two-body S--S interaction term for both the original and reduced models.
In addition to having fewer identifiable parameters, reduced models include the physics that is informed by the data.
In this case, details about the repulsive core were removed at the boundary, resulting in a stronger repulsion at short distances that were not constrained by the fit.
This example illustrates that reduced order models change the nature of the physics that is encoded in the potential.
While the simple model may be just as accurate as the original model on the training data, it may be less accurate for downstream applications.
Additional work is needed to develop these information geometric insights into concrete tools for selecting and developing interatomic potentials for target applications.
Alternatively, by identifying the short-range interactions as the missing information in the training set, information geometry can potentially guide the extension of training data to constrain relevant parameter combinations.

It is interesting to compare the geodesic to the purely local analysis of the Fisher information.
The initial direction of the geodesic is given by the sloppiest eigendirection of the Fisher information.
That eigendirection, as can be seen in Fig.~\ref{fig:fim_SW_eigenvectors}, is dominated by the parameters $B_{\text{S--S}}$ and $\sigma_{\text{S--S}}$.
Initially, the geodesic decreases $B_{\text{S--S}}$ and increases $\sigma_{\text{S--S}}$; however, as $\sigma_{\text{S--S}}$ increases, it plays a more important role in the model.
It is no longer part of the sloppy combination, so the geodesic rotates to align with $A_{\text{S--S}}$ and $B_{\text{S--S}}$, the parameters that eventually participate in the boundary (see Fig. \ref{fig:geo_mcmc_SW_evaporation}).
This simple comparison illustrates how the geodesic naturally extends the local analysis.



\section{DISCUSSION}
\label{sec:discussion}
Sloppy models are often identified by their characteristic FIM spectra with eigenvalues spanning many orders of magnitude.\cite{Transtrum_Machta_Brown_Daniels_Myers_Sethna_2015, Transtrum_Machta_Sethna_2010, Gutenkunst_Waterfall_Casey_Brown_Myers_Sethna_2007, Waterfall_Casey_Gutenkunst_Brown_Myers_Brouwer_Elser_Sethna_2006}
Previous work has noted sloppiness in many contexts,\cite{Jeong_Zhuang_Transtrum_Zhou_Qiu_2018, White_Tolman_Thames_Withers_Mason_Transtrum_2016, Transtrum_Qiu_2016, Mannakee_Ragsdale_Transtrum_Gutenkunst_2016,  Transtrum_Machta_Brown_Daniels_Myers_Sethna_2015, Machta_Chachra_Transtrum_Sethna_2013, Transtrum_Qiu_2012} including IPs,\cite{Frederiksen_Jacobsen_Brown_Sethna_2004, Wen_Li_Brommer_Elliott_Sethna_Tadmor_2016, Wen_Li_Brommer_Elliott_Sethna_Tadmor_2016, Longbottom_Brommer_2019} and our results corroborate this conclusion (see Fig.~\ref{fig:fim_eigenvalues}).
Subsequent work using information geometry showed sloppiness to be a consequence of global properties of the model, specifically that the model manifold is bounded with a hierarchy of widths.\cite{Transtrum_Machta_Sethna_2010}
In this work, we have extended the local sloppy model analysis using (Bayesian) MCMC and (frequentist) profile likelihoods.
Each of these methods gives a unique perspective on the ``global sloppiness" of the model.
For example, MCMC samples evaporate sloppy parameters, and cost profiles have flat plateaus resulting in diverging uncertainties.
We connect these traditional statistical tools to sloppy model analysis using information geometry and geodesics.
We show that the problems associated with both of these methods are features of the same underlying phenomena, sloppiness in the form of bounded model manifolds.
We now discuss each of these observations in more detail.

Figure~\ref{fig:resultsLJ} illustrates that cost contours of sloppy models have broad plateaus in parameter space.
These regions can be thought of as high-entropy states, i.e., large volumes of parameter space with approximately equal cost.
Although these regions may have sub-optimal cost, they can dominate the Bayesian posterior because of their high entropy.
Previous work in sloppy models has noted this ``parameter evaporation'' phenomenon when the posterior becomes dominated by sub-optimal, high-entropy samples.\cite{Gutenkunst_2007}
Geometrically, the model maps these high-entropy regions to compressed areas near the boundaries of the model manifold.
Parameter evaporation is thus the Bayesian manifestation of sloppiness that is a consequence of model boundaries.
As such, it is a nonlinear effect and distinct from other well-documented statistical phenomena related to high dimension.

Model boundaries have different distances from the data.
This implies that the height of the cost plateau varies in each parameter direction.
Consequently, specific parameters evaporate at different sampling temperatures.
This is analogous to particles in a classical finite potential well; only those particles (MCMC walkers) with high enough energy can escape the well (cost surface).
Figure~\ref{fig:results_SW_compare} shows that different parameter combinations evaporate at different temperatures.
Indeed, previous studies in IPs have lowered the sampling temperature specifically to avoid parameter evaporation.\cite{Frederiksen_Jacobsen_Brown_Sethna_2004, Longbottom_Brommer_2019}
However, for a sufficiently broad prior, parameters evaporate at any non-zero temperature, although the evaporation time may be very long for large cost barriers, making it difficult to assess convergence.

Figure~\ref{fig:eigval_mcmc_compare} reports the number of evaporated parameters for different sampling temperatures for our SW potential.
Note that the number of identifiable parameter combinations, i.e., non-evaporated at a given temperature, correlates with the number of eigenvalues above that temperature.
Elsewhere, it has been shown that the eigenvalues of the FIM are a good approximation for the widths of the model manifold.\cite{Transtrum_Machta_Sethna_2010, Transtrum_Machta_Sethna_2011}
Since the sampling temperature corresponds to a distance in data space, Fig~\ref{fig:eigval_mcmc_compare} confirms that the eigenvalues are a reasonable (local) approximation for the widths of the model manifold of IPs.

Most of the challenges associated with formulating and performing a Bayesian analysis of a sloppy IP result from the interplay of entropy and energy in the posterior.
Recall that the Bayesian prior acts as a measure on the parameter space (see \Subsec~\ref{subsec:method_likelihood}), i.e., it quantifies the entropy associated with volumes of parameter space.
The ambiguity in the choice of prior leads to the issues we report here.
We advocate comparing MCMC with a frequentist method to assess the effect the prior has on the posterior, as we have done using profile likelihoods.

The profile likelihood analysis, being a frequentist method, does not use a prior.
As such, global sloppiness manifests itself differently.
The cost profiles exhibit plateaus that asymptotically approach constant values, as seen in Figs.~\ref{fig:resultsLJMorse} and \ref{fig:results_SW_compare}.
Uncertainty in a parameter is set by selecting a level of statistical significance, e.g., 95\% confidence interval, and identifying those parameter values with cost less than the corresponding cost threshold.
As the cost threshold approaches that of the plateau, the uncertainty diverges.
This leads to uncertainty metrics that are very sensitive to the level of statistical significance, making it difficult to draw conclusions from the UQ analysis.

Another complication due to sloppiness is related to parameter correlations.
Sloppy canyons and plateaus do not naturally align with the parameter axes due to correlations in the parameters.
These correlations can bend sloppy canyons, as seen in Fig.~\hyperref[fig:resultsLJMorse]{12(a)} for LJ.
As the profile likelihood projects bending canyons onto parameter axes, correlation is lost and the results are misleading.
Using a more natural parameterization, motivated by information geometry, weakens parameter correlations and unwinds the canyons asymptotically aligning them with the parameter axes.

Geodesics extend the local FIM analysis to a global regime.
For example, consider Fig.~\ref{fig:geo_mcmc_SW_evaporation}.
The geodesic initially pointed in the sloppiest direction, as indicated by the FIM, but changed directions to follow the global sloppiness as it approached the manifold boundary.
This behavior is due to non-linearity in the model and is known as parameter-effects curvature.\cite{Bates_Watts_198}

The global nature of geodesics is used to find boundaries of the model manifold, revealing the cost plateaus and suggesting natural parameterizations of the model.
Again, consider Fig.~\ref{fig:geo_mcmc_SW_evaporation} where the geodesics found the manifold boundary represented by the parameter limits $A_{\text{S--S}}\rightarrow 0$ and $B_{\text{S--S}}\rightarrow \infty$.
This limit demonstrates a more natural parameterization of the model with parameters $\epsilon = 1/B_{\text{S--S}}$ and $\theta = A_{\text{S--S}} B_{\text{S--S}}$, where $\epsilon$ is strictly non-negative.
In this parameterization, only one parameter participates in the boundary, i.e., $\epsilon \rightarrow 0$ while $\theta \sim \mathcal{O}(1)$.
The sloppy direction aligns with a single parameter $\epsilon$ that is (mostly) uncorrelated from $\theta$.
Furthermore, rather than a diverging confidence interval for $B_{\text{S--S}}$, the confidence intervals for $\epsilon$ and $\theta$ remain finite.

This new parameterization suggests a simplified model in which $\epsilon$ has been removed.
The Manifold Boundary Approximation Method (MBAM) is a tool that utilizes information geometry to find these boundary limits and construct reduced models.
In this specific example, the reduced model would correspond to $\epsilon \rightarrow 0$ while holding $\theta \sim \mathcal{O}(1)$.
This limit removes the sloppy parameter from the model, leaving the identifiable combination, $\theta = A_{\text{S--S}} B_{\text{S--S}}$.
Performing MBAM before UQ, i.e., finding less-sloppy models, would remove the challenges we have discussed.
In this paper, we have performed the first step of MBAM by using geodesics to find manifold boundaries.

Reduced models often do not transfer well to new predictive regimes.
However, they make new predictions with higher levels of certainty.
When the large parametric uncertainties of sloppy models are propagated to new predictions, the resulting uncertainties can be large or infinite.
Reducing the sloppiness of models decreases parametric uncertainty as well as the propagated uncertainty in new predictive regimes.
Future work will continue this process and perform UQ on reduced models.

We have shown that sloppy models lead to ill-posed UQ problems.
For Bayesians, the challenge is how to unambiguously select a prior that does not lead to large-entropy contributions in the posterior.
For frequentists, the challenge is sensitivity to the confidence level and plateaus that do not naturally align with the bare parameters (i.e., occur due to correlations among parameters).
By identifying the root cause of these problems, we hope this work will lead to more transparency in the future UQ studies for IPs.
In particular, information geometry suggests solutions to these issues by identifying natural parameterizations near boundaries that provide simplified, less-sloppy models.

In conclusion, we provide suggestions both for model developers and UQ practitioners alike.
For developers of empirical potentials, we recommend using the FIM to assess how parameters locally affect calculated quantities.
To extend this analysis to a global regime, we recommend using geodesics to identify more natural parameterizations as well as additional training data that are needed to identify model parameters.
For example, geodesic calculations for the LJ potential above identified that the training data only contained information about the attractive part of the potential.
To fully identify the LJ parameters additional data that probe the repulsive regime are needed.
Alternatively, unidentifiable parameters may be removed.
Using these geodesics to reduce the model with an MBAM step will decrease model sloppiness and improve future UQ.
This process can then be iterated starting from the reduced model until a simple, yet accurate model is attained.
However, caution must be used as these parameters may be relevant for downstream applications.
Future work may focus on how to selectively remove parameters for target applications.

For performing UQ of IPs, we recommend starting with the FIM analysis to assess the sloppiness of the model.
This analysis also provides a local estimate of which parameters evaporate at a given sampling temperature.
If performing UQ with MCMC, we recommend using several different sampling temperatures, including the natural temperature.\cite{Brown_Sethna_2003} and some alternative priors.
For example, we have sampled the SW at multiple temperatures (see Fig~\ref{fig:results_SW_compare}) and for different priors (see the \href{https://aip.scitation.org/doi/suppl/10.1063/5.0084988}{supplementary material}).
Then, we advocate comparing the sampling results to geodesics, a frequentist method, to assess the effect of the Bayesian prior on parameter uncertainty.
Additionally, researchers can perform other frequentist analysis, e.g., profile likelihood.
Finally, for an extended UQ analysis, researchers can apply MBAM to perform model reduction and/or collect additional training data, in which case we recommend iterating the steps in the previous paragraph.
We are working on implementing these analysis tools within the KIM-based Learning-Integrated Fitting Framework (KLIFF) Python package.\cite{Wen_Afshar_Elliott_Tadmor_2022}


\section*{SUPPLEMENTARY MATERIAL}
In the \href{https://aip.scitation.org/doi/suppl/10.1063/5.0084988}{supplementary materials}, we first show plots of the profile likelihood and the MCMC samples at several sampling temperatures for all parameters of the SW potential for MoS$_2$ system.
The sampling temperatures are given with respect to the natural temperature $T_0 \approx 1.85 \times 10^5 $.
We then present the MCMC samples for SW potential, where the sampling is done using linear parameterization with a corresponding uniform prior in parameter space.

\begin{acknowledgments}
    This work was supported by the National Science Foundation under Awards Nos. DMR-1834251 and DMR-1834332.
    Some of the calculations were done on computational facilities provided by the Brigham Young University Office of Research Computing.
\end{acknowledgments}

\section*{AUTHOR DECLARATION}
\subsection*{Conflict of Interest}
The authors have no conflicts to disclose.

\section*{DATA AVAILABILITY}
The data that support the findings of this study are available from the corresponding author upon reasonable request.


\bibliography{refs.bib}

\end{document}


\title{Supplementary Material: Bayesian, frequentist, and information geometric approaches to parametric uncertainty quantification of classical empirical interatomic potentials}
\author{Yonatan Kurniawan}
\author{Cody L. Petrie}
\author{Kinamo J. Williams Jr.}
\author{Mark K. Transtrum}
\email{mktranstrum@byu.edu}
\affiliation{Department of Physics and Astronomy, Brigham Young University, Provo, UT 84604, United States}

\author{Ellad B. Tadmor}
\author{Ryan S. Elliott}
\author{Daniel S. Karls}
\affiliation{Department of Aerospace Engineering and Mechanics, University of Minnesota, Minneapolis, MN 55455, United States}

\author{Mingjian Wen}
\affiliation{Energy Technologies Area, Lawrence Berkeley National Laboratory, Berkeley, CA 94720, United States}

\date{\today}

\maketitle

\tableofcontents

\section{UQ results for SW MoS$_2$ potential}
\label{sec:sup_uq_sw_log}

In this section, we present the profile likelihood and the MCMC samples at several sampling temperatures for all parameters of the SW potential.
The parameters are calibrated to DFT data of the atomic forces in an MoS$_2$ monolayer at 750 K.
We set the Bayesian prior to be a uniform distribution on a rectangular region, bounded by $\left|\log(\theta_i)\right| < 24$, where $\theta_i$ are the parameters in this potential.
The sampling temperatures are given with respect to the natural temperature $T_0 \approx 1.85 \times 10^5 $.

Sec.~\ref{subsec:Mo-Mo}--\ref{subsec:3-body} show the UQ results for parameters corresponding to each interaction type.
On the lower triangle panes, the samples are plotted as the black points while the red and blue curves show the profile likelihood paths for the parameters on the horizontal and vertical axes, respectively.
On the diagonal, we superimpose the cost profiles (red curves) on top of the marginal distribution of the MCMC samples.
The results for parameters corresponding to different interaction types are given in Sec.~\ref{subsec:Mo-Mo_Mo-S}--\ref{subsec:S-S_3-body}.
On these latter sections, we do not plot the probability density of the MCMC samples nor the cost profile, for the purpose of visual clarity, but these plots can be found in the earlier sections.
Additionally, these plots show a lack of correlation between parameters corresponding to different interaction types.

\subsection{Mo--Mo parameters}
\label{subsec:Mo-Mo}
Profile likelihood and MCMC samples for parameters corresponding to 2-body Mo--Mo interaction.

\begin{figure*}[!h]
    \centering
    \includegraphics[width=0.6\textwidth]{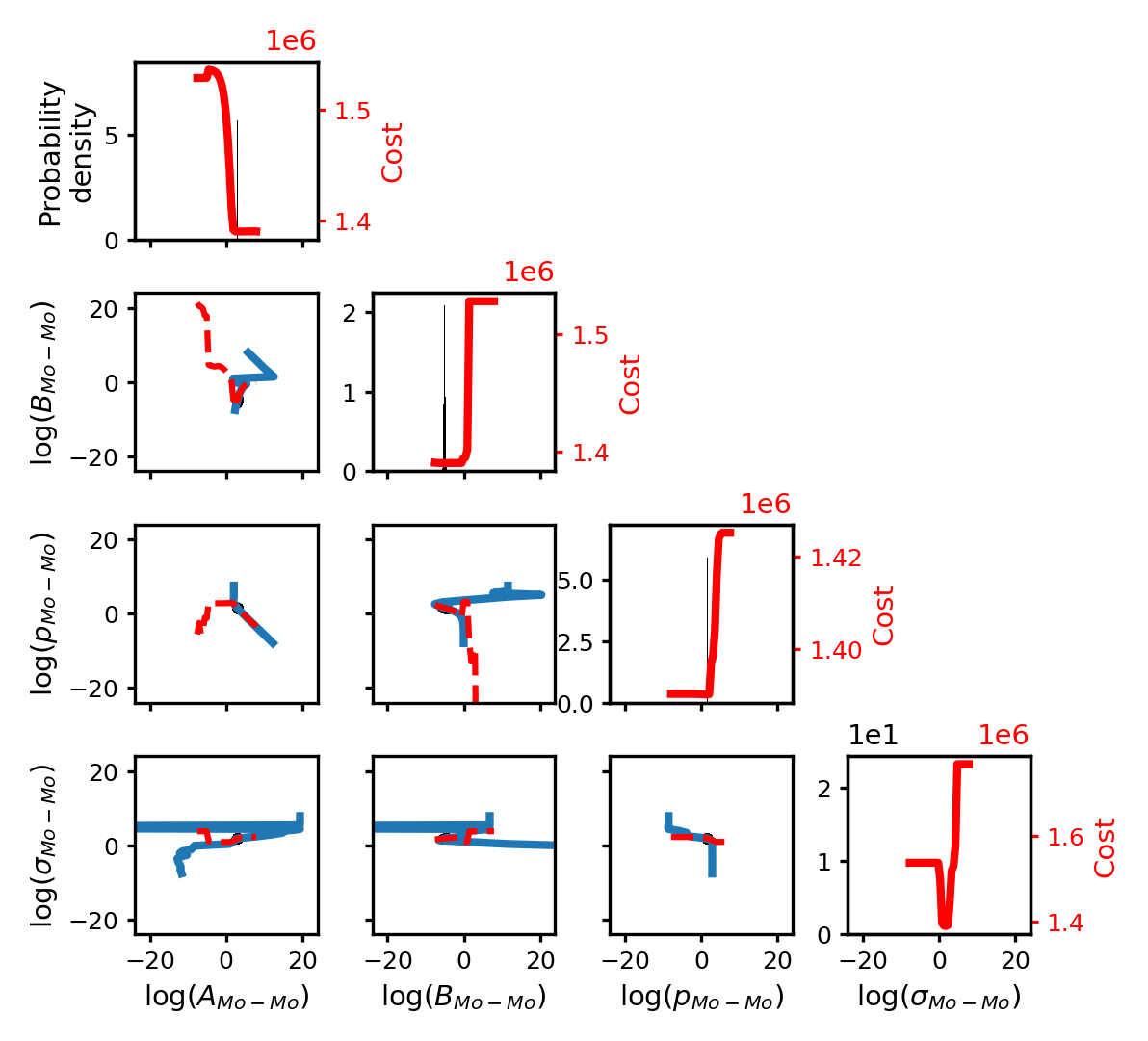}
    \caption[UQ results for SW potential Mo--Mo parameters at $T = 5.40 \times 10^{-6}~T_0$]{
        Profile likelihood and MCMC samples for Mo--Mo parameters at sampling temperature $5.40 \times 10^{-6}~T_0$ for the SW MoS$_2$ potential.
    }
\end{figure*}

\begin{figure*}[!h]
    \centering
    \includegraphics[width=0.6\textwidth]{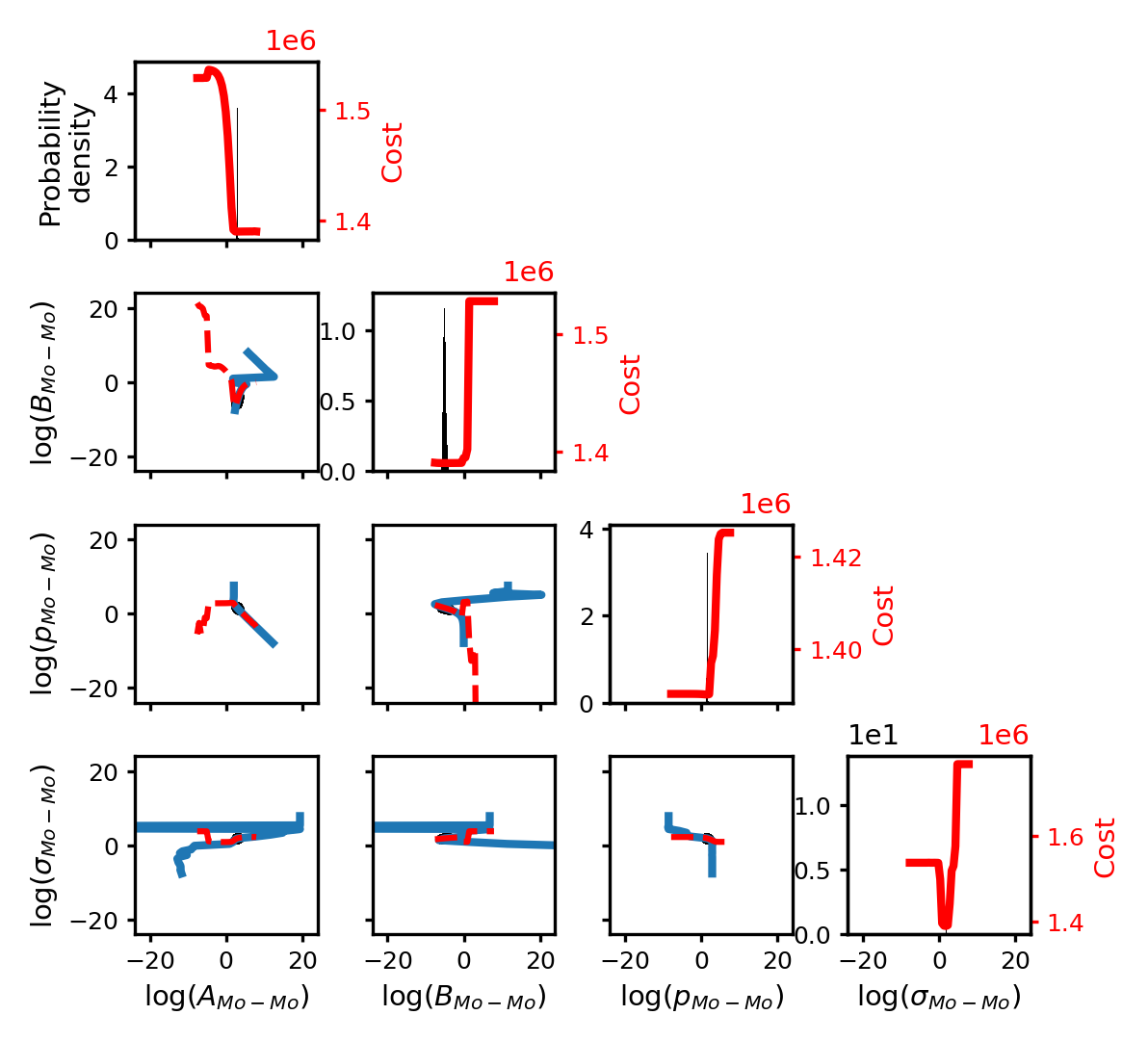}
    \caption[UQ results for SW potential Mo--Mo parameters at $T = 1.71 \times 10^{-5}~T_0$]{
        Profile likelihood and MCMC samples for Mo--Mo parameters at sampling temperature $1.71 \times 10^{-5}~T_0$ for the SW MoS$_2$ potential.
    }
\end{figure*}

\begin{figure*}[!h]
    \centering
    \includegraphics[width=0.6\textwidth]{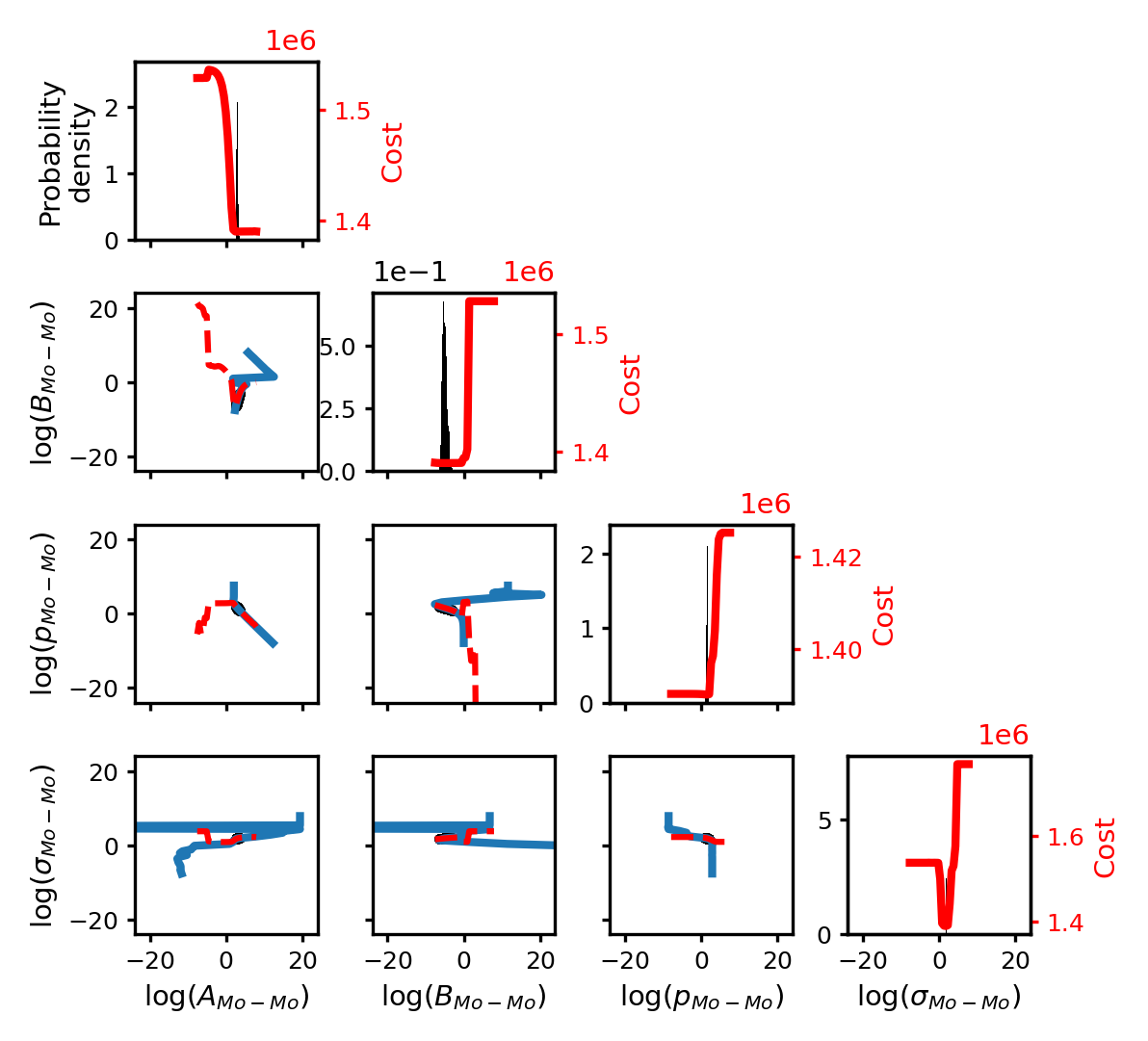}
    \caption[UQ results for SW potential Mo--Mo parameters at $T = 5.40 \times 10^{-5}~T_0$]{
        Profile likelihood and MCMC samples for Mo--Mo parameters at sampling temperature $5.40 \times 10^{-5}~T_0$ for the SW MoS$_2$ potential.
    }
\end{figure*}

\begin{figure*}[!h]
    \centering
    \includegraphics[width=0.6\textwidth]{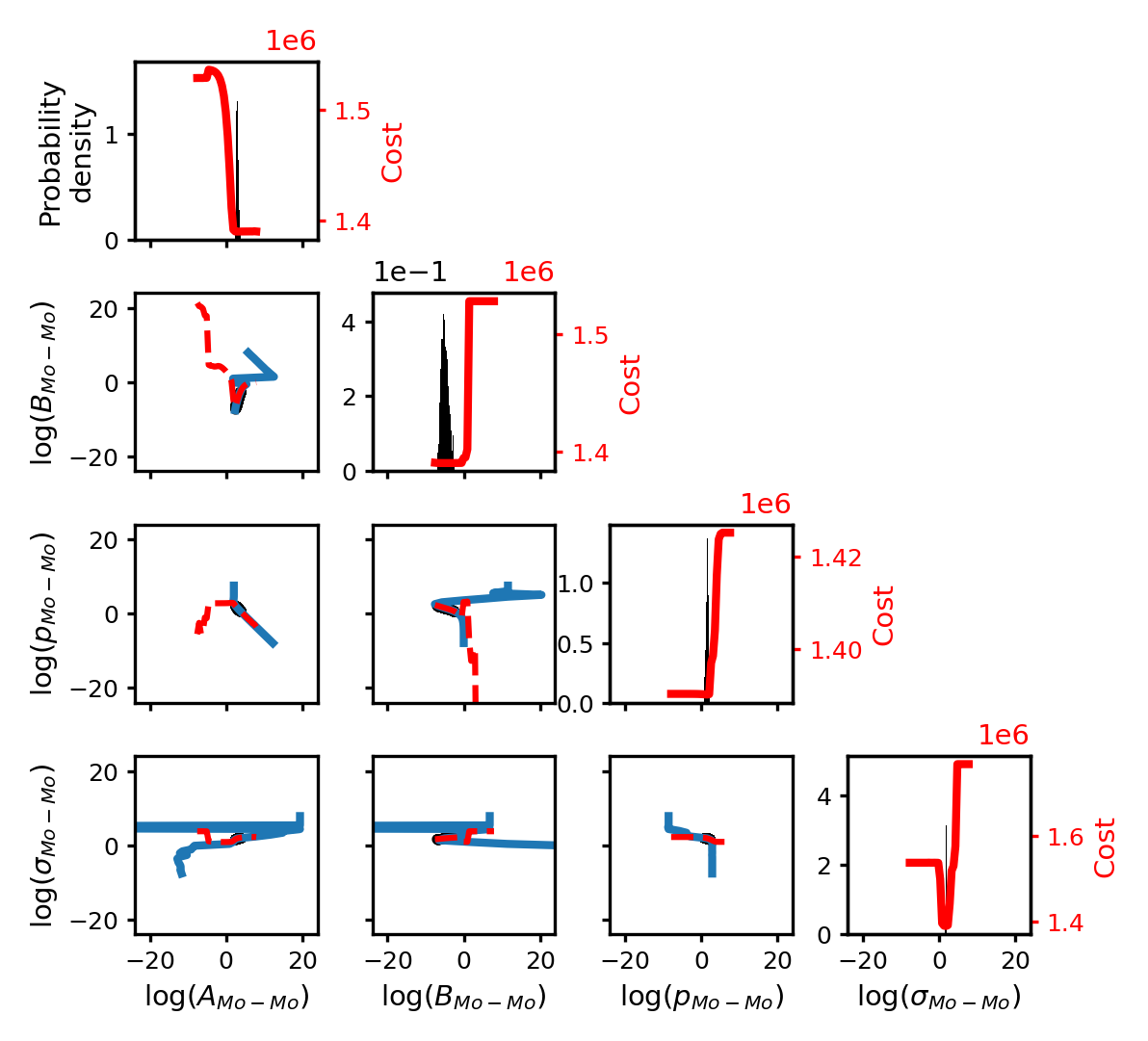}
    \caption[UQ results for SW potential Mo--Mo parameters at $T = 1.71 \times 10^{-4}~T_0$]{
        Profile likelihood and MCMC samples for Mo--Mo parameters at sampling temperature $1.71 \times 10^{-4}~T_0$ for the SW MoS$_2$ potential.
    }
\end{figure*}

\begin{figure*}[!h]
    \centering
    \includegraphics[width=0.6\textwidth]{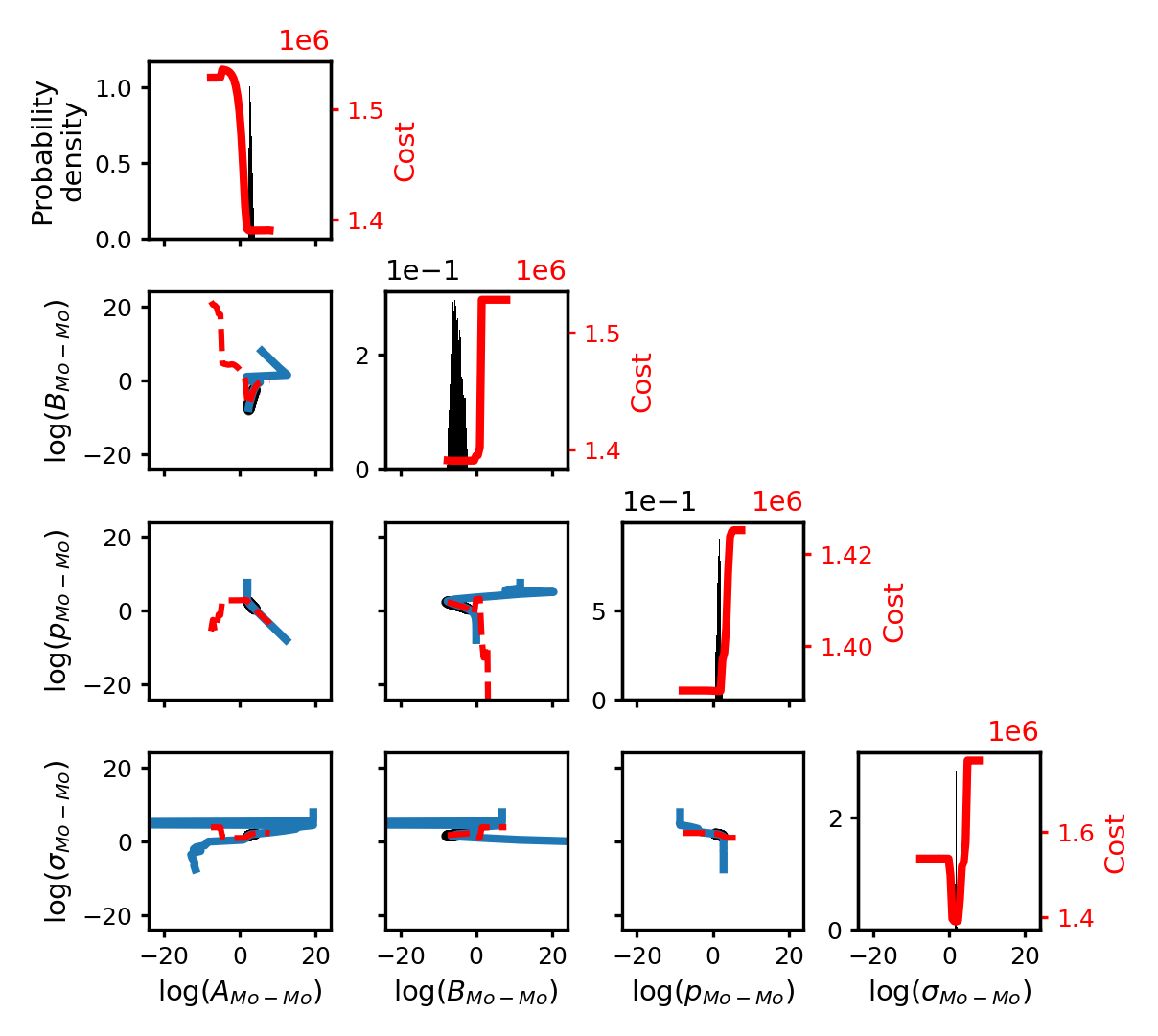}
    \caption[UQ results for SW potential Mo--Mo parameters at $T = 5.40 \times 10^{-4}~T_0$]{
        Profile likelihood and MCMC samples for Mo--Mo parameters at sampling temperature $5.40 \times 10^{-4}~T_0$ for the SW MoS$_2$ potential.
    }
\end{figure*}

\begin{figure*}[!h]
    \centering
    \includegraphics[width=0.6\textwidth]{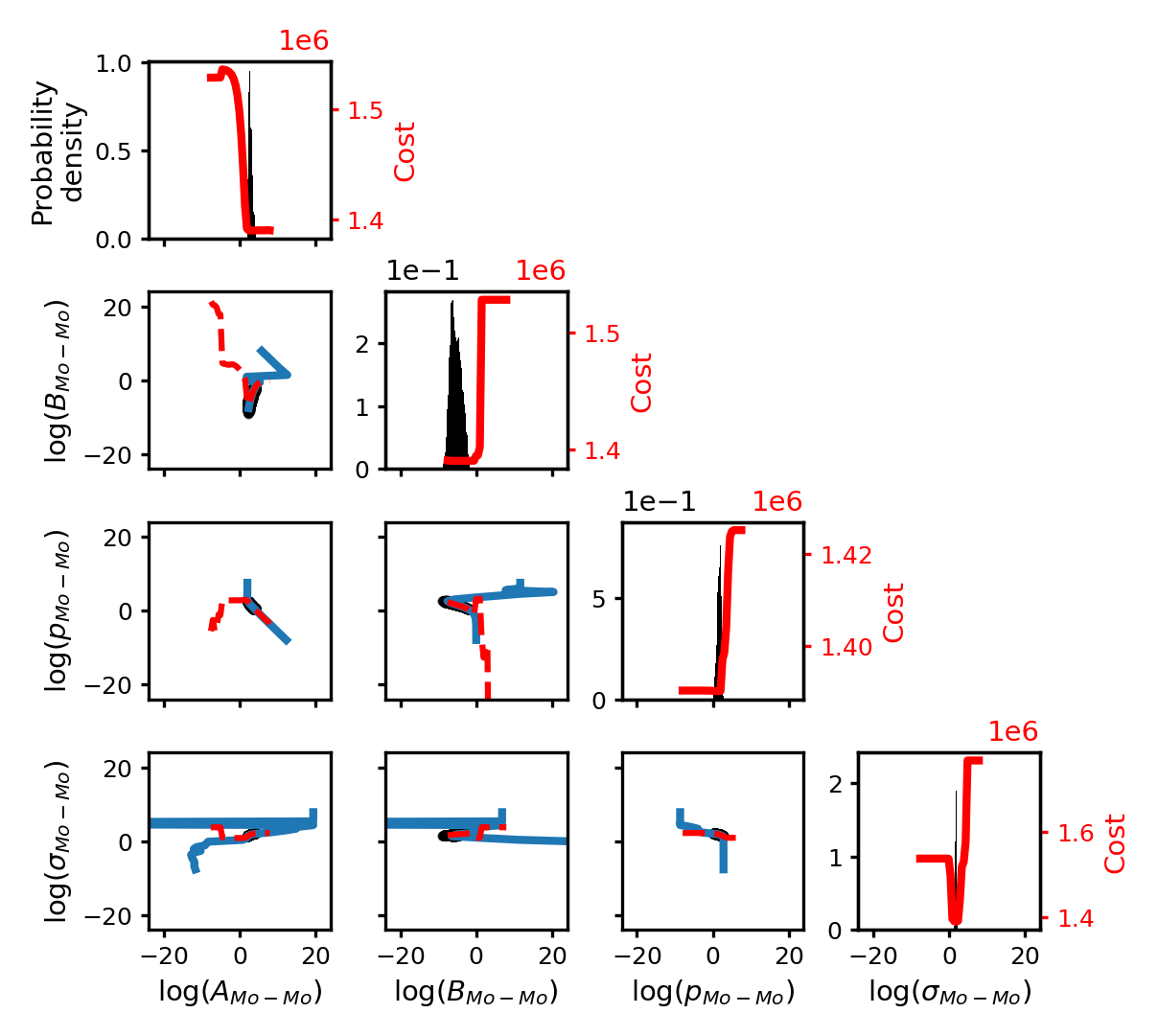}
    \caption[UQ results for SW potential Mo--Mo parameters at $T = 1.71 \times 10^{-3}~T_0$]{
        Profile likelihood and MCMC samples for Mo--Mo parameters at sampling temperature $1.71 \times 10^{-3}~T_0$ for the SW MoS$_2$ potential.
    }
\end{figure*}

\begin{figure*}[!h]
    \centering
    \includegraphics[width=0.6\textwidth]{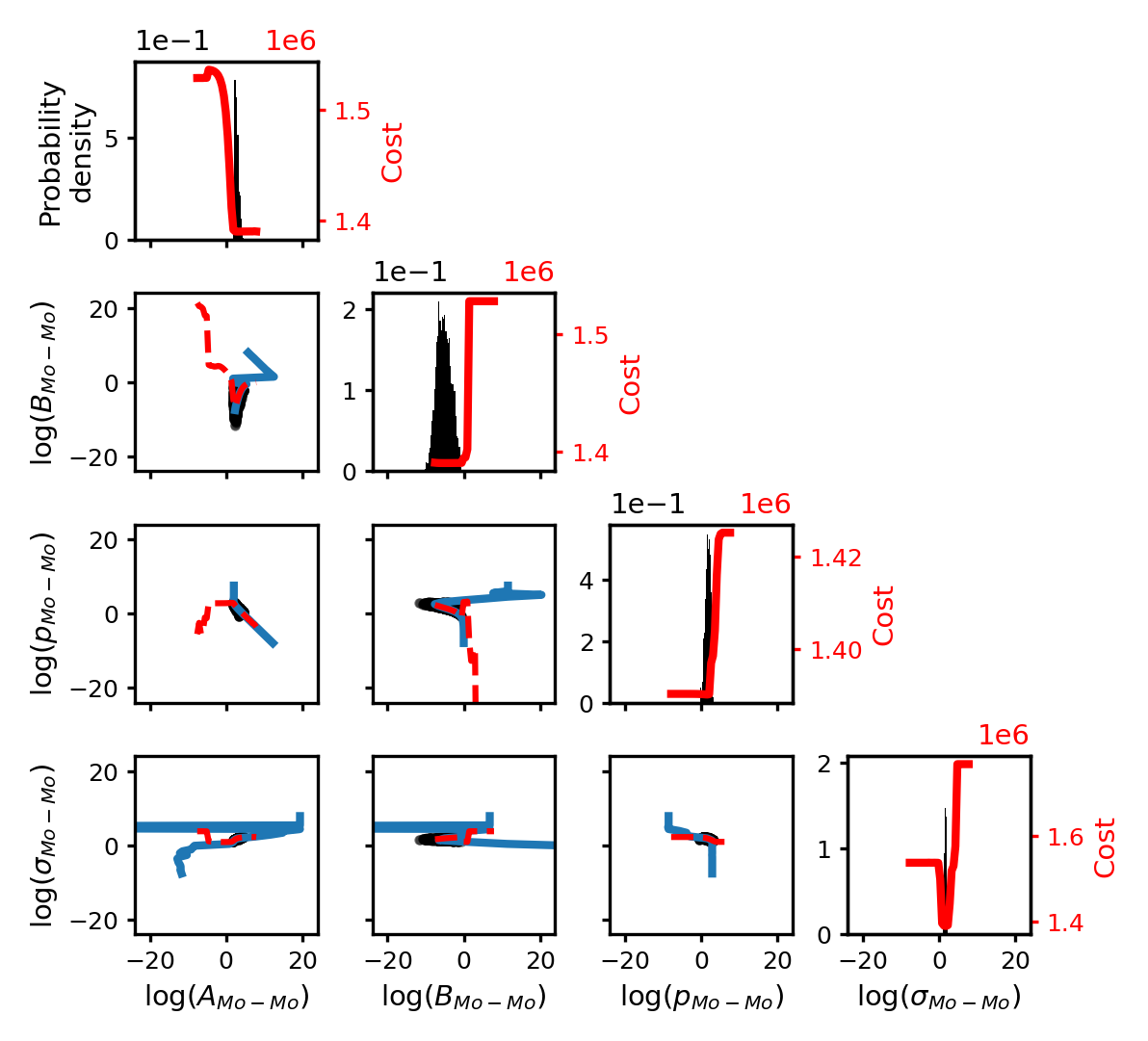}
    \caption[UQ results for SW potential Mo--Mo parameters at $T = 5.40 \times 10^{-3}~T_0$]{
        Profile likelihood and MCMC samples for Mo--Mo parameters at sampling temperature $5.40 \times 10^{-3}~T_0$ for the SW MoS$_2$ potential.
    }
\end{figure*}

\begin{figure*}[!h]
    \centering
    \includegraphics[width=0.6\textwidth]{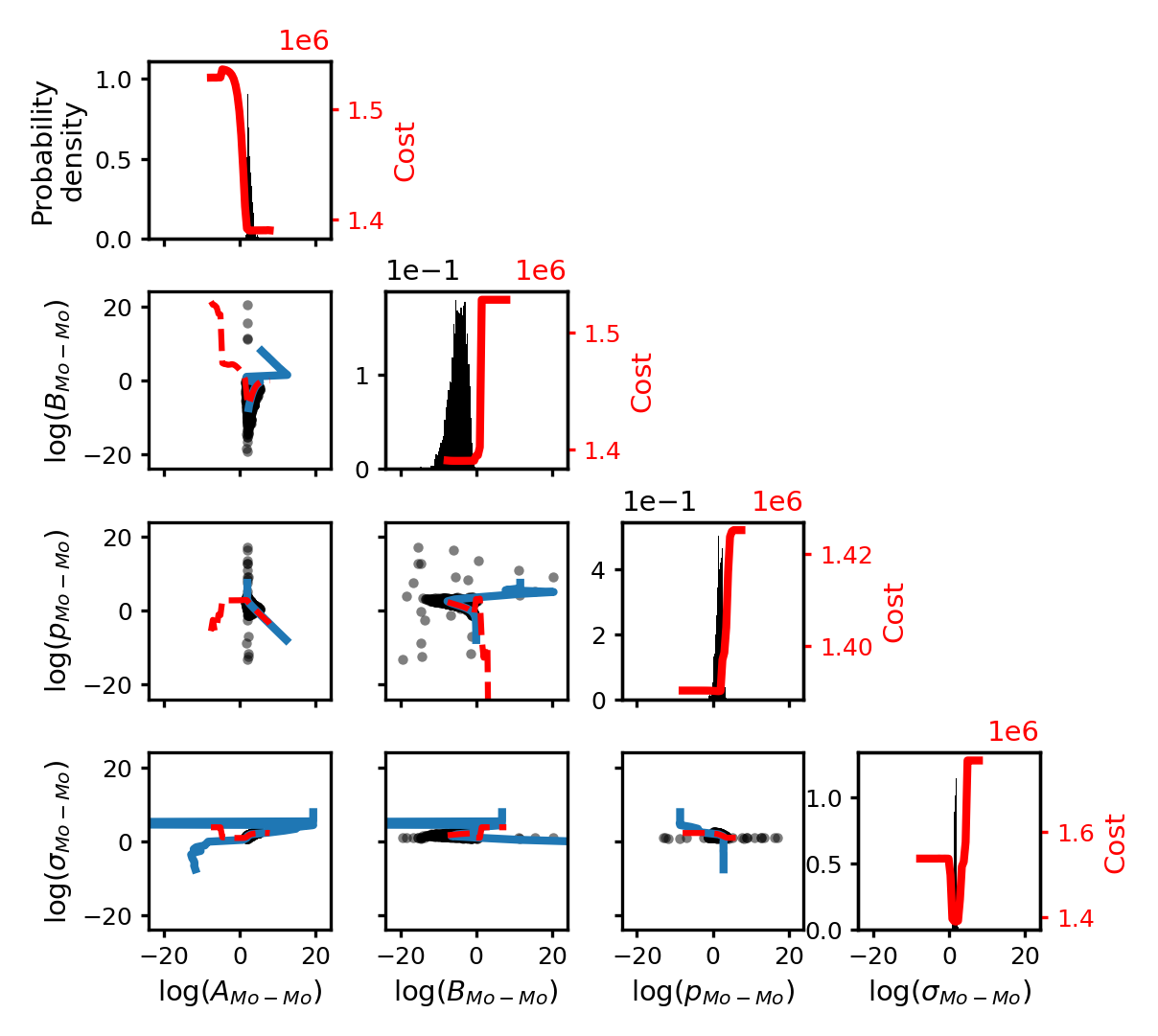}
    \caption[UQ results for SW potential Mo--Mo parameters at $T = 1.71 \times 10^{-2}~T_0$]{
        Profile likelihood and MCMC samples for Mo--Mo parameters at sampling temperature $1.71 \times 10^{-2}~T_0$ for the SW MoS$_2$ potential.
    }
\end{figure*}

\begin{figure*}[!h]
    \centering
    \includegraphics[width=0.6\textwidth]{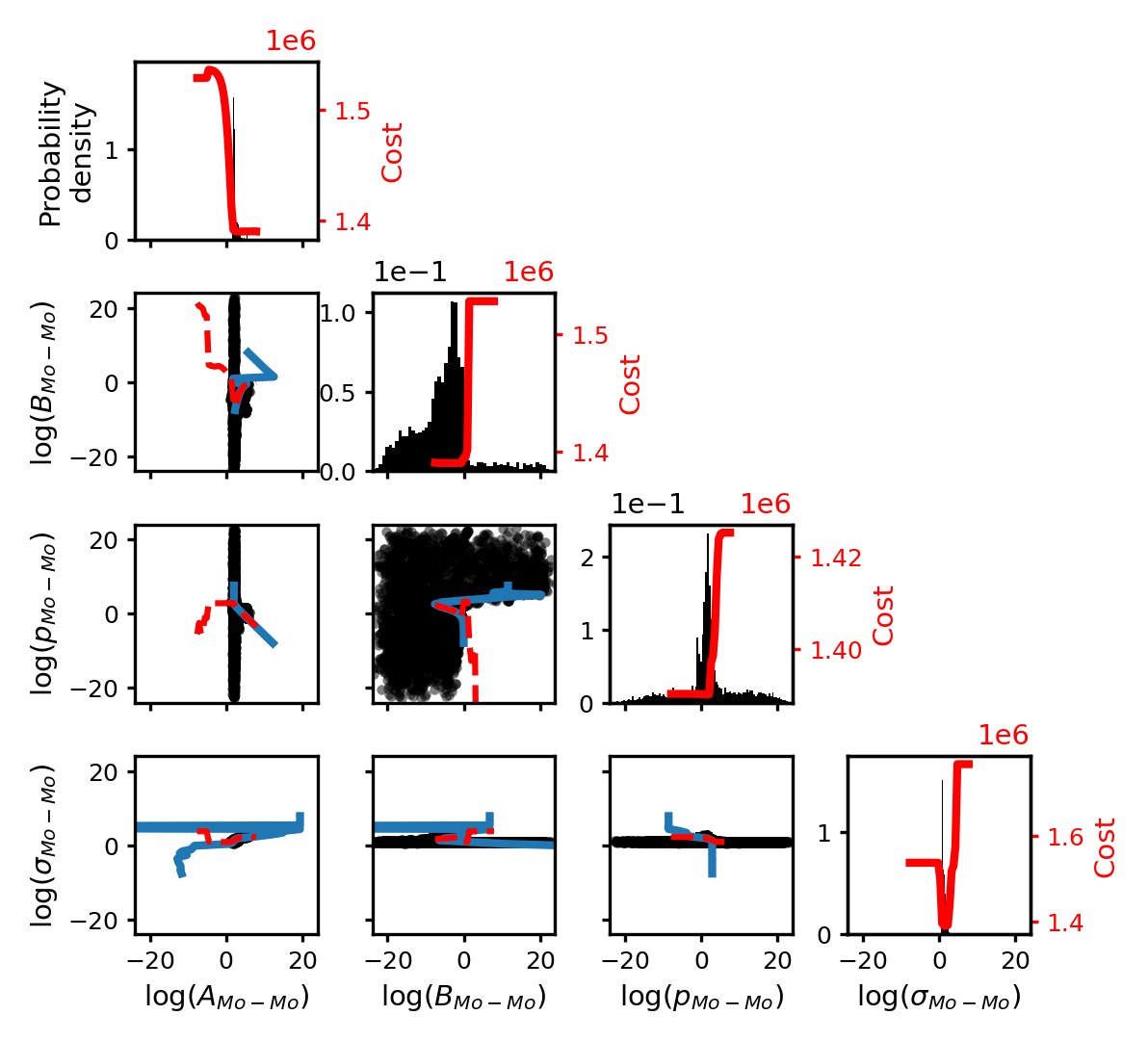}
    \caption[UQ results for SW potential Mo--Mo parameters at $T = 5.40 \times 10^{-2}~T_0$]{
        Profile likelihood and MCMC samples for Mo--Mo parameters at sampling temperature $5.40 \times 10^{-2}~T_0$ for the SW MoS$_2$ potential.
    }
\end{figure*}

\begin{figure*}[!h]
    \centering
    \includegraphics[width=0.6\textwidth]{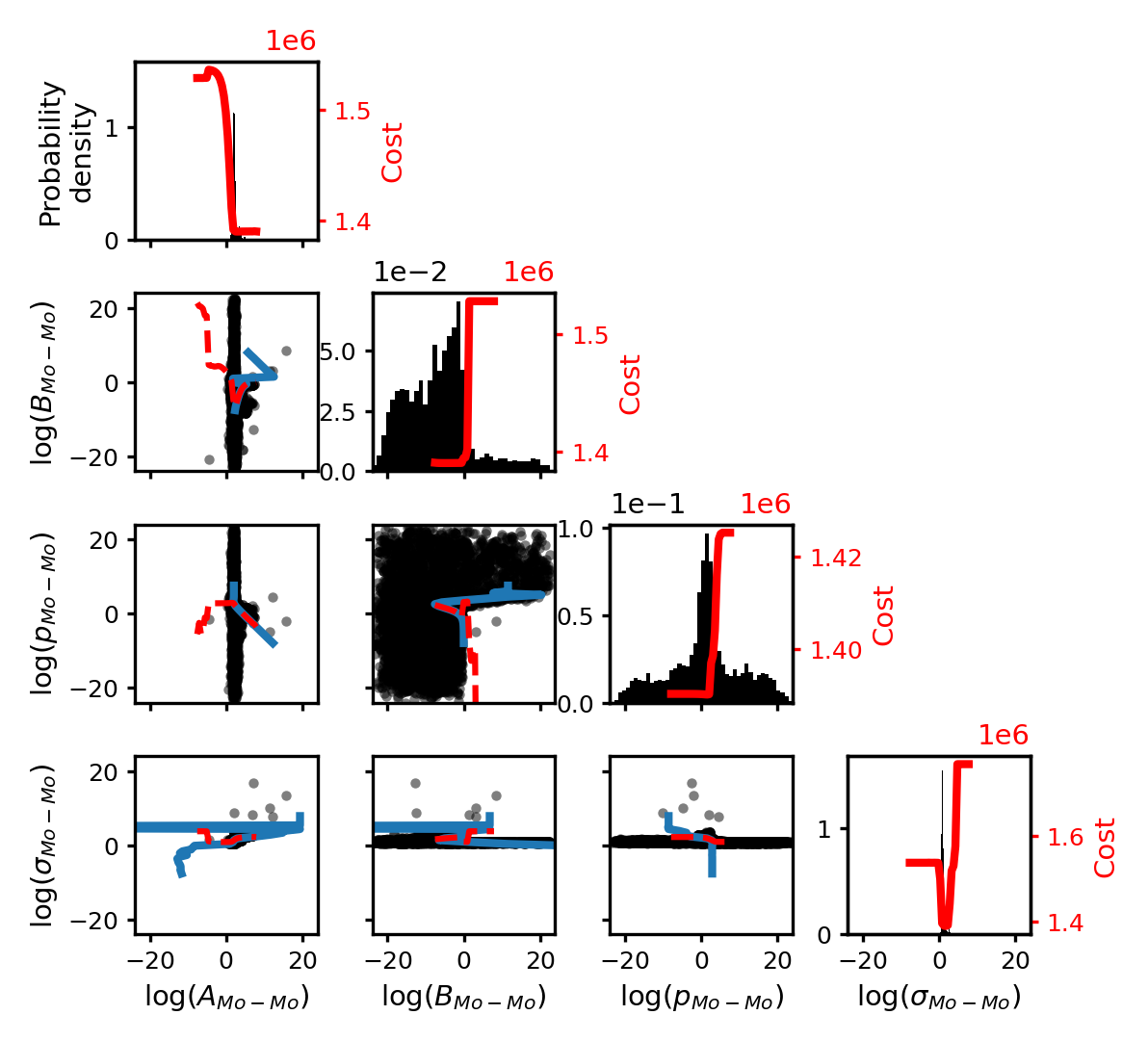}
    \caption[UQ results for SW potential Mo--Mo parameters at $T = 1.71 \times 10^{-1}~T_0$]{
        Profile likelihood and MCMC samples for Mo--Mo parameters at sampling temperature $1.71 \times 10^{-1}~T_0$ for the SW MoS$_2$ potential.
    }
\end{figure*}

\begin{figure*}[!h]
    \centering
    \includegraphics[width=0.6\textwidth]{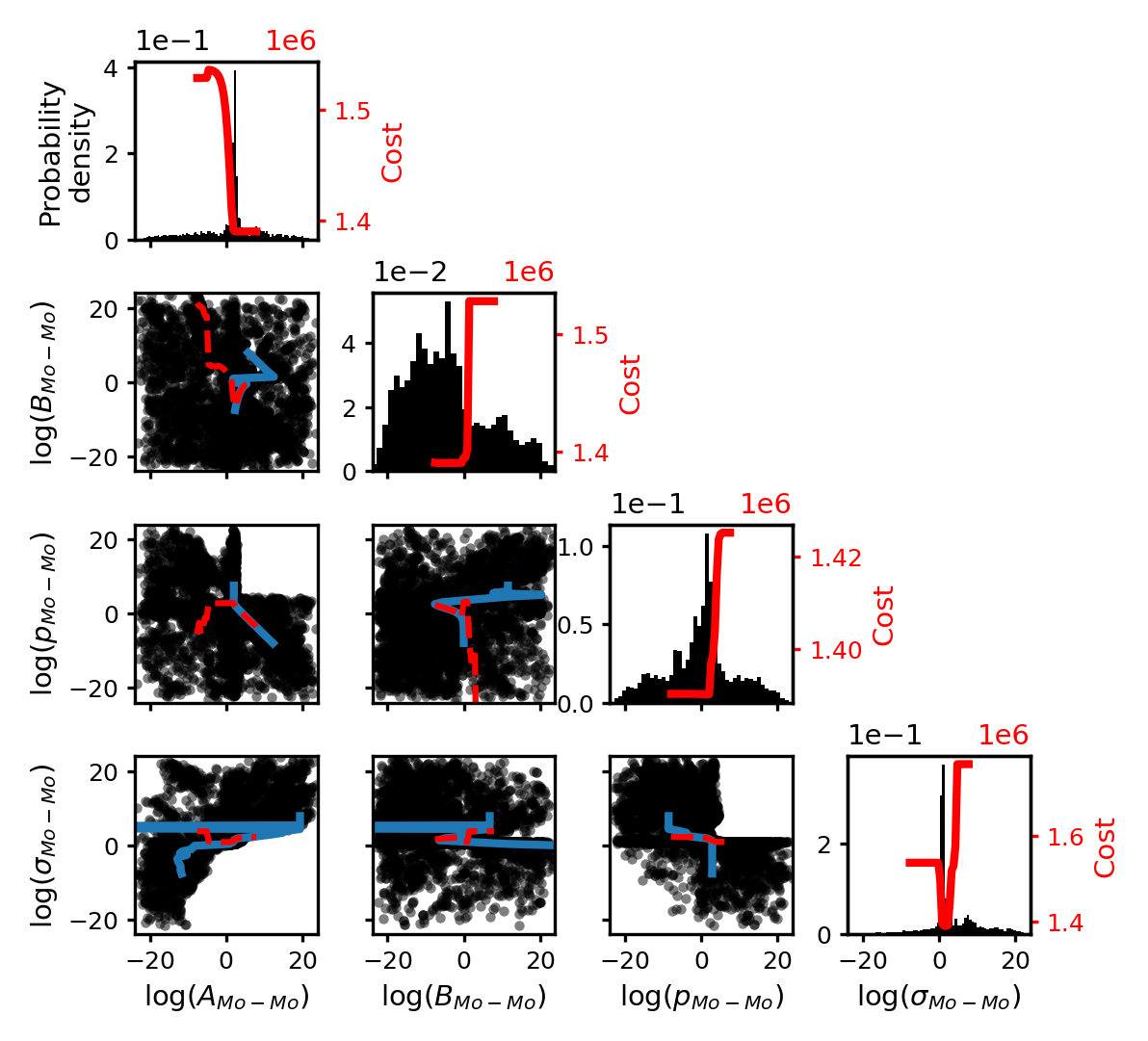}
    \caption[UQ results for SW potential Mo--Mo parameters at $T = 5.40 \times 10^{-1}~T_0$]{
        Profile likelihood and MCMC samples for Mo--Mo parameters at sampling temperature $5.40 \times 10^{-1}~T_0$ for the SW MoS$_2$ potential.
    }
\end{figure*}

\ifincludeTo
    \begin{figure*}[!h]
        \centering
        \includegraphics[width=0.6\textwidth]{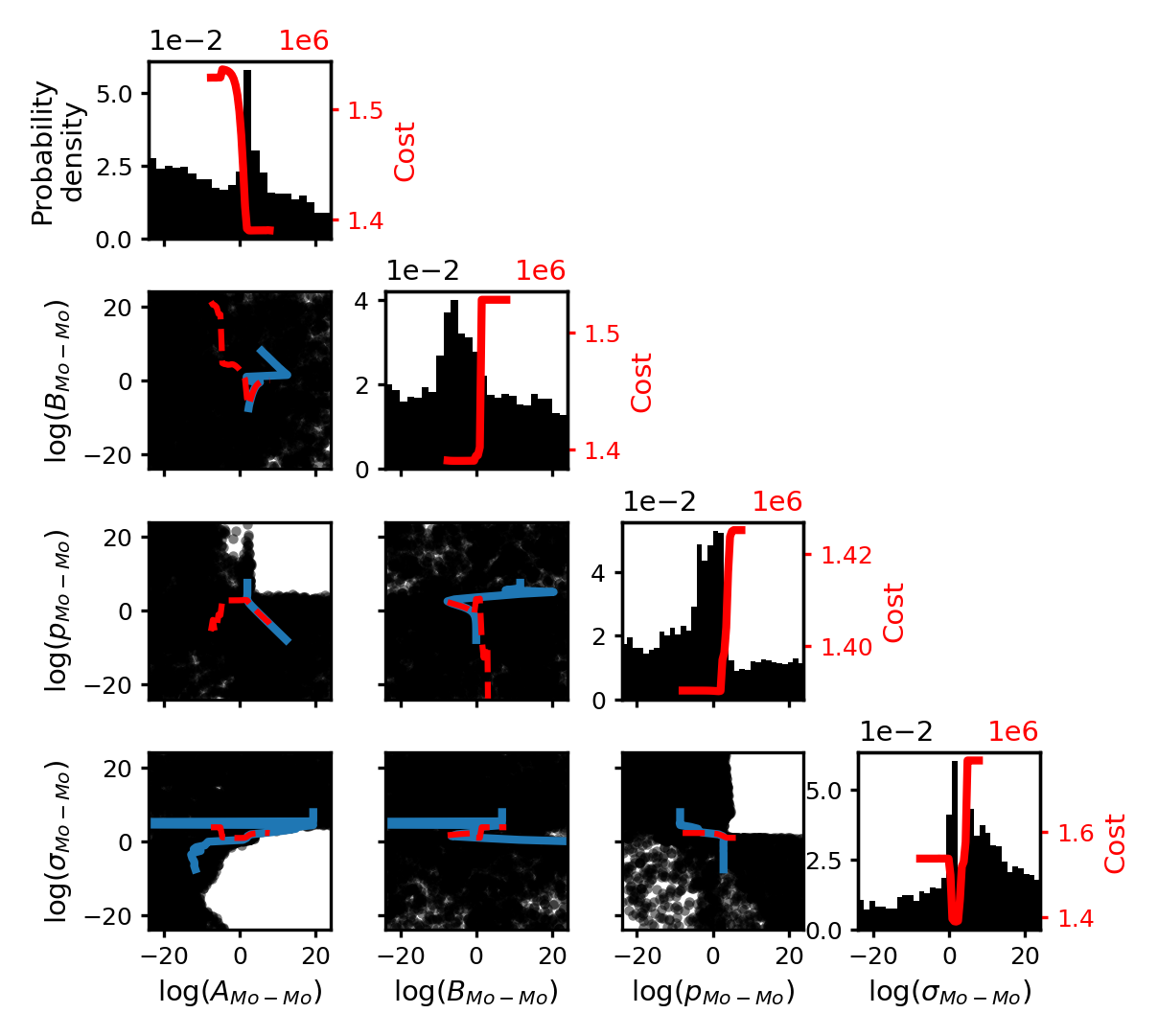}
        \caption[UQ results for SW potential Mo--Mo parameters at $T = T_0$]{
            Profile likelihood and MCMC samples for Mo--Mo parameters at sampling temperature $T_0$ for the SW MoS$_2$ potential.
        }
    \end{figure*}
\fi

\begin{figure*}[!h]
    \centering
    \includegraphics[width=0.6\textwidth]{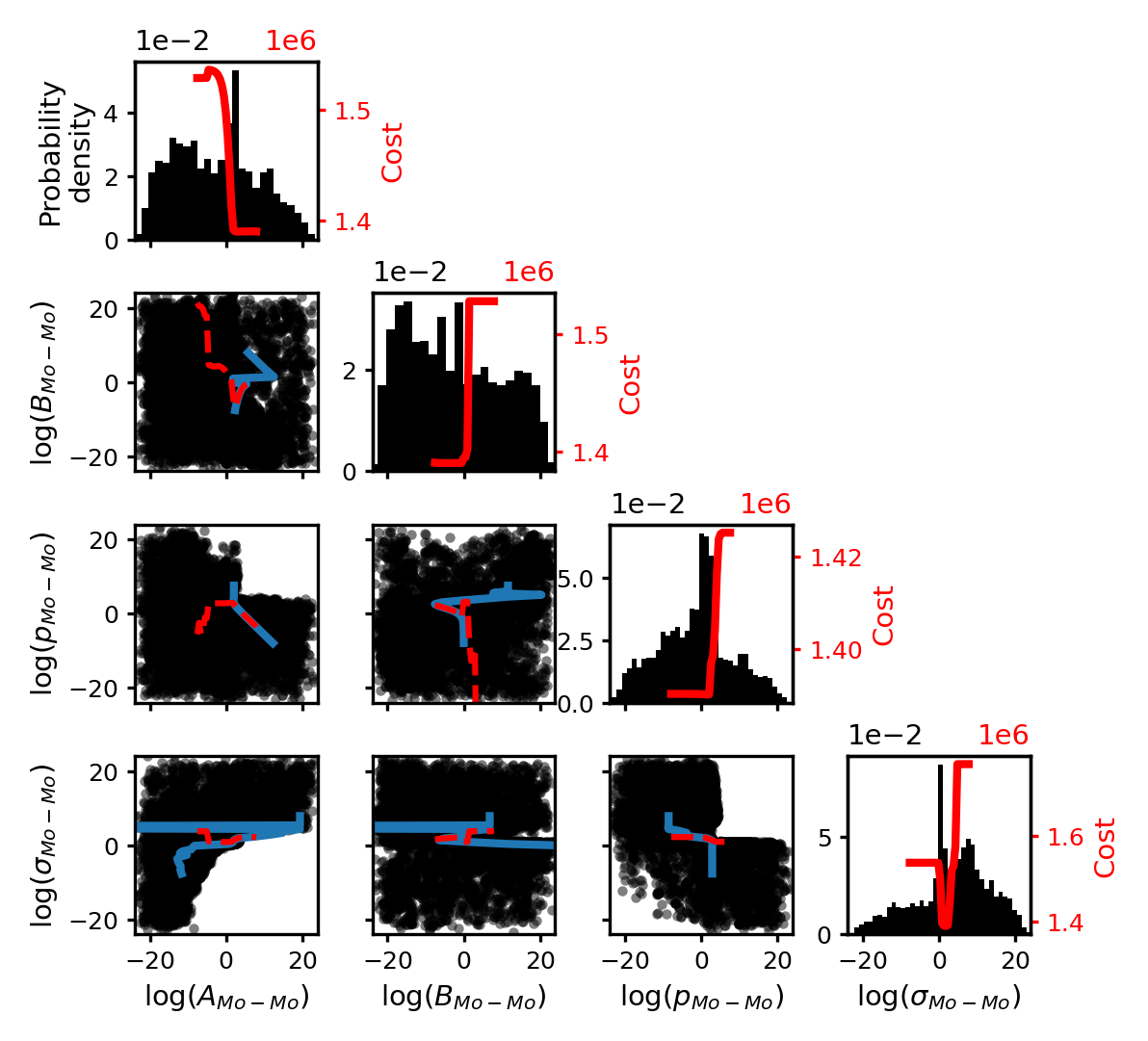}
    \caption[UQ results for SW potential Mo--Mo parameters at $T = 1.71~T_0$]{
        Profile likelihood and MCMC samples for Mo--Mo parameters at sampling temperature $1.71~T_0$ for the SW MoS$_2$ potential.
    }
\end{figure*}

\begin{figure*}[!h]
    \centering
    \includegraphics[width=0.6\textwidth]{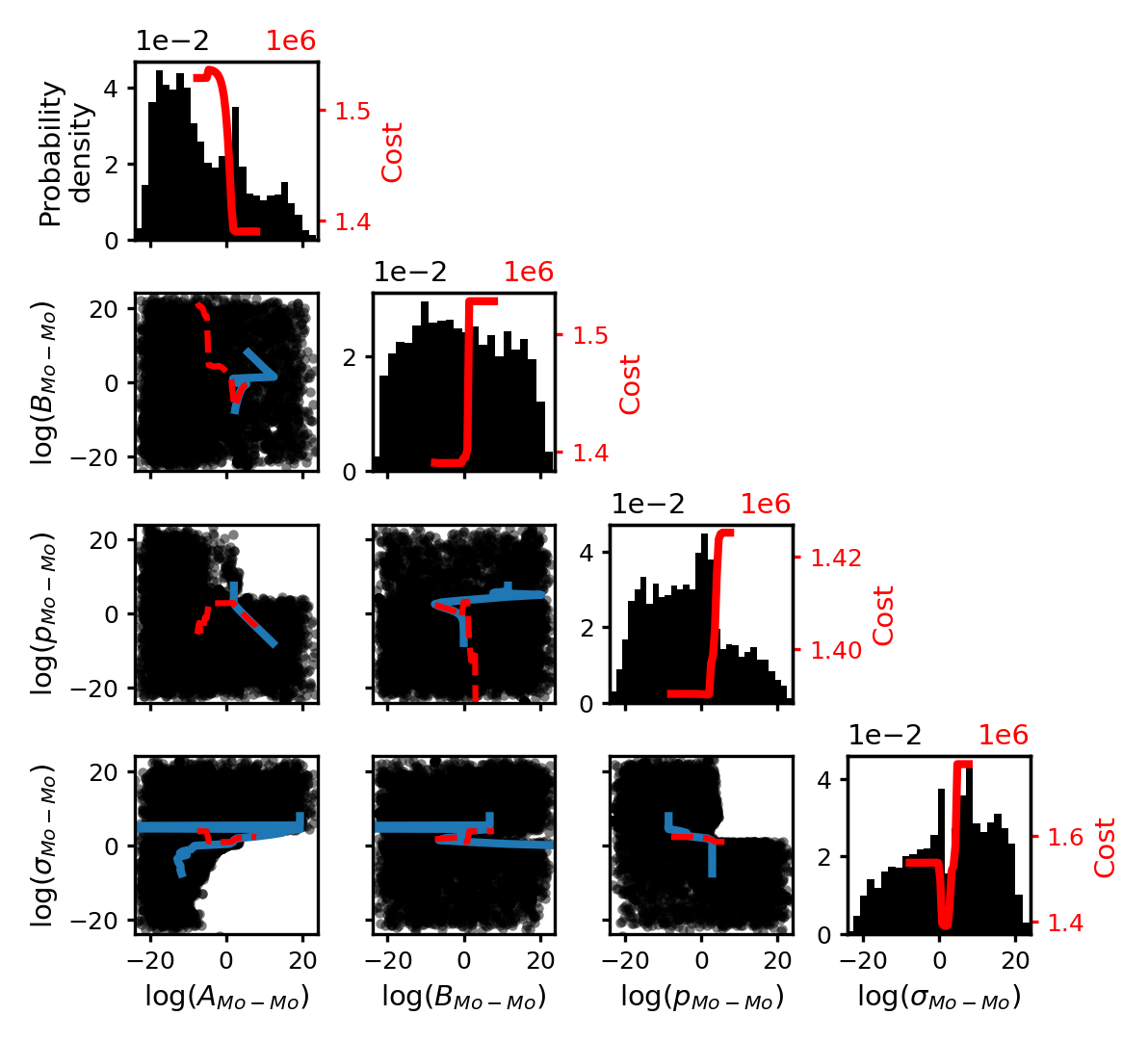}
    \caption[UQ results for SW potential Mo--Mo parameters at $T = 5.40~T_0$]{
        Profile likelihood and MCMC samples for Mo--Mo parameters at sampling temperature $5.40~T_0$ for the SW MoS$_2$ potential.
    }
\end{figure*}

\cleardoublepage

\subsection{Mo--S parameters}
\label{subsec:Mo-S}
Profile likelihood and MCMC samples for parameters corresponding to 2-body Mo--S interaction.

\begin{figure*}[!h]
    \centering
    \includegraphics[width=0.6\textwidth]{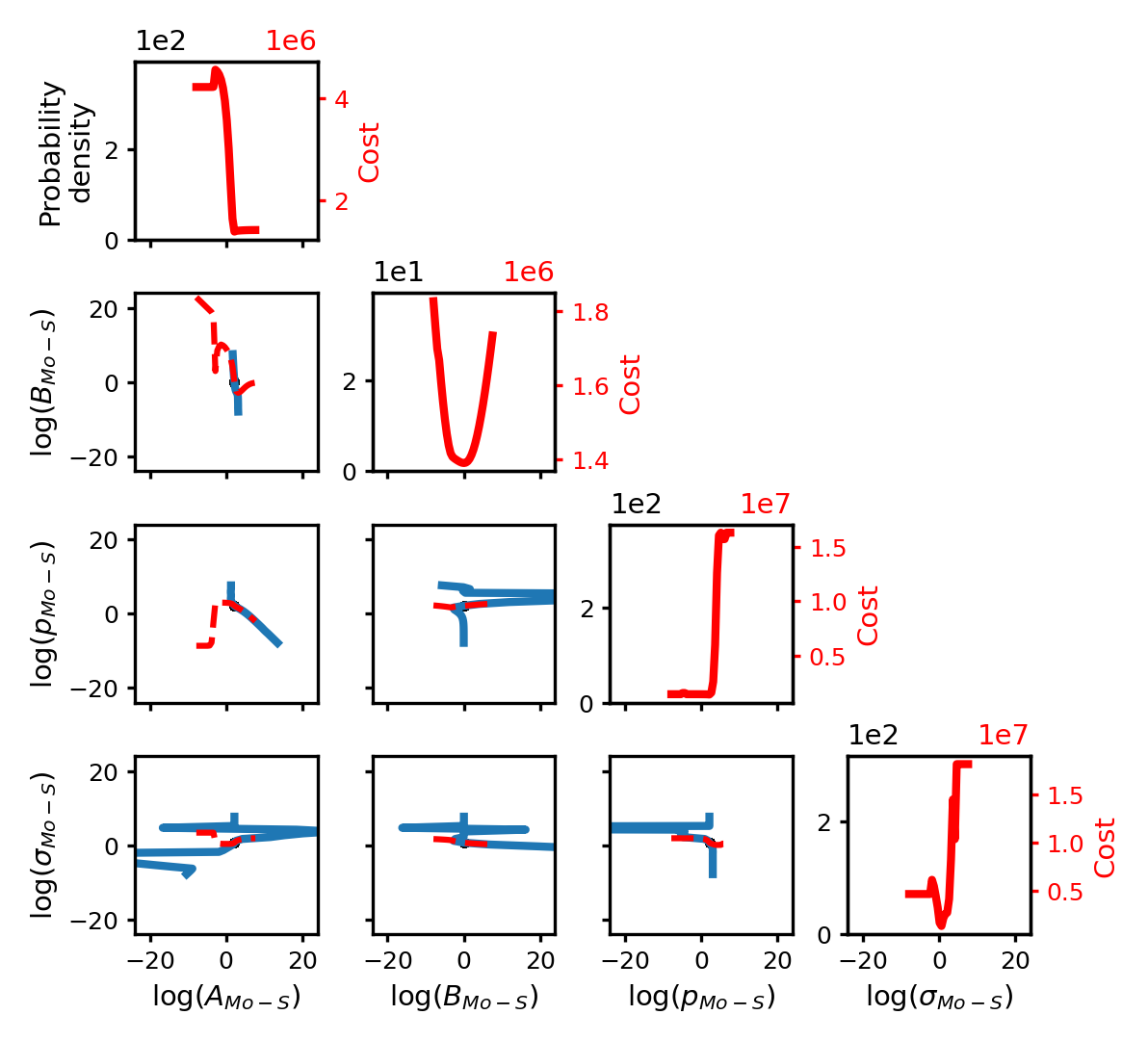}
    \caption[UQ results for SW potential Mo--S parameters at $T = 5.40 \times 10^{-6}~T_0$]{
        Profile likelihood and MCMC samples for Mo--S parameters at sampling temperature $5.40 \times 10^{-6}~T_0$ for the SW MoS$_2$ potential.
    }
\end{figure*}

\begin{figure*}[!h]
    \centering
    \includegraphics[width=0.6\textwidth]{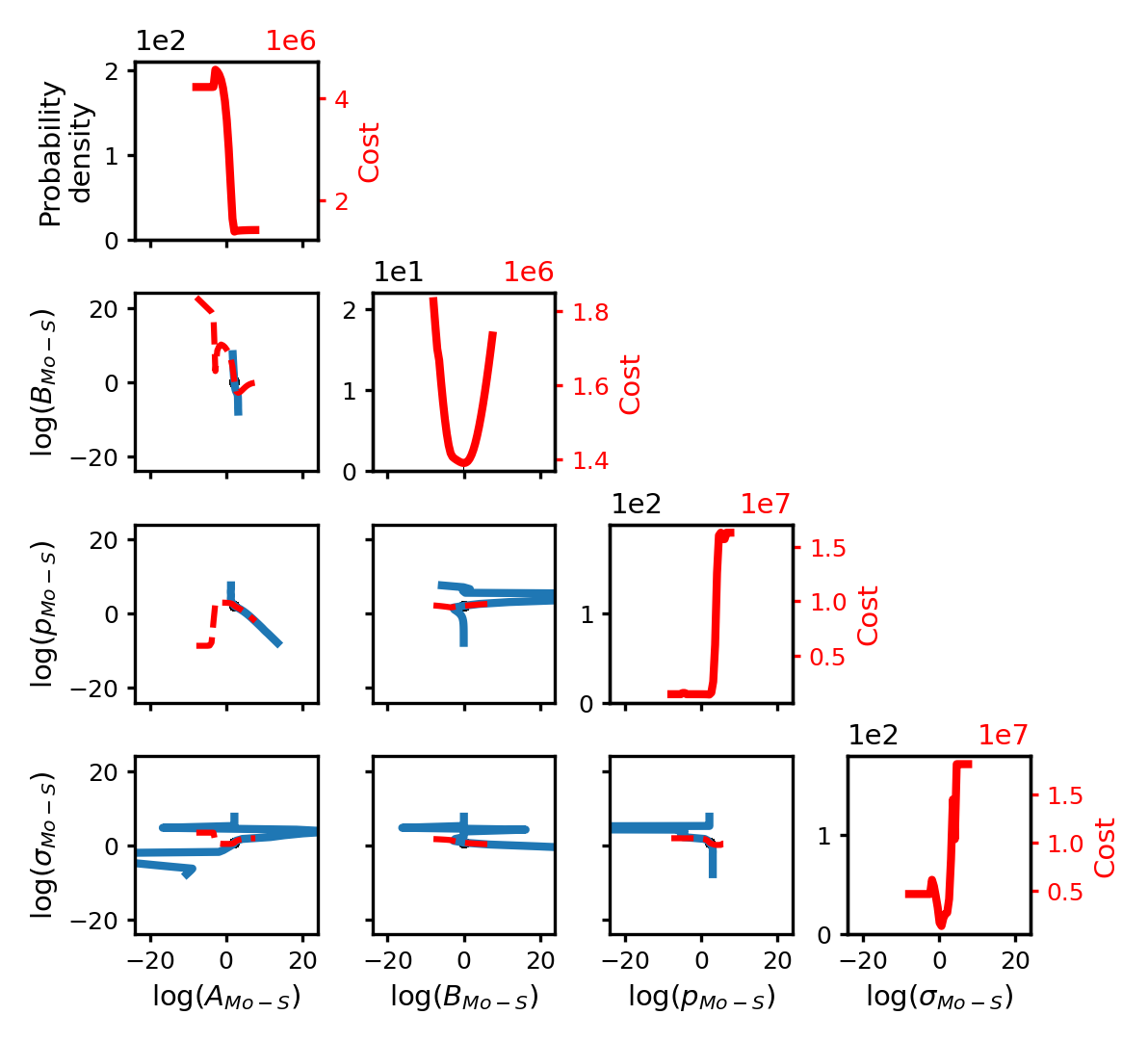}
    \caption[UQ results for SW potential Mo--S parameters at $T = 1.71 \times 10^{-5}~T_0$]{
        Profile likelihood and MCMC samples for Mo--S parameters at sampling temperature $1.71 \times 10^{-5}~T_0$ for the SW MoS$_2$ potential.
    }
\end{figure*}

\begin{figure*}[!h]
    \centering
    \includegraphics[width=0.6\textwidth]{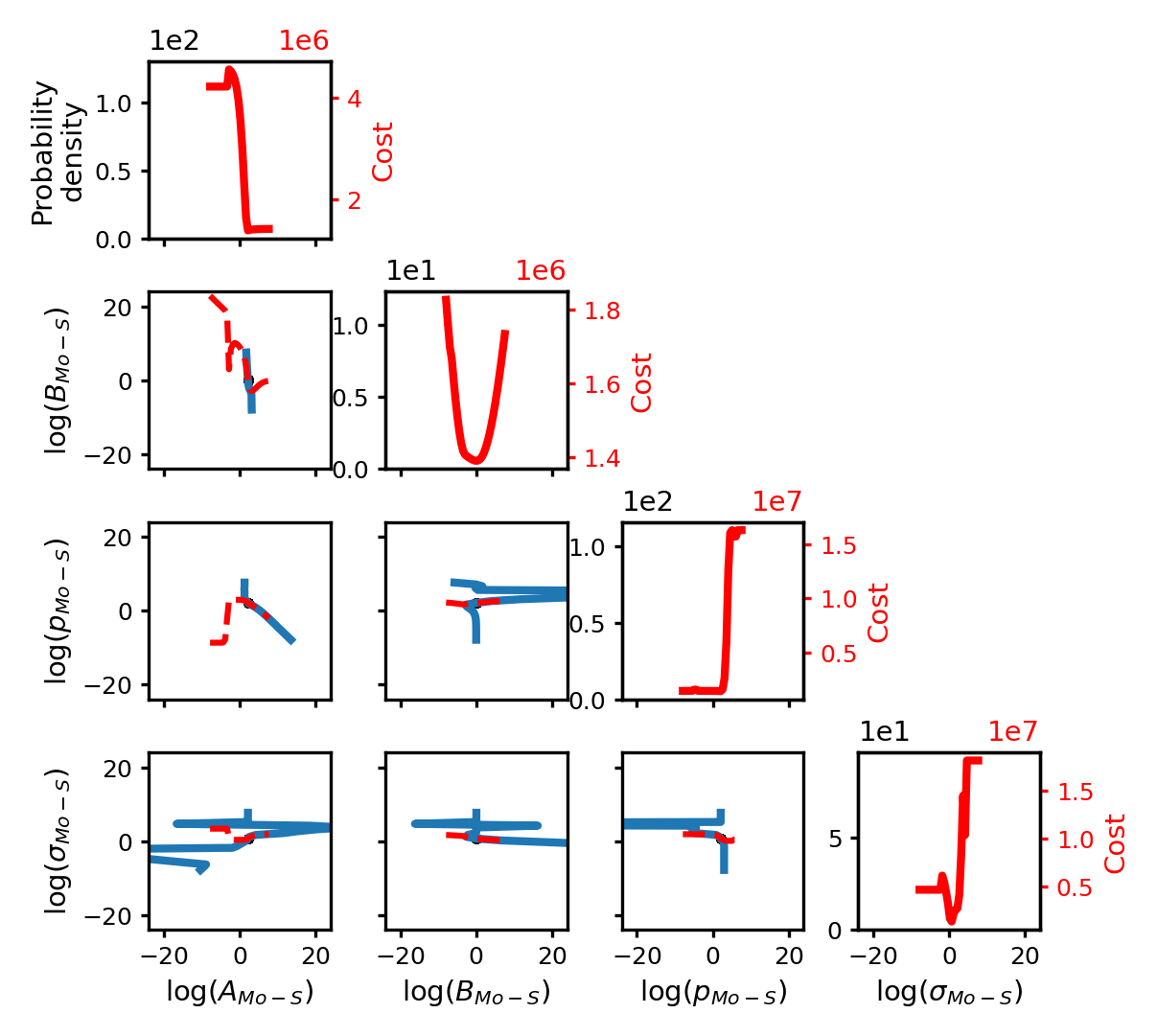}
    \caption[UQ results for SW potential Mo--S parameters at $T = 5.40 \times 10^{-5}~T_0$]{
        Profile likelihood and MCMC samples for Mo--S parameters at sampling temperature $5.40 \times 10^{-5}~T_0$ for the SW MoS$_2$ potential.
    }
\end{figure*}

\begin{figure*}[!h]
    \centering
    \includegraphics[width=0.6\textwidth]{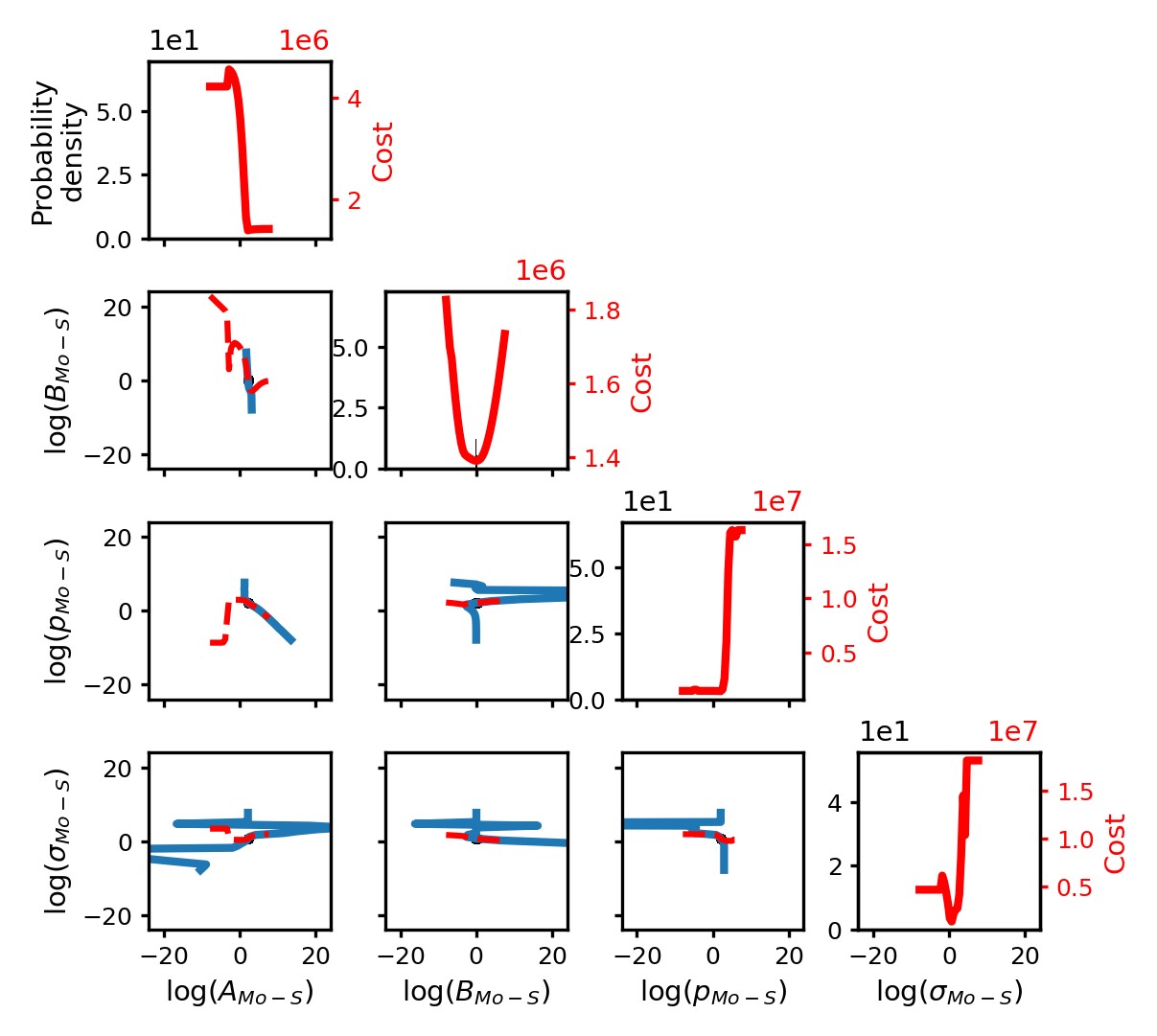}
    \caption[UQ results for SW potential Mo--S parameters at $T = 1.71 \times 10^{-4}~T_0$]{
        Profile likelihood and MCMC samples for Mo--S parameters at sampling temperature $1.71 \times 10^{-4}~T_0$ for the SW MoS$_2$ potential.
    }
\end{figure*}

\begin{figure*}[!h]
    \centering
    \includegraphics[width=0.6\textwidth]{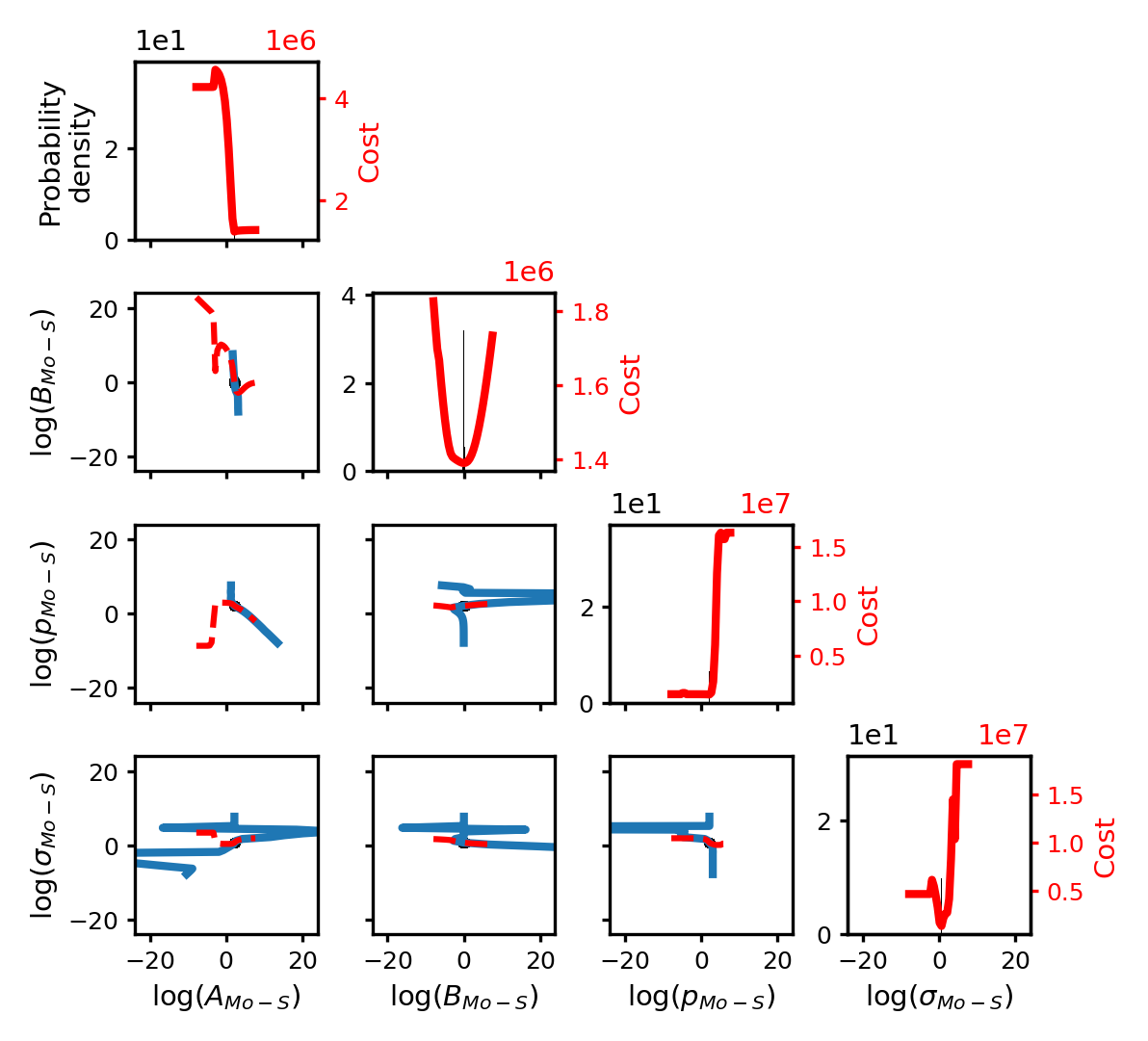}
    \caption[UQ results for SW potential Mo--S parameters at $T = 5.40 \times 10^{-4}~T_0$]{
        Profile likelihood and MCMC samples for Mo--S parameters at sampling temperature $5.40 \times 10^{-4}~T_0$ for the SW MoS$_2$ potential.
    }
\end{figure*}

\begin{figure*}[!h]
    \centering
    \includegraphics[width=0.6\textwidth]{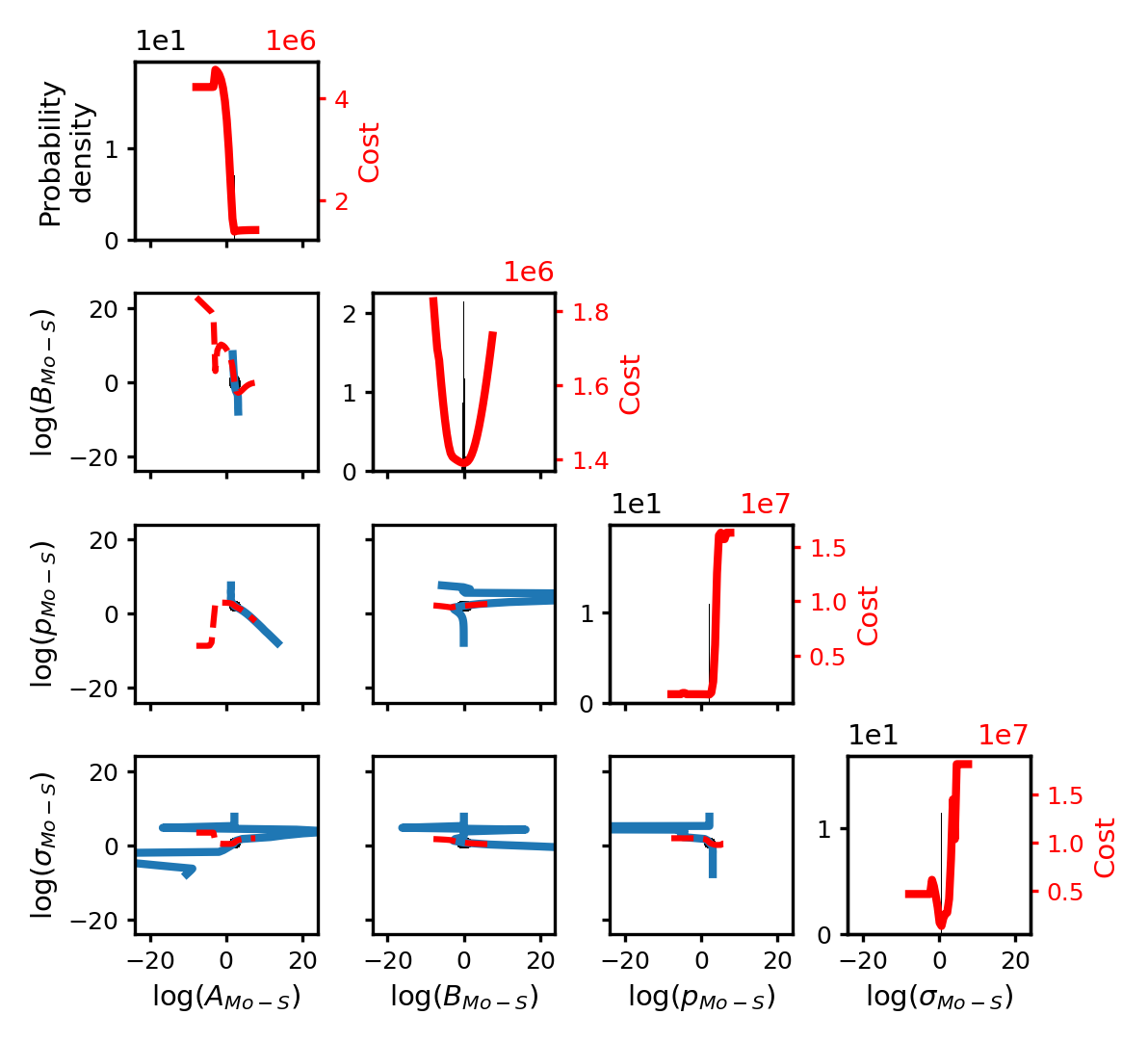}
    \caption[UQ results for SW potential Mo--S parameters at $T = 1.71 \times 10^{-3}~T_0$]{
        Profile likelihood and MCMC samples for Mo--S parameters at sampling temperature $1.71 \times 10^{-3}~T_0$ for the SW MoS$_2$ potential.
    }
\end{figure*}

\begin{figure*}[!h]
    \centering
    \includegraphics[width=0.6\textwidth]{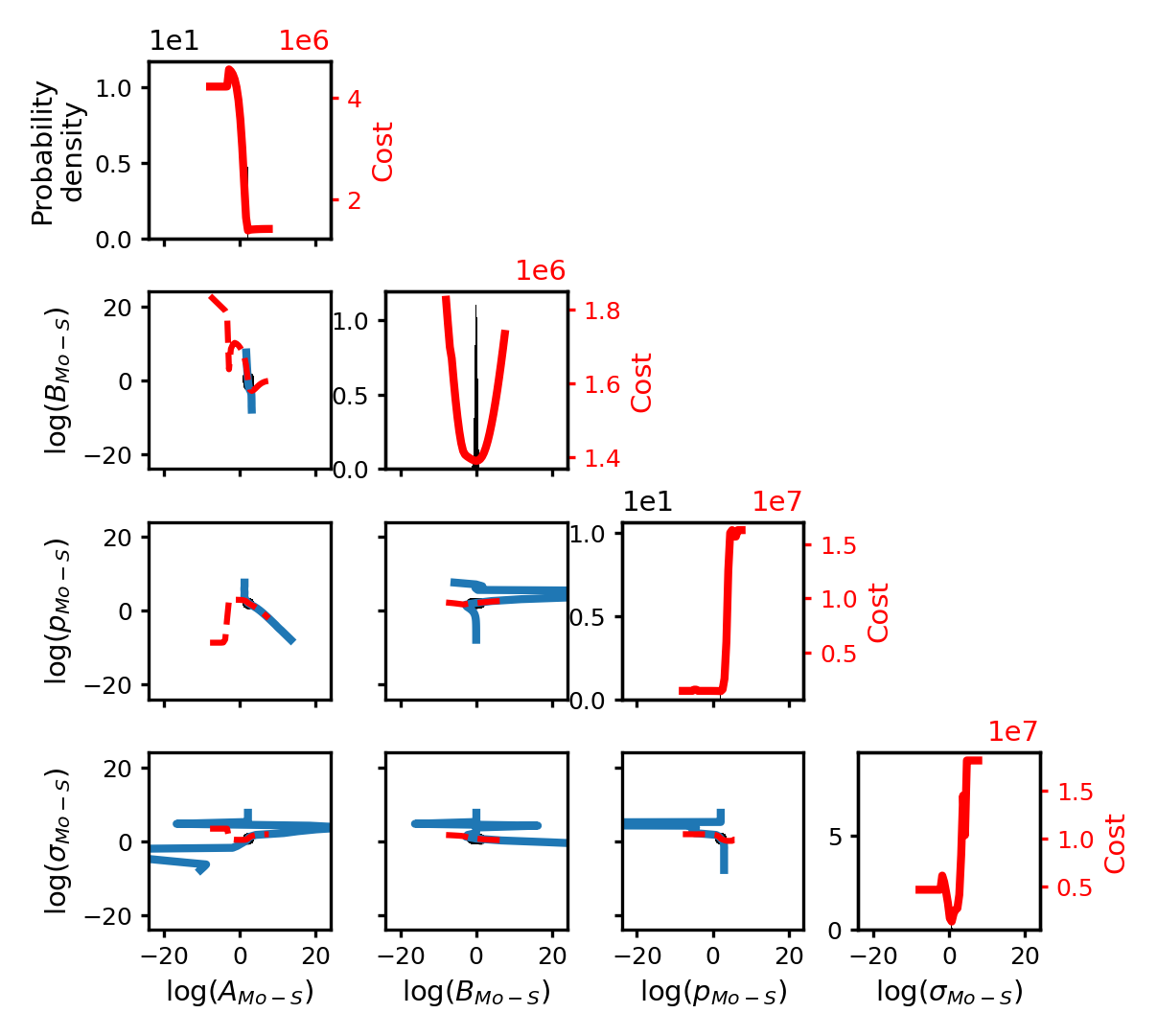}
    \caption[UQ results for SW potential Mo--S parameters at $T = 5.40 \times 10^{-3}~T_0$]{
        Profile likelihood and MCMC samples for Mo--S parameters at sampling temperature $5.40 \times 10^{-3}~T_0$ for the SW MoS$_2$ potential.
    }
\end{figure*}

\begin{figure*}[!h]
    \centering
    \includegraphics[width=0.6\textwidth]{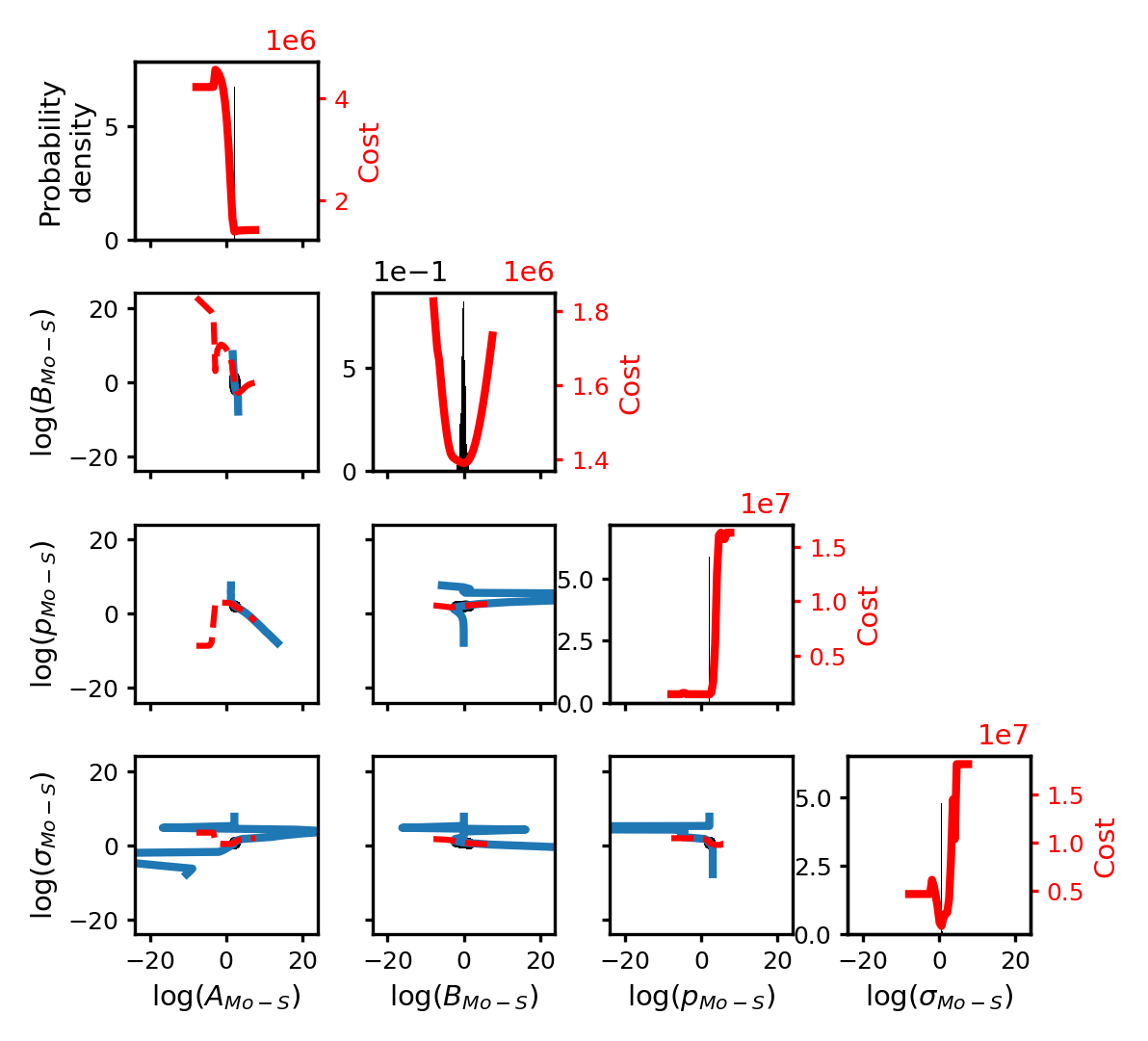}
    \caption[UQ results for SW potential Mo--S parameters at $T = 1.71 \times 10^{-2}~T_0$]{
        Profile likelihood and MCMC samples for Mo--S parameters at sampling temperature $1.71 \times 10^{-2}~T_0$ for the SW MoS$_2$ potential.
    }
\end{figure*}

\begin{figure*}[!h]
    \centering
    \includegraphics[width=0.6\textwidth]{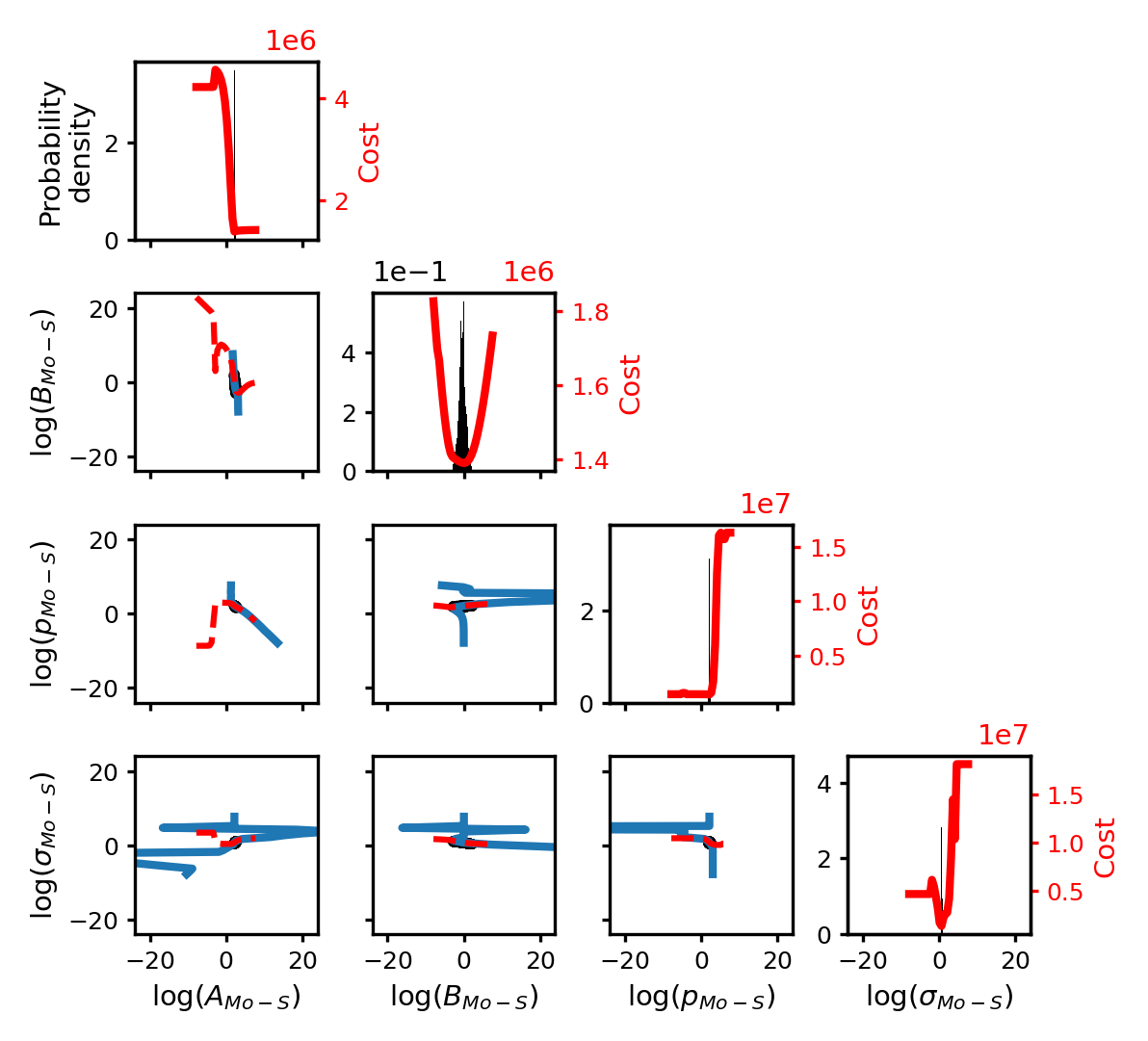}
    \caption[UQ results for SW potential Mo--S parameters at $T = 5.40 \times 10^{-2}~T_0$]{
        Profile likelihood and MCMC samples for Mo--S parameters at sampling temperature $5.40 \times 10^{-2}~T_0$ for the SW MoS$_2$ potential.
    }
\end{figure*}

\begin{figure*}[!h]
    \centering
    \includegraphics[width=0.6\textwidth]{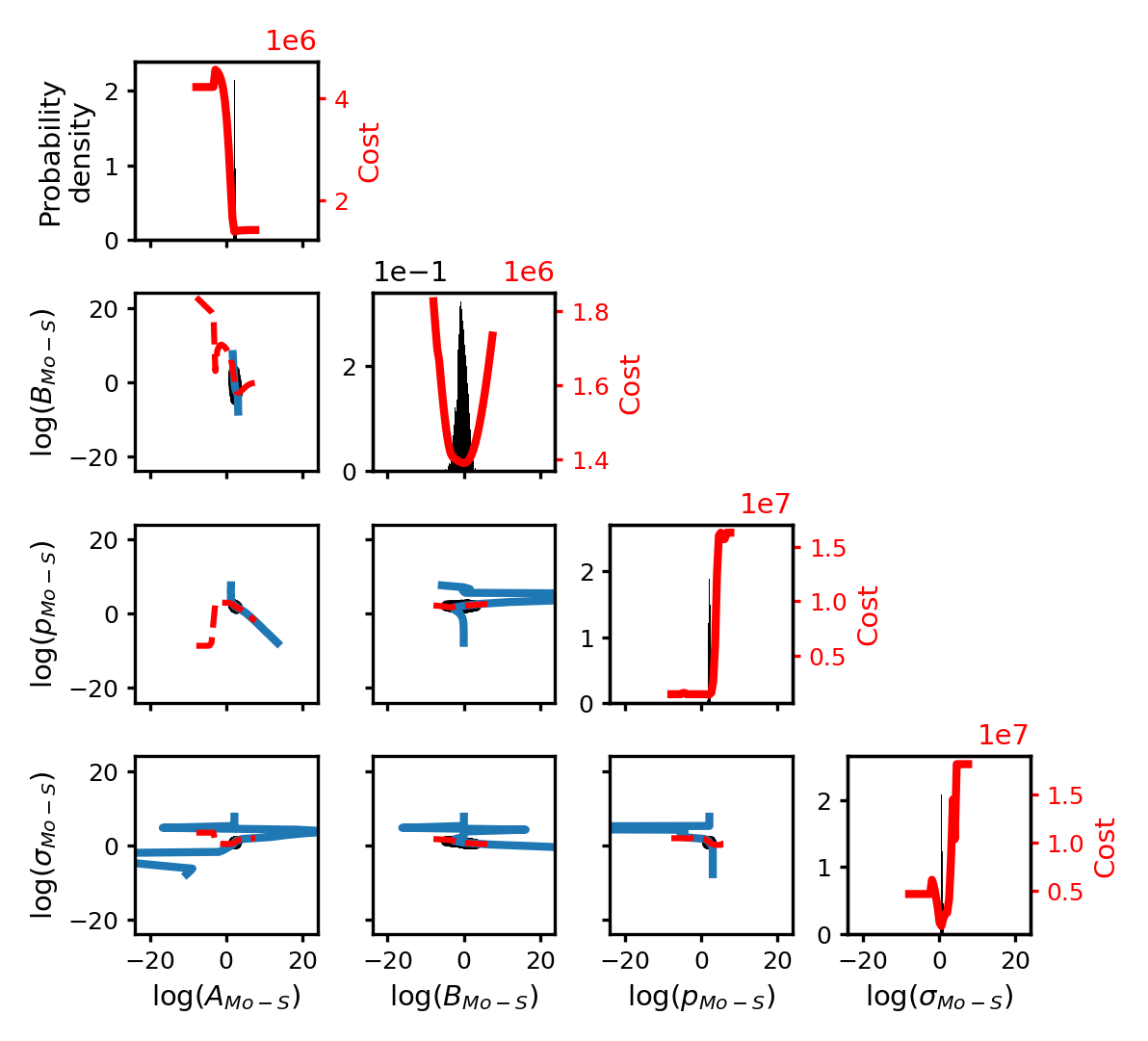}
    \caption[UQ results for SW potential Mo--S parameters at $T = 1.71 \times 10^{-1}~T_0$]{
        Profile likelihood and MCMC samples for Mo--S parameters at sampling temperature $1.71 \times 10^{-1}~T_0$ for the SW MoS$_2$ potential.
    }
\end{figure*}

\begin{figure*}[!h]
    \centering
    \includegraphics[width=0.6\textwidth]{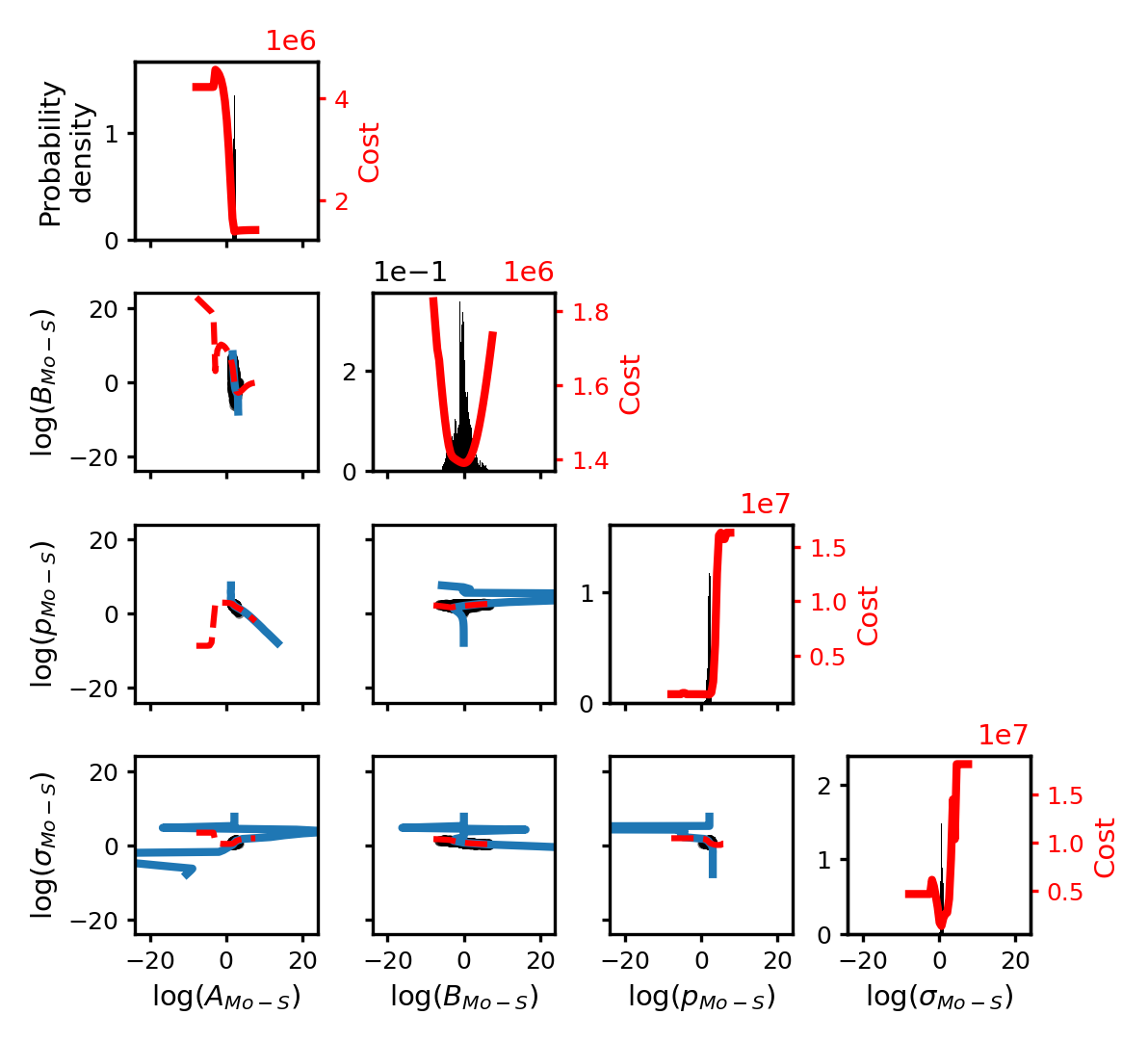}
    \caption[UQ results for SW potential Mo--S parameters at $T = 5.40 \times 10^{-1}~T_0$]{
        Profile likelihood and MCMC samples for Mo--S parameters at sampling temperature $5.40 \times 10^{-1}~T_0$ for the SW MoS$_2$ potential.
    }
\end{figure*}

\ifincludeTo
    \begin{figure*}[!h]
        \centering
        \includegraphics[width=0.6\textwidth]{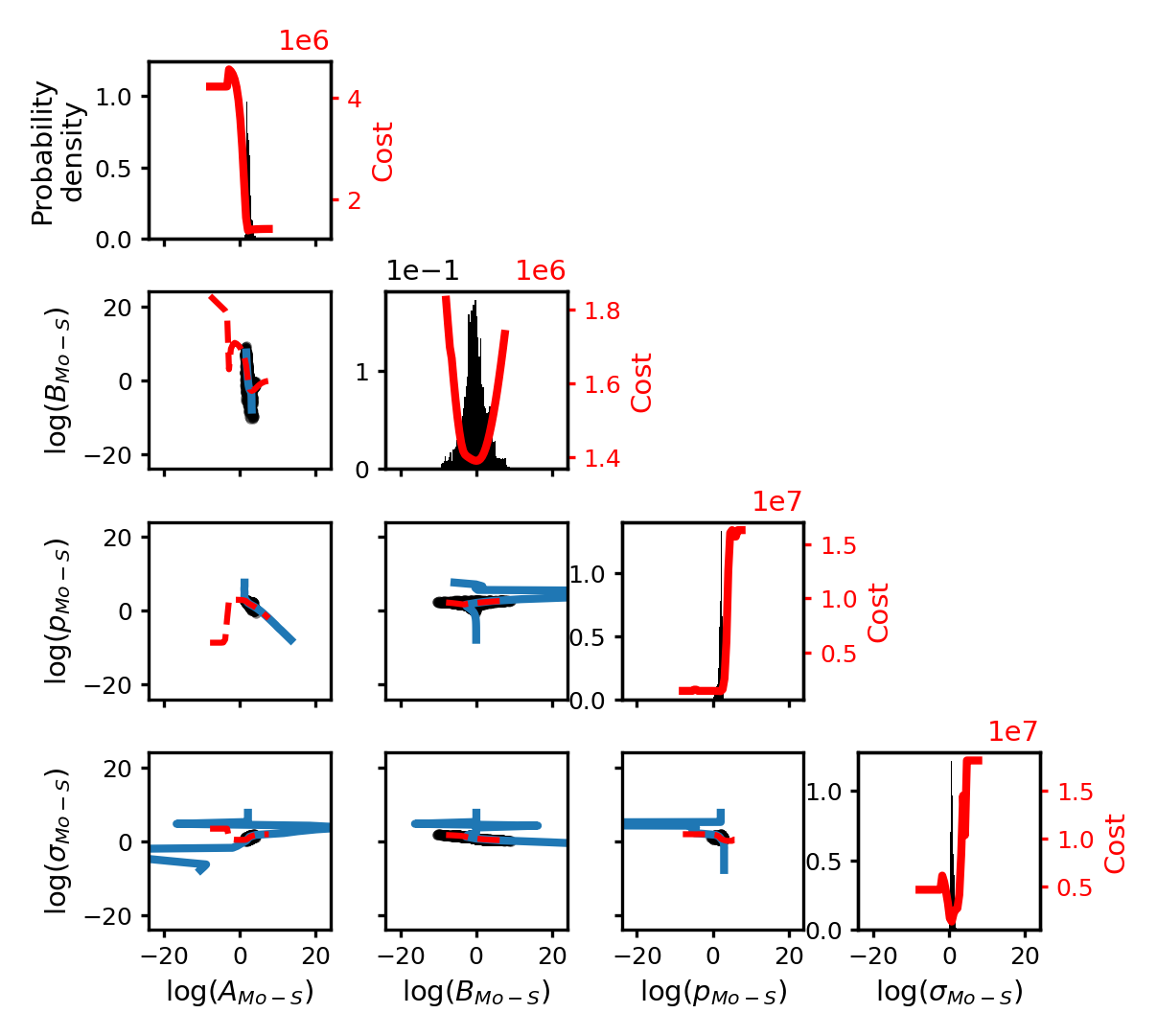}
        \caption[UQ results for SW potential Mo--S parameters at $T = T_0$]{
            Profile likelihood and MCMC samples for Mo--S parameters at sampling temperature $T_0$ for the SW MoS$_2$ potential.
        }
    \end{figure*}
\fi

\begin{figure*}[!h]
    \centering
    \includegraphics[width=0.6\textwidth]{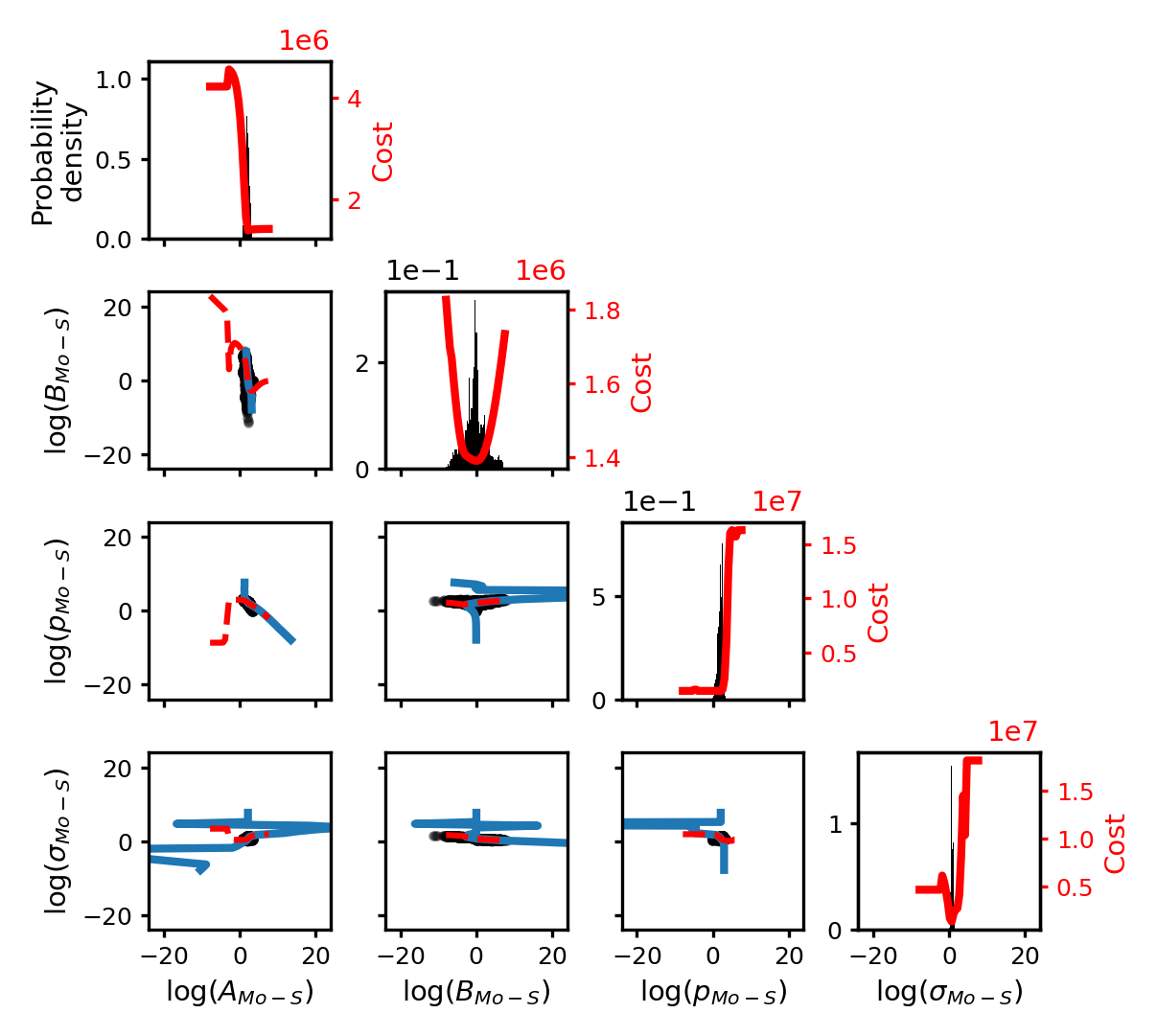}
    \caption[UQ results for SW potential Mo--S parameters at $T = 1.71~T_0$]{
        Profile likelihood and MCMC samples for Mo--S parameters at sampling temperature $1.71~T_0$ for the SW MoS$_2$ potential.
    }
\end{figure*}

\begin{figure*}[!h]
    \centering
    \includegraphics[width=0.6\textwidth]{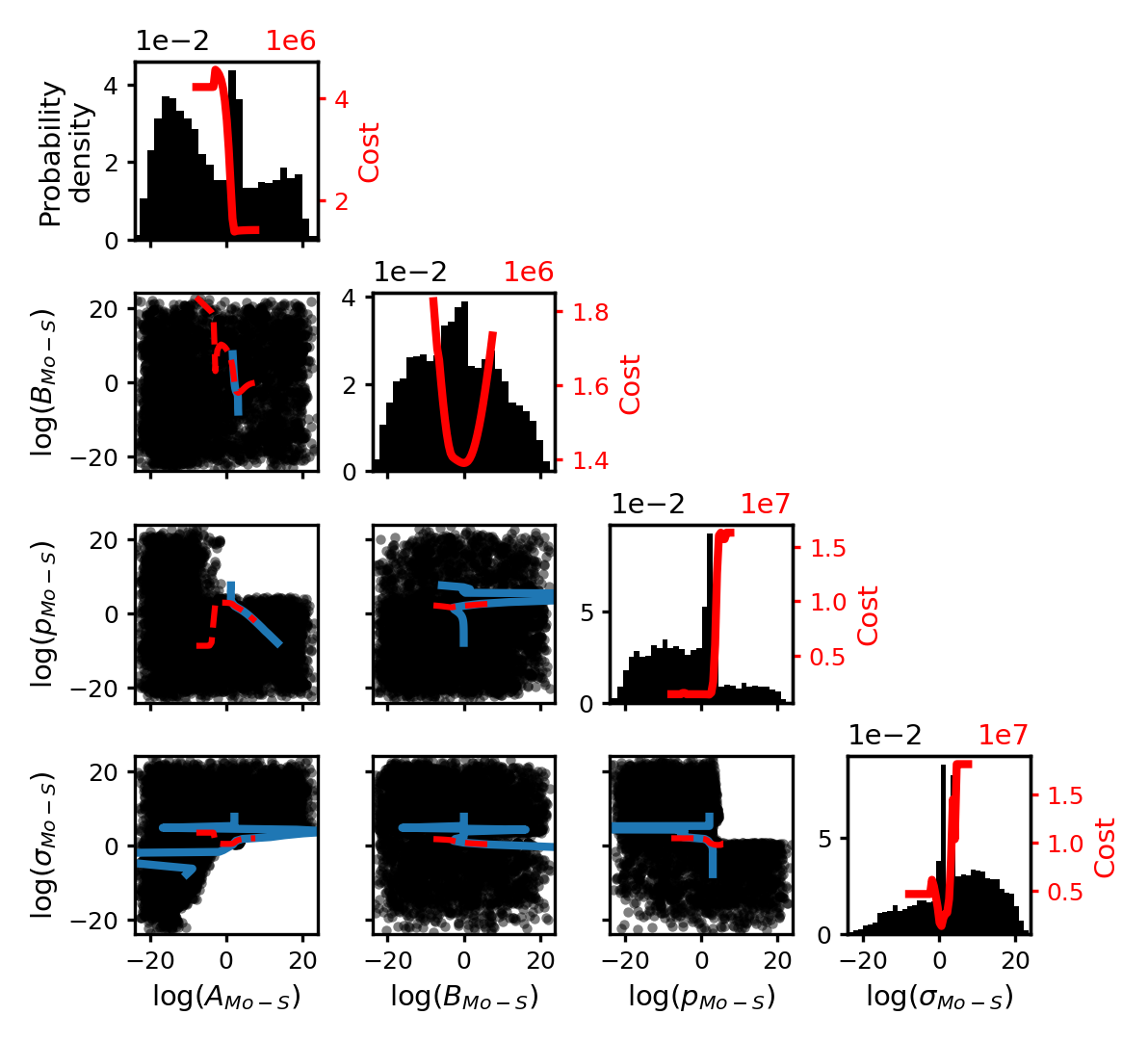}
    \caption[UQ results for SW potential Mo--S parameters at $T = 5.40~T_0$]{
        Profile likelihood and MCMC samples for Mo--S parameters at sampling temperature $5.40~T_0$ for the SW MoS$_2$ potential.
    }
\end{figure*}

\cleardoublepage

\subsection{S--S parameters}
\label{subsec:S-S}
Profile likelihood and MCMC samples for parameters corresponding to 2-body S--S interaction.

\begin{figure*}[!h]
    \centering
    \includegraphics[width=0.6\textwidth]{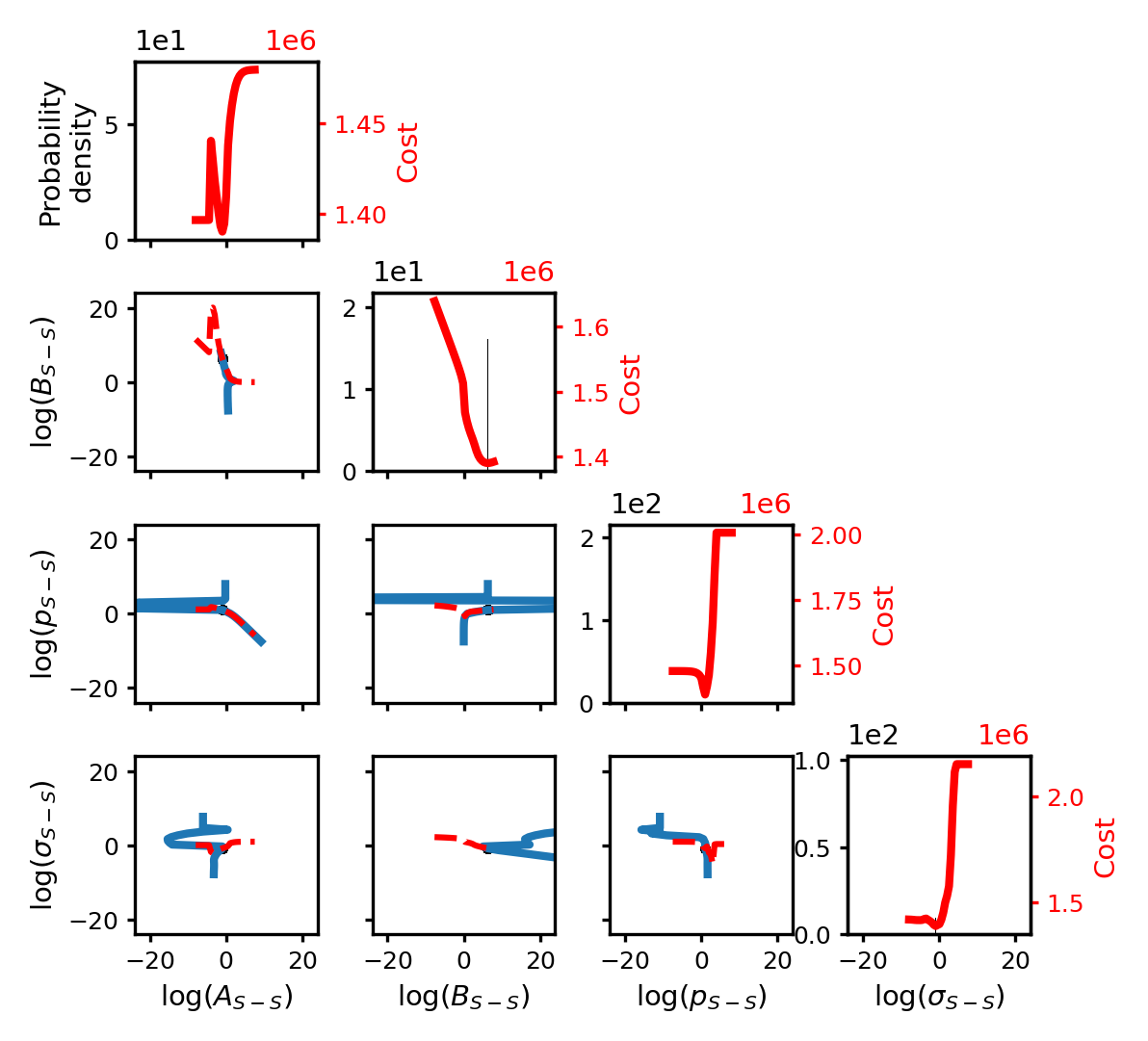}
    \caption[UQ results for SW potential S--S parameters at $T = 5.40 \times 10^{-6}~T_0$]{
        Profile likelihood and MCMC samples for S--S parameters at sampling temperature $5.40 \times 10^{-6}~T_0$ for the SW MoS$_2$ potential.
    }
\end{figure*}

\begin{figure*}[!h]
    \centering
    \includegraphics[width=0.6\textwidth]{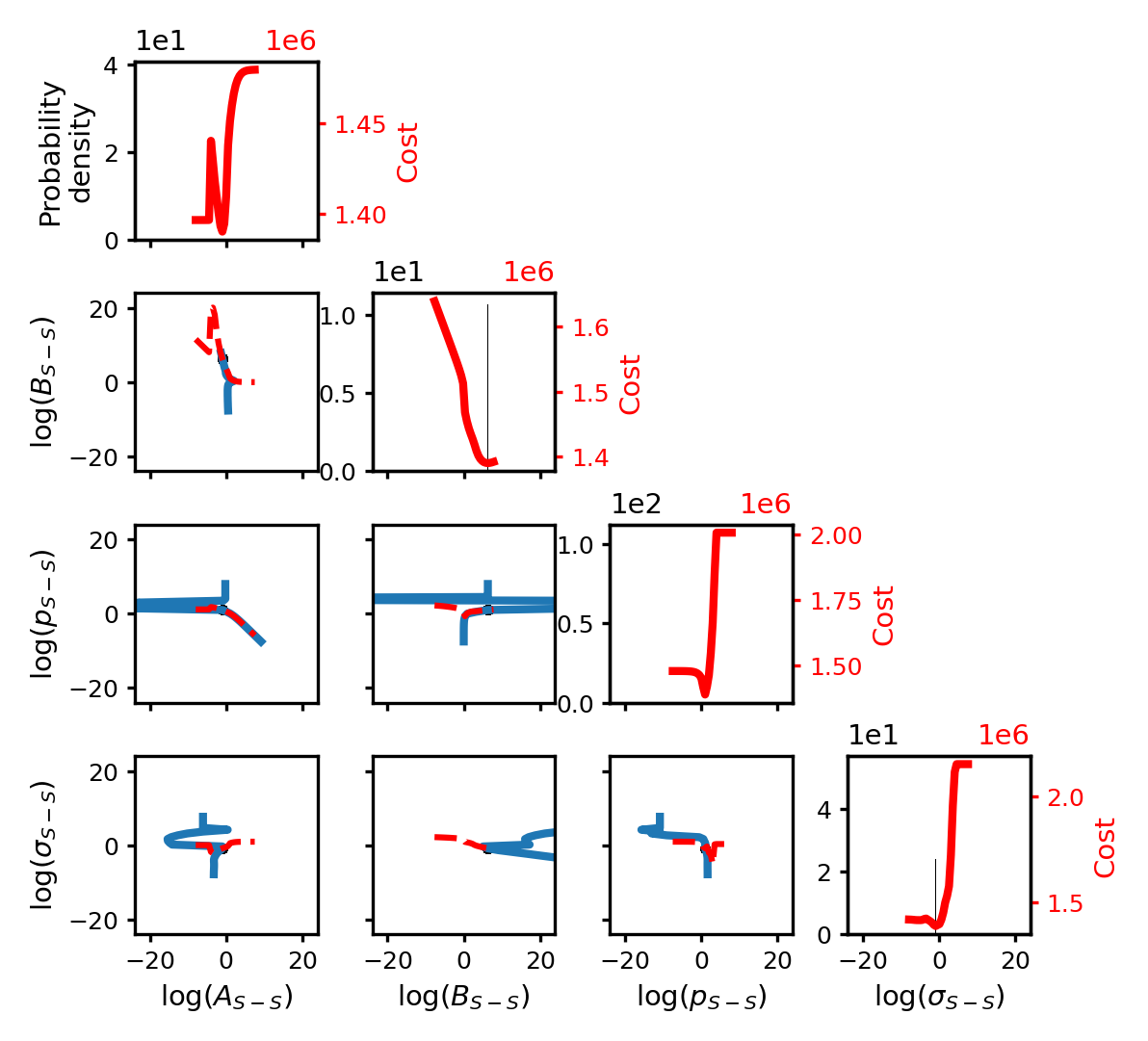}
    \caption[UQ results for SW potential S--S parameters at $T = 1.71 \times 10^{-5}~T_0$]{
        Profile likelihood and MCMC samples for S--S parameters at sampling temperature $1.71 \times 10^{-5}~T_0$ for the SW MoS$_2$ potential.
    }
\end{figure*}

\begin{figure*}[!h]
    \centering
    \includegraphics[width=0.6\textwidth]{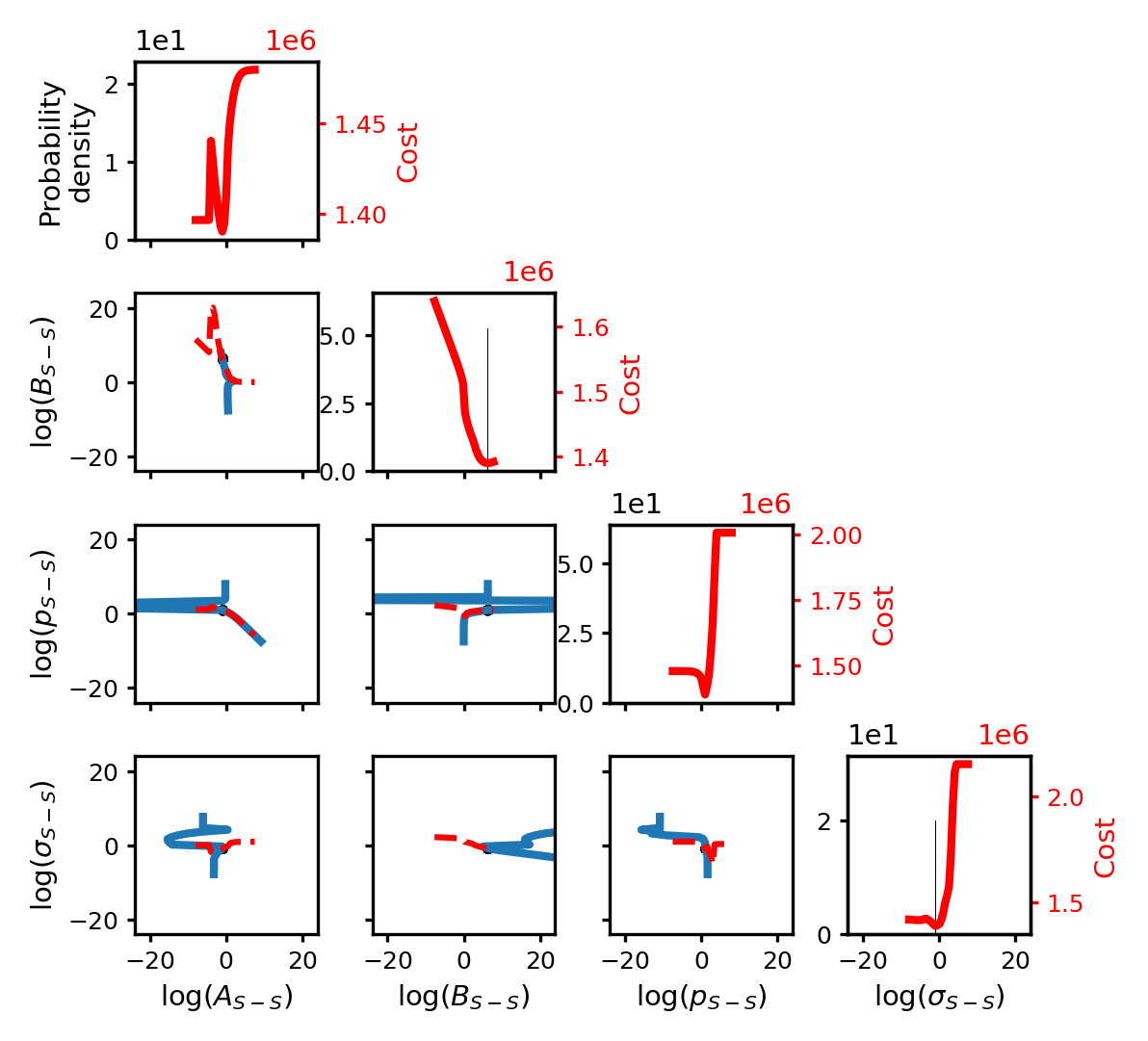}
    \caption[UQ results for SW potential S--S parameters at $T = 5.40 \times 10^{-5}~T_0$]{
        Profile likelihood and MCMC samples for S--S parameters at sampling temperature $5.40 \times 10^{-5}~T_0$ for the SW MoS$_2$ potential.
    }
\end{figure*}

\begin{figure*}[!h]
    \centering
    \includegraphics[width=0.6\textwidth]{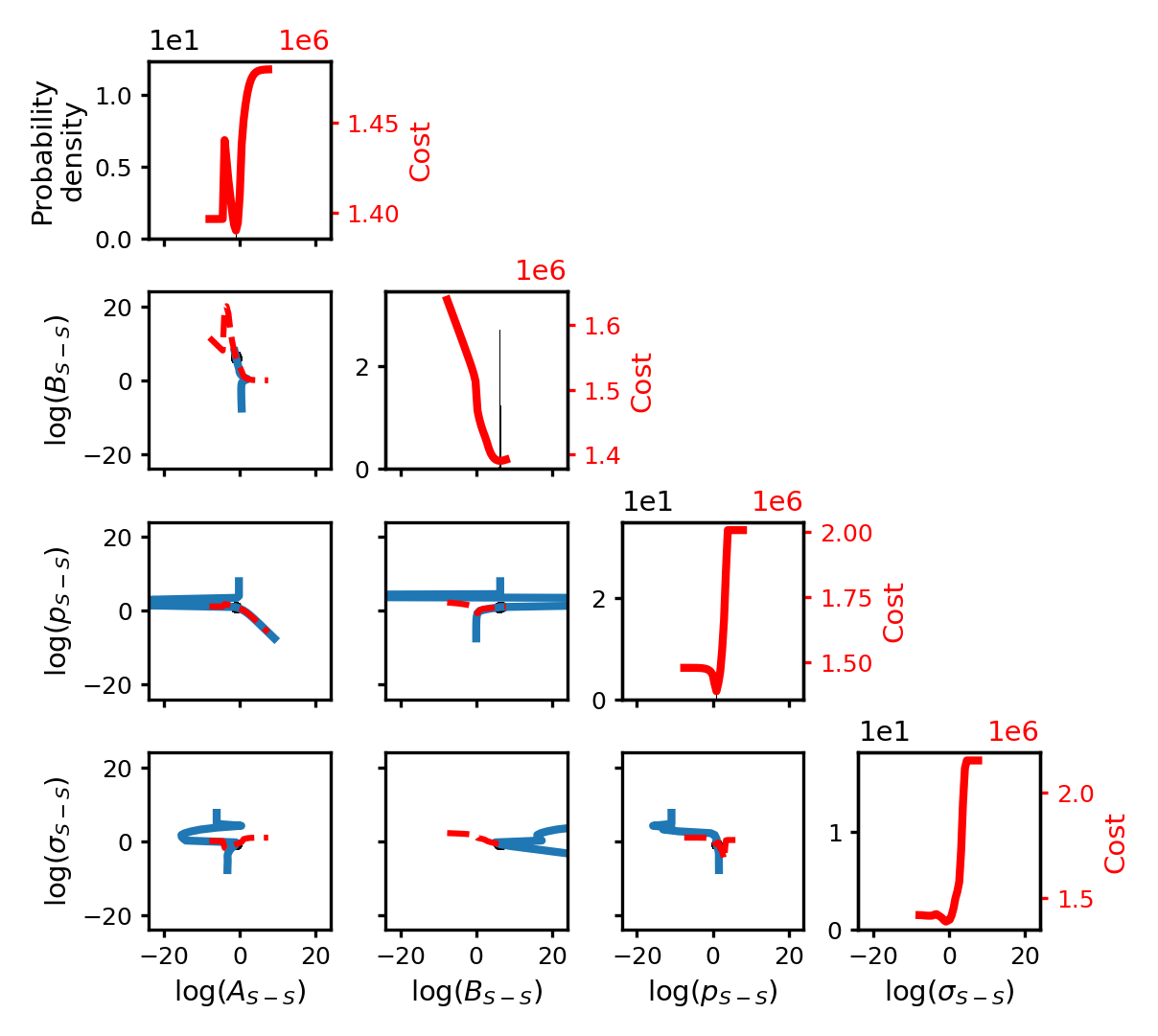}
    \caption[UQ results for SW potential S--S parameters at $T = 1.71 \times 10^{-4}~T_0$]{
        Profile likelihood and MCMC samples for S--S parameters at sampling temperature $1.71 \times 10^{-4}~T_0$ for the SW MoS$_2$ potential.
    }
\end{figure*}

\begin{figure*}[!h]
    \centering
    \includegraphics[width=0.6\textwidth]{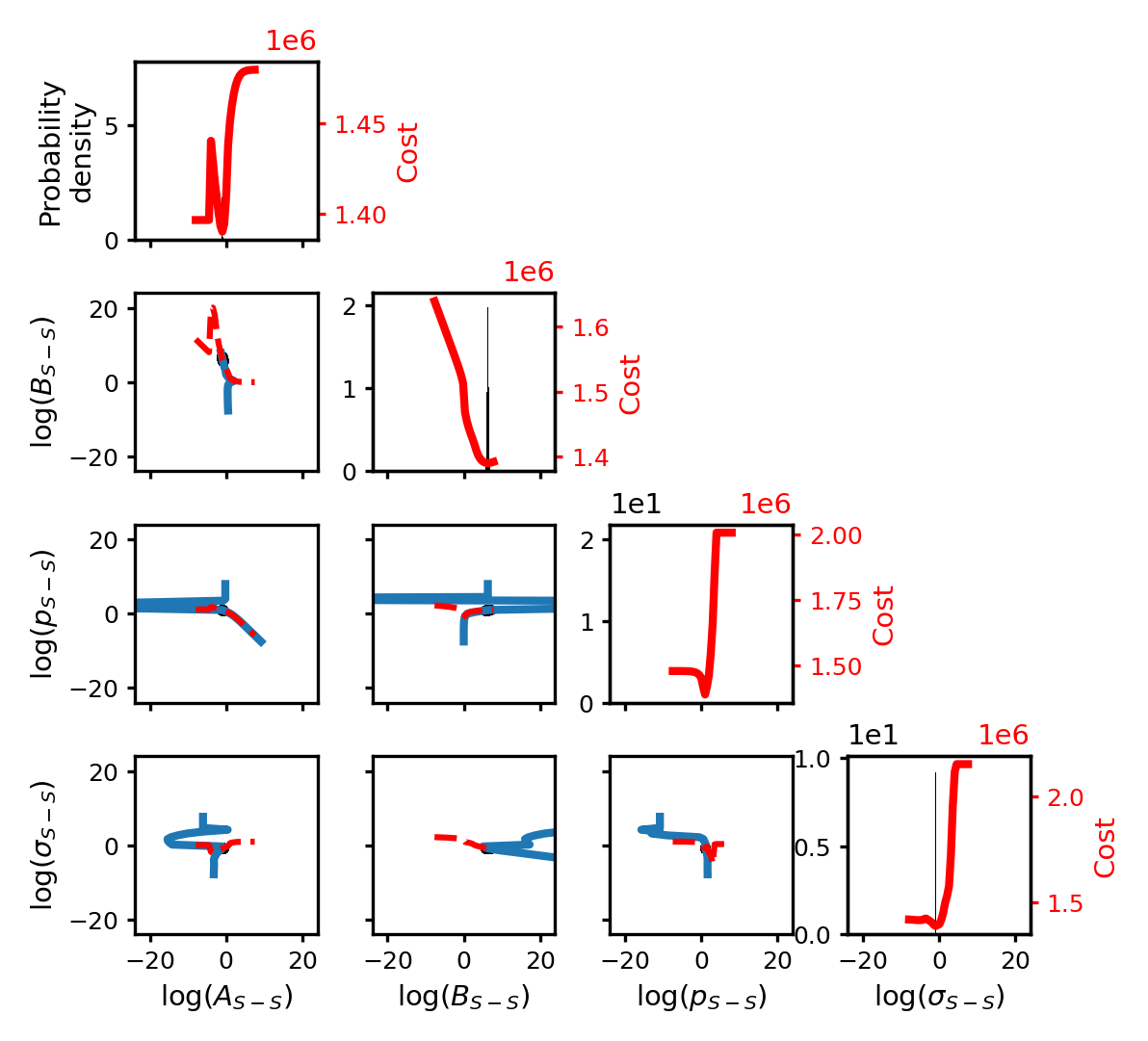}
    \caption[UQ results for SW potential S--S parameters at $T = 5.40 \times 10^{-4}~T_0$]{
        Profile likelihood and MCMC samples for S--S parameters at sampling temperature $5.40 \times 10^{-4}~T_0$ for the SW MoS$_2$ potential.
    }
\end{figure*}

\begin{figure*}[!h]
    \centering
    \includegraphics[width=0.6\textwidth]{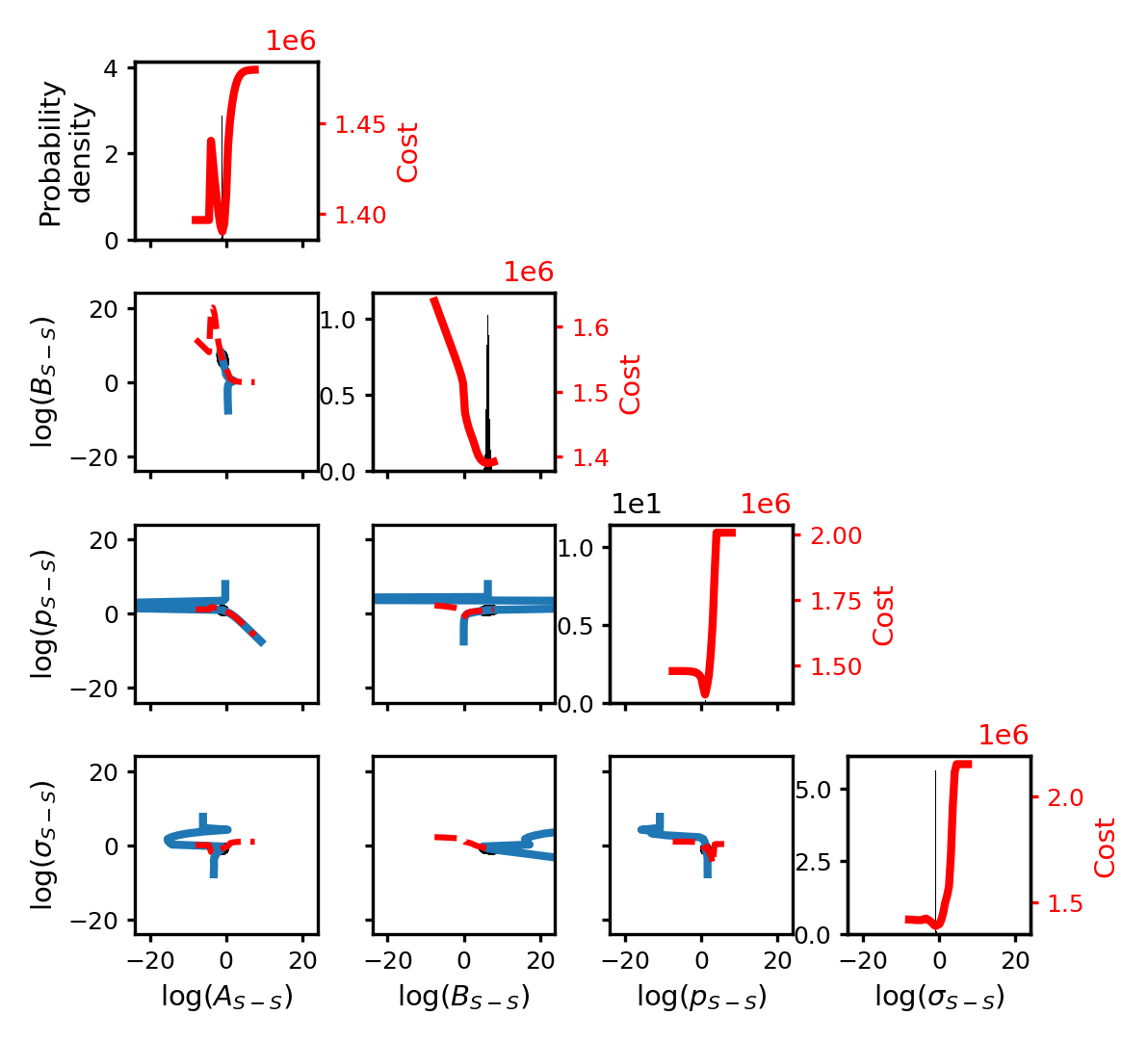}
    \caption[UQ results for SW potential S--S parameters at $T = 1.71 \times 10^{-3}~T_0$]{
        Profile likelihood and MCMC samples for S--S parameters at sampling temperature $1.71 \times 10^{-3}~T_0$ for the SW MoS$_2$ potential.
    }
\end{figure*}

\begin{figure*}[!h]
    \centering
    \includegraphics[width=0.6\textwidth]{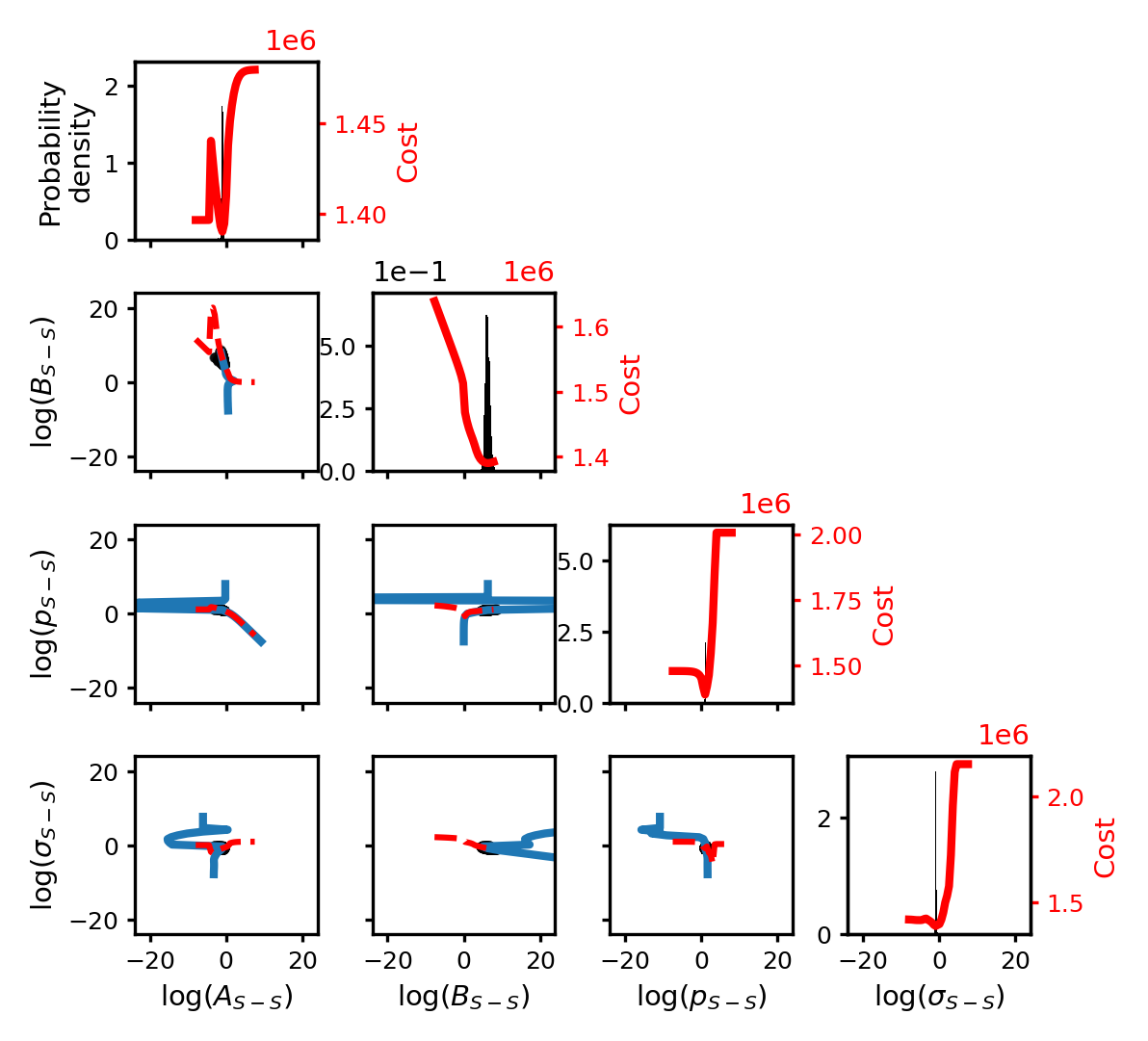}
    \caption[UQ results for SW potential S--S parameters at $T = 5.40 \times 10^{-3}~T_0$]{
        Profile likelihood and MCMC samples for S--S parameters at sampling temperature $5.40 \times 10^{-3}~T_0$ for the SW MoS$_2$ potential.
    }
\end{figure*}

\begin{figure*}[!h]
    \centering
    \includegraphics[width=0.6\textwidth]{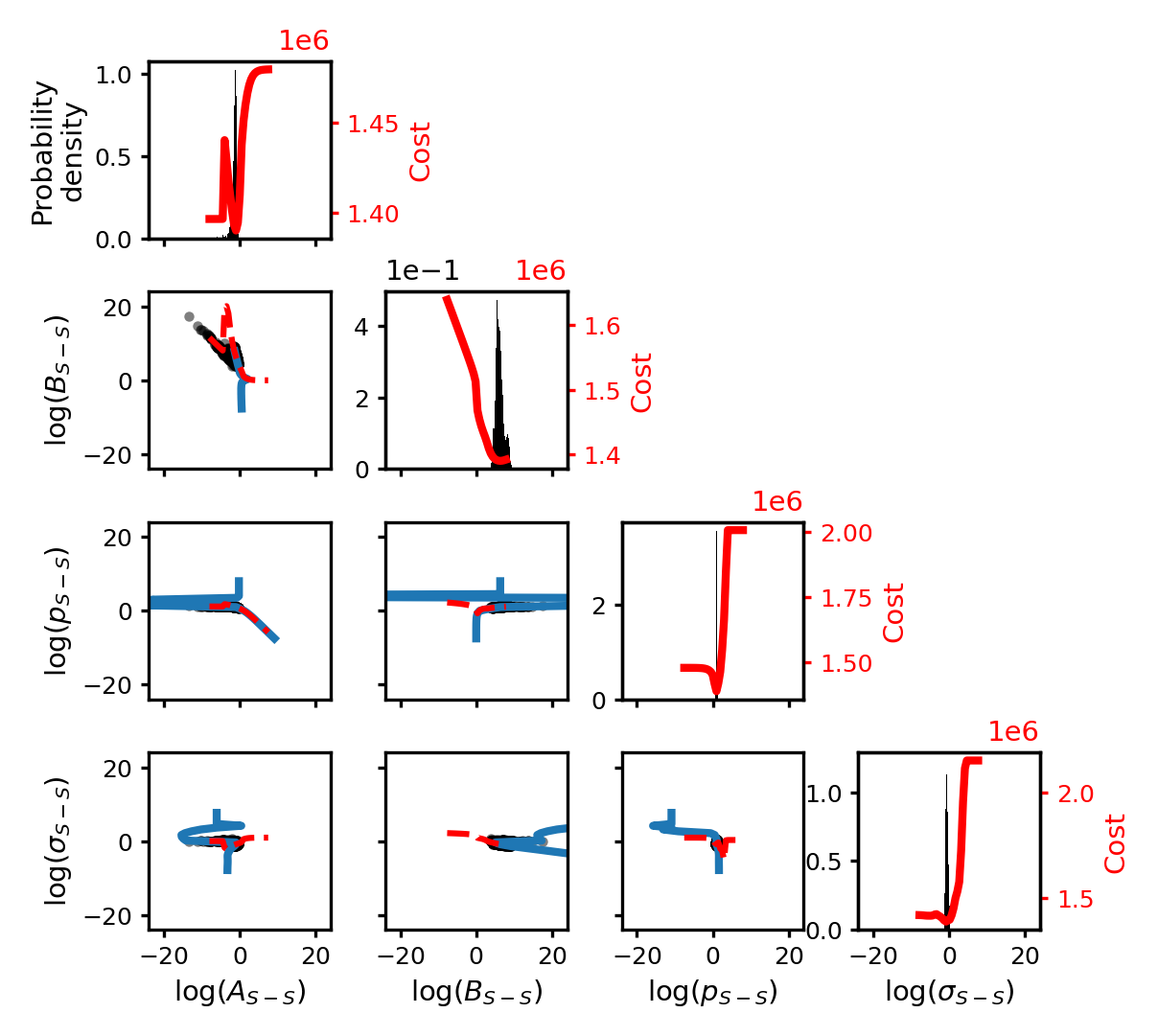}
    \caption[UQ results for SW potential S--S parameters at $T = 1.71 \times 10^{-2}~T_0$]{
        Profile likelihood and MCMC samples for S--S parameters at sampling temperature $1.71 \times 10^{-2}~T_0$ for the SW MoS$_2$ potential.
    }
\end{figure*}

\begin{figure*}[!h]
    \centering
    \includegraphics[width=0.6\textwidth]{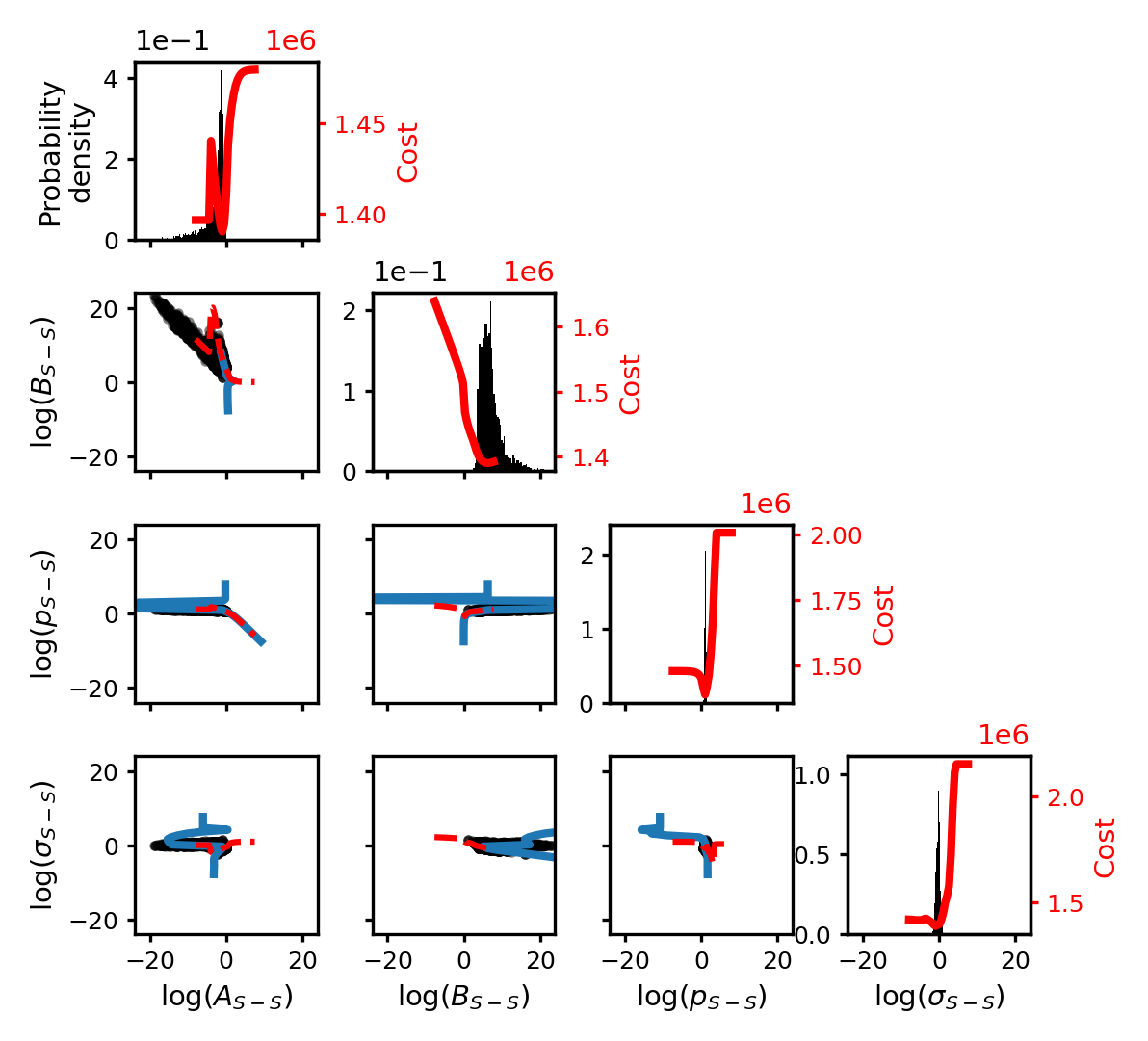}
    \caption[UQ results for SW potential S--S parameters at $T = 5.40 \times 10^{-2}~T_0$]{
        Profile likelihood and MCMC samples for S--S parameters at sampling temperature $5.40 \times 10^{-2}~T_0$ for the SW MoS$_2$ potential.
    }
\end{figure*}

\begin{figure*}[!h]
    \centering
    \includegraphics[width=0.6\textwidth]{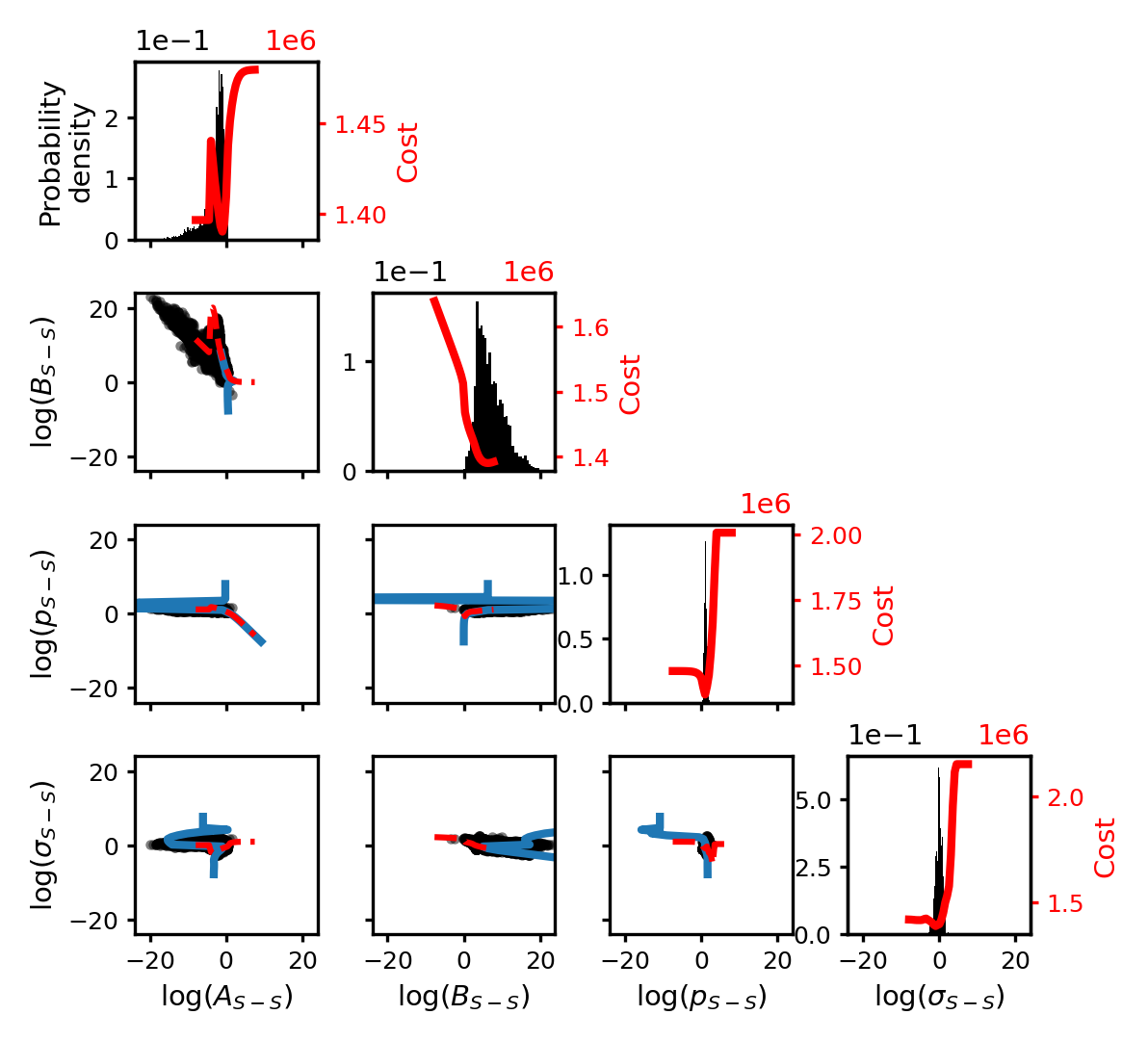}
    \caption[UQ results for SW potential S--S parameters at $T = 1.71 \times 10^{-1}~T_0$]{
        Profile likelihood and MCMC samples for S--S parameters at sampling temperature $1.71 \times 10^{-1}~T_0$ for the SW MoS$_2$ potential.
    }
\end{figure*}

\begin{figure*}[!h]
    \centering
    \includegraphics[width=0.6\textwidth]{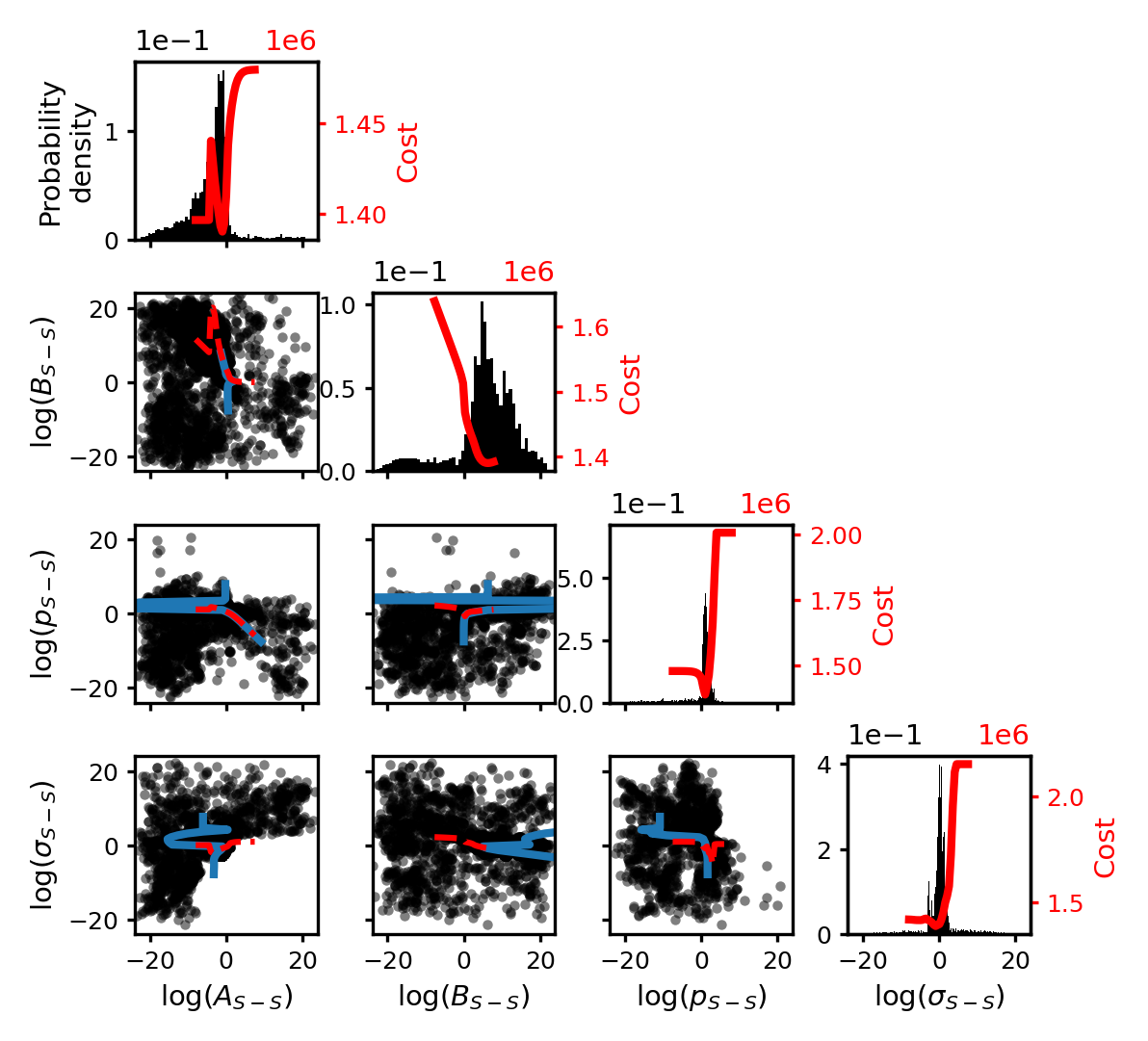}
    \caption[UQ results for SW potential S--S parameters at $T = 5.40 \times 10^{-1}~T_0$]{
        Profile likelihood and MCMC samples for S--S parameters at sampling temperature $5.40 \times 10^{-1}~T_0$ for the SW MoS$_2$ potential.
    }
\end{figure*}

\ifincludeTo
    \begin{figure*}[!h]
        \centering
        \includegraphics[width=0.6\textwidth]{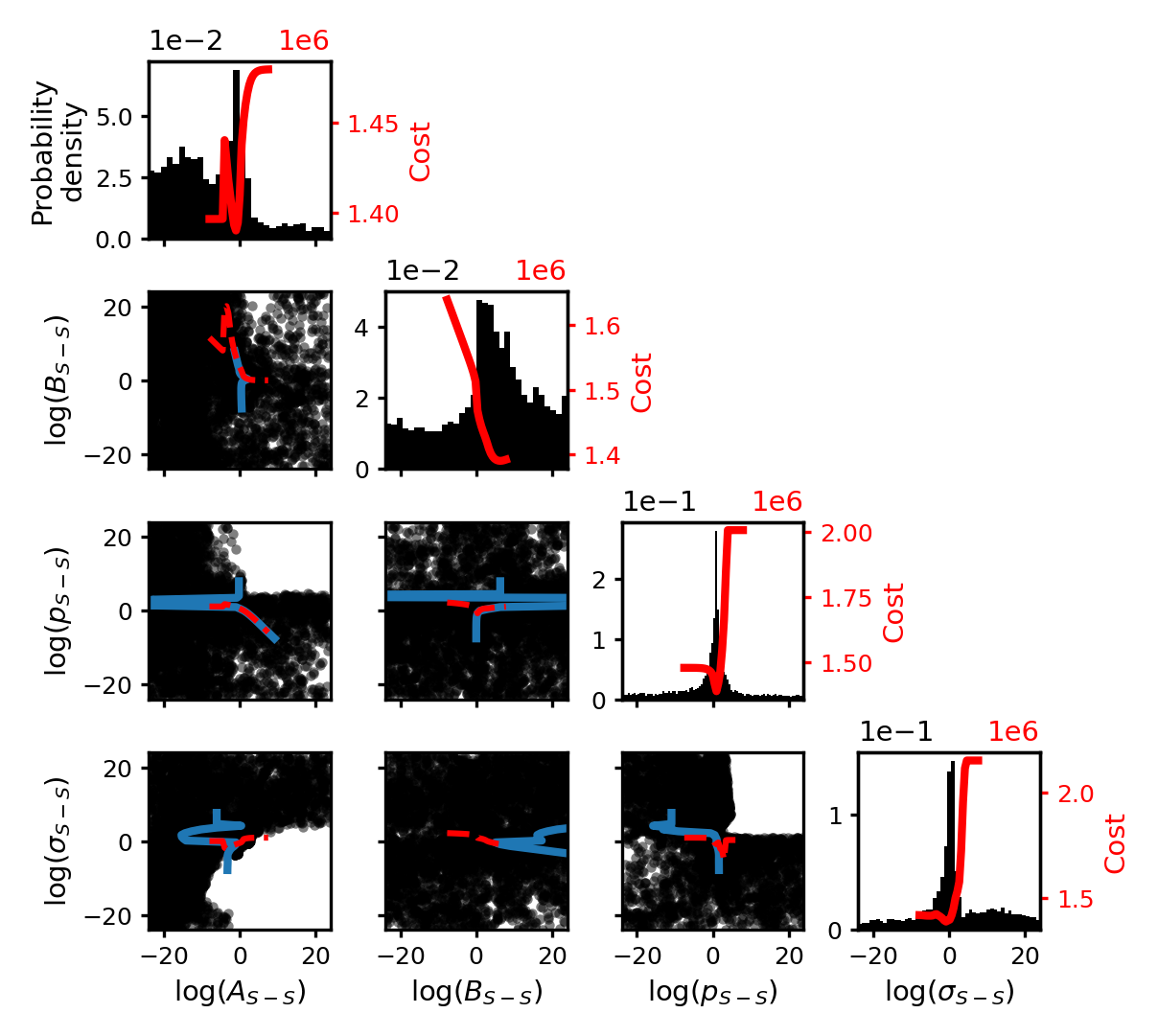}
        \caption[UQ results for SW potential S--S parameters at $T = T_0$]{
            Profile likelihood and MCMC samples for S--S parameters at sampling temperature $T_0$ for the SW MoS$_2$ potential.
        }
    \end{figure*}
\fi

\begin{figure*}[!h]
    \centering
    \includegraphics[width=0.6\textwidth]{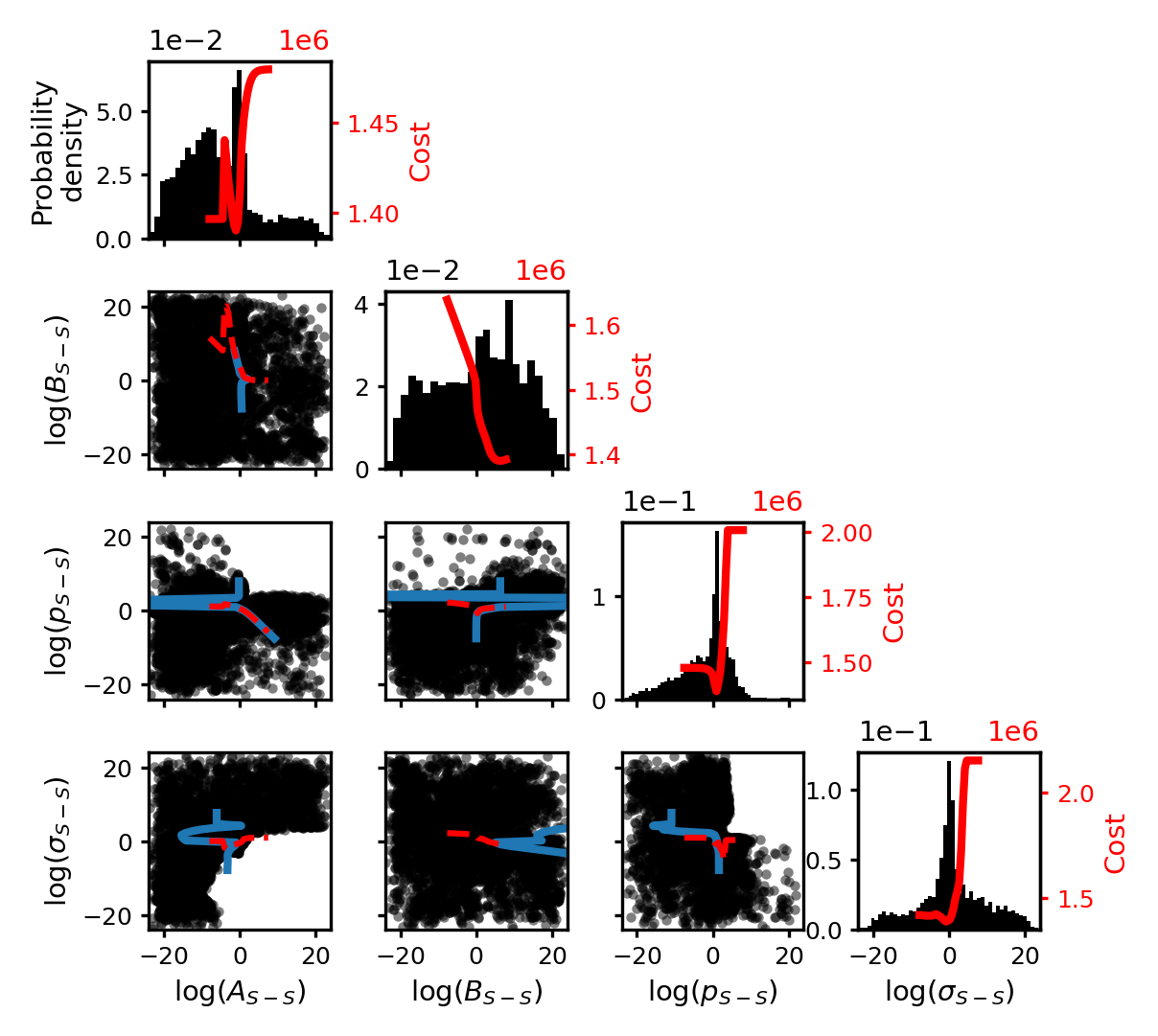}
    \caption[UQ results for SW potential S--S parameters at $T = 1.71~T_0$]{
        Profile likelihood and MCMC samples for S--S parameters at sampling temperature $1.71~T_0$ for the SW MoS$_2$ potential.
    }
\end{figure*}

\begin{figure*}[!h]
    \centering
    \includegraphics[width=0.6\textwidth]{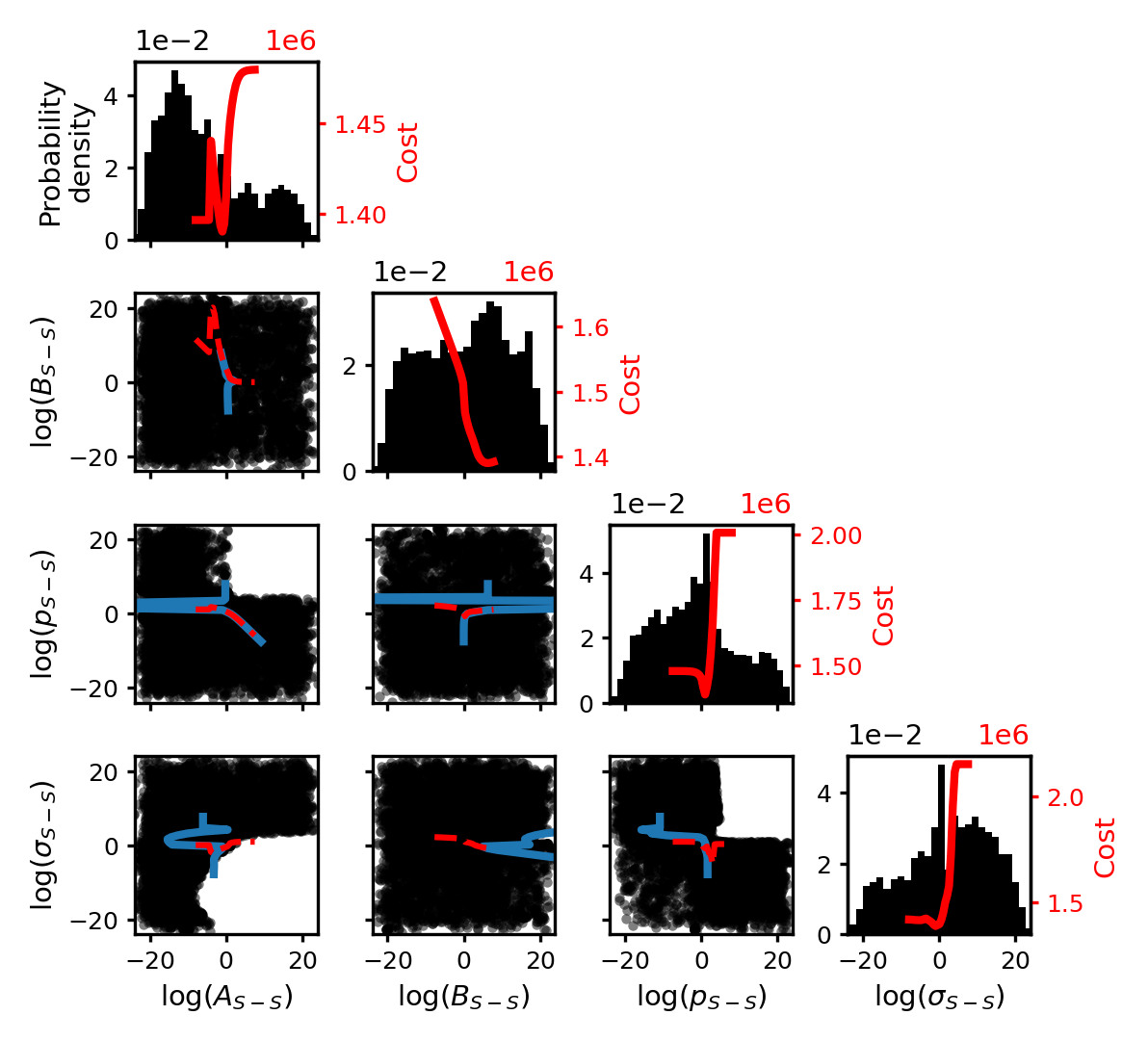}
    \caption[UQ results for SW potential S--S parameters at $T = 5.40~T_0$]{
        Profile likelihood and MCMC samples for S--S parameters at sampling temperature $5.40~T_0$ for the SW MoS$_2$ potential.
    }
\end{figure*}

\cleardoublepage

\subsection{3-body parameters}
\label{subsec:3-body}
Profile likelihood and MCMC samples for parameters corresponding to 3-body interactions.

\begin{figure*}[!h]
    \centering
    \includegraphics[width=0.6\textwidth]{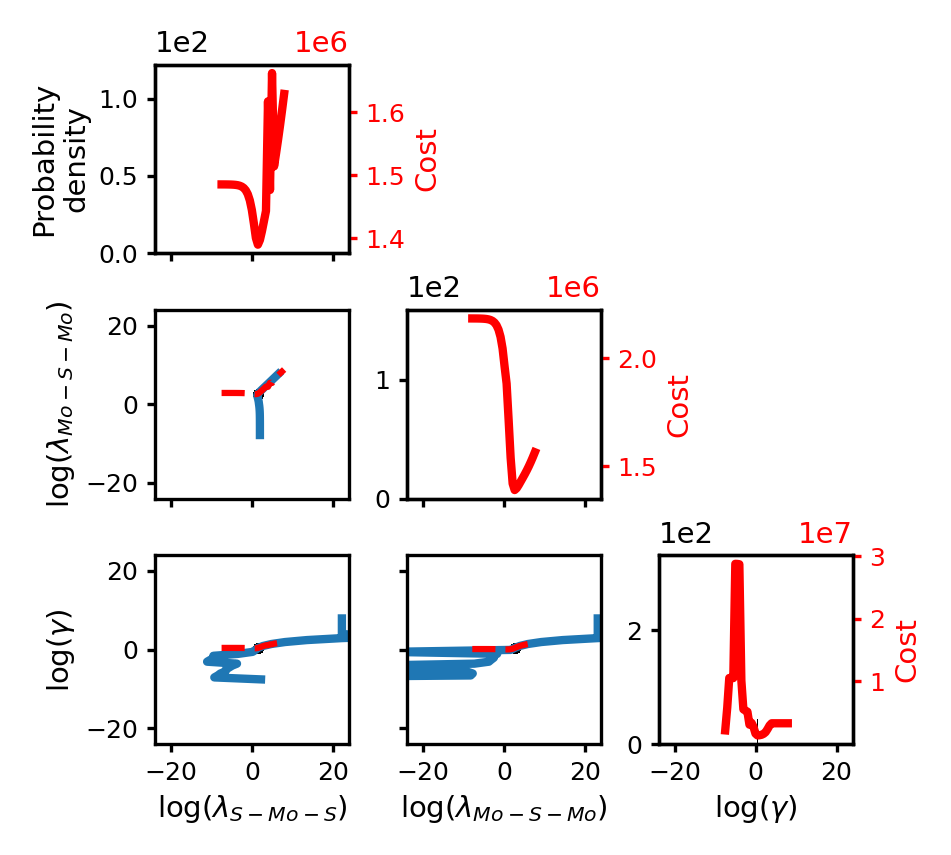}
    \caption[UQ results for SW potential 3-body parameters at $T = 5.40 \times 10^{-6}~T_0$]{
        Profile likelihood and MCMC samples for 3-body parameters at sampling temperature $5.40 \times 10^{-6}~T_0$ for the SW MoS$_2$ potential.
    }
\end{figure*}

\begin{figure*}[!h]
    \centering
    \includegraphics[width=0.6\textwidth]{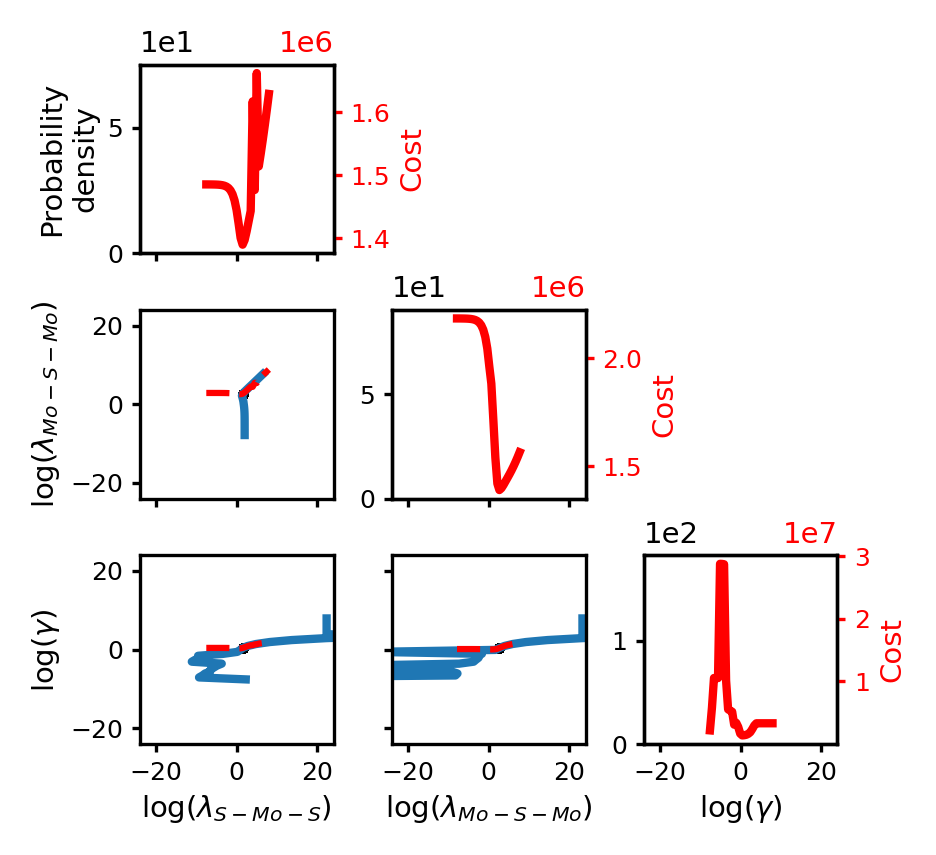}
    \caption[UQ results for SW potential 3-body parameters at $T = 1.71 \times 10^{-5}~T_0$]{
        Profile likelihood and MCMC samples for 3-body parameters at sampling temperature $1.71 \times 10^{-5}~T_0$ for the SW MoS$_2$ potential.
    }
\end{figure*}

\begin{figure*}[!h]
    \centering
    \includegraphics[width=0.6\textwidth]{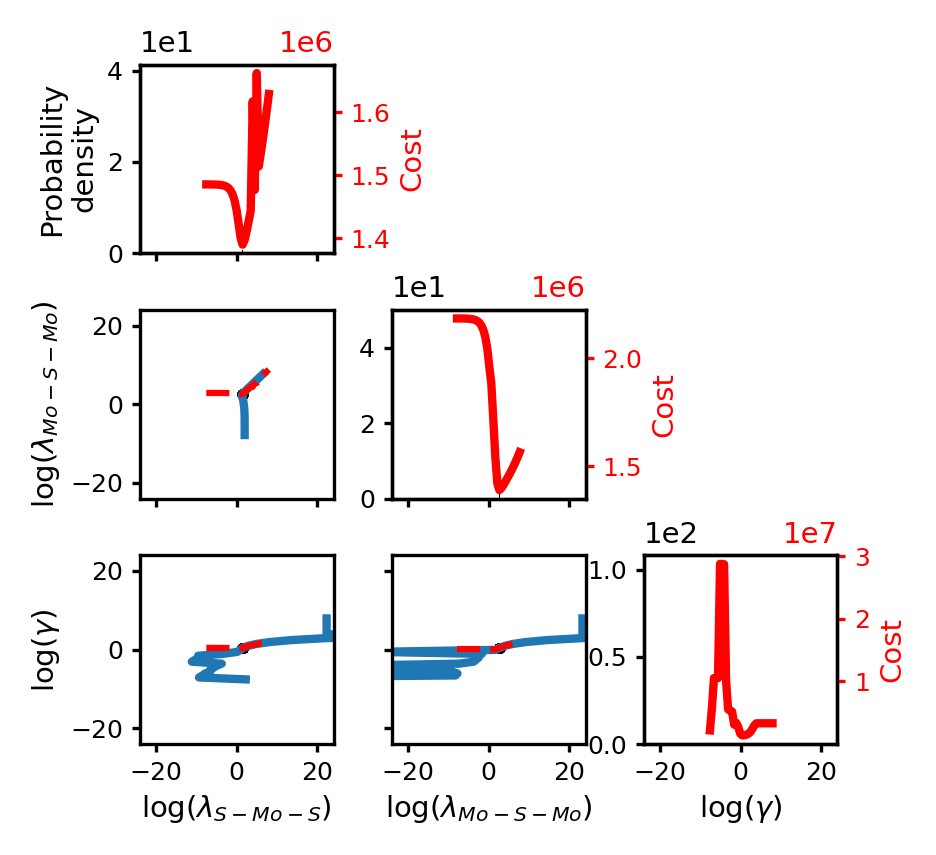}
    \caption[UQ results for SW potential 3-body parameters at $T = 5.40 \times 10^{-5}~T_0$]{
        Profile likelihood and MCMC samples for 3-body parameters at sampling temperature $5.40 \times 10^{-5}~T_0$ for the SW MoS$_2$ potential.
    }
\end{figure*}

\begin{figure*}[!h]
    \centering
    \includegraphics[width=0.6\textwidth]{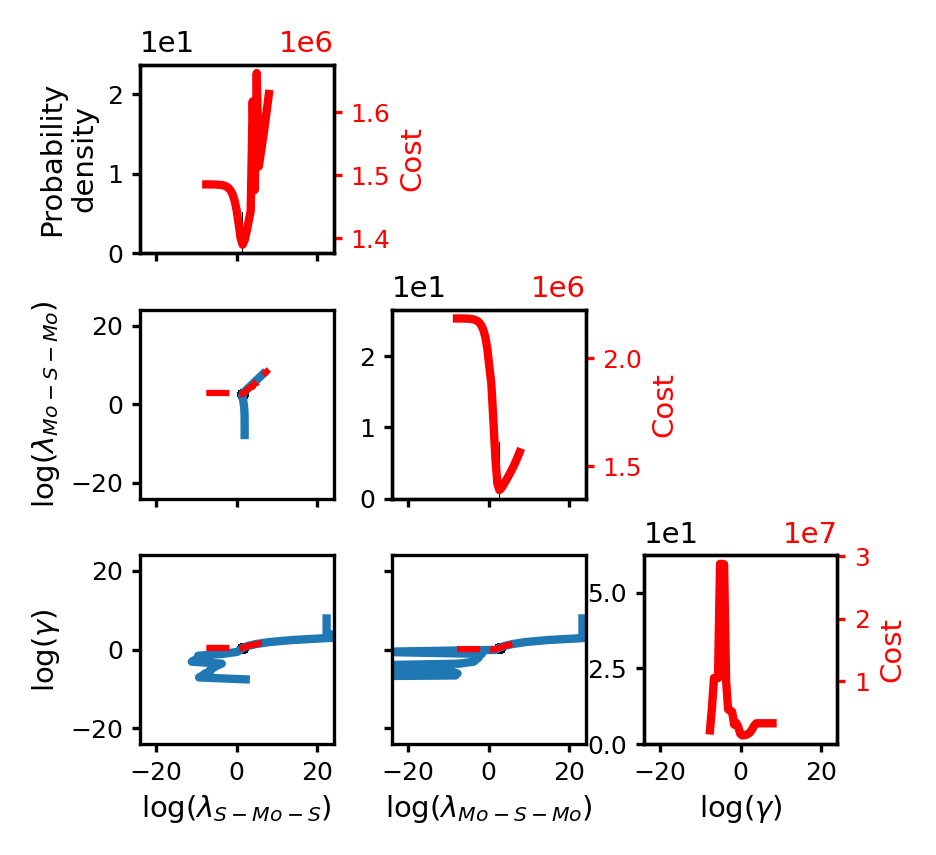}
    \caption[UQ results for SW potential 3-body parameters at $T = 1.71 \times 10^{-4}~T_0$]{
        Profile likelihood and MCMC samples for 3-body parameters at sampling temperature $1.71 \times 10^{-4}~T_0$ for the SW MoS$_2$ potential.
    }
\end{figure*}

\begin{figure*}[!h]
    \centering
    \includegraphics[width=0.6\textwidth]{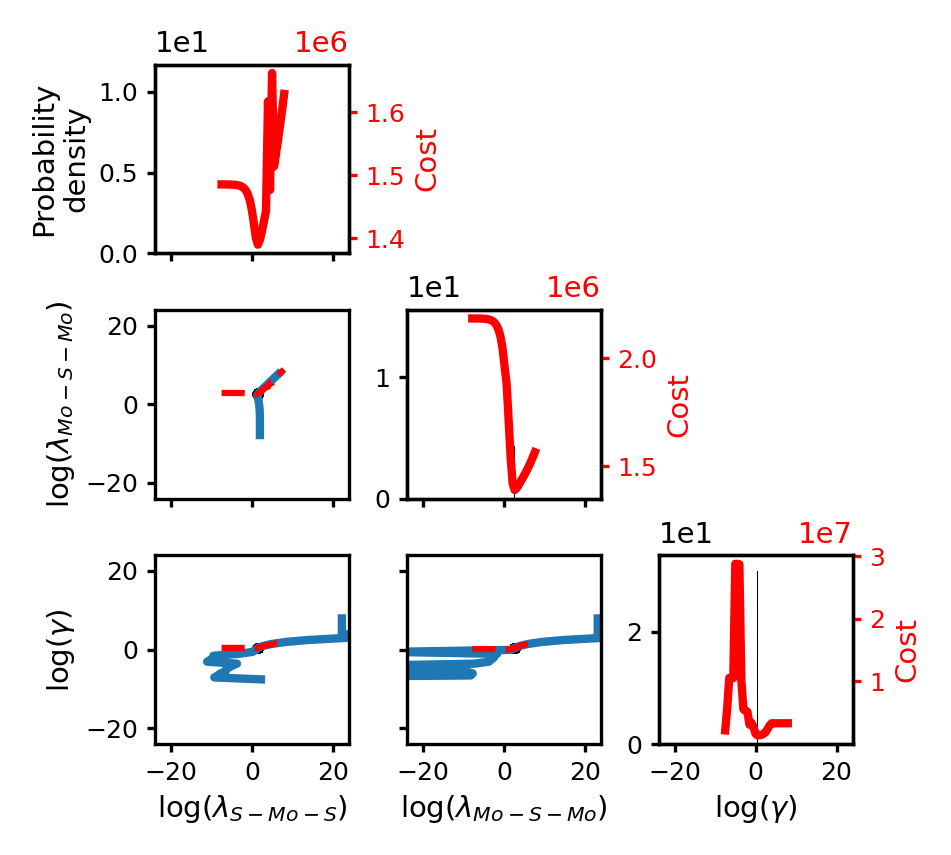}
    \caption[UQ results for SW potential 3-body parameters at $T = 5.40 \times 10^{-4}~T_0$]{
        Profile likelihood and MCMC samples for 3-body parameters at sampling temperature $5.40 \times 10^{-4}~T_0$ for the SW MoS$_2$ potential.
    }
\end{figure*}

\begin{figure*}[!h]
    \centering
    \includegraphics[width=0.6\textwidth]{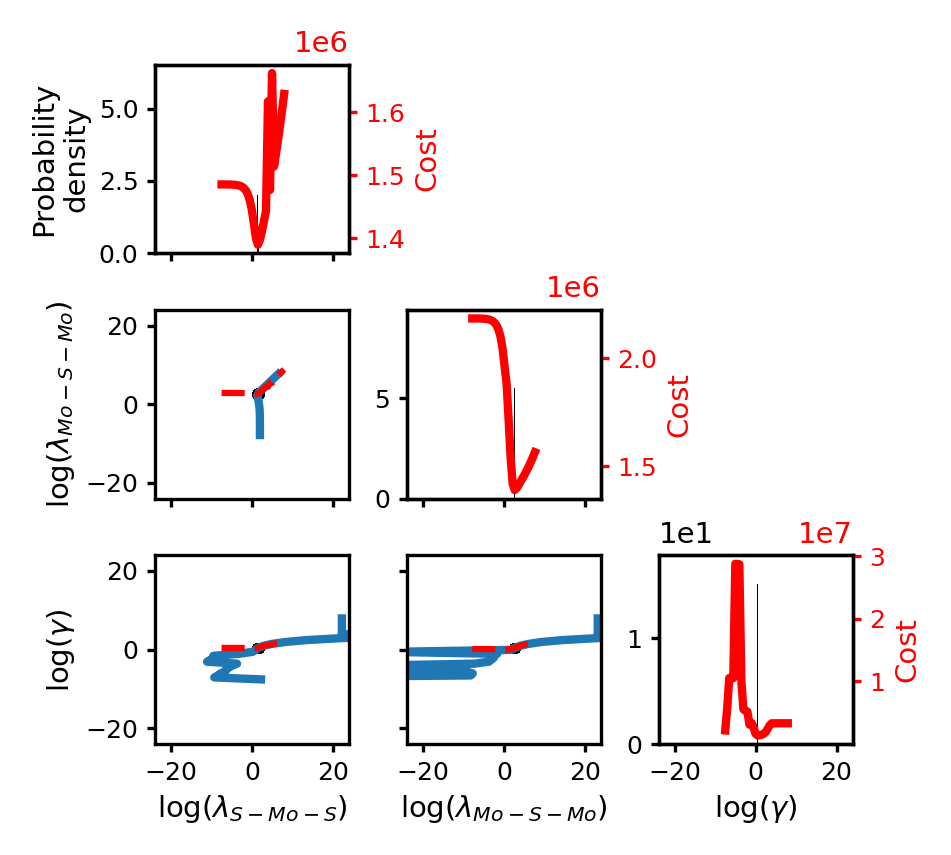}
    \caption[UQ results for SW potential 3-body parameters at $T = 1.71 \times 10^{-3}~T_0$]{
        Profile likelihood and MCMC samples for 3-body parameters at sampling temperature $1.71 \times 10^{-3}~T_0$ for the SW MoS$_2$ potential.
    }
\end{figure*}

\begin{figure*}[!h]
    \centering
    \includegraphics[width=0.6\textwidth]{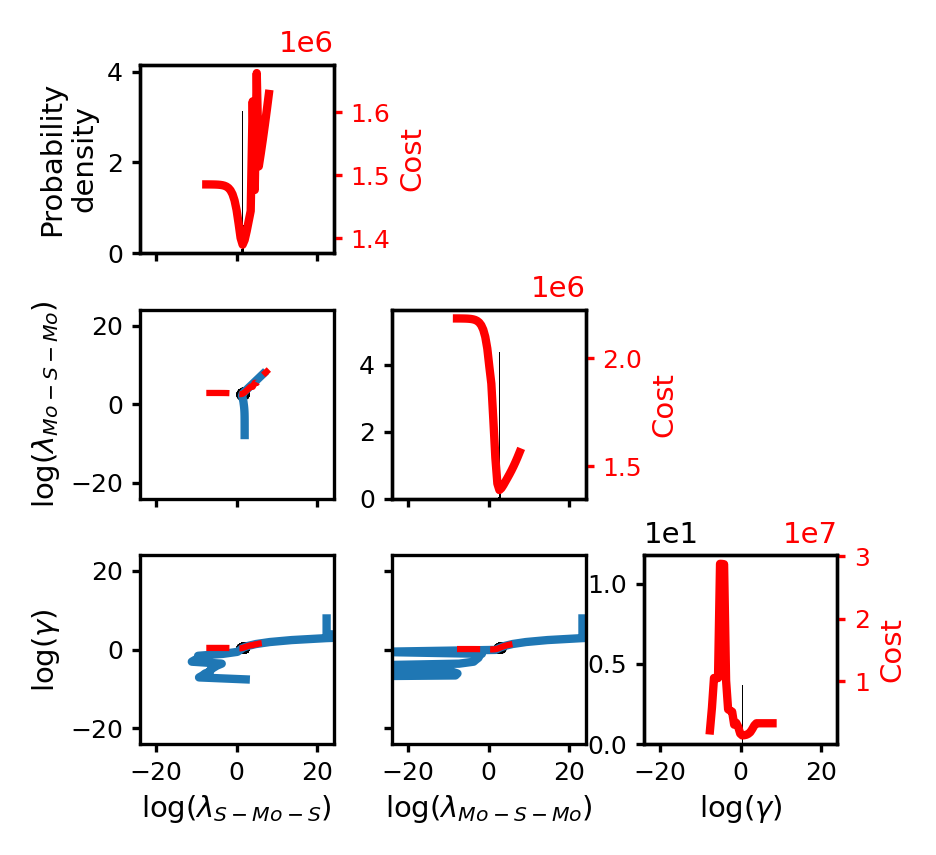}
    \caption[UQ results for SW potential 3-body parameters at $T = 5.40 \times 10^{-3}~T_0$]{
        Profile likelihood and MCMC samples for 3-body parameters at sampling temperature $5.40 \times 10^{-3}~T_0$ for the SW MoS$_2$ potential.
    }
\end{figure*}

\begin{figure*}[!h]
    \centering
    \includegraphics[width=0.6\textwidth]{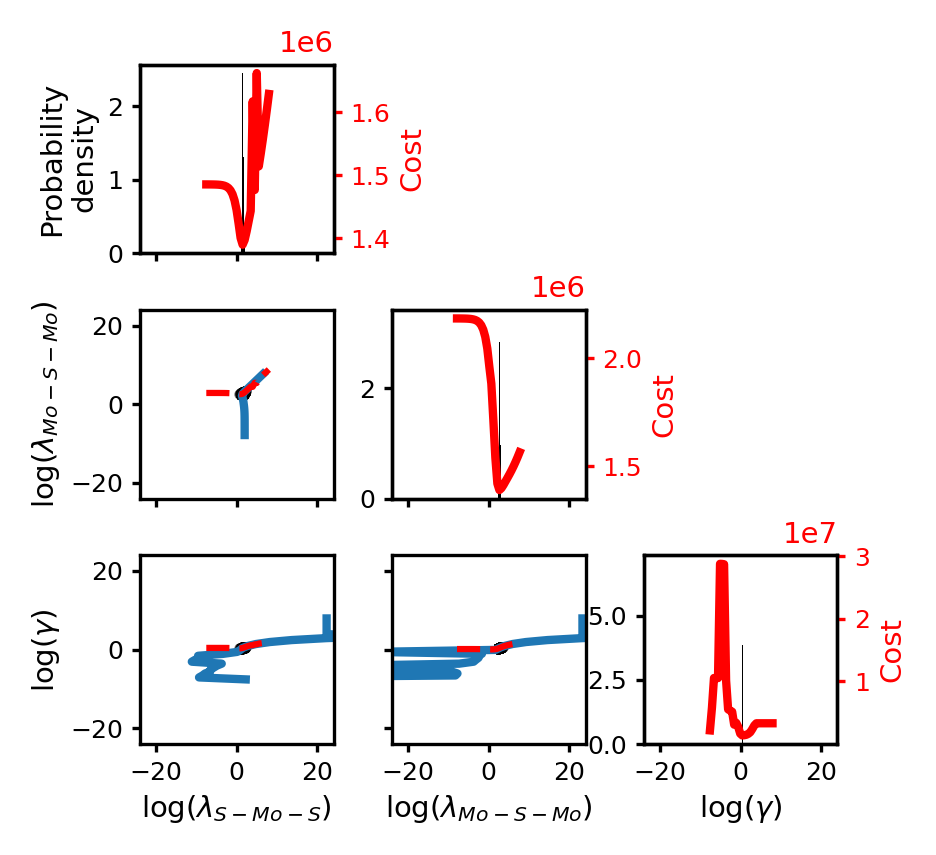}
    \caption[UQ results for SW potential 3-body parameters at $T = 1.71 \times 10^{-2}~T_0$]{
        Profile likelihood and MCMC samples for 3-body parameters at sampling temperature $1.71 \times 10^{-2}~T_0$ for the SW MoS$_2$ potential.
    }
\end{figure*}

\begin{figure*}[!h]
    \centering
    \includegraphics[width=0.6\textwidth]{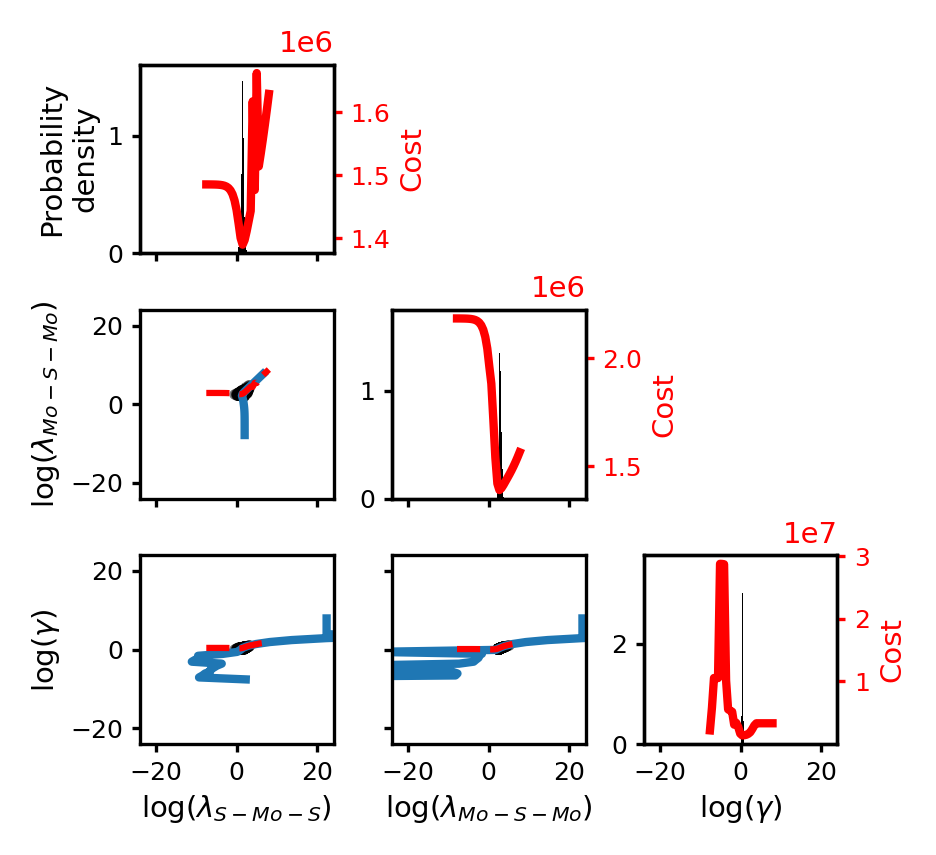}
    \caption[UQ results for SW potential 3-body parameters at $T = 5.40 \times 10^{-2}~T_0$]{
        Profile likelihood and MCMC samples for 3-body parameters at sampling temperature $5.40 \times 10^{-2}~T_0$ for the SW MoS$_2$ potential.
    }
\end{figure*}

\begin{figure*}[!h]
    \centering
    \includegraphics[width=0.6\textwidth]{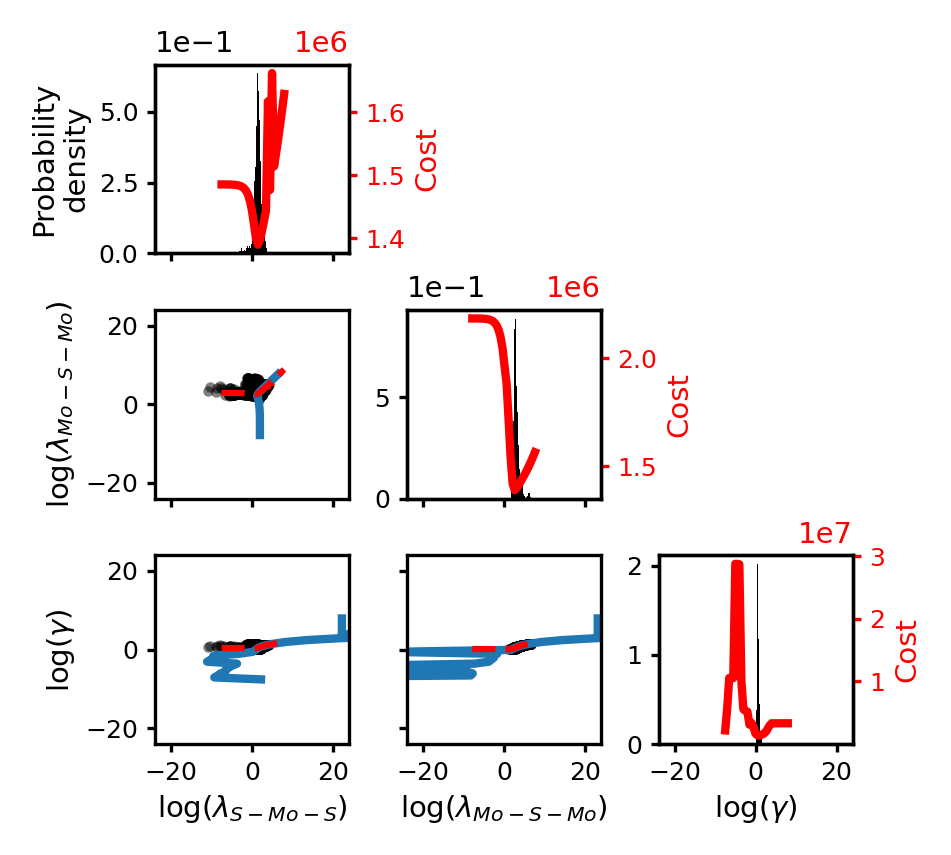}
    \caption[UQ results for SW potential 3-body parameters at $T = 1.71 \times 10^{-1}~T_0$]{
        Profile likelihood and MCMC samples for 3-body parameters at sampling temperature $1.71 \times 10^{-1}~T_0$ for the SW MoS$_2$ potential.
    }
\end{figure*}

\begin{figure*}[!h]
    \centering
    \includegraphics[width=0.6\textwidth]{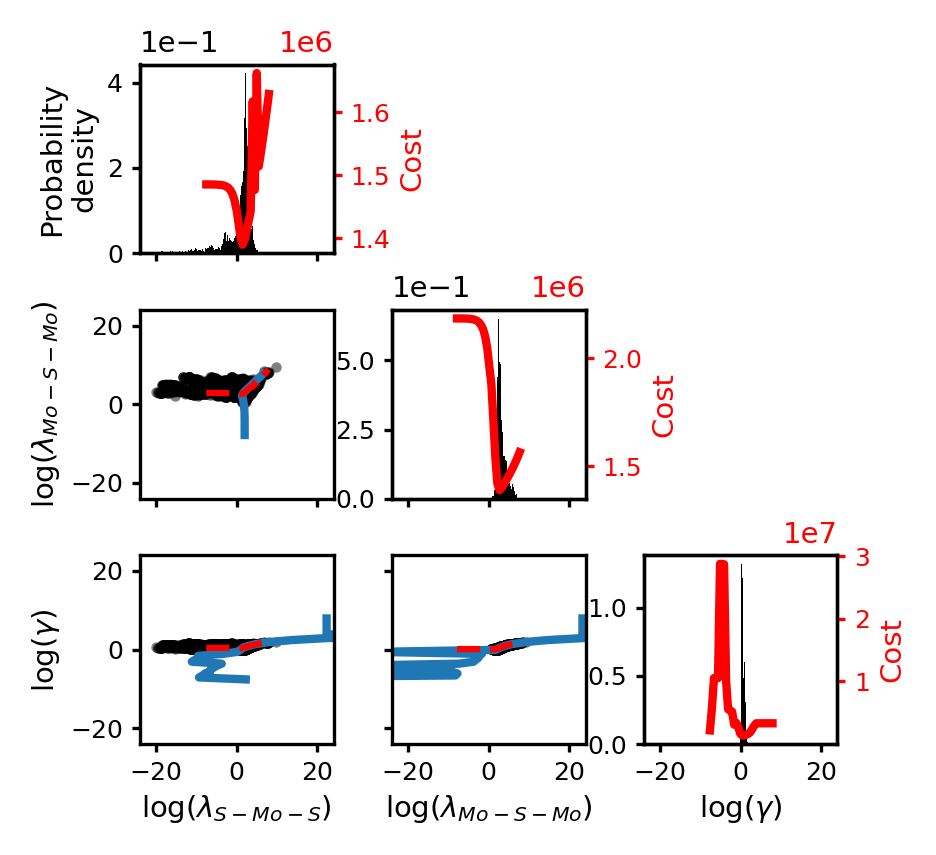}
    \caption[UQ results for SW potential 3-body parameters at $T = 5.40 \times 10^{-1}~T_0$]{
        Profile likelihood and MCMC samples for 3-body parameters at sampling temperature $5.40 \times 10^{-1}~T_0$ for the SW MoS$_2$ potential.
    }
\end{figure*}

\ifincludeTo
    \begin{figure*}[!h]
        \centering
        \includegraphics[width=0.6\textwidth]{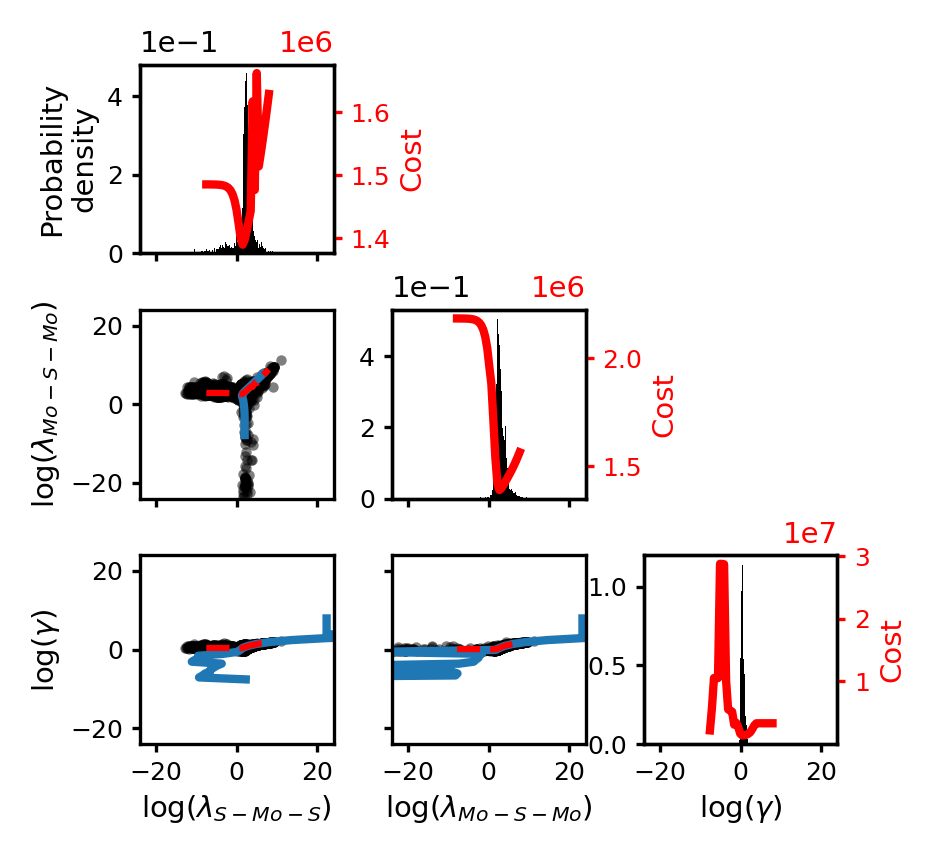}
        \caption[UQ results for SW potential 3-body parameters at $T = T_0$]{
            Profile likelihood and MCMC samples for 3-body parameters at sampling temperature $T_0$ for the SW MoS$_2$ potential.
        }
    \end{figure*}
\fi

\begin{figure*}[!h]
    \centering
    \includegraphics[width=0.6\textwidth]{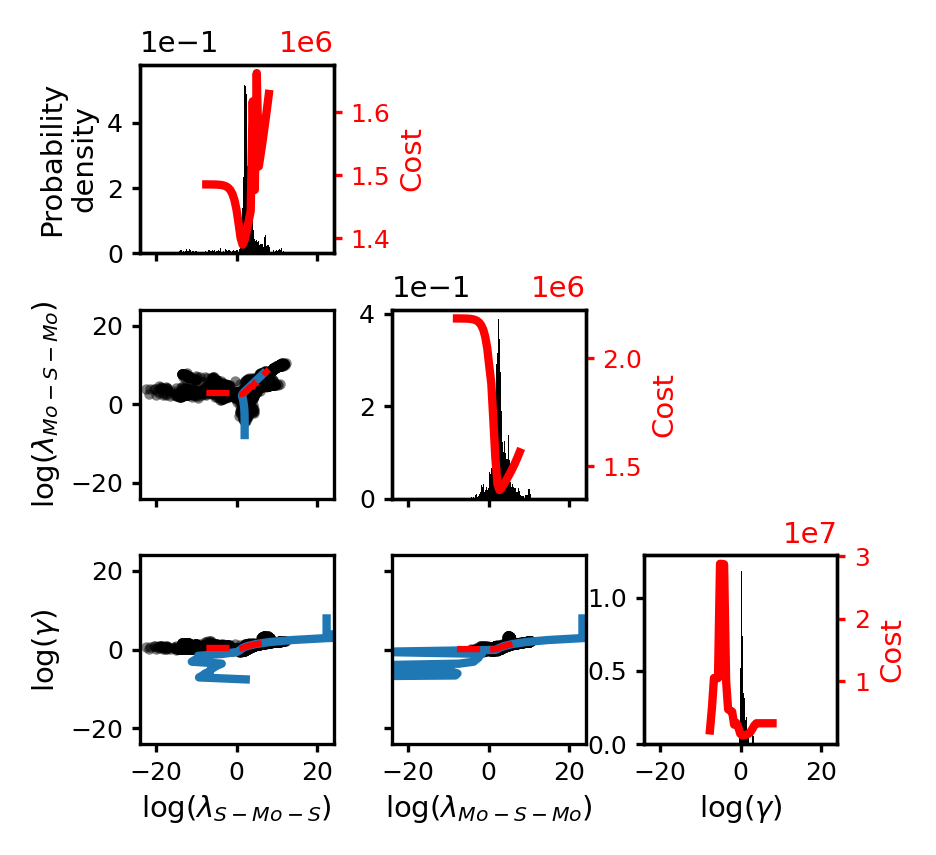}
    \caption[UQ results for SW potential 3-body parameters at $T = 1.71~T_0$]{
        Profile likelihood and MCMC samples for 3-body parameters at sampling temperature $1.71~T_0$ for the SW MoS$_2$ potential.
    }
\end{figure*}

\begin{figure*}[!h]
    \centering
    \includegraphics[width=0.6\textwidth]{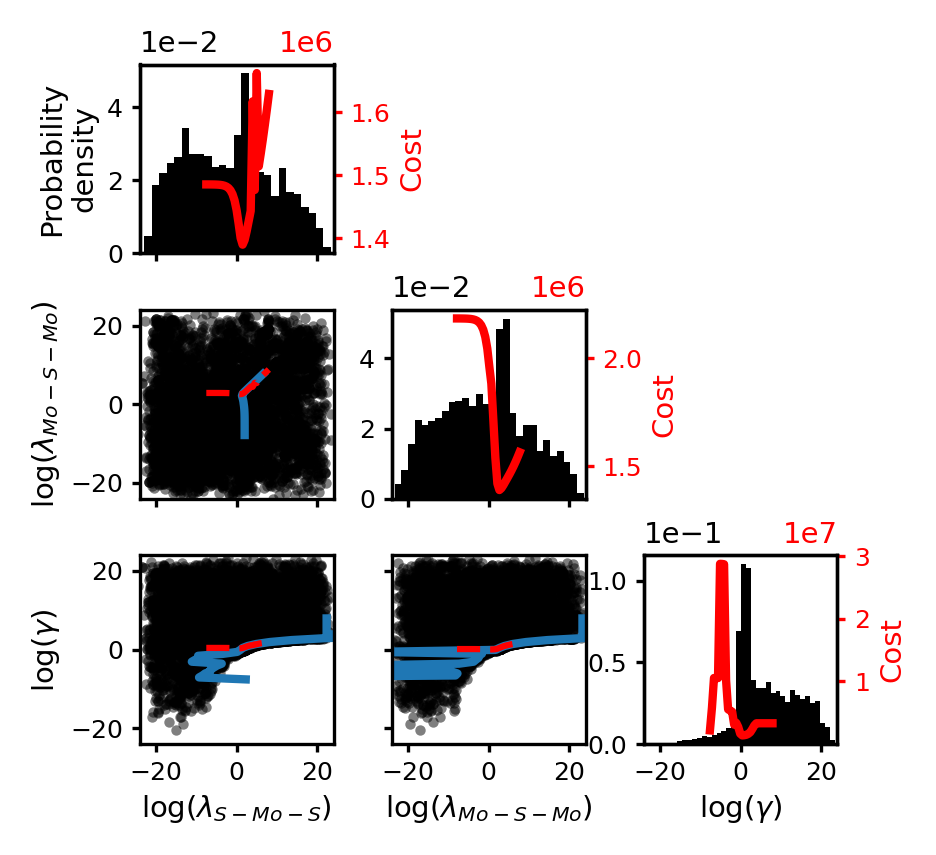}
    \caption[UQ results for SW potential 3-body parameters at $T = 5.40~T_0$]{
        Profile likelihood and MCMC samples for 3-body parameters at sampling temperature $5.40~T_0$ for the SW MoS$_2$ potential.
    }
\end{figure*}

\cleardoublepage

\subsection{Mo--Mo and Mo--S parameters}
\label{subsec:Mo-Mo_Mo-S}
Profile likelihood and MCMC samples between Mo--Mo and Mo--S parameters.
Notice that there is a lack of correlation between parameters corresponding to different interaction types.

\begin{figure*}[!h]
    \centering
    \includegraphics[width=0.6\textwidth]{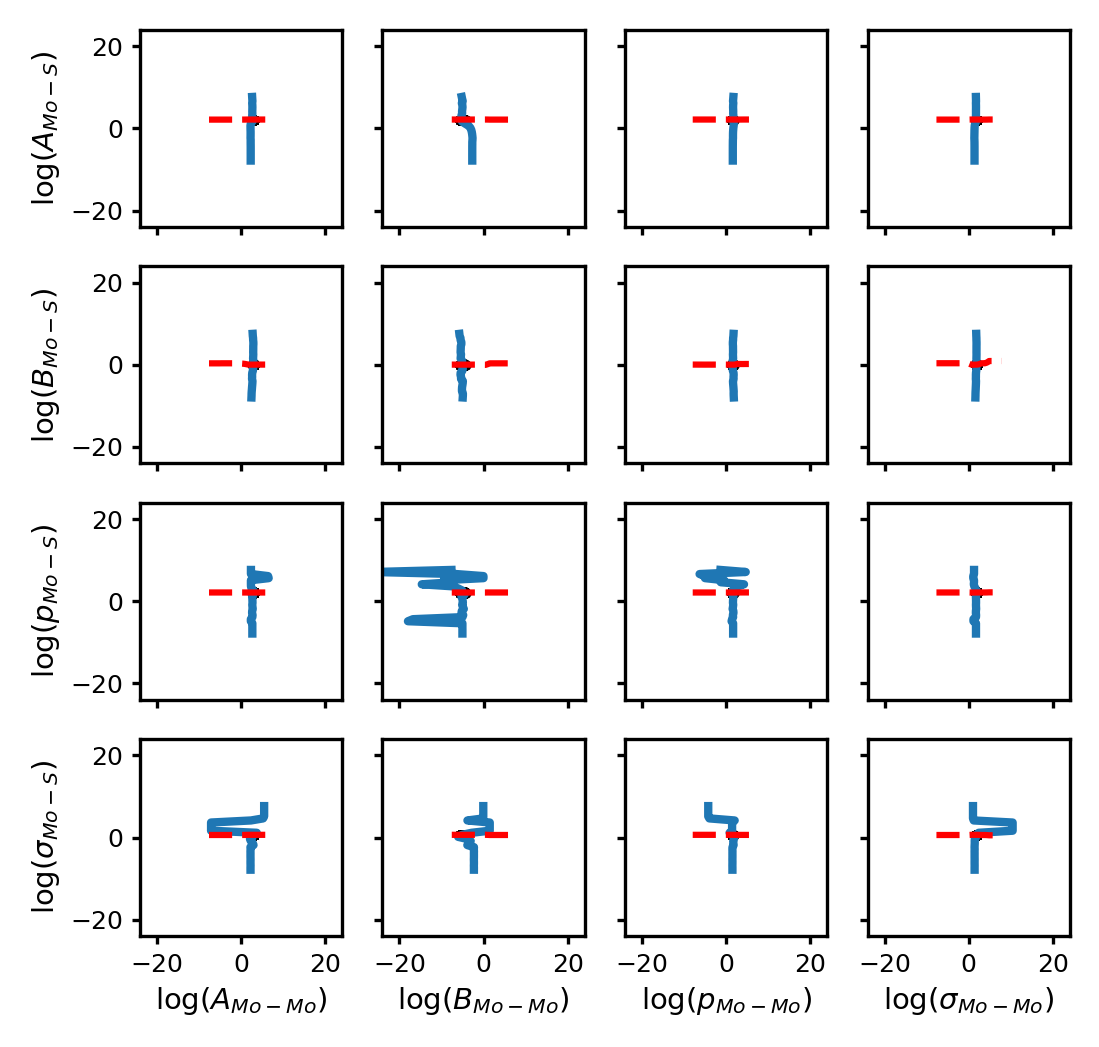}
    \caption[UQ results for SW potential Mo--Mo and Mo--S parameters at $T = 5.40\times10^{-6}~T_0$]{
        Profile likelihood and MCMC samples for Mo--Mo and Mo--S parameters at sampling temperature $5.40\times10^{-6}~T_0$ for the SW MoS$_2$ potential.
    }
\end{figure*}

\begin{figure*}[!h]
    \centering
    \includegraphics[width=0.6\textwidth]{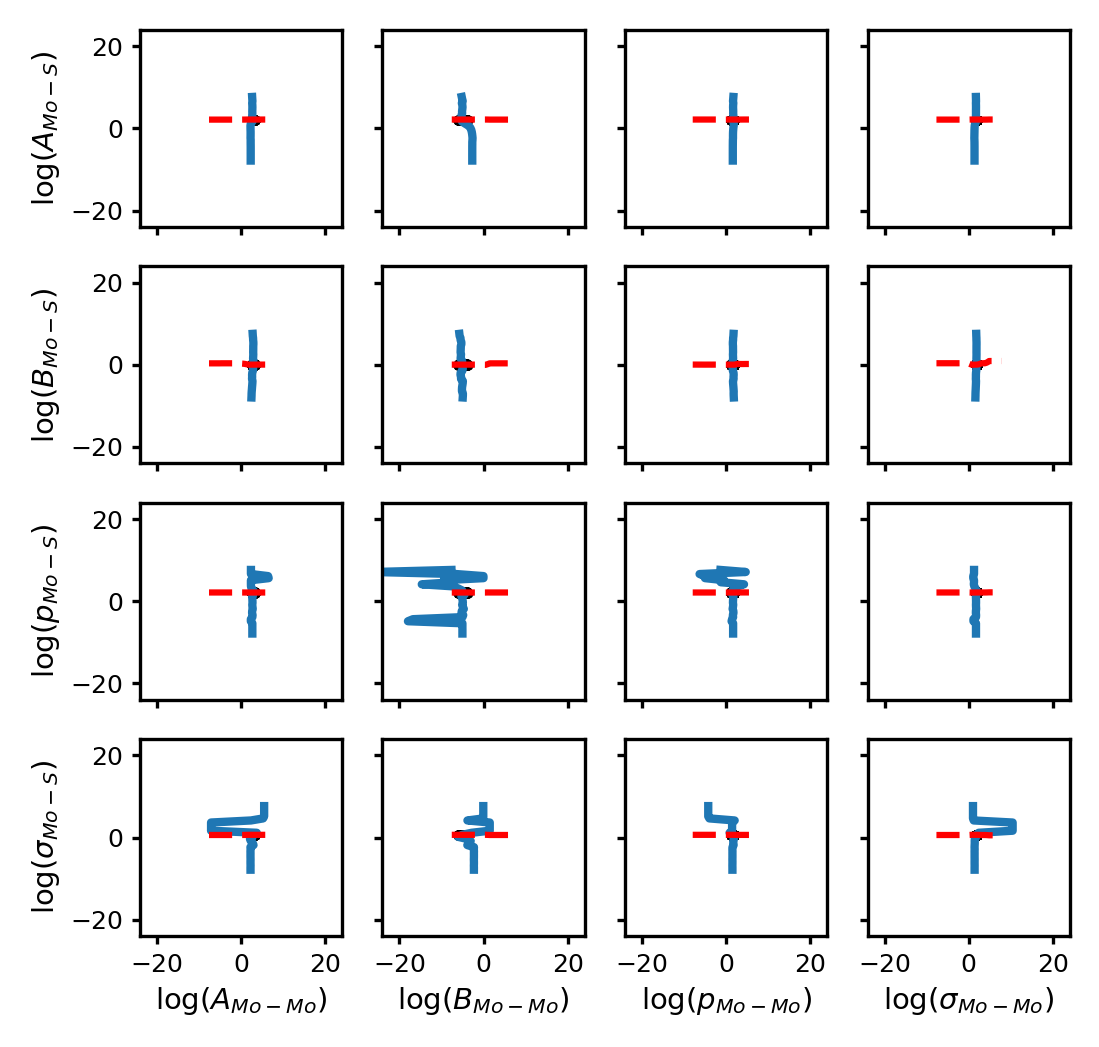}
    \caption[UQ results for SW potential Mo--Mo and Mo--S parameters at $T = 1.71\times10^{-5}~T_0$]{
        Profile likelihood and MCMC samples for Mo--Mo and Mo--S parameters at sampling temperature $1.71\times10^{-5}~T_0$ for the SW MoS$_2$ potential.
    }
\end{figure*}

\begin{figure*}[!h]
    \centering
    \includegraphics[width=0.6\textwidth]{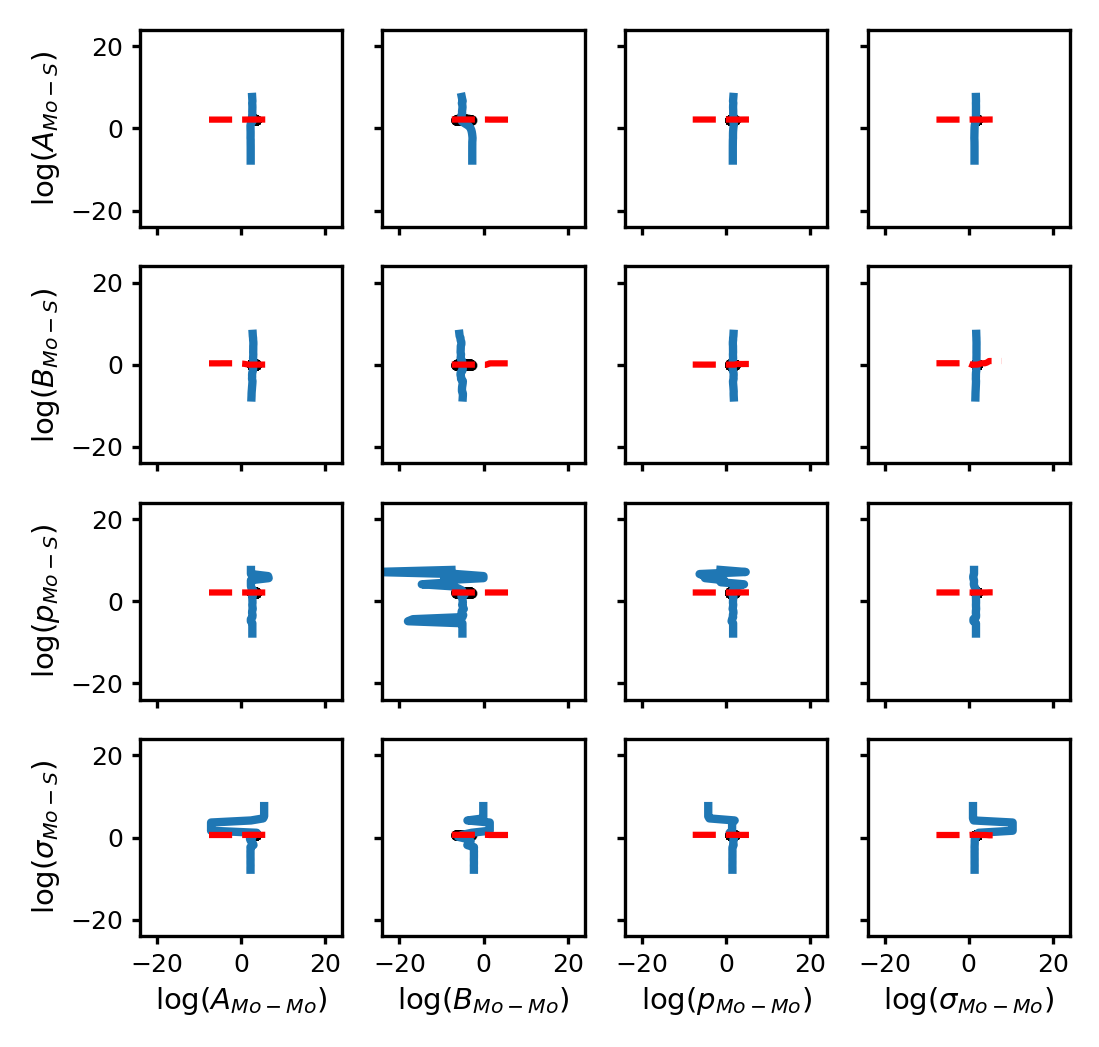}
    \caption[UQ results for SW potential Mo--Mo and Mo--S parameters at $T = 5.40\times10^{-5}~T_0$]{
        Profile likelihood and MCMC samples for Mo--Mo and Mo--S parameters at sampling temperature $5.40\times10^{-5}~T_0$ for the SW MoS$_2$ potential.
    }
\end{figure*}

\begin{figure*}[!h]
    \centering
    \includegraphics[width=0.6\textwidth]{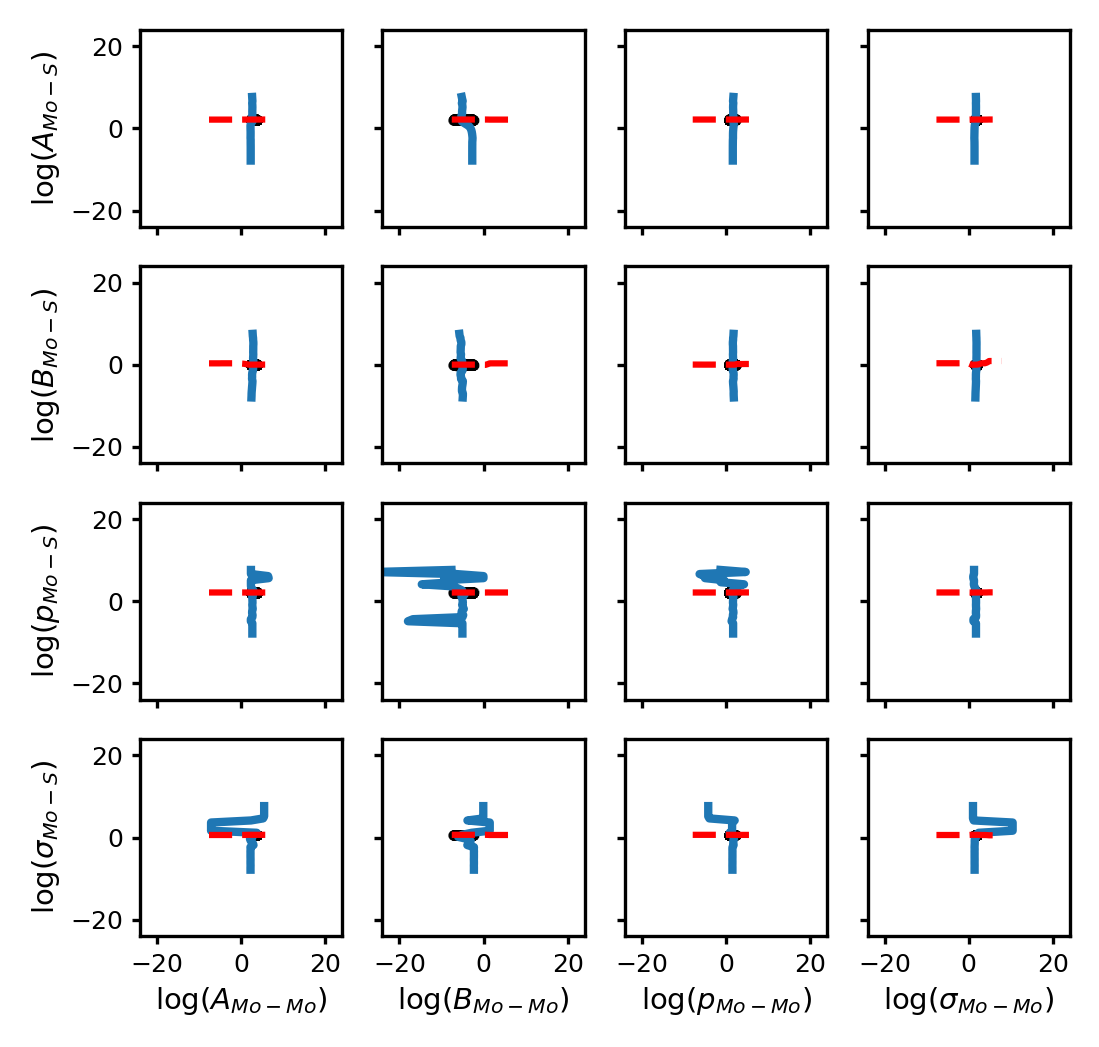}
    \caption[UQ results for SW potential Mo--Mo and Mo--S parameters at $T = 1.71\times10^{-4}~T_0$]{
        Profile likelihood and MCMC samples for Mo--Mo and Mo--S parameters at sampling temperature $1.71\times10^{-4}~T_0$ for the SW MoS$_2$ potential.
    }
\end{figure*}

\begin{figure*}[!h]
    \centering
    \includegraphics[width=0.6\textwidth]{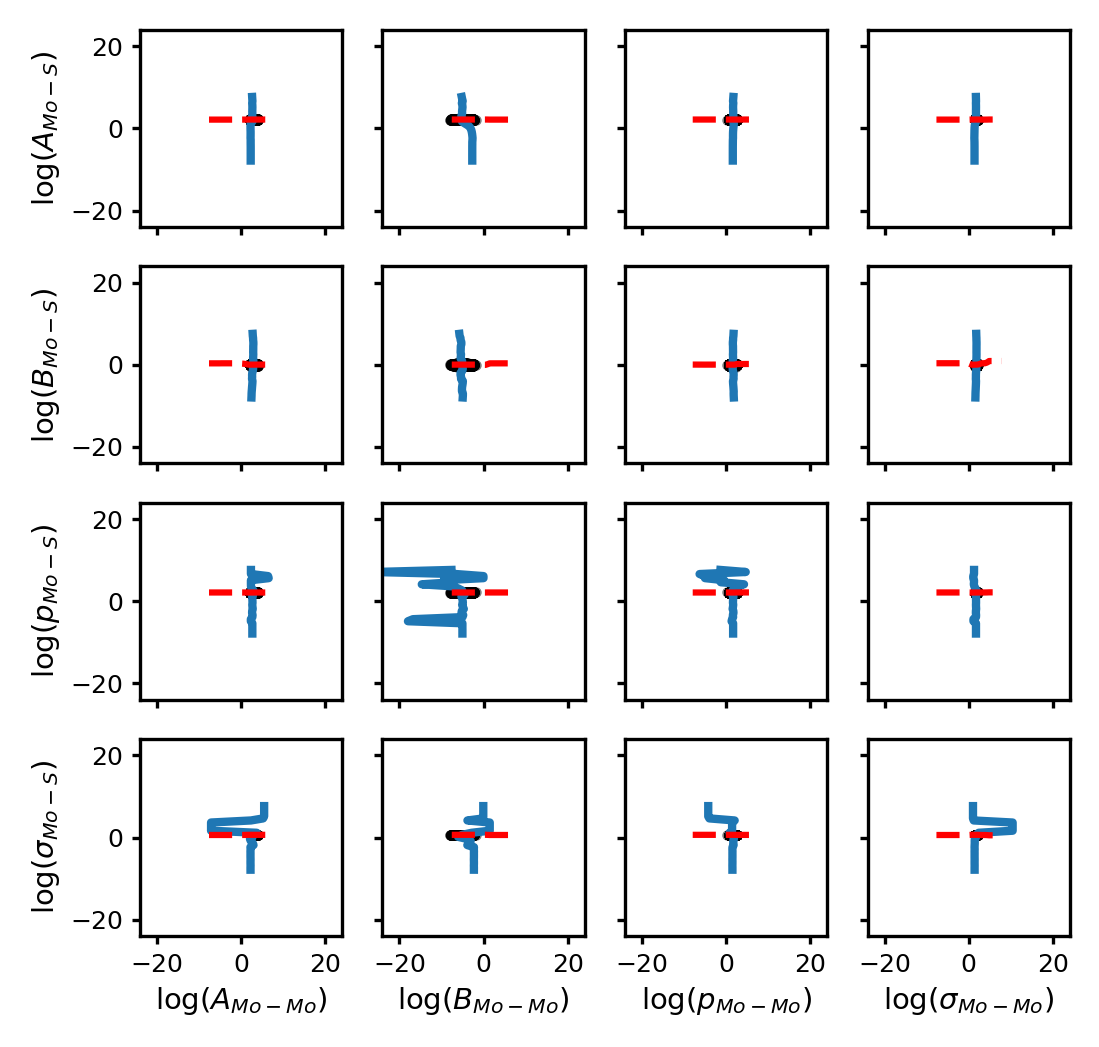}
    \caption[UQ results for SW potential Mo--Mo and Mo--S parameters at $T = 5.40\times10^{-4}~T_0$]{
        Profile likelihood and MCMC samples for Mo--Mo and Mo--S parameters at sampling temperature $5.40\times10^{-4}~T_0$ for the SW MoS$_2$ potential.
    }
\end{figure*}

\begin{figure*}[!h]
    \centering
    \includegraphics[width=0.6\textwidth]{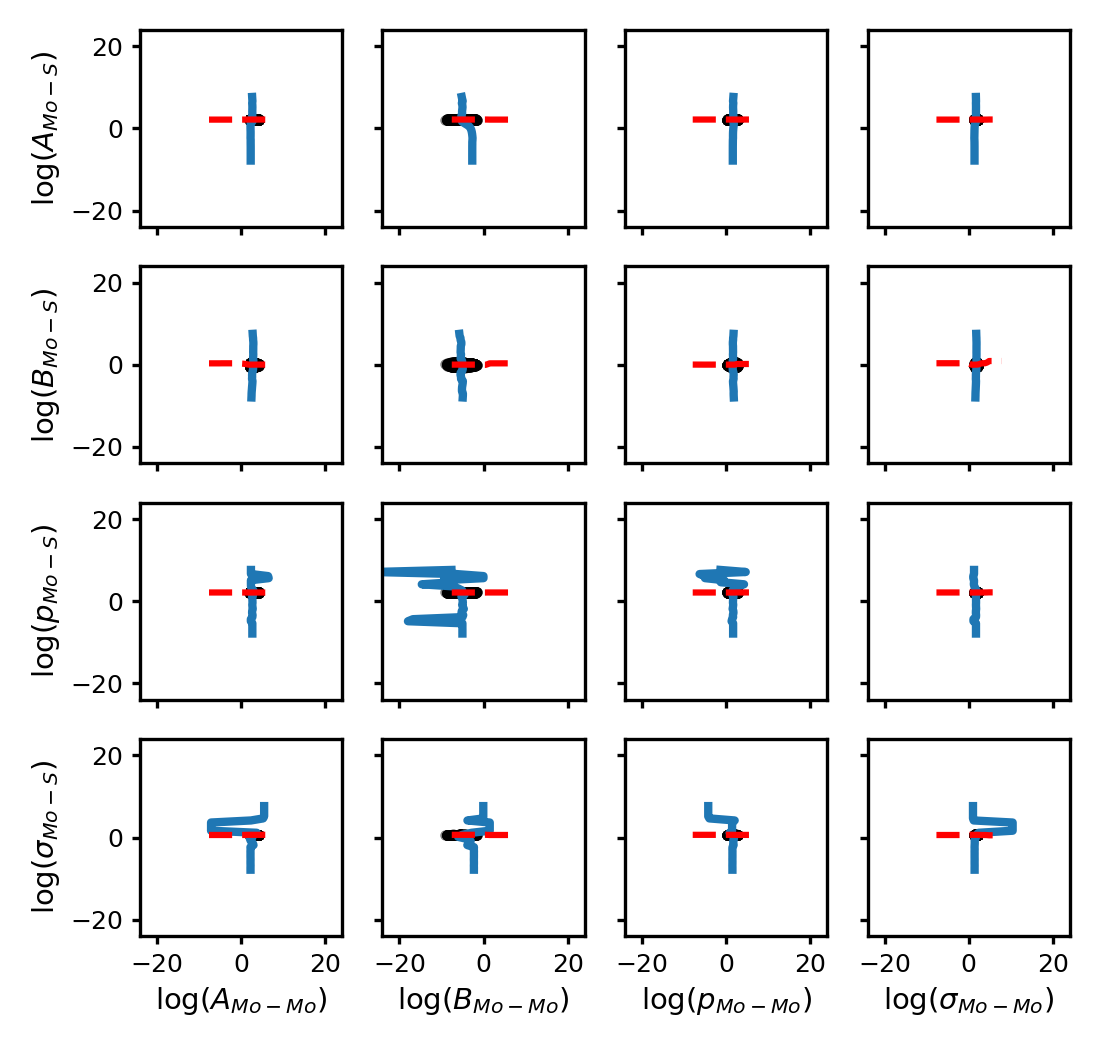}
    \caption[UQ results for SW potential Mo--Mo and Mo--S parameters at $T = 1.71\times10^{-3}~T_0$]{
        Profile likelihood and MCMC samples for Mo--Mo and Mo--S parameters at sampling temperature $1.71\times10^{-3}~T_0$ for the SW MoS$_2$ potential.
    }
\end{figure*}

\begin{figure*}[!h]
    \centering
    \includegraphics[width=0.6\textwidth]{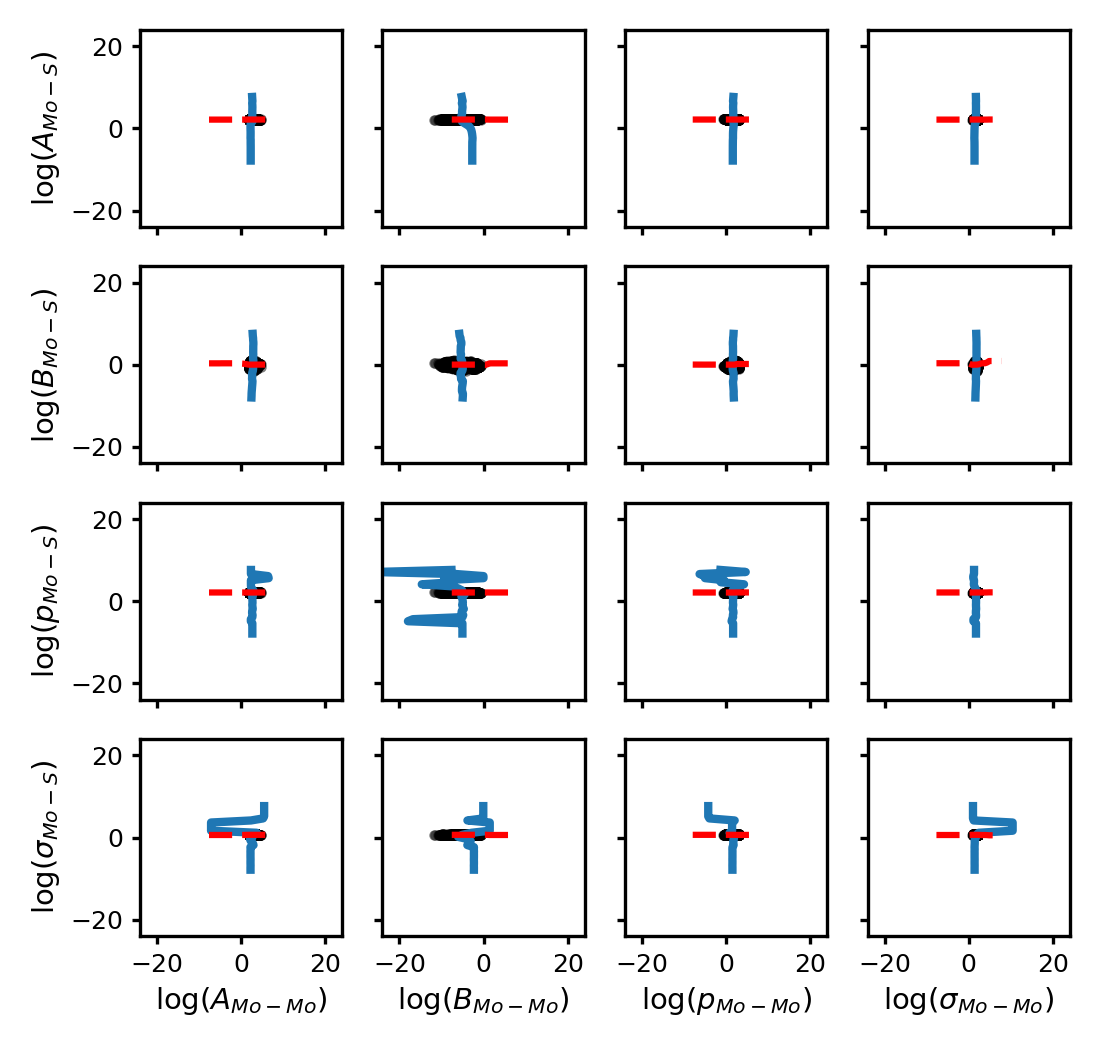}
    \caption[UQ results for SW potential Mo--Mo and Mo--S parameters at $T = 5.40\times10^{-3}~T_0$]{
        Profile likelihood and MCMC samples for Mo--Mo and Mo--S parameters at sampling temperature $5.40\times10^{-3}~T_0$ for the SW MoS$_2$ potential.
    }
\end{figure*}

\begin{figure*}[!h]
    \centering
    \includegraphics[width=0.6\textwidth]{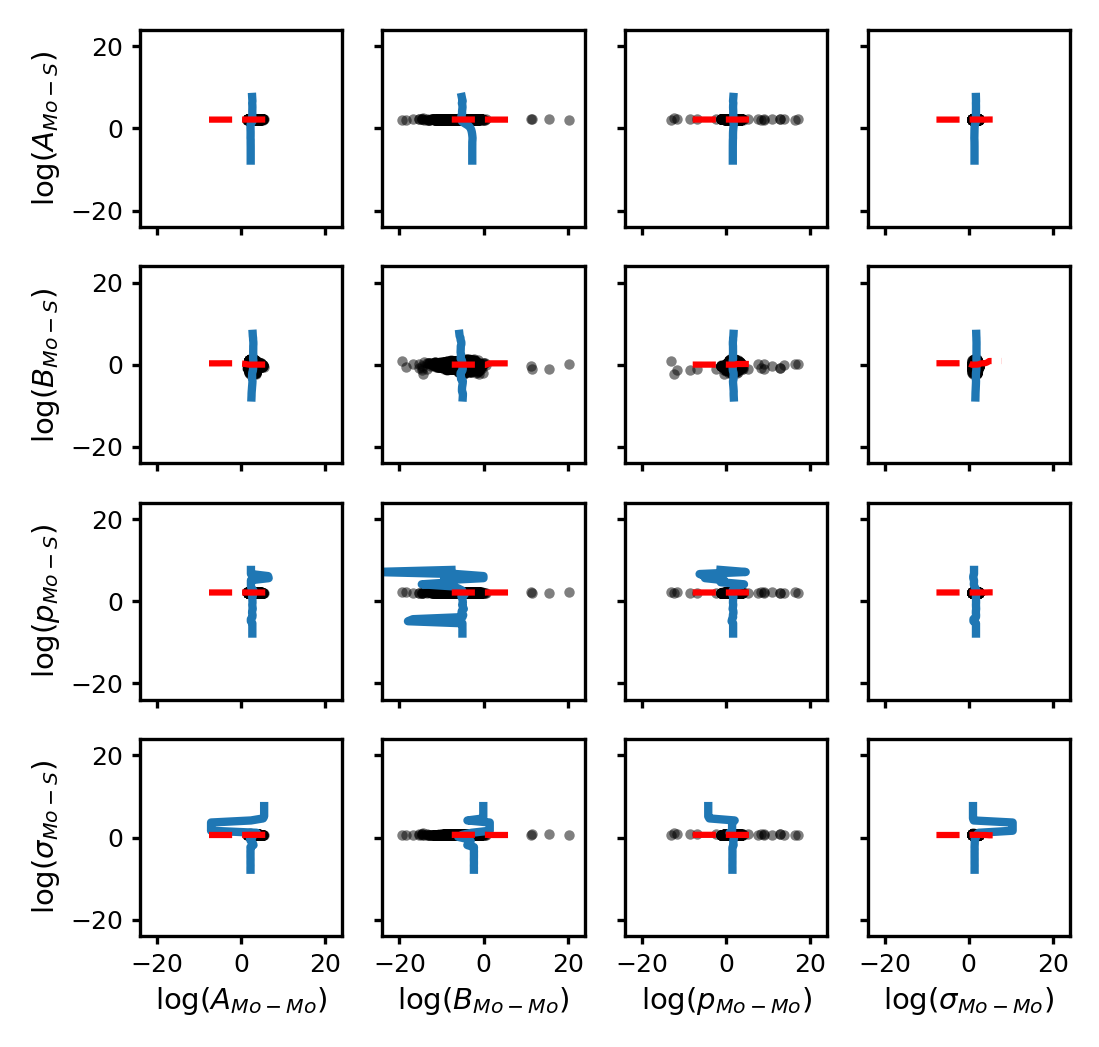}
    \caption[UQ results for SW potential Mo--Mo and Mo--S parameters at $T = 1.71\times10^{-2}~T_0$]{
        Profile likelihood and MCMC samples for Mo--Mo and Mo--S parameters at sampling temperature $1.71\times10^{-2}~T_0$ for the SW MoS$_2$ potential.
    }
\end{figure*}

\begin{figure*}[!h]
    \centering
    \includegraphics[width=0.6\textwidth]{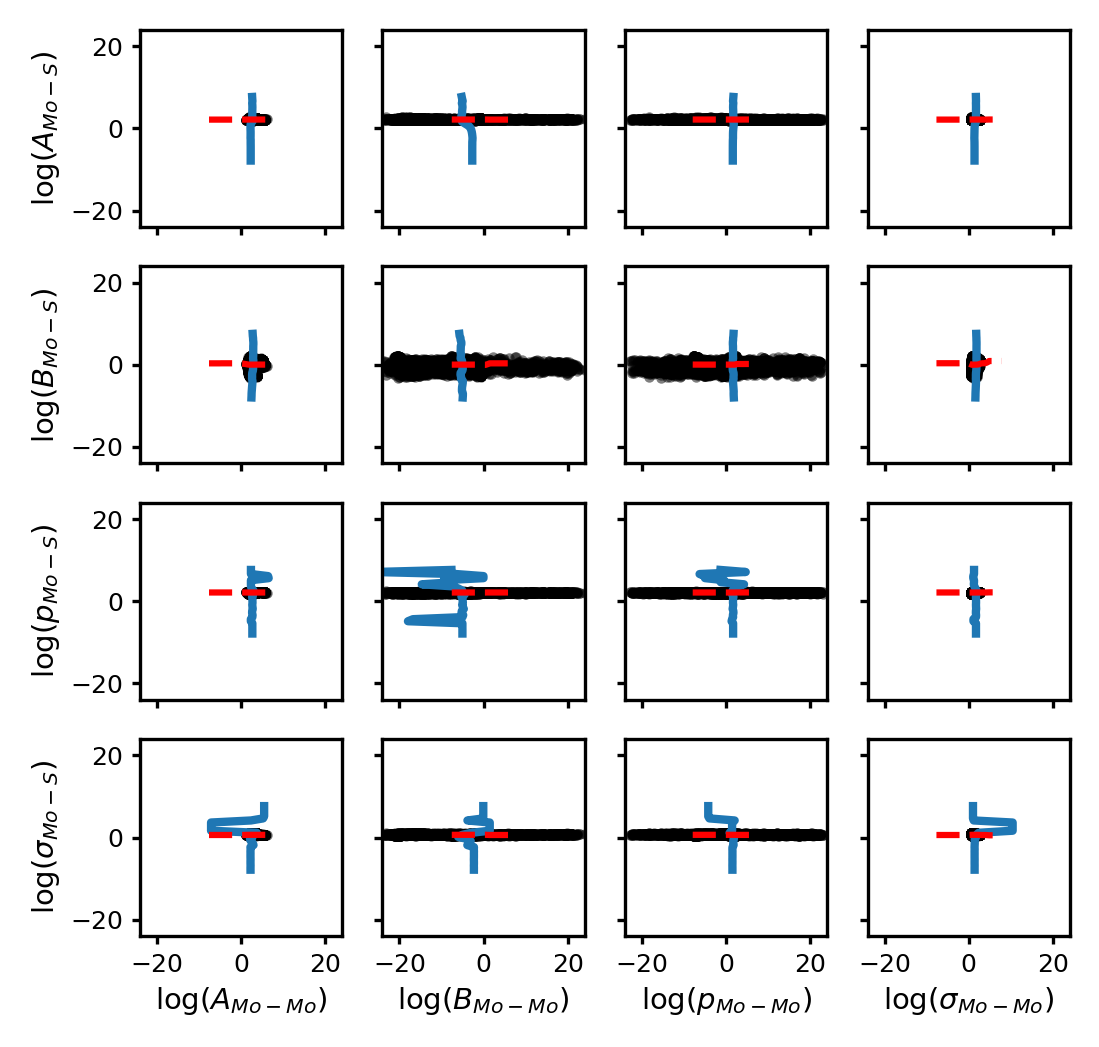}
    \caption[UQ results for SW potential Mo--Mo and Mo--S parameters at $T = 5.40\times10^{-2}~T_0$]{
        Profile likelihood and MCMC samples for Mo--Mo and Mo--S parameters at sampling temperature $5.40\times10^{-2}~T_0$ for the SW MoS$_2$ potential.
    }
\end{figure*}

\begin{figure*}[!h]
    \centering
    \includegraphics[width=0.6\textwidth]{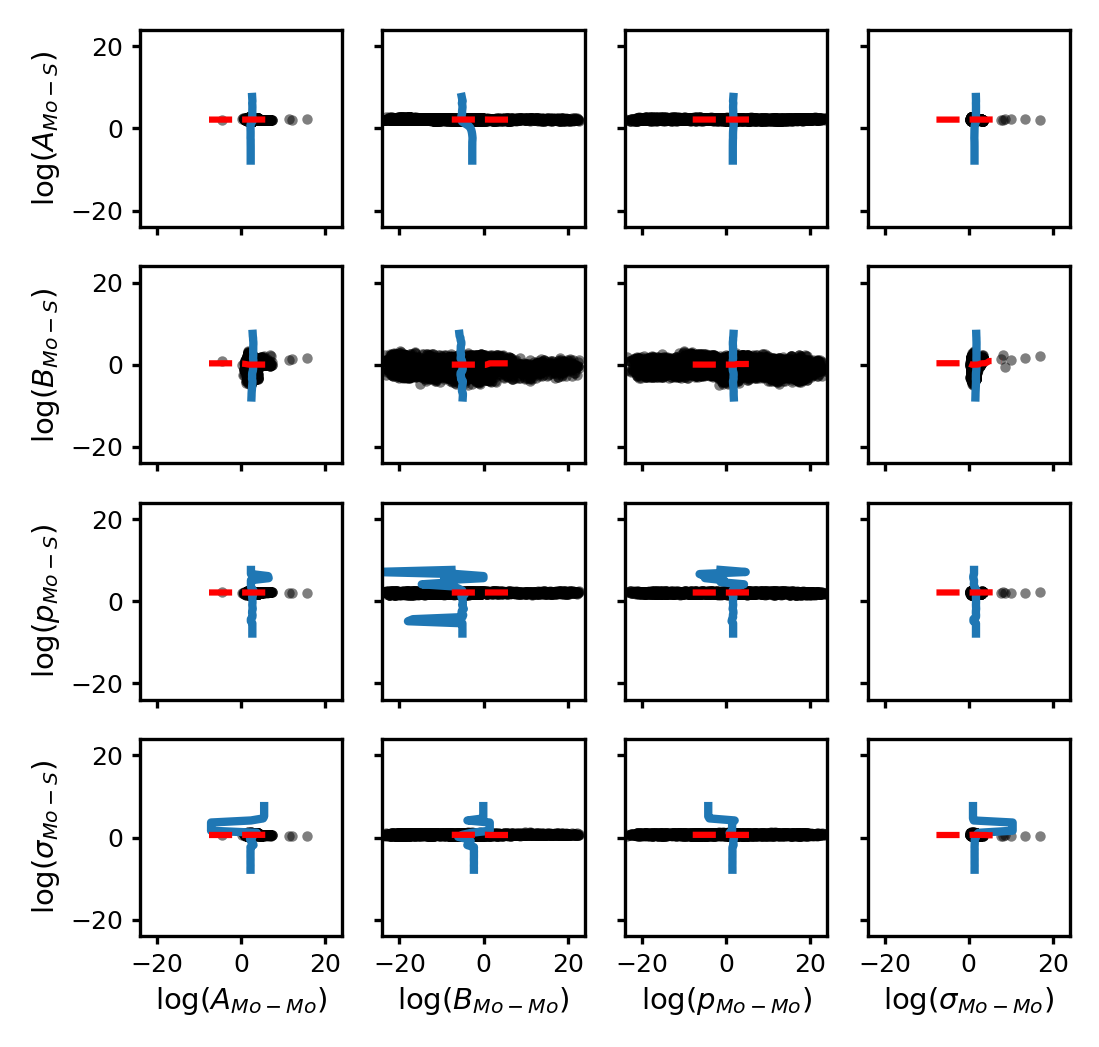}
    \caption[UQ results for SW potential Mo--Mo and Mo--S parameters at $T = 1.71\times10^{-1}~T_0$]{
        Profile likelihood and MCMC samples for Mo--Mo and Mo--S parameters at sampling temperature $1.71\times10^{-1}~T_0$ for the SW MoS$_2$ potential.
    }
\end{figure*}

\begin{figure*}[!h]
    \centering
    \includegraphics[width=0.6\textwidth]{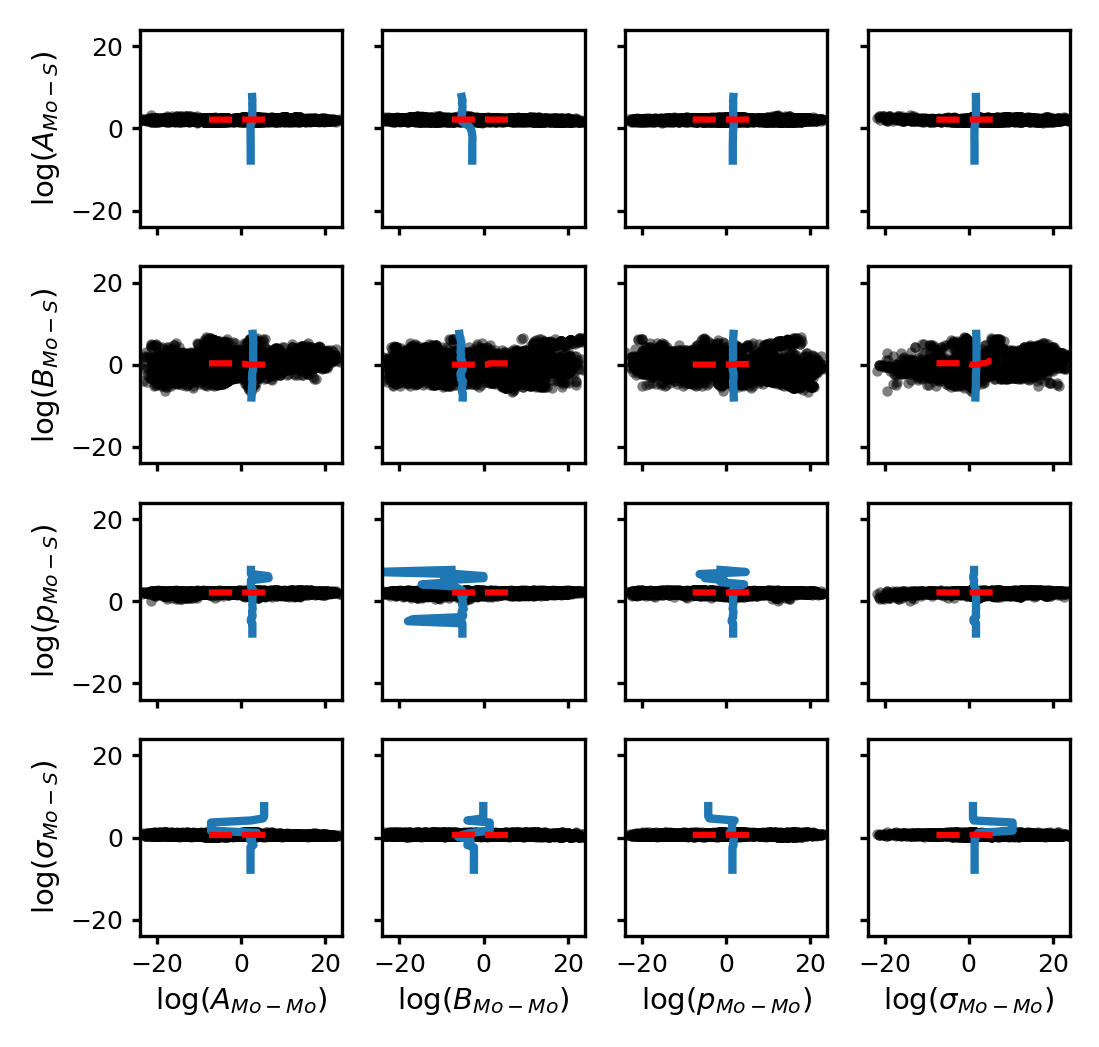}
    \caption[UQ results for SW potential Mo--Mo and Mo--S parameters at $T = 5.40\times10^{-1}~T_0$]{
        Profile likelihood and MCMC samples for Mo--Mo and Mo--S parameters at sampling temperature $5.40\times10^{-1}~T_0$ for the SW MoS$_2$ potential.
    }
\end{figure*}

\ifincludeTo
    \begin{figure*}[!h]
        \centering
        \includegraphics[width=0.6\textwidth]{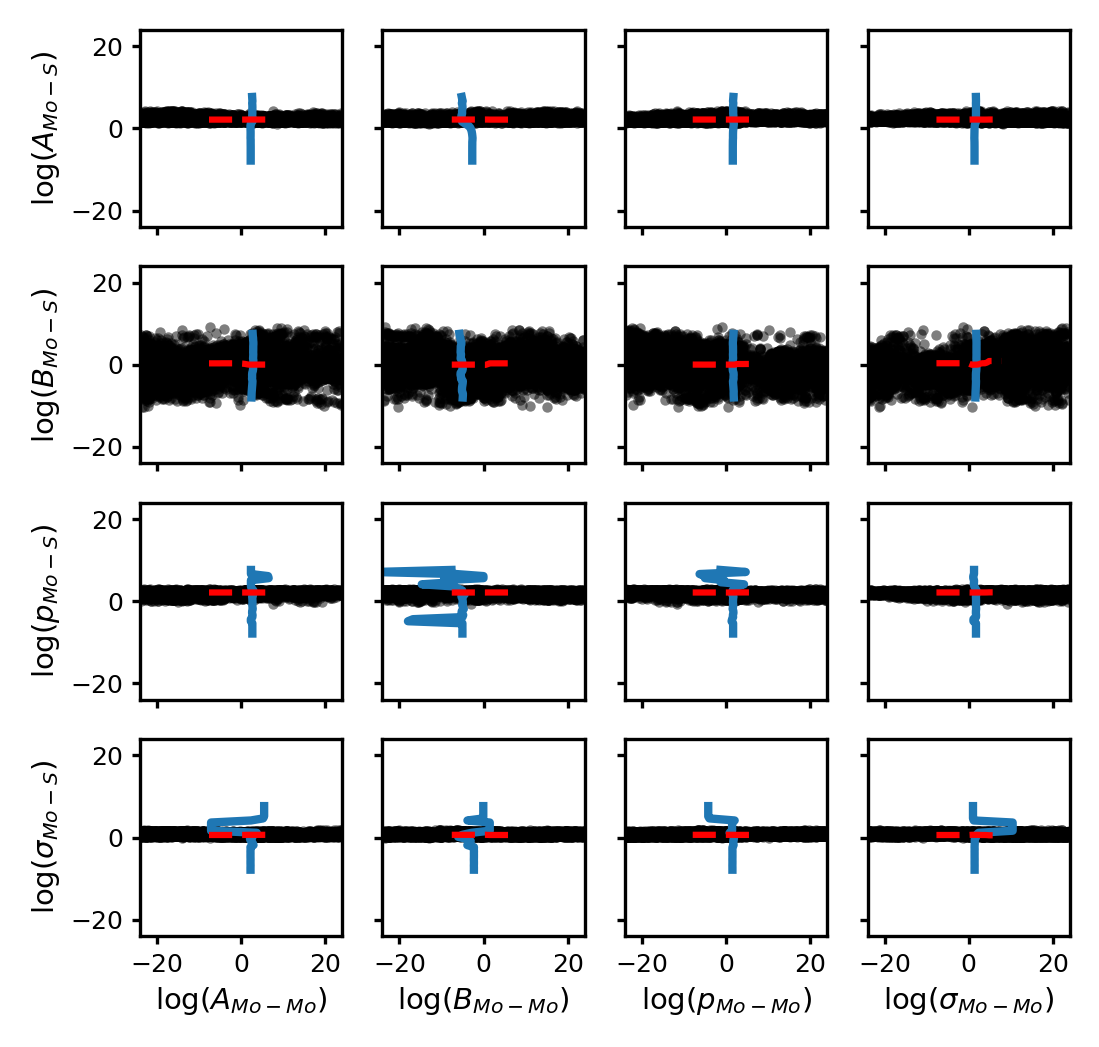}
        \caption[UQ results for SW potential Mo--Mo and Mo--S parameters at $T = T_0$]{
            Profile likelihood and MCMC samples for Mo--Mo and Mo--S parameters at sampling temperature $T_0$ for the SW MoS$_2$ potential.
        }
    \end{figure*}
\fi

\begin{figure*}[!h]
    \centering
    \includegraphics[width=0.6\textwidth]{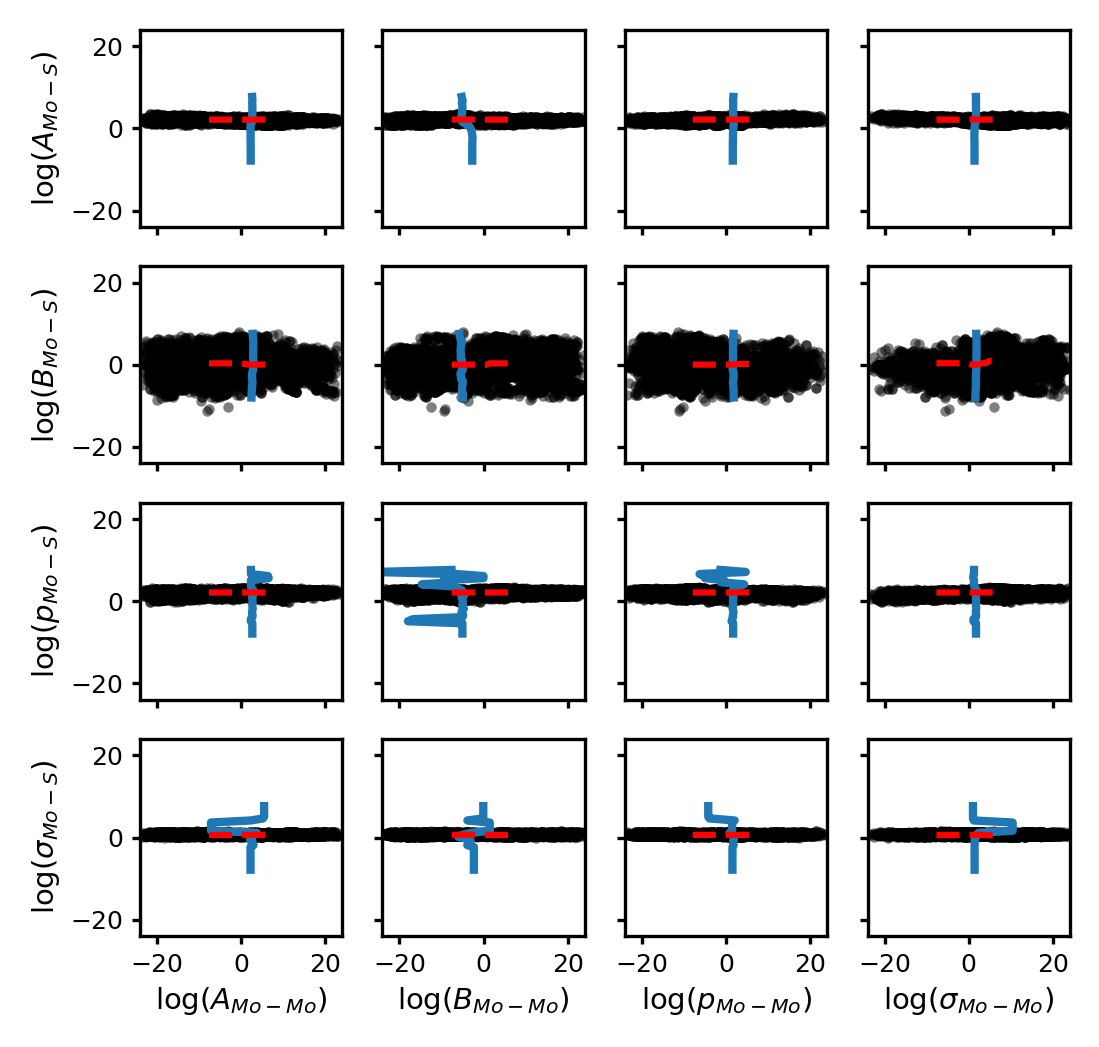}
    \caption[UQ results for SW potential Mo--Mo and Mo--S parameters at $T = 1.71~T_0$]{
        Profile likelihood and MCMC samples for Mo--Mo and Mo--S parameters at sampling temperature $1.71~T_0$ for the SW MoS$_2$ potential.
    }
\end{figure*}

\begin{figure*}[!h]
    \centering
    \includegraphics[width=0.6\textwidth]{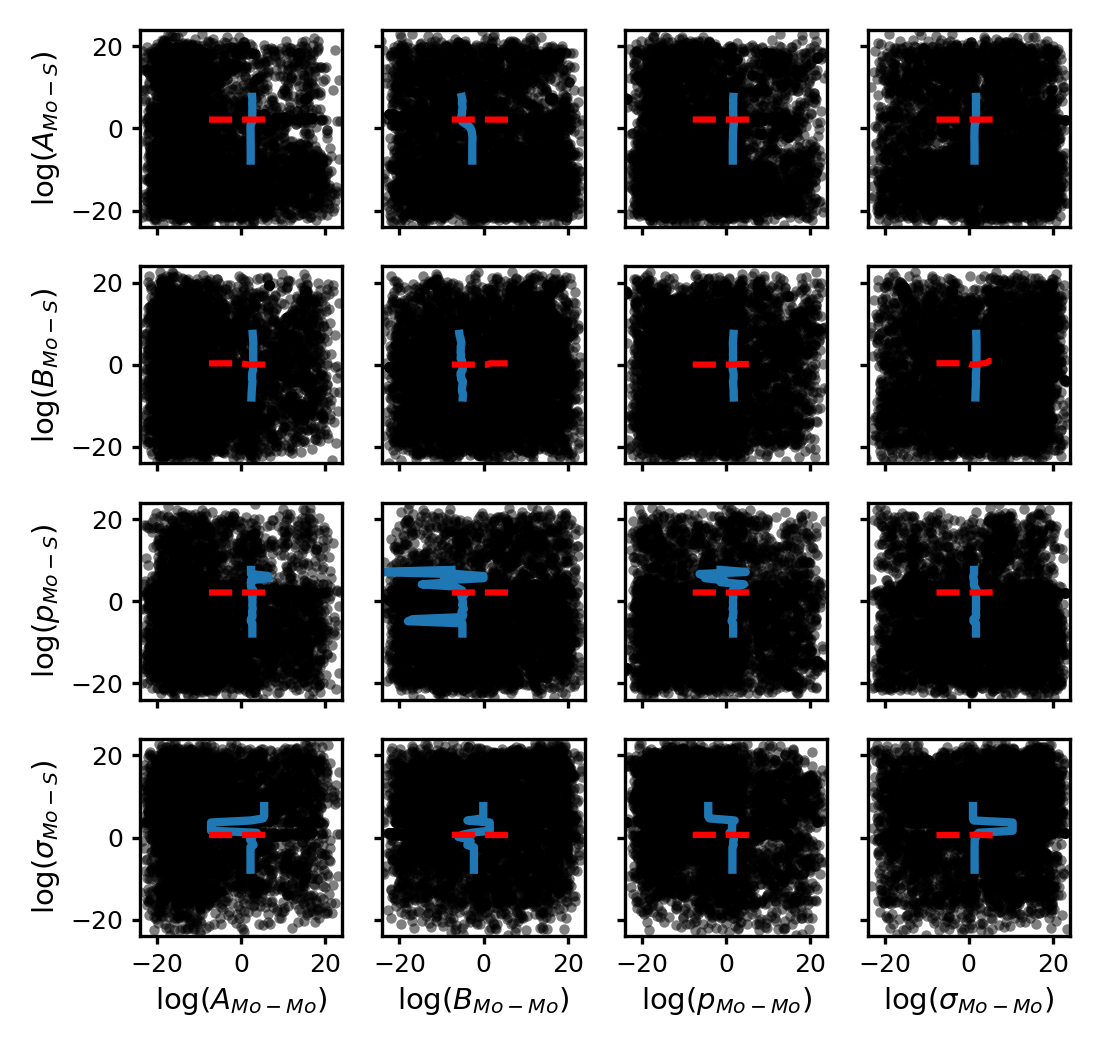}
    \caption[UQ results for SW potential Mo--Mo and Mo--S parameters at $T = 5.40~T_0$]{
        Profile likelihood and MCMC samples for Mo--Mo and Mo--S parameters at sampling temperature $5.40~T_0$ for the SW MoS$_2$ potential.
    }
\end{figure*}

\cleardoublepage

\subsection{Mo--Mo and S--S parameters}
\label{subsec:Mo-Mo_S-S}
Profile likelihood and MCMC samples between Mo--Mo and S--S parameters.
Notice that there is a lack of correlation between between parameters corresponding to different interaction types.

\begin{figure*}[!h]
    \centering
    \includegraphics[width=0.6\textwidth]{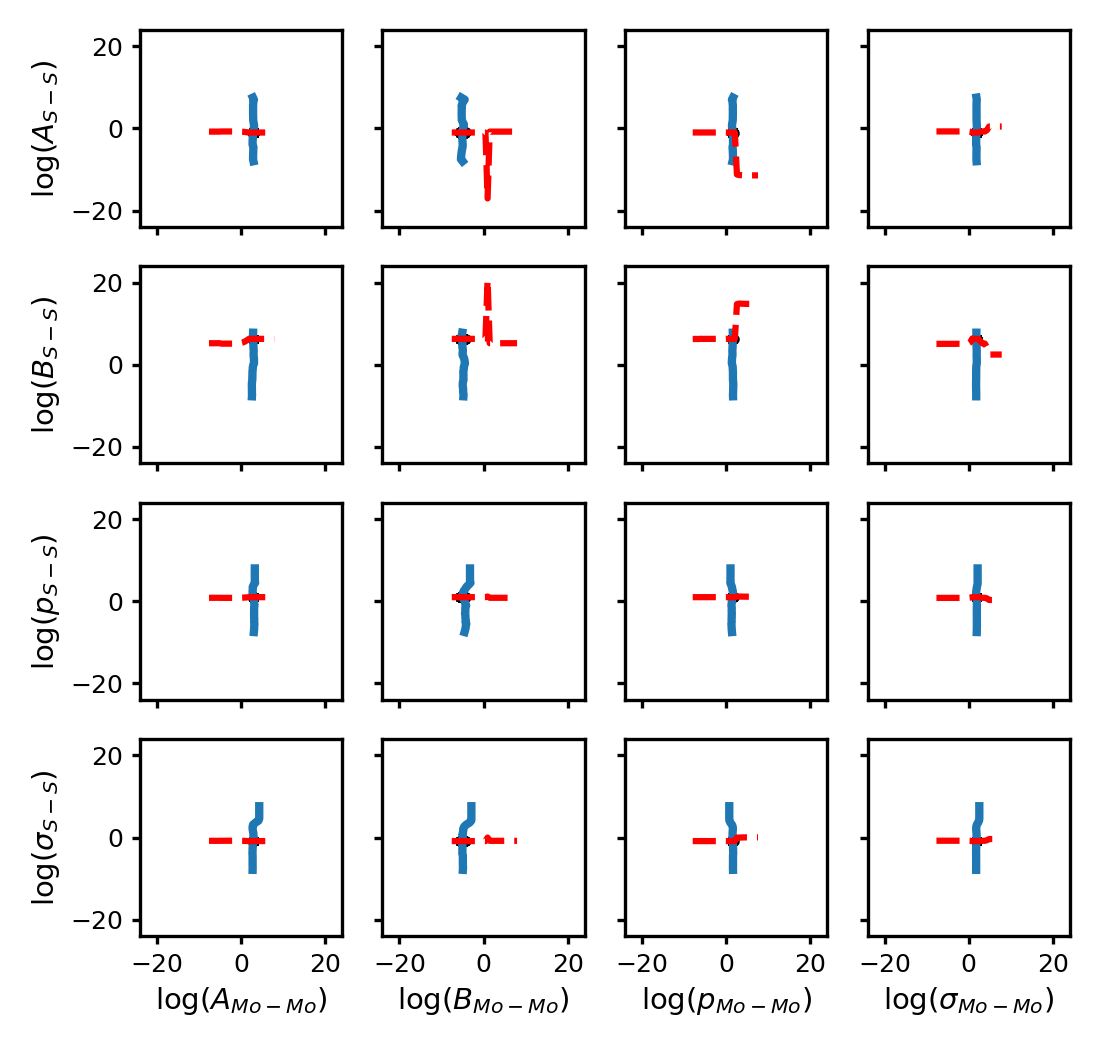}
    \caption[UQ results for SW potential Mo--Mo and S--S parameters at $T = 5.40\times10^{-6}~T_0$]{
        Profile likelihood and MCMC samples for Mo--Mo and S--S parameters at sampling temperature $5.40\times10^{-6}~T_0$ for the SW MoS$_2$ potential.
    }
\end{figure*}

\begin{figure*}[!h]
    \centering
    \includegraphics[width=0.6\textwidth]{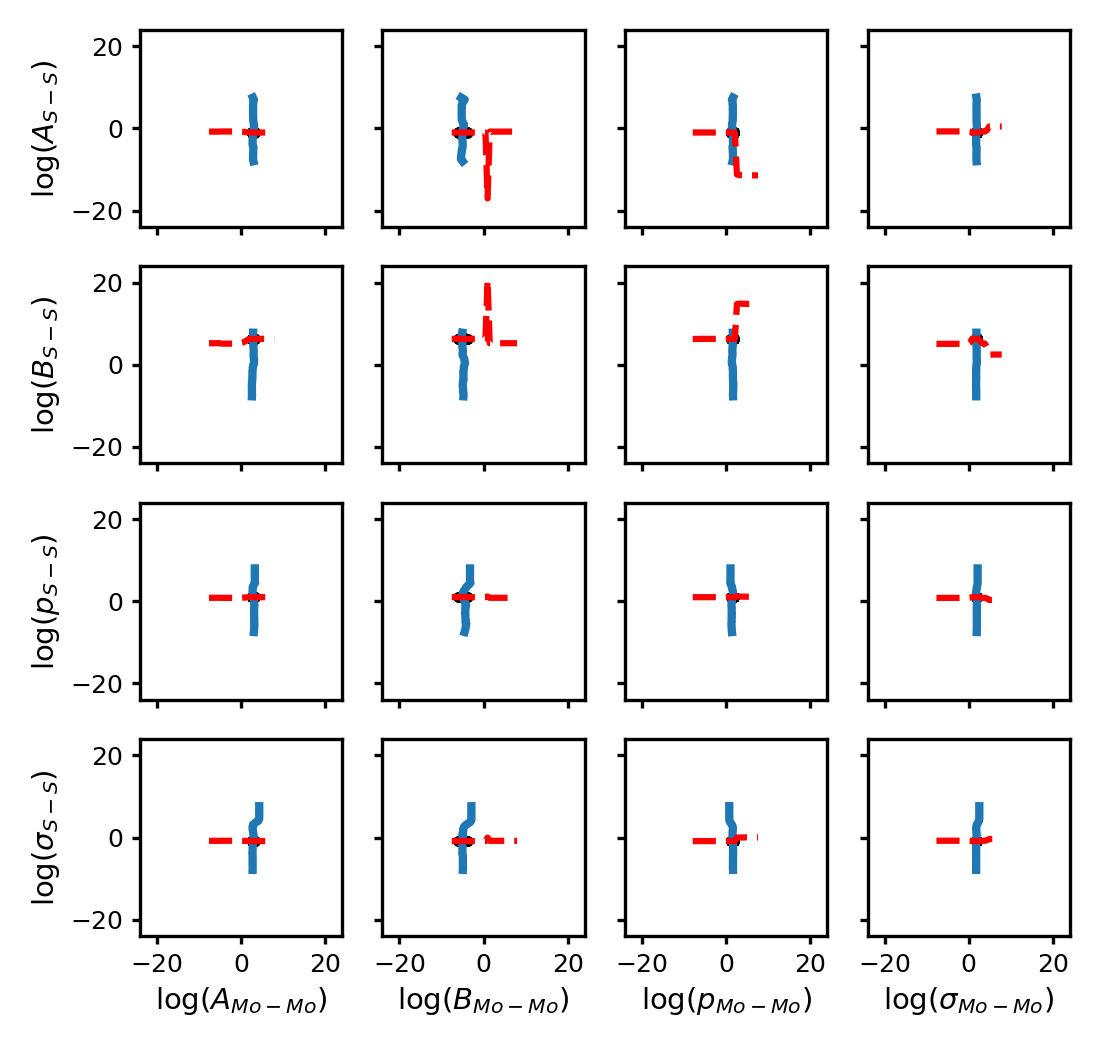}
    \caption[UQ results for SW potential Mo--Mo and S--S parameters at $T = 1.71\times10^{-5}~T_0$]{
        Profile likelihood and MCMC samples for Mo--Mo and S--S parameters at sampling temperature $1.71\times10^{-5}~T_0$ for the SW MoS$_2$ potential.
    }
\end{figure*}

\begin{figure*}[!h]
    \centering
    \includegraphics[width=0.6\textwidth]{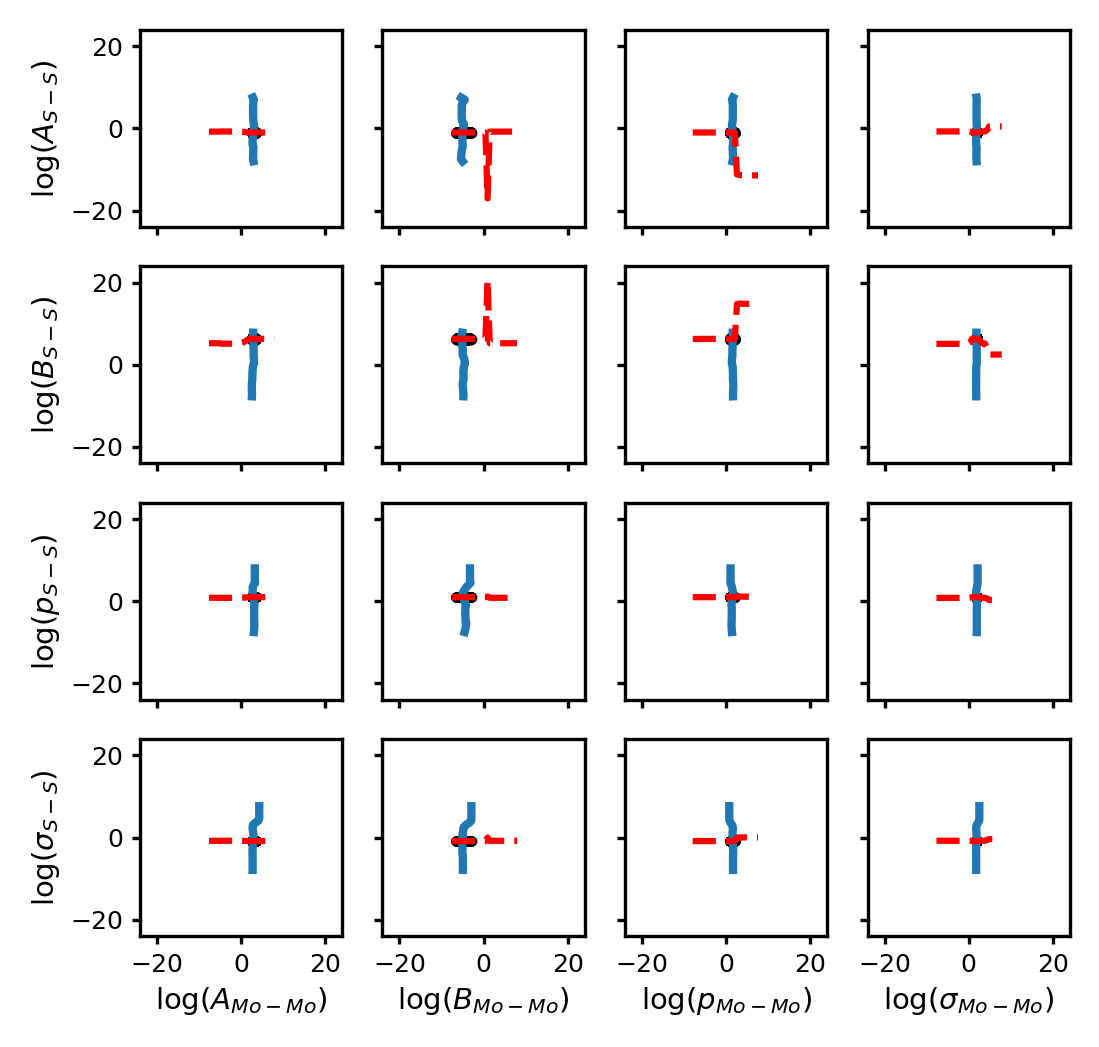}
    \caption[UQ results for SW potential Mo--Mo and S--S parameters at $T = 5.40\times10^{-5}~T_0$]{
        Profile likelihood and MCMC samples for Mo--Mo and S--S parameters at sampling temperature $5.40\times10^{-5}~T_0$ for the SW MoS$_2$ potential.
    }
\end{figure*}

\begin{figure*}[!h]
    \centering
    \includegraphics[width=0.6\textwidth]{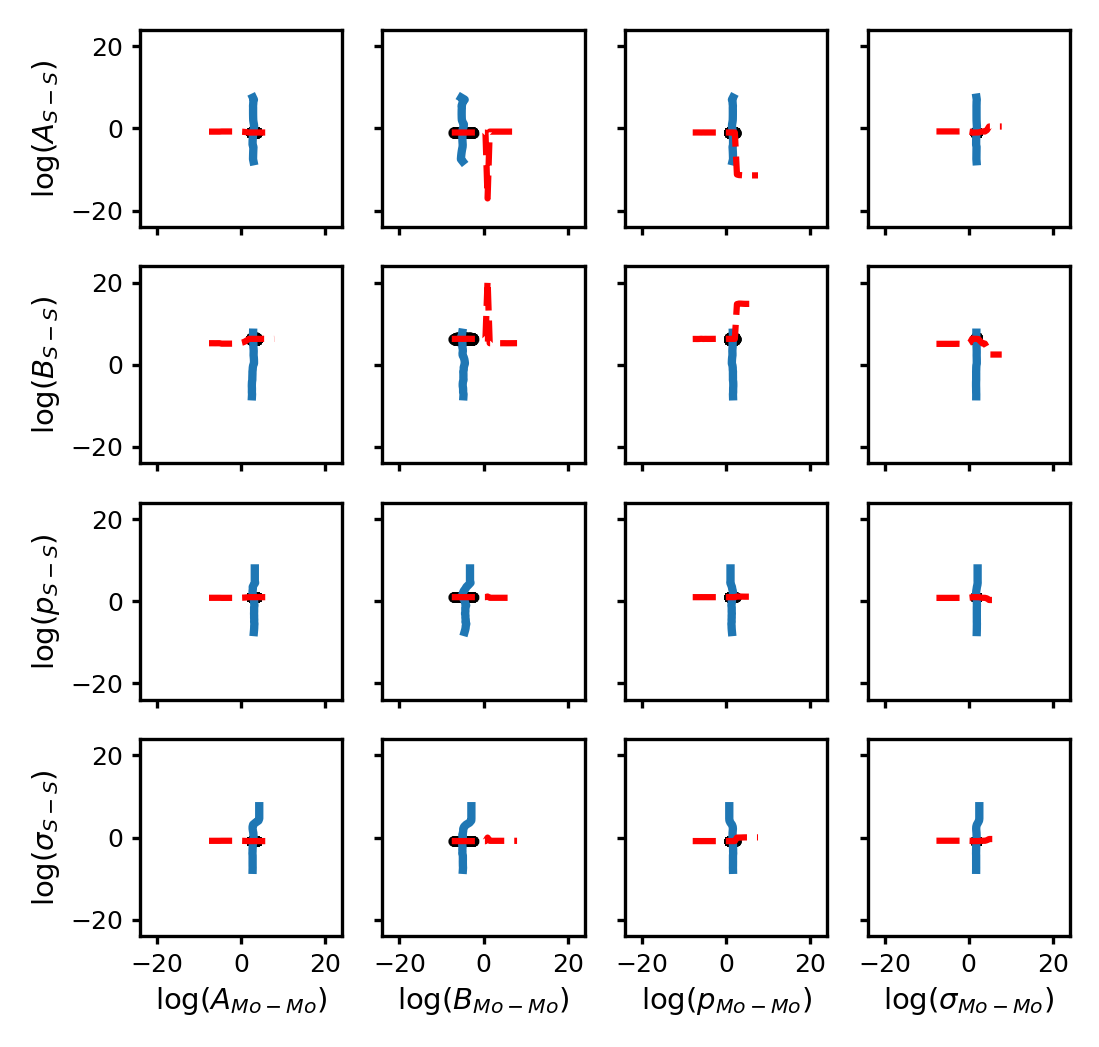}
    \caption[UQ results for SW potential Mo--Mo and S--S parameters at $T = 1.71\times10^{-4}~T_0$]{
        Profile likelihood and MCMC samples for Mo--Mo and S--S parameters at sampling temperature $1.71\times10^{-4}~T_0$ for the SW MoS$_2$ potential.
    }
\end{figure*}

\begin{figure*}[!h]
    \centering
    \includegraphics[width=0.6\textwidth]{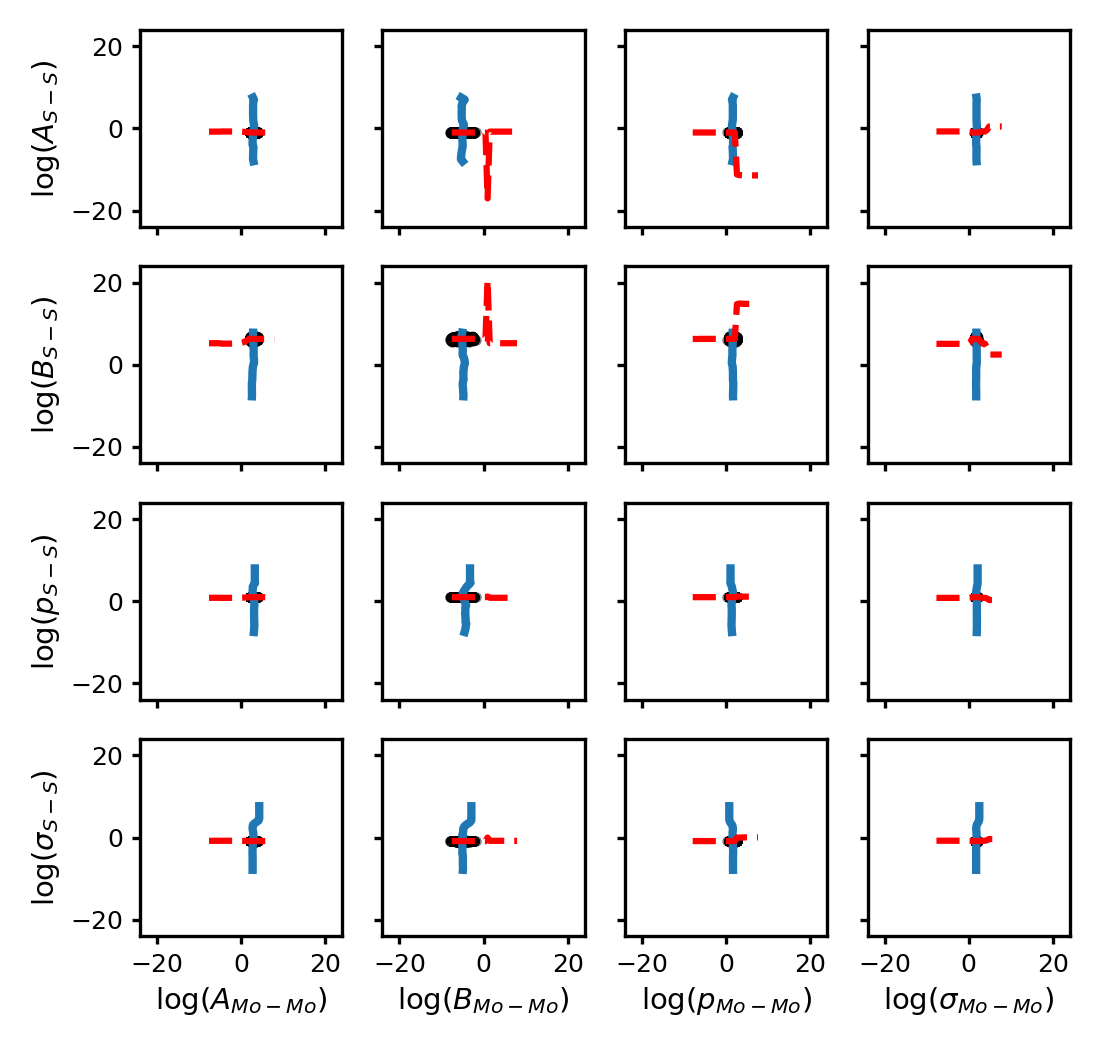}
    \caption[UQ results for SW potential Mo--Mo and S--S parameters at $T = 5.40\times10^{-4}~T_0$]{
        Profile likelihood and MCMC samples for Mo--Mo and S--S parameters at sampling temperature $5.40\times10^{-4}~T_0$ for the SW MoS$_2$ potential.
    }
\end{figure*}

\begin{figure*}[!h]
    \centering
    \includegraphics[width=0.6\textwidth]{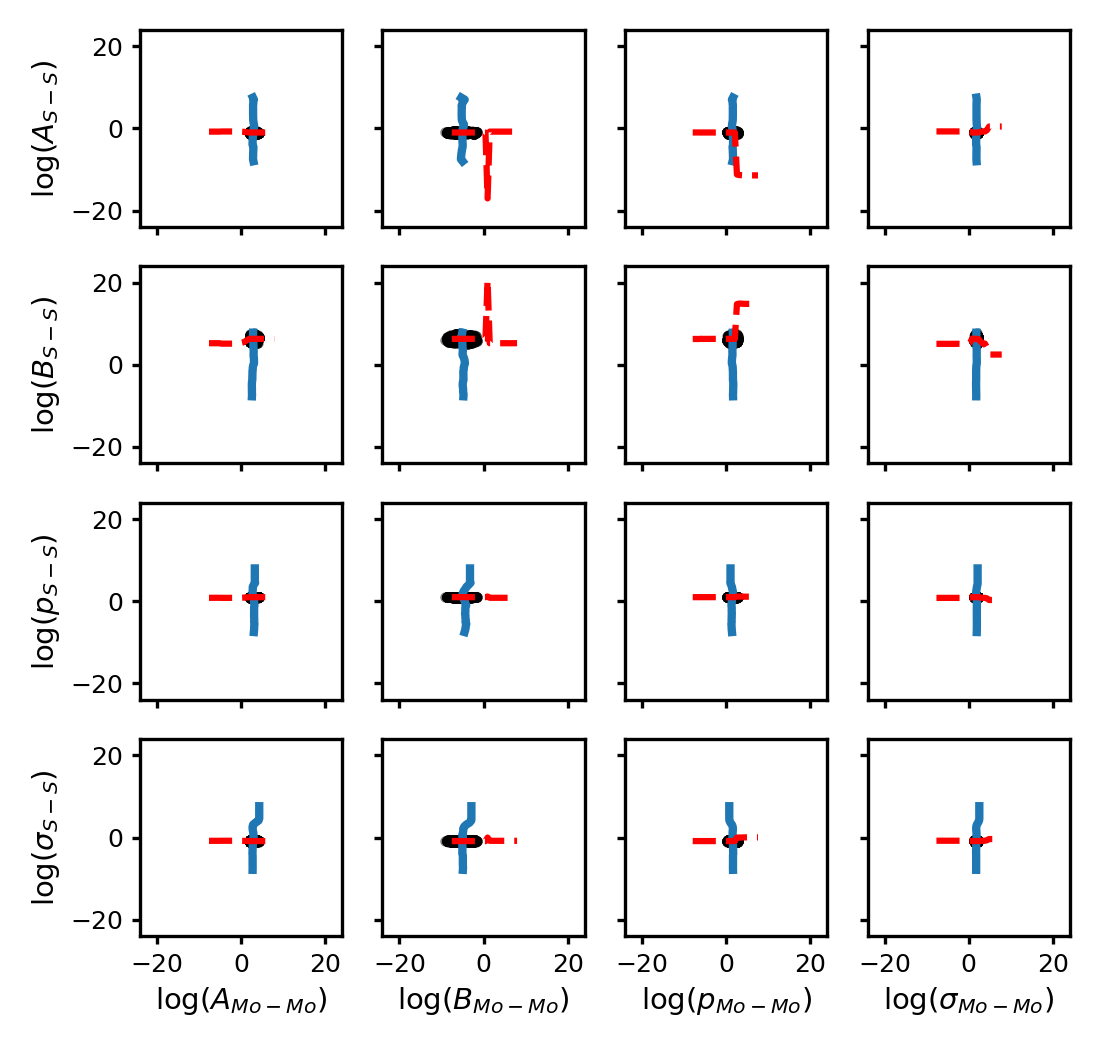}
    \caption[UQ results for SW potential Mo--Mo and S--S parameters at $T = 1.71\times10^{-3}~T_0$]{
        Profile likelihood and MCMC samples for Mo--Mo and S--S parameters at sampling temperature $1.71\times10^{-3}~T_0$ for the SW MoS$_2$ potential.
    }
\end{figure*}

\begin{figure*}[!h]
    \centering
    \includegraphics[width=0.6\textwidth]{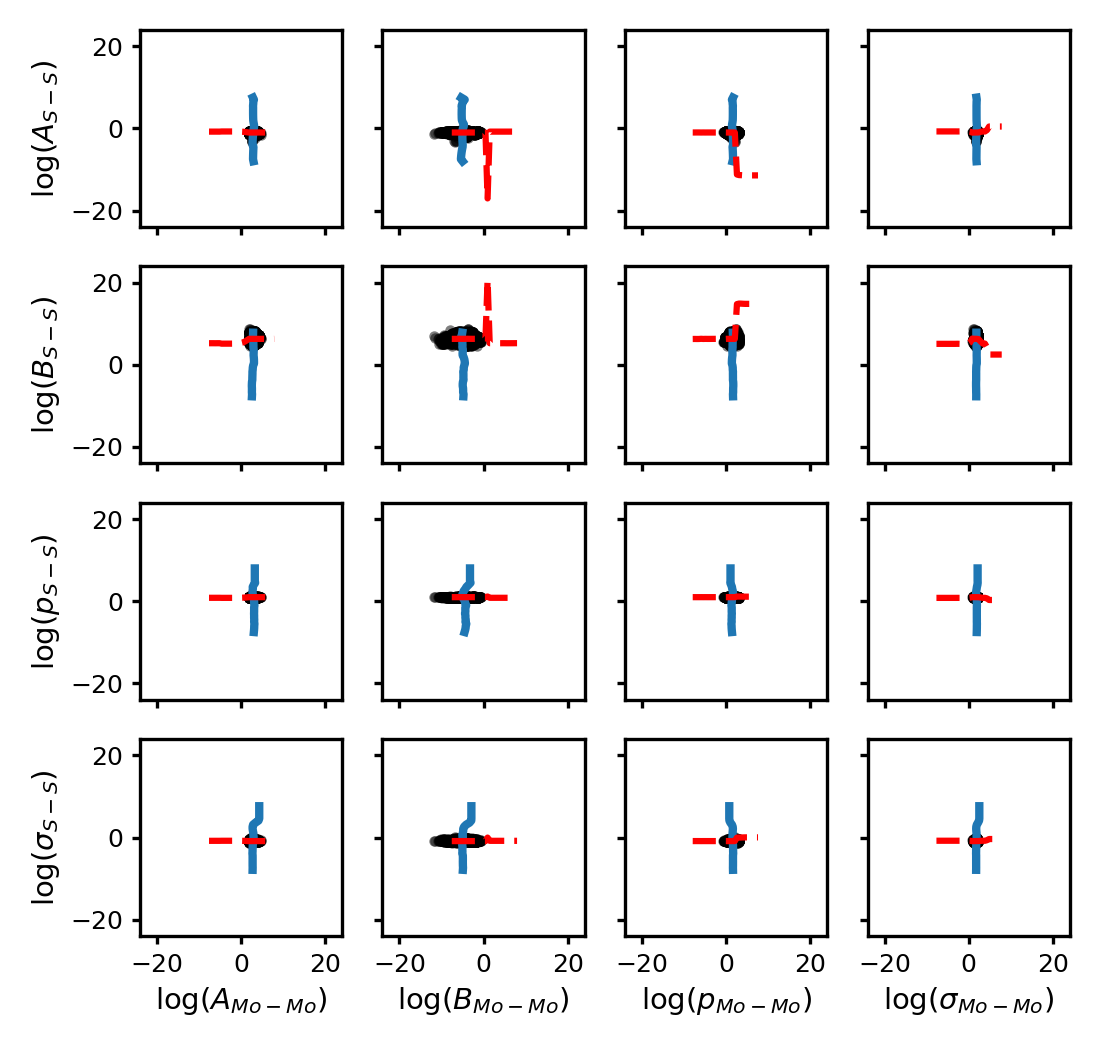}
    \caption[UQ results for SW potential Mo--Mo and S--S parameters at $T = 5.40\times10^{-3}~T_0$]{
        Profile likelihood and MCMC samples for Mo--Mo and S--S parameters at sampling temperature $5.40\times10^{-3}~T_0$ for the SW MoS$_2$ potential.
    }
\end{figure*}

\begin{figure*}[!h]
    \centering
    \includegraphics[width=0.6\textwidth]{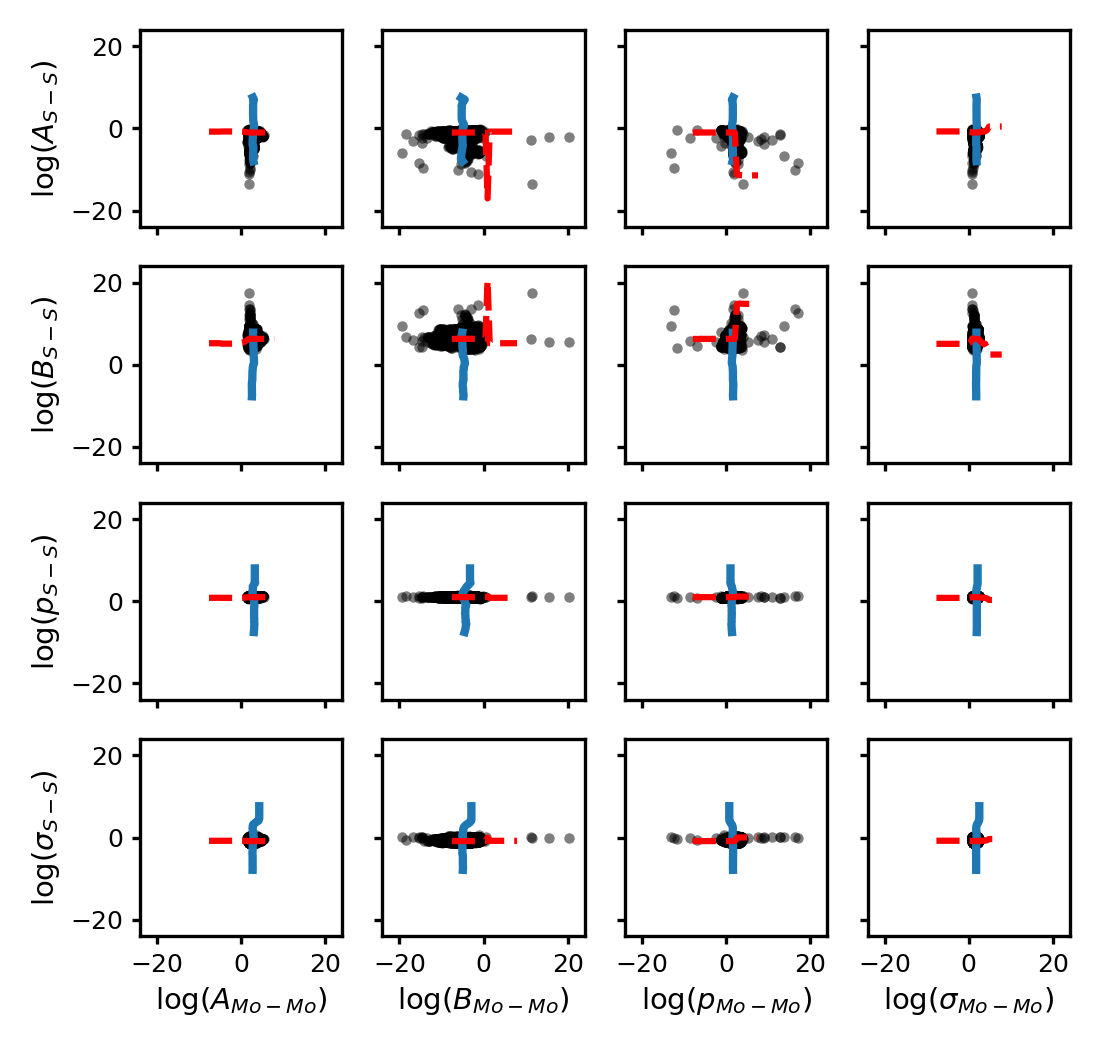}
    \caption[UQ results for SW potential Mo--Mo and S--S parameters at $T = 1.71\times10^{-2}~T_0$]{
        Profile likelihood and MCMC samples for Mo--Mo and S--S parameters at sampling temperature $1.71\times10^{-2}~T_0$ for the SW MoS$_2$ potential.
    }
\end{figure*}

\begin{figure*}[!h]
    \centering
    \includegraphics[width=0.6\textwidth]{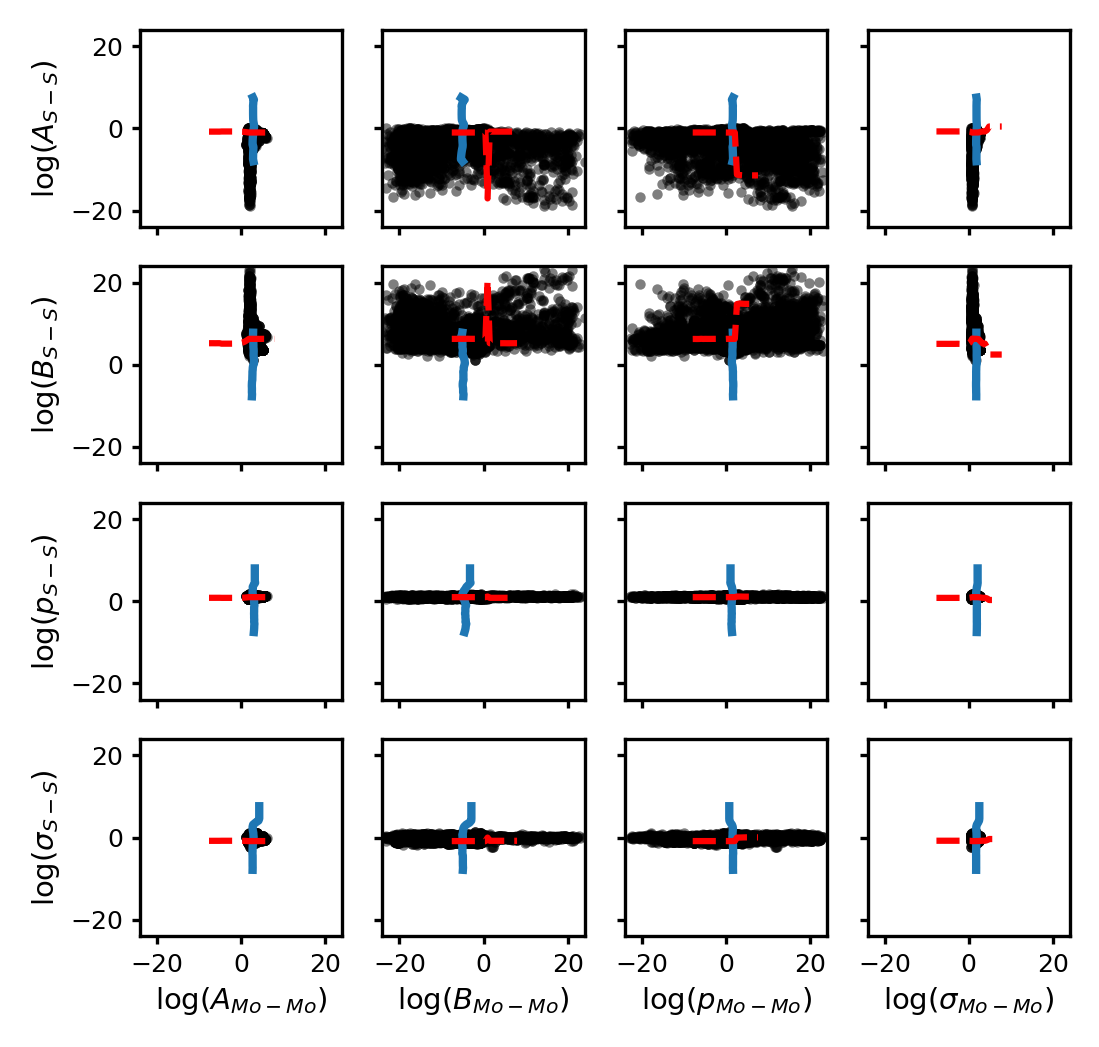}
    \caption[UQ results for SW potential Mo--Mo and S--S parameters at $T = 5.40\times10^{-2}~T_0$]{
        Profile likelihood and MCMC samples for Mo--Mo and S--S parameters at sampling temperature $5.40\times10^{-2}~T_0$ for the SW MoS$_2$ potential.
    }
\end{figure*}

\begin{figure*}[!h]
    \centering
    \includegraphics[width=0.6\textwidth]{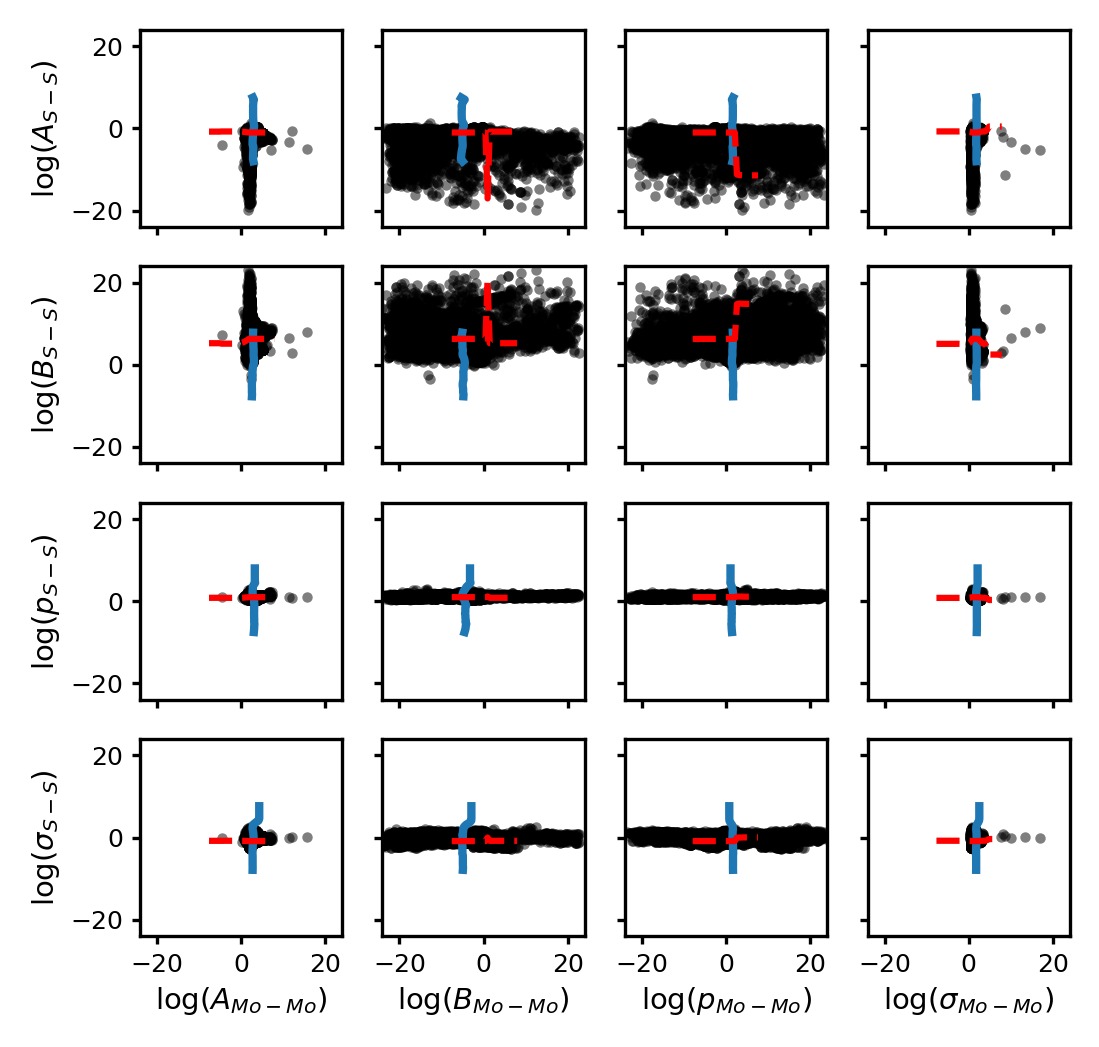}
    \caption[UQ results for SW potential Mo--Mo and S--S parameters at $T = 1.71\times10^{-1}~T_0$]{
        Profile likelihood and MCMC samples for Mo--Mo and S--S parameters at sampling temperature $1.71\times10^{-1}~T_0$ for the SW MoS$_2$ potential.
    }
\end{figure*}

\begin{figure*}[!h]
    \centering
    \includegraphics[width=0.6\textwidth]{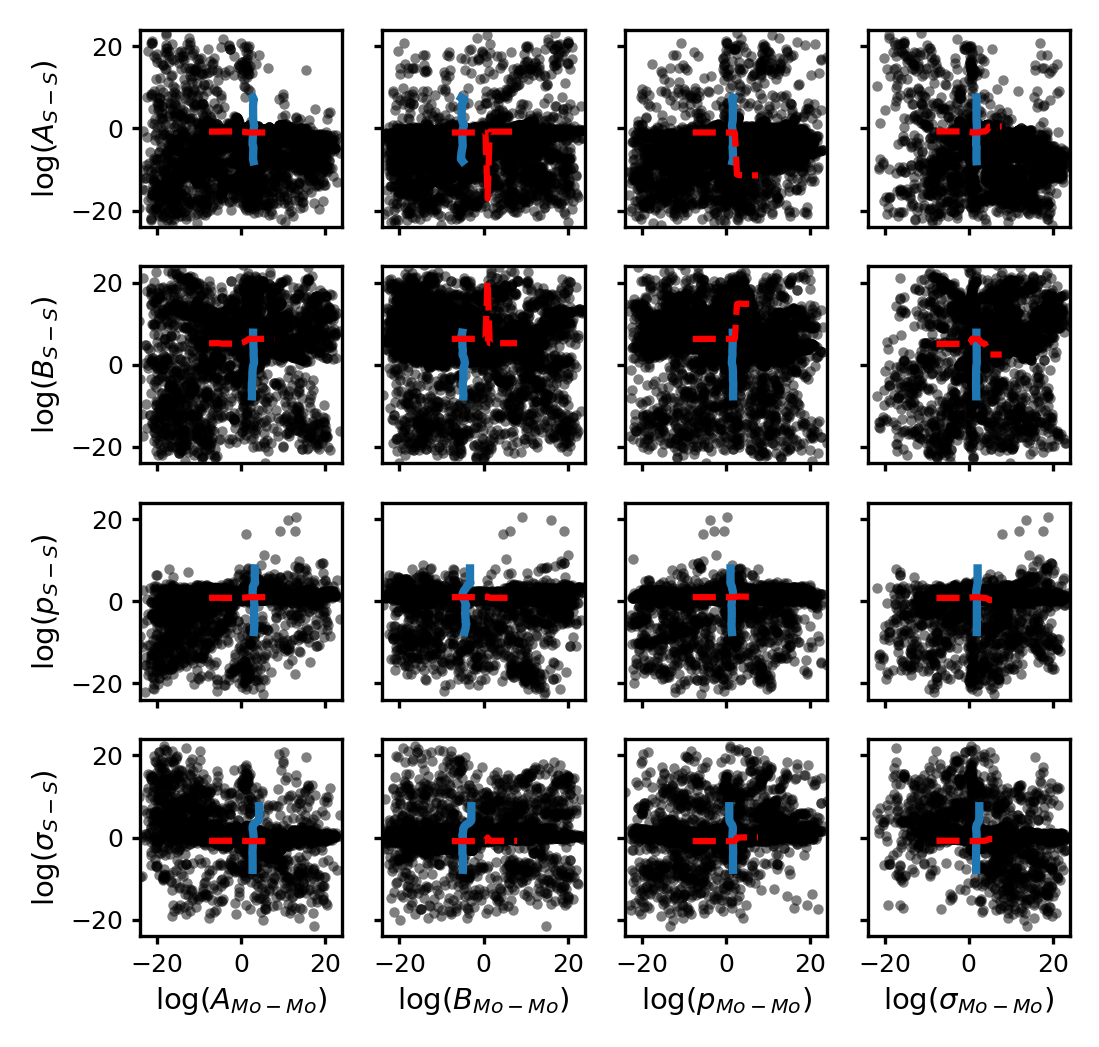}
    \caption[UQ results for SW potential Mo--Mo and S--S parameters at $T = 5.40\times10^{-1}~T_0$]{
        Profile likelihood and MCMC samples for Mo--Mo and S--S parameters at sampling temperature $5.40\times10^{-1}~T_0$ for the SW MoS$_2$ potential.
    }
\end{figure*}

\ifincludeTo
    \begin{figure*}[!h]
        \centering
        \includegraphics[width=0.6\textwidth]{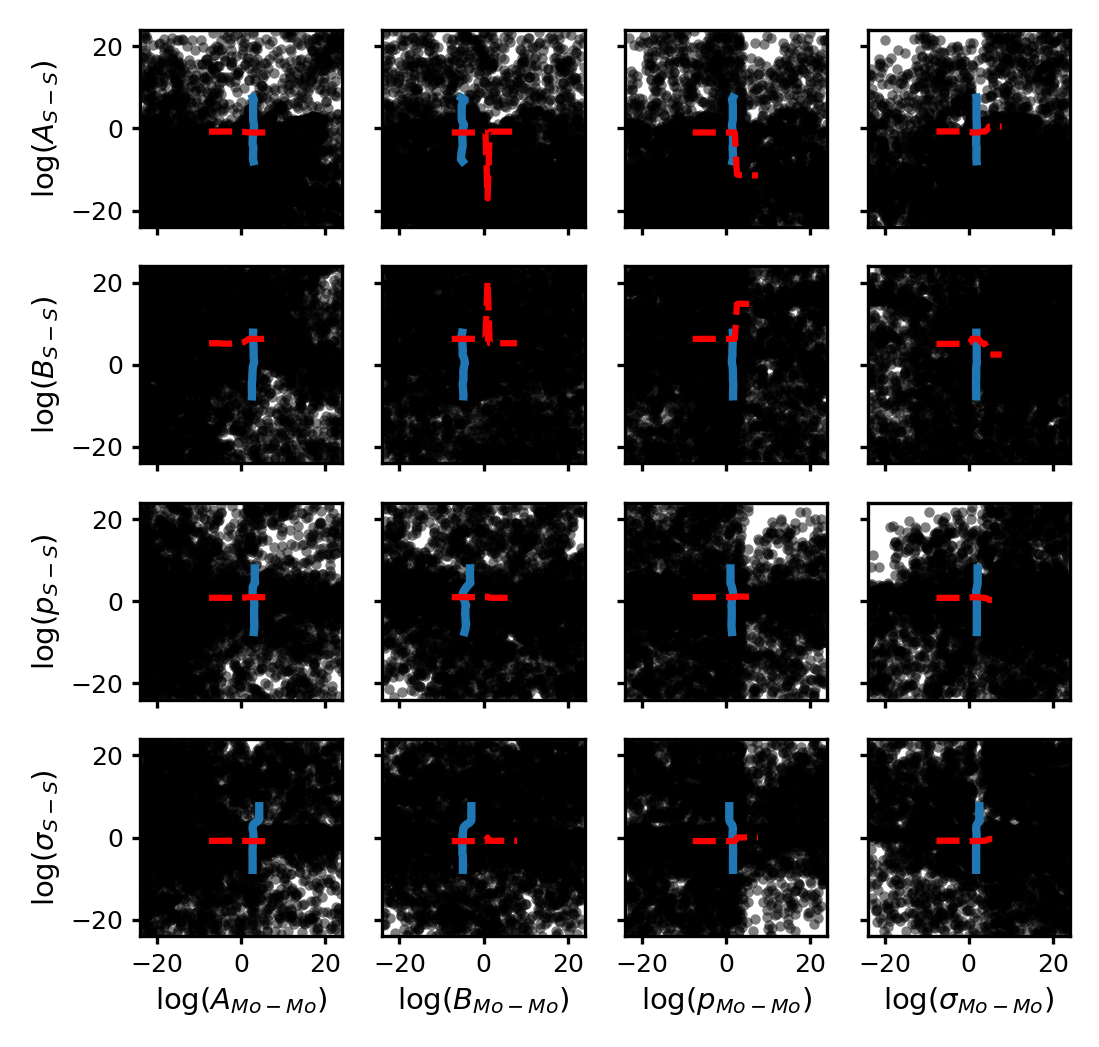}
        \caption[UQ results for SW potential Mo--Mo and S--S parameters at $T = T_0$]{
            Profile likelihood and MCMC samples for Mo--Mo and S--S parameters at sampling temperature $T_0$ for the SW MoS$_2$ potential.
        }
    \end{figure*}
\fi

\begin{figure*}[!h]
    \centering
    \includegraphics[width=0.6\textwidth]{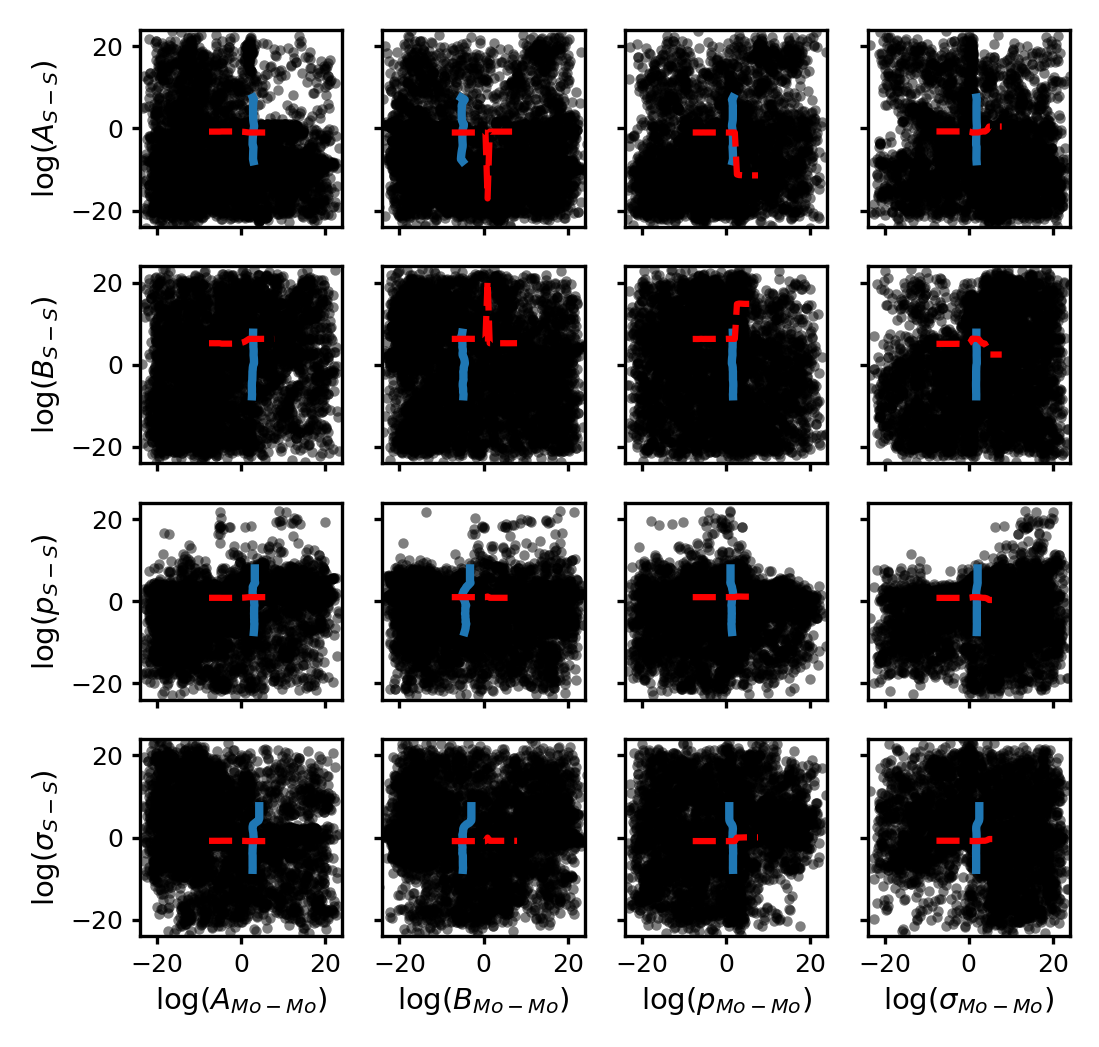}
    \caption[UQ results for SW potential Mo--Mo and S--S parameters at $T = 1.71~T_0$]{
        Profile likelihood and MCMC samples for Mo--Mo and S--S parameters at sampling temperature $1.71~T_0$ for the SW MoS$_2$ potential.
    }
\end{figure*}

\begin{figure*}[!h]
    \centering
    \includegraphics[width=0.6\textwidth]{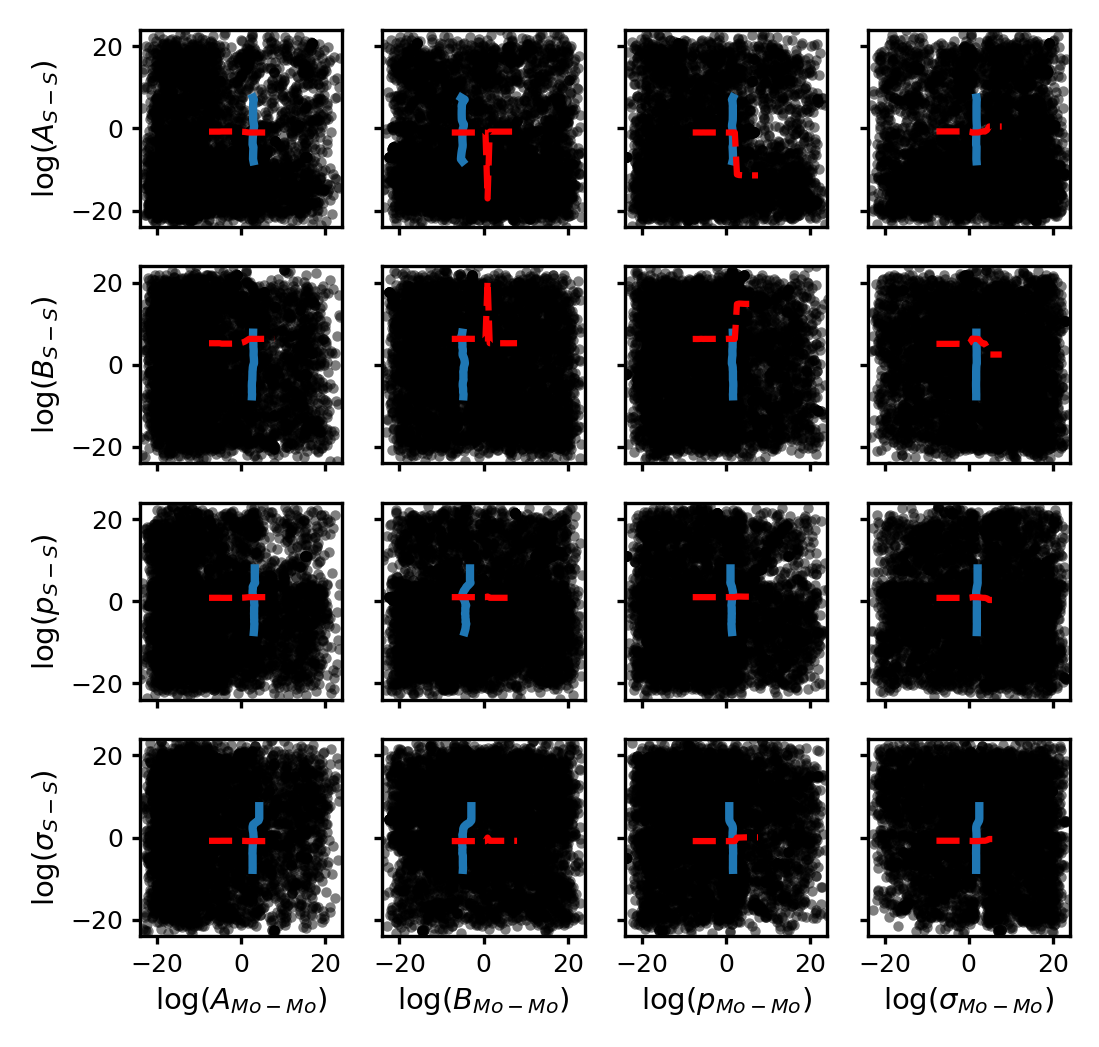}
    \caption[UQ results for SW potential Mo--Mo and S--S parameters at $T = 5.40~T_0$]{
        Profile likelihood and MCMC samples for Mo--Mo and S--S parameters at sampling temperature $5.40~T_0$ for the SW MoS$_2$ potential.
    }
\end{figure*}

\cleardoublepage

\subsection{Mo--Mo and 3-body parameters}
\label{subsec:Mo-Mo_3-body}
Profile likelihood and MCMC samples between Mo--Mo and 3-body parameters.
Notice that there is a lack of correlation between between parameters corresponding to different interaction types.

\begin{figure*}[!h]
    \centering
    \includegraphics[width=0.6\textwidth]{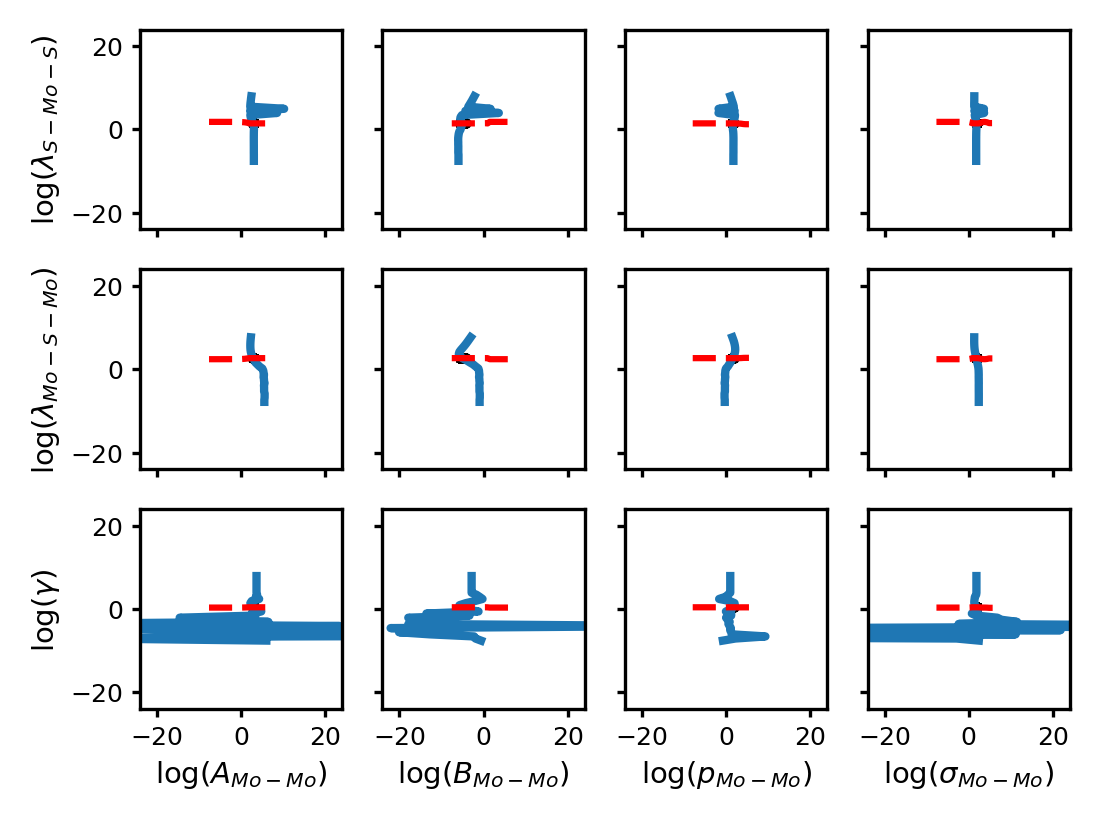}
    \caption[UQ results for SW potential Mo--Mo and 3-body parameters at $T = 5.40\times10^{-6}~T_0$]{
        Profile likelihood and MCMC samples for Mo--Mo and 3-body parameters at sampling temperature $5.40\times10^{-6}~T_0$ for the SW MoS$_2$ potential.
    }
\end{figure*}

\begin{figure*}[!h]
    \centering
    \includegraphics[width=0.6\textwidth]{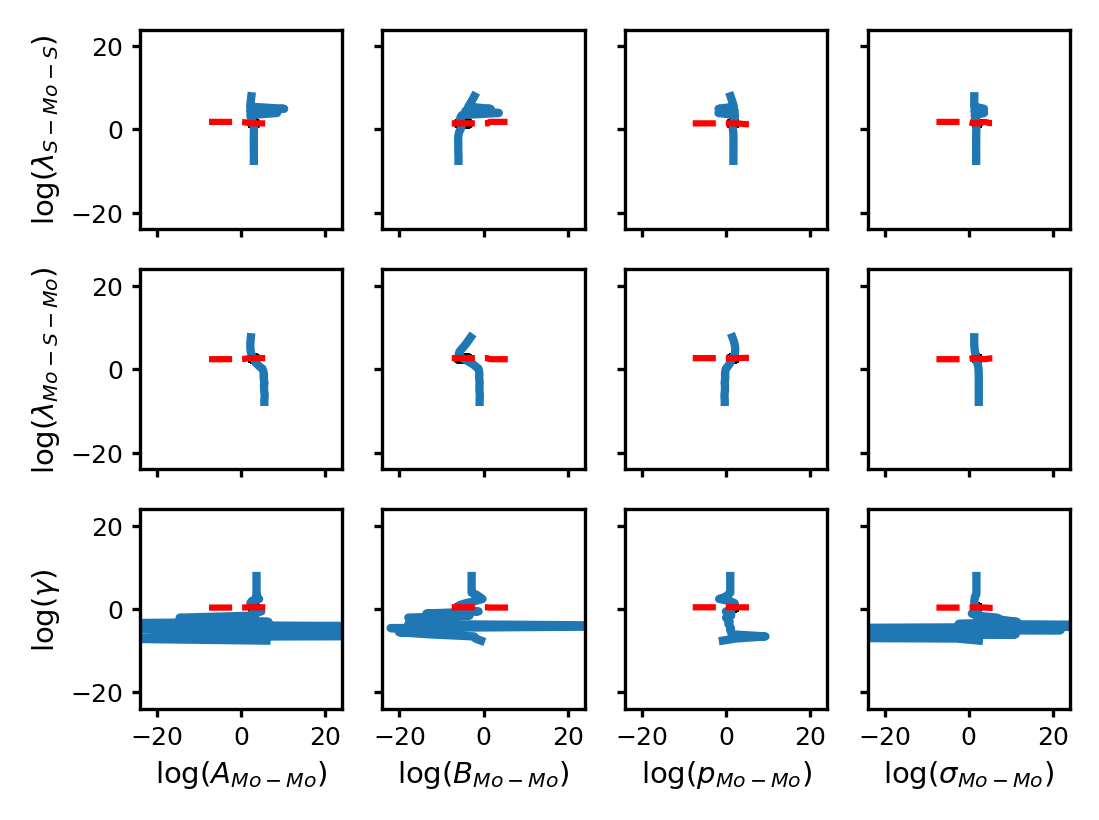}
    \caption[UQ results for SW potential Mo--Mo and 3-body parameters at $T = 1.71\times10^{-5}~T_0$]{
        Profile likelihood and MCMC samples for Mo--Mo and 3-body parameters at sampling temperature $1.71\times10^{-5}~T_0$ for the SW MoS$_2$ potential.
    }
\end{figure*}

\begin{figure*}[!h]
    \centering
    \includegraphics[width=0.6\textwidth]{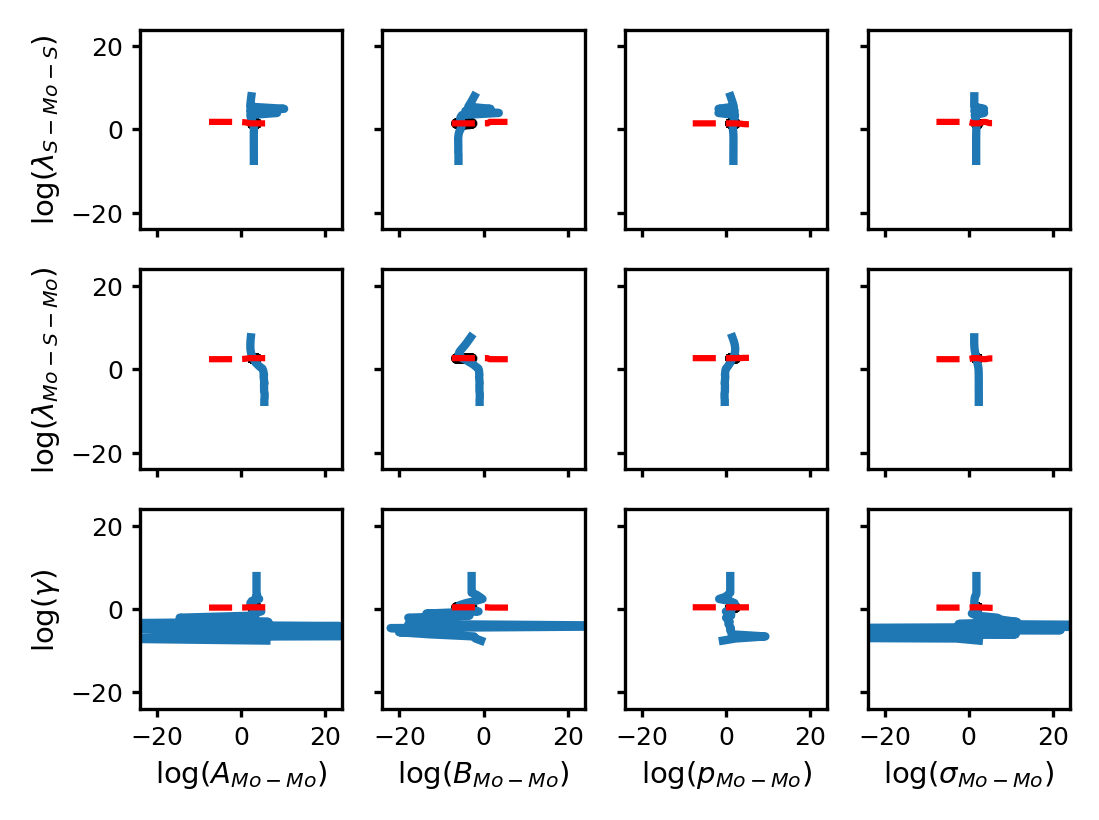}
    \caption[UQ results for SW potential Mo--Mo and 3-body parameters at $T = 5.40\times10^{-5}~T_0$]{
        Profile likelihood and MCMC samples for Mo--Mo and 3-body parameters at sampling temperature $5.40\times10^{-5}~T_0$ for the SW MoS$_2$ potential.
    }
\end{figure*}

\begin{figure*}[!h]
    \centering
    \includegraphics[width=0.6\textwidth]{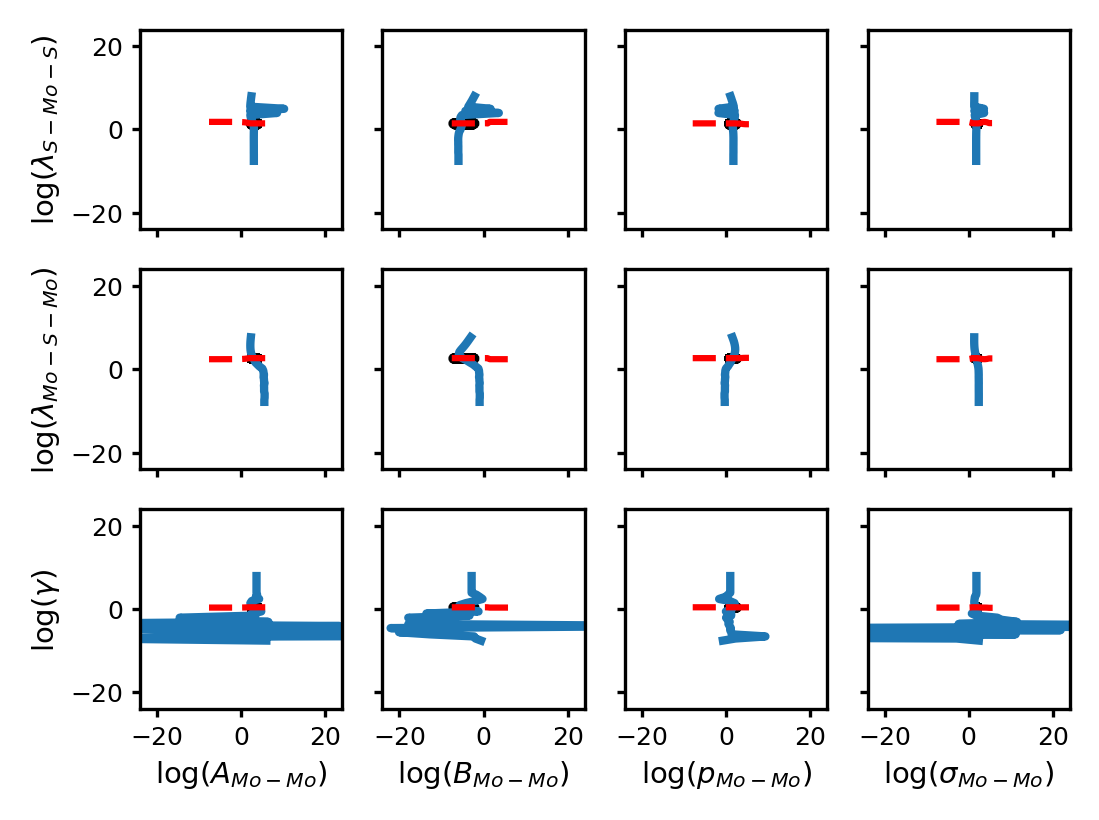}
    \caption[UQ results for SW potential Mo--Mo and 3-body parameters at $T = 1.71\times10^{-4}~T_0$]{
        Profile likelihood and MCMC samples for Mo--Mo and 3-body parameters at sampling temperature $1.71\times10^{-4}~T_0$ for the SW MoS$_2$ potential.
    }
\end{figure*}

\begin{figure*}[!h]
    \centering
    \includegraphics[width=0.6\textwidth]{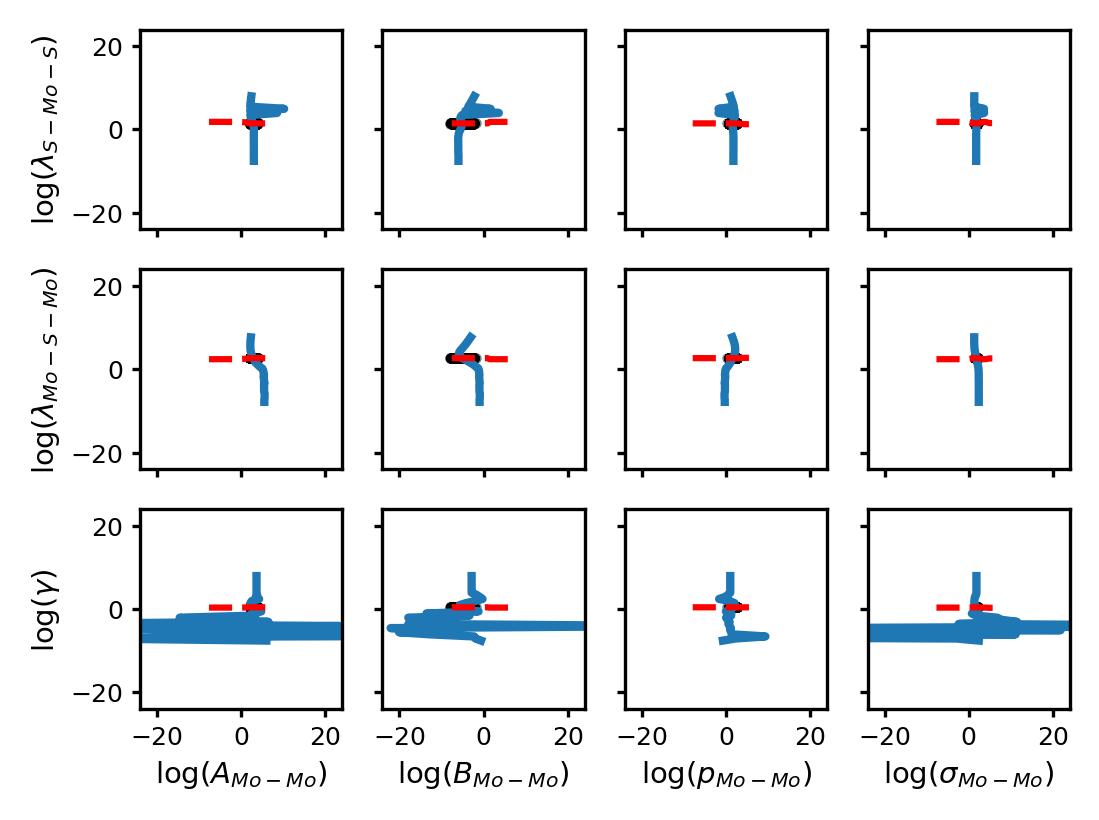}
    \caption[UQ results for SW potential Mo--Mo and 3-body parameters at $T = 5.40\times10^{-4}~T_0$]{
        Profile likelihood and MCMC samples for Mo--Mo and 3-body parameters at sampling temperature $5.40\times10^{-4}~T_0$ for the SW MoS$_2$ potential.
    }
\end{figure*}

\begin{figure*}[!h]
    \centering
    \includegraphics[width=0.6\textwidth]{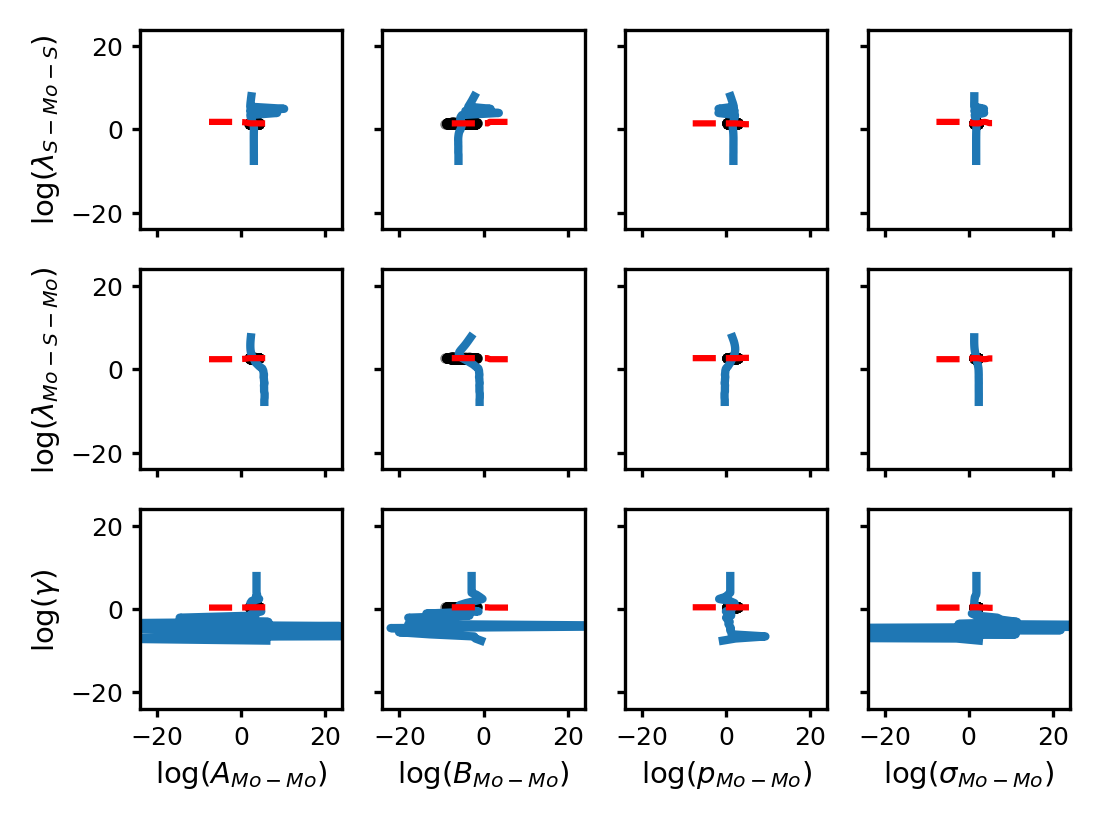}
    \caption[UQ results for SW potential Mo--Mo and 3-body parameters at $T = 1.71\times10^{-3}~T_0$]{
        Profile likelihood and MCMC samples for Mo--Mo and 3-body parameters at sampling temperature $1.71\times10^{-3}~T_0$ for the SW MoS$_2$ potential.
    }
\end{figure*}

\begin{figure*}[!h]
    \centering
    \includegraphics[width=0.6\textwidth]{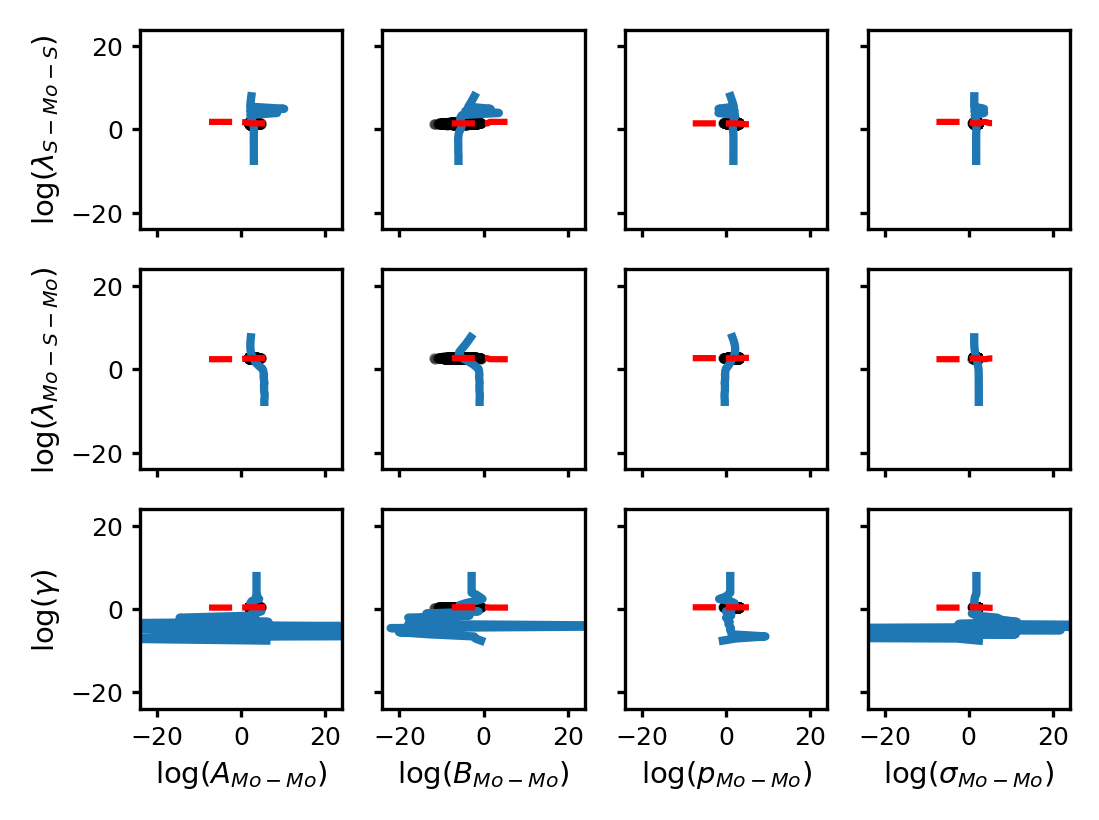}
    \caption[UQ results for SW potential Mo--Mo and 3-body parameters at $T = 5.40\times10^{-3}~T_0$]{
        Profile likelihood and MCMC samples for Mo--Mo and 3-body parameters at sampling temperature $5.40\times10^{-3}~T_0$ for the SW MoS$_2$ potential.
    }
\end{figure*}

\begin{figure*}[!h]
    \centering
    \includegraphics[width=0.6\textwidth]{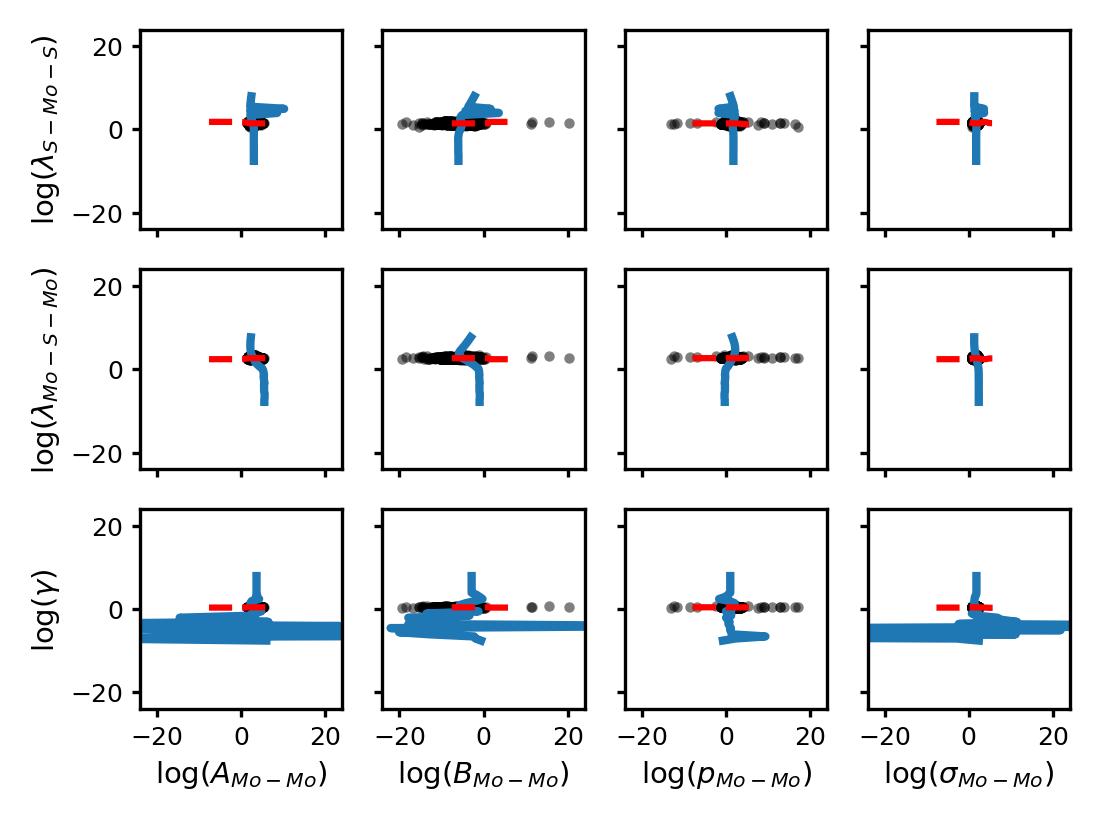}
    \caption[UQ results for SW potential Mo--Mo and 3-body parameters at $T = 1.71\times10^{-2}~T_0$]{
        Profile likelihood and MCMC samples for Mo--Mo and 3-body parameters at sampling temperature $1.71\times10^{-2}~T_0$ for the SW MoS$_2$ potential.
    }
\end{figure*}

\begin{figure*}[!h]
    \centering
    \includegraphics[width=0.6\textwidth]{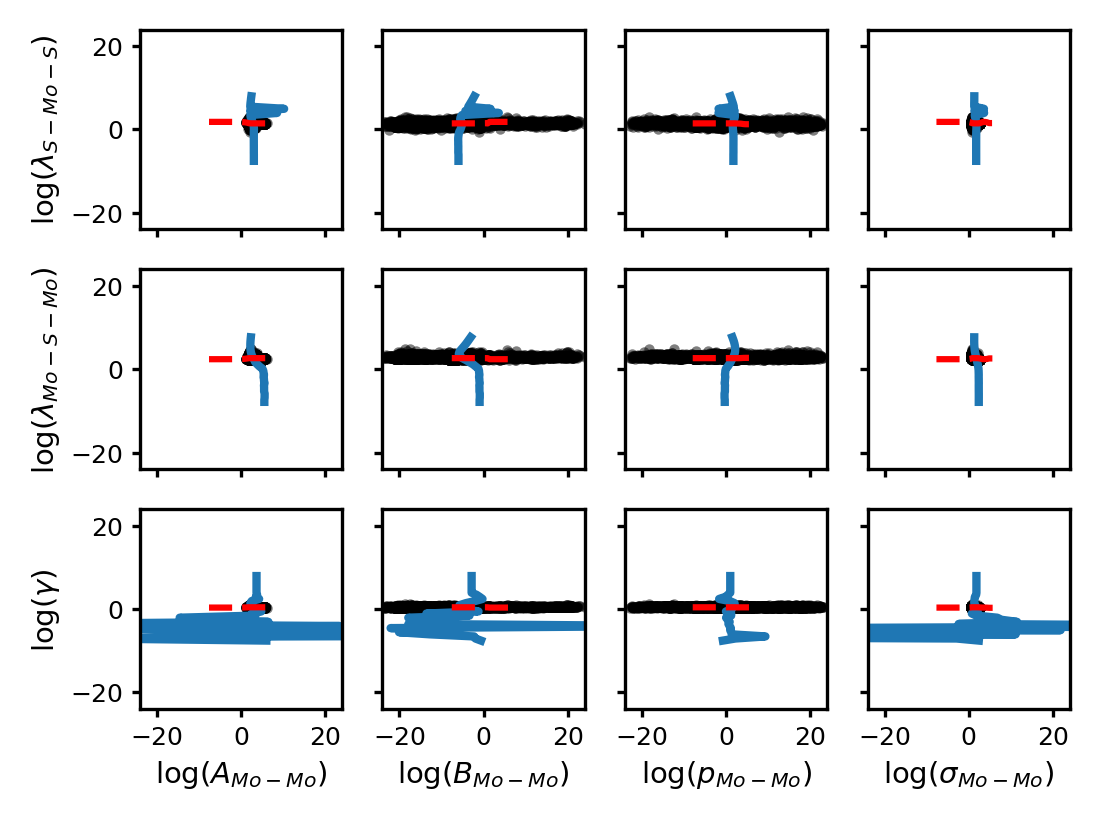}
    \caption[UQ results for SW potential Mo--Mo and 3-body parameters at $T = 5.40\times10^{-2}~T_0$]{
        Profile likelihood and MCMC samples for Mo--Mo and 3-body parameters at sampling temperature $5.40\times10^{-2}~T_0$ for the SW MoS$_2$ potential.
    }
\end{figure*}

\begin{figure*}[!h]
    \centering
    \includegraphics[width=0.6\textwidth]{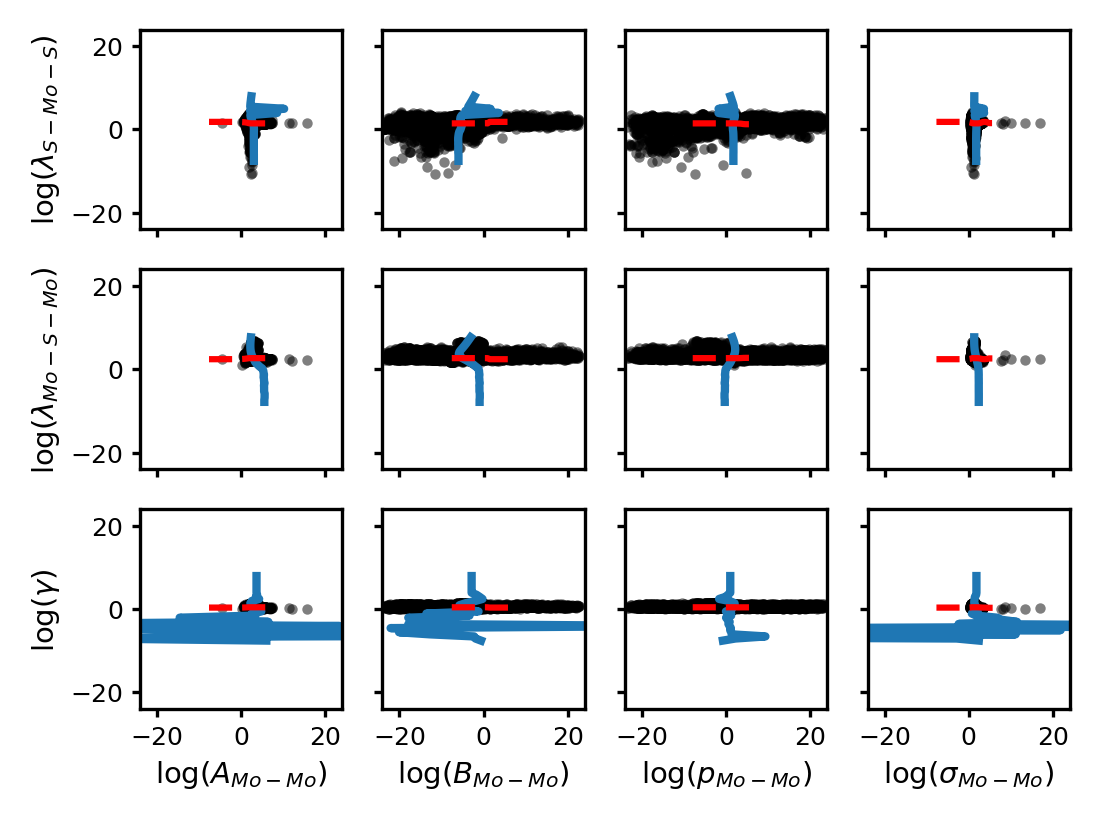}
    \caption[UQ results for SW potential Mo--Mo and 3-body parameters at $T = 1.71\times10^{-1}~T_0$]{
        Profile likelihood and MCMC samples for Mo--Mo and 3-body parameters at sampling temperature $1.71\times10^{-1}~T_0$ for the SW MoS$_2$ potential.
    }
\end{figure*}

\begin{figure*}[!h]
    \centering
    \includegraphics[width=0.6\textwidth]{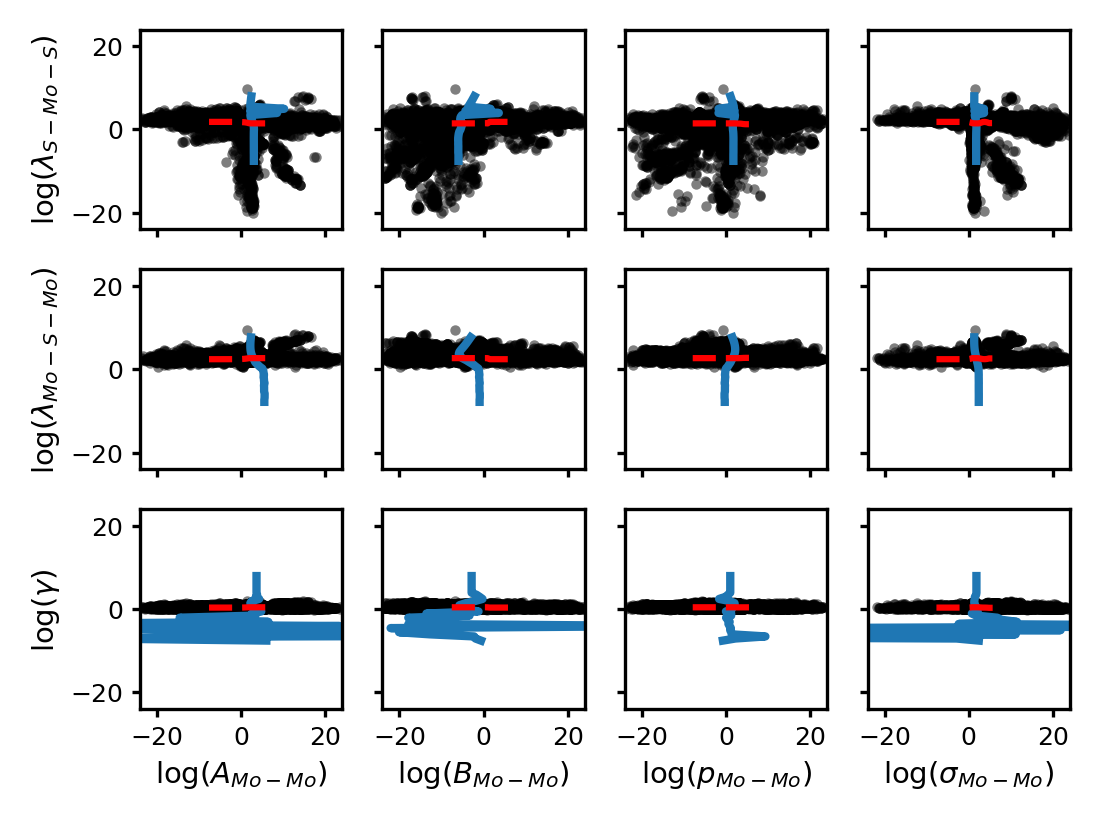}
    \caption[UQ results for SW potential Mo--Mo and 3-body parameters at $T = 5.40\times10^{-1}~T_0$]{
        Profile likelihood and MCMC samples for Mo--Mo and 3-body parameters at sampling temperature $5.40\times10^{-1}~T_0$ for the SW MoS$_2$ potential.
    }
\end{figure*}

\ifincludeTo
    \begin{figure*}[!h]
        \centering
        \includegraphics[width=0.6\textwidth]{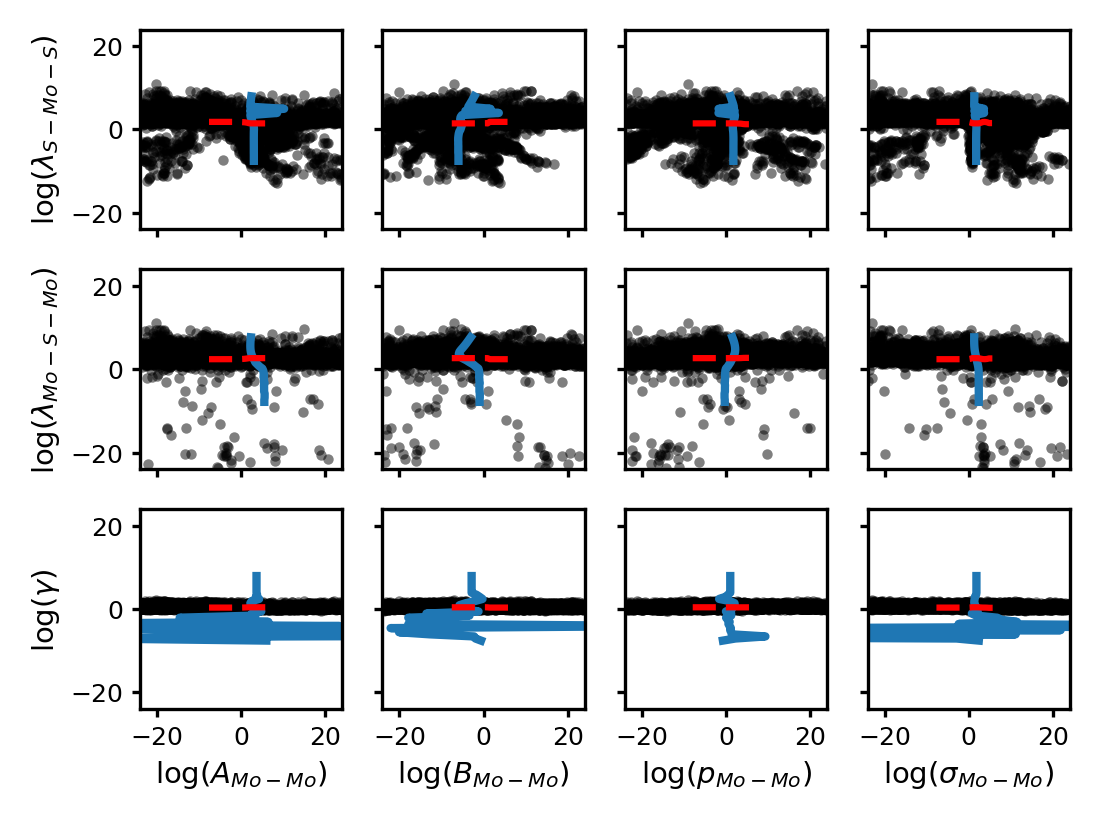}
        \caption[UQ results for SW potential Mo--Mo and 3-body parameters at $T = T_0$]{
            Profile likelihood and MCMC samples for Mo--Mo and 3-body parameters at sampling temperature $T_0$ for the SW MoS$_2$ potential.
        }
    \end{figure*}
\fi

\begin{figure*}[!h]
    \centering
    \includegraphics[width=0.6\textwidth]{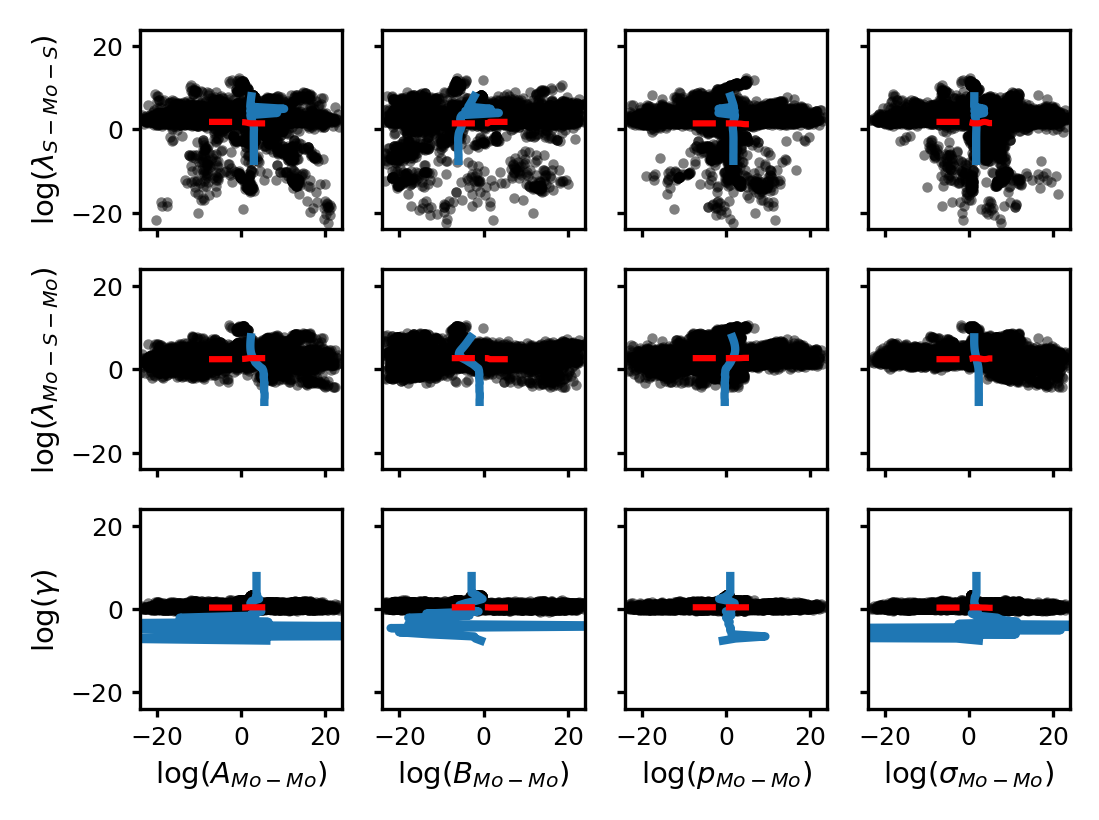}
    \caption[UQ results for SW potential Mo--Mo and 3-body parameters at $T = 1.71~T_0$]{
        Profile likelihood and MCMC samples for Mo--Mo and 3-body parameters at sampling temperature $1.71~T_0$ for the SW MoS$_2$ potential.
    }
\end{figure*}

\begin{figure*}[!h]
    \centering
    \includegraphics[width=0.6\textwidth]{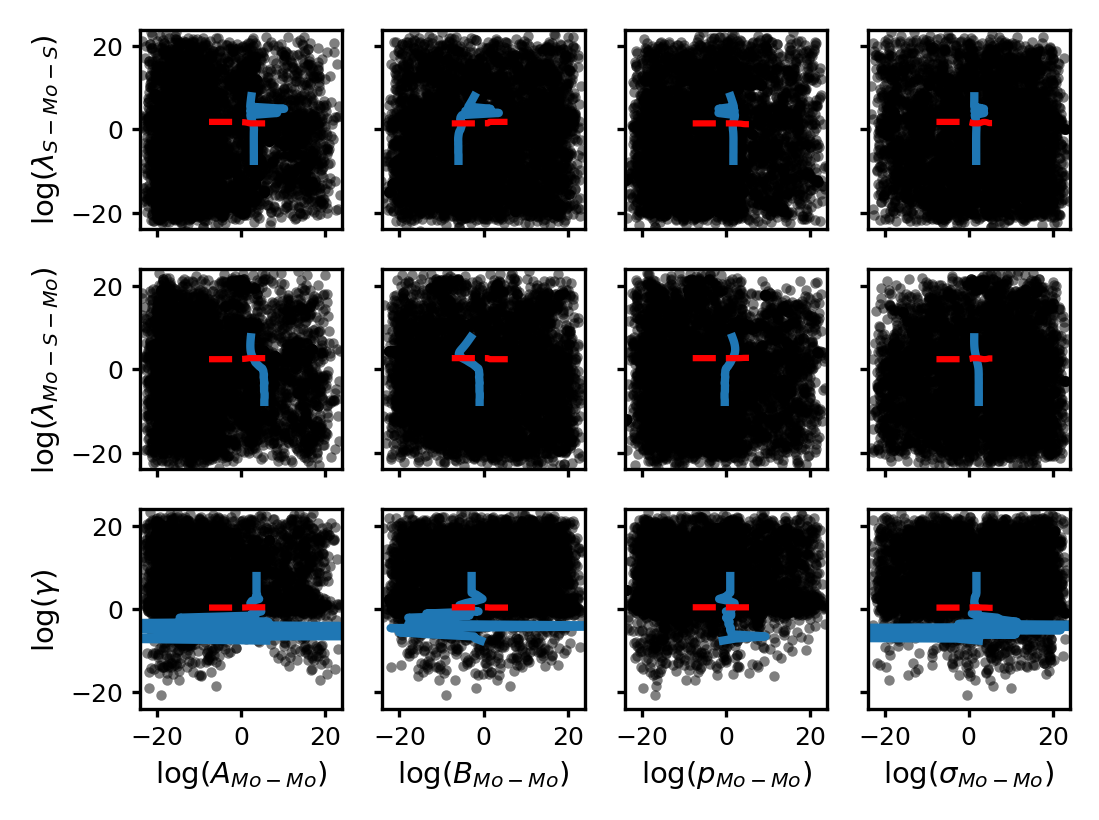}
    \caption[UQ results for SW potential Mo--Mo and 3-body parameters at $T = 5.40~T_0$]{
        Profile likelihood and MCMC samples for Mo--Mo and 3-body parameters at sampling temperature $5.40~T_0$ for the SW MoS$_2$ potential.
    }
\end{figure*}

\cleardoublepage

\subsection{Mo--S and S--S parameters}
\label{subsec:Mo-S_S-S}
Profile likelihood and MCMC samples between Mo--S and S--S parameters.
Notice that there is a lack of correlation between between parameters corresponding to different interaction types.

\begin{figure*}[!h]
    \centering
    \includegraphics[width=0.6\textwidth]{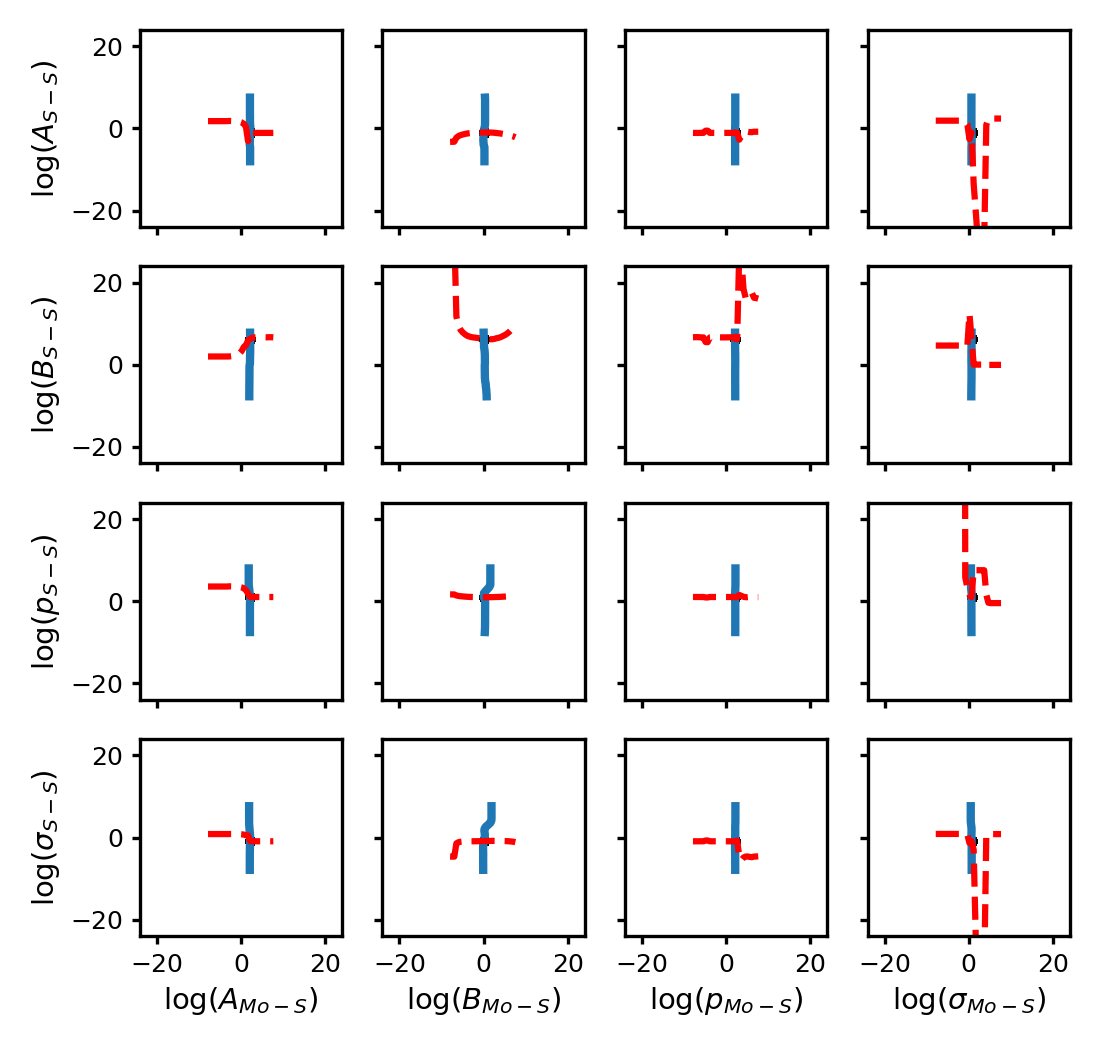}
    \caption[UQ results for SW potential Mo--S and S--S parameters at $T = 5.40\times10^{-6}~T_0$]{
        Profile likelihood and MCMC samples for Mo--S and S--S parameters at sampling temperature $5.40\times10^{-6}~T_0$ for the SW MoS$_2$ potential.
    }
\end{figure*}

\begin{figure*}[!h]
    \centering
    \includegraphics[width=0.6\textwidth]{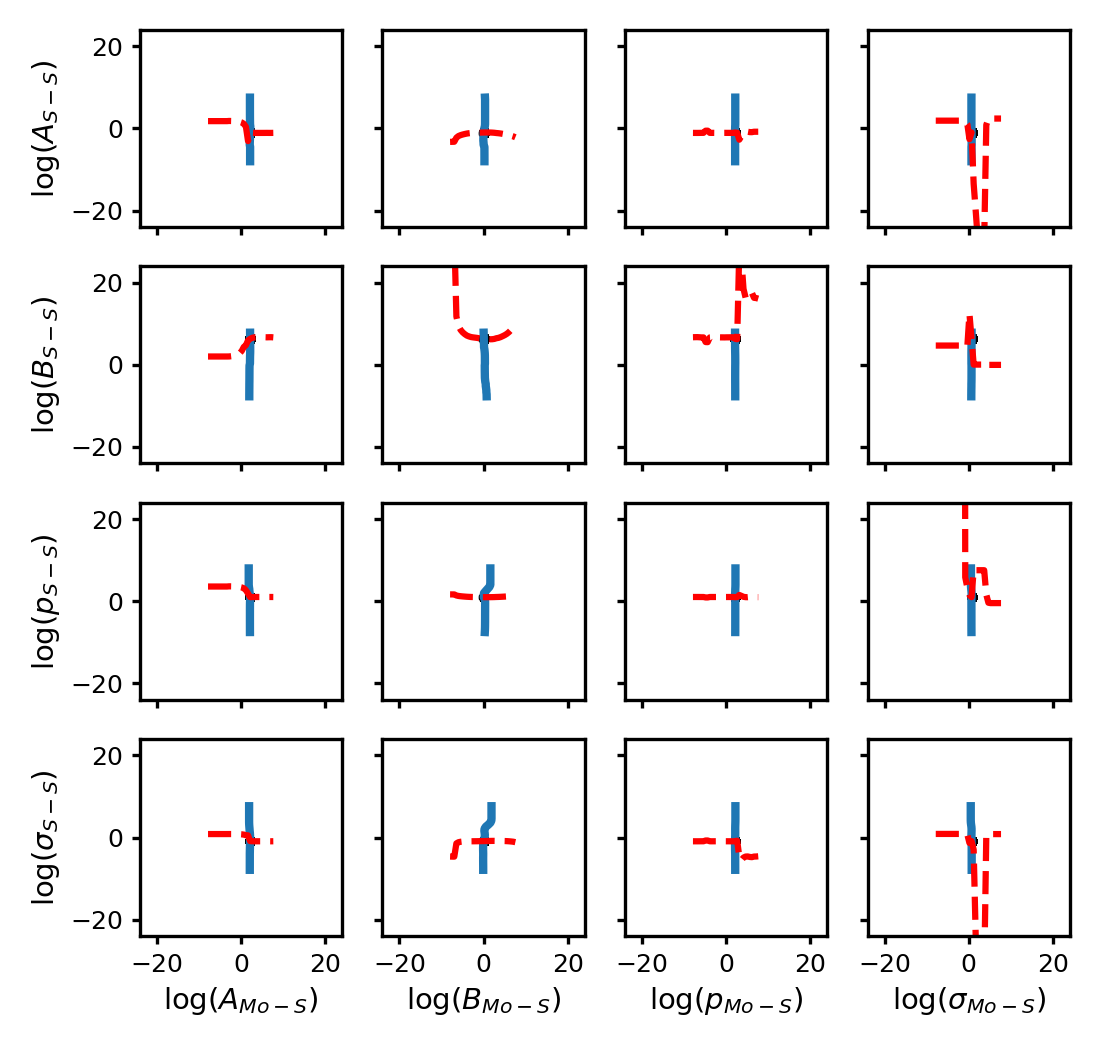}
    \caption[UQ results for SW potential Mo--S and S--S parameters at $T = 1.71\times10^{-5}~T_0$]{
        Profile likelihood and MCMC samples for Mo--S and S--S parameters at sampling temperature $1.71\times10^{-5}~T_0$ for the SW MoS$_2$ potential.
    }
\end{figure*}

\begin{figure*}[!h]
    \centering
    \includegraphics[width=0.6\textwidth]{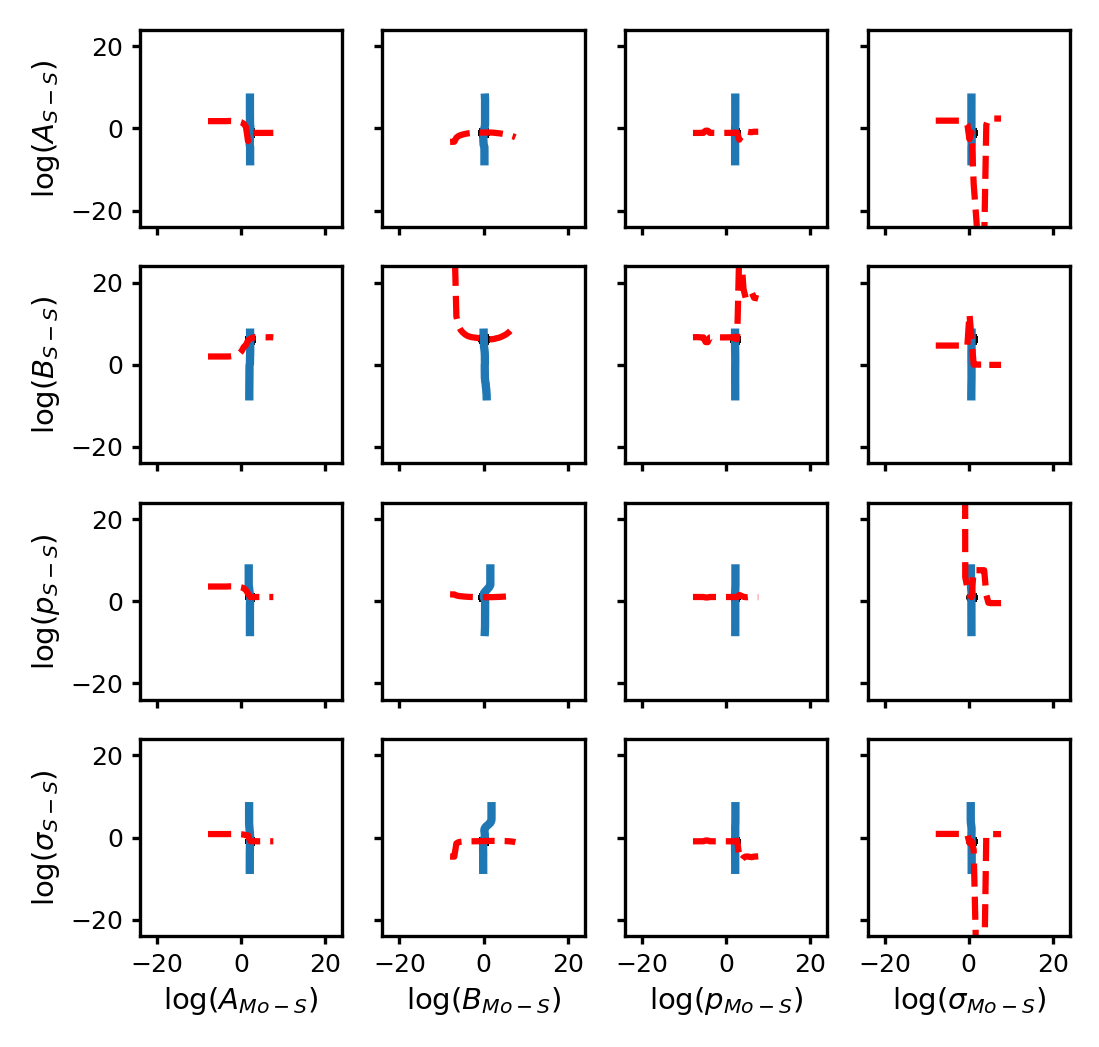}
    \caption[UQ results for SW potential Mo--S and S--S parameters at $T = 5.40\times10^{-5}~T_0$]{
        Profile likelihood and MCMC samples for Mo--S and S--S parameters at sampling temperature $5.40\times10^{-5}~T_0$ for the SW MoS$_2$ potential.
    }
\end{figure*}

\begin{figure*}[!h]
    \centering
    \includegraphics[width=0.6\textwidth]{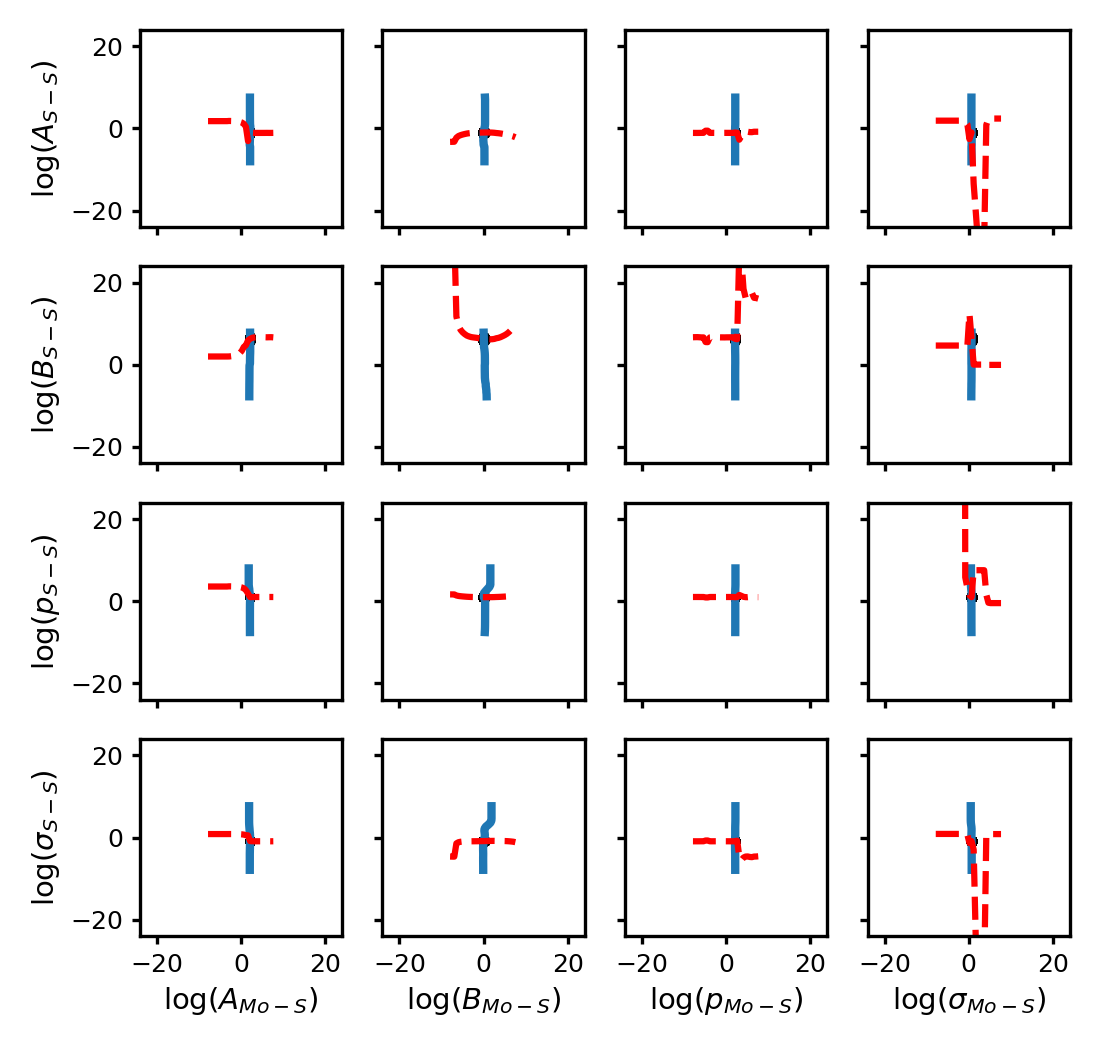}
    \caption[UQ results for SW potential Mo--S and S--S parameters at $T = 1.71\times10^{-4}~T_0$]{
        Profile likelihood and MCMC samples for Mo--S and S--S parameters at sampling temperature $1.71\times10^{-4}~T_0$ for the SW MoS$_2$ potential.
    }
\end{figure*}

\begin{figure*}[!h]
    \centering
    \includegraphics[width=0.6\textwidth]{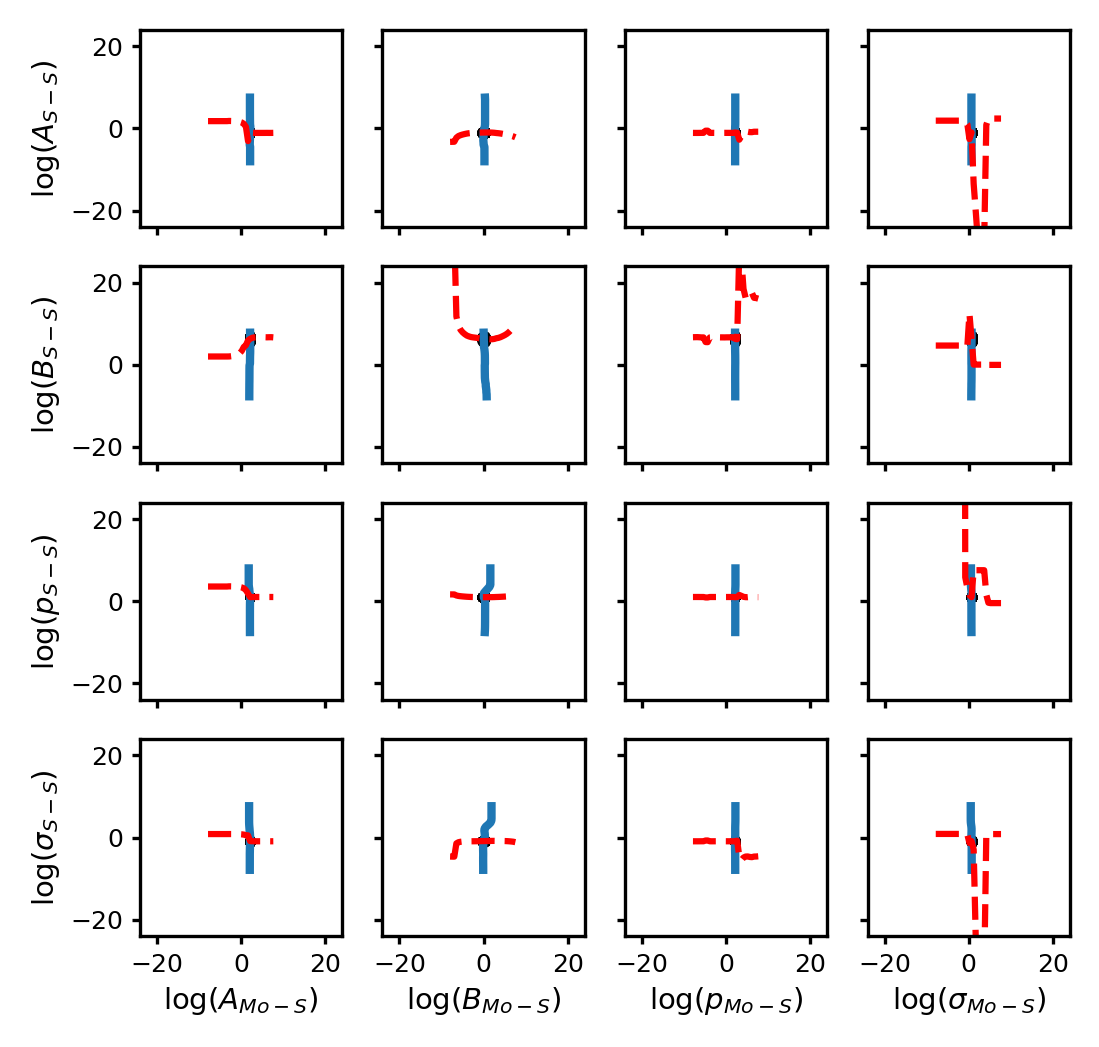}
    \caption[UQ results for SW potential Mo--S and S--S parameters at $T = 5.40\times10^{-4}~T_0$]{
        Profile likelihood and MCMC samples for Mo--S and S--S parameters at sampling temperature $5.40\times10^{-4}~T_0$ for the SW MoS$_2$ potential.
    }
\end{figure*}

\begin{figure*}[!h]
    \centering
    \includegraphics[width=0.6\textwidth]{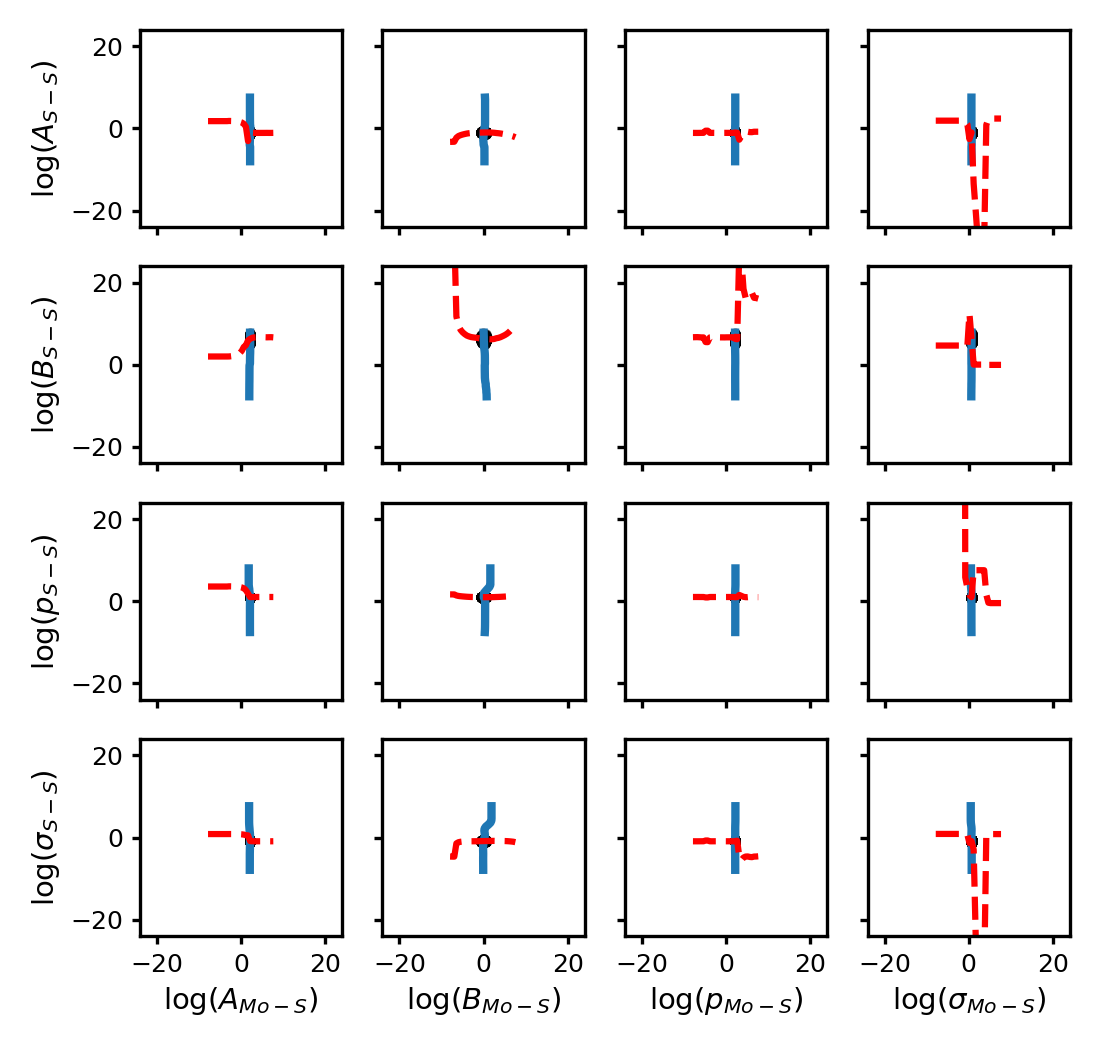}
    \caption[UQ results for SW potential Mo--S and S--S parameters at $T = 1.71\times10^{-3}~T_0$]{
        Profile likelihood and MCMC samples for Mo--S and S--S parameters at sampling temperature $1.71\times10^{-3}~T_0$ for the SW MoS$_2$ potential.
    }
\end{figure*}

\begin{figure*}[!h]
    \centering
    \includegraphics[width=0.6\textwidth]{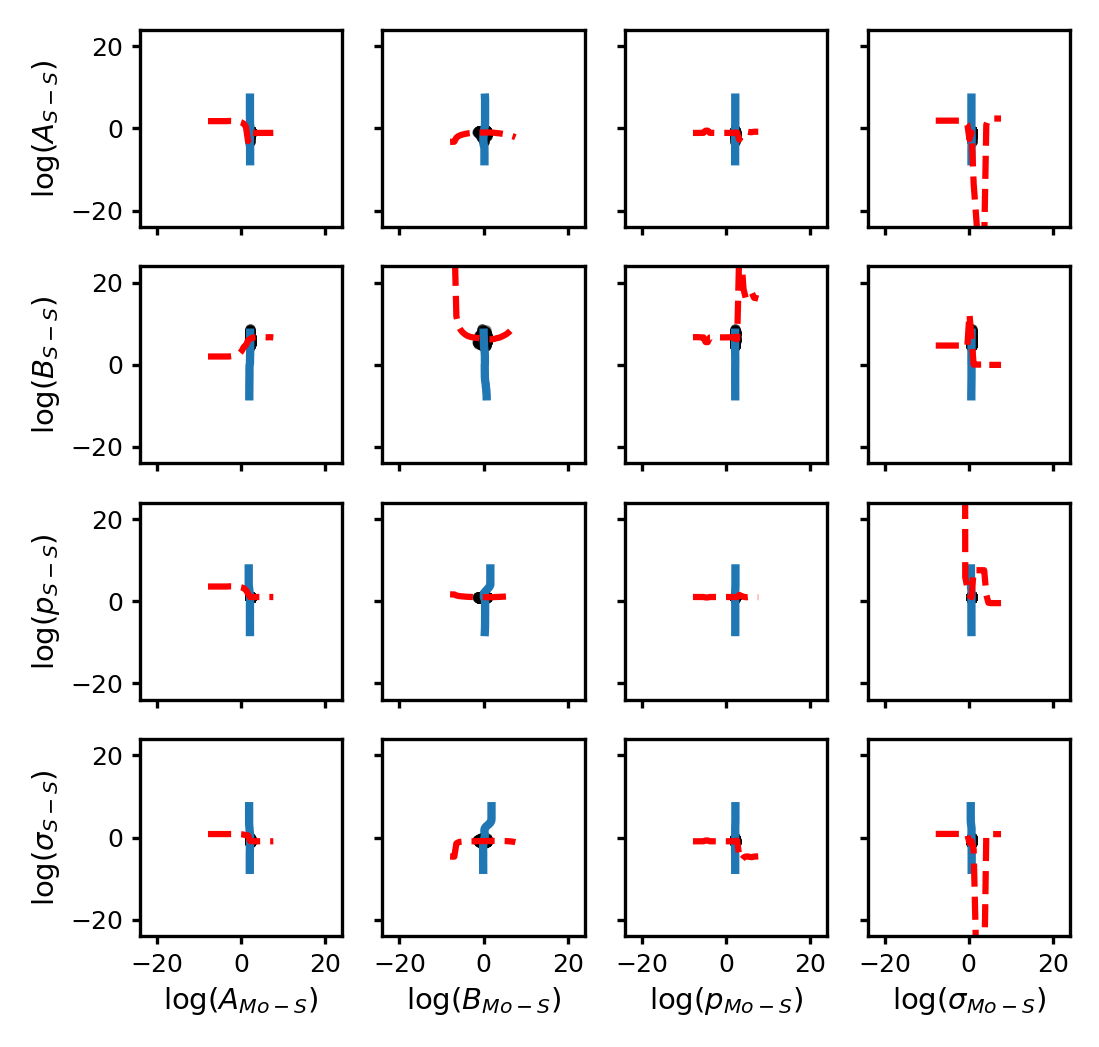}
    \caption[UQ results for SW potential Mo--S and S--S parameters at $T = 5.40\times10^{-3}~T_0$]{
        Profile likelihood and MCMC samples for Mo--S and S--S parameters at sampling temperature $5.40\times10^{-3}~T_0$ for the SW MoS$_2$ potential.
    }
\end{figure*}

\begin{figure*}[!h]
    \centering
    \includegraphics[width=0.6\textwidth]{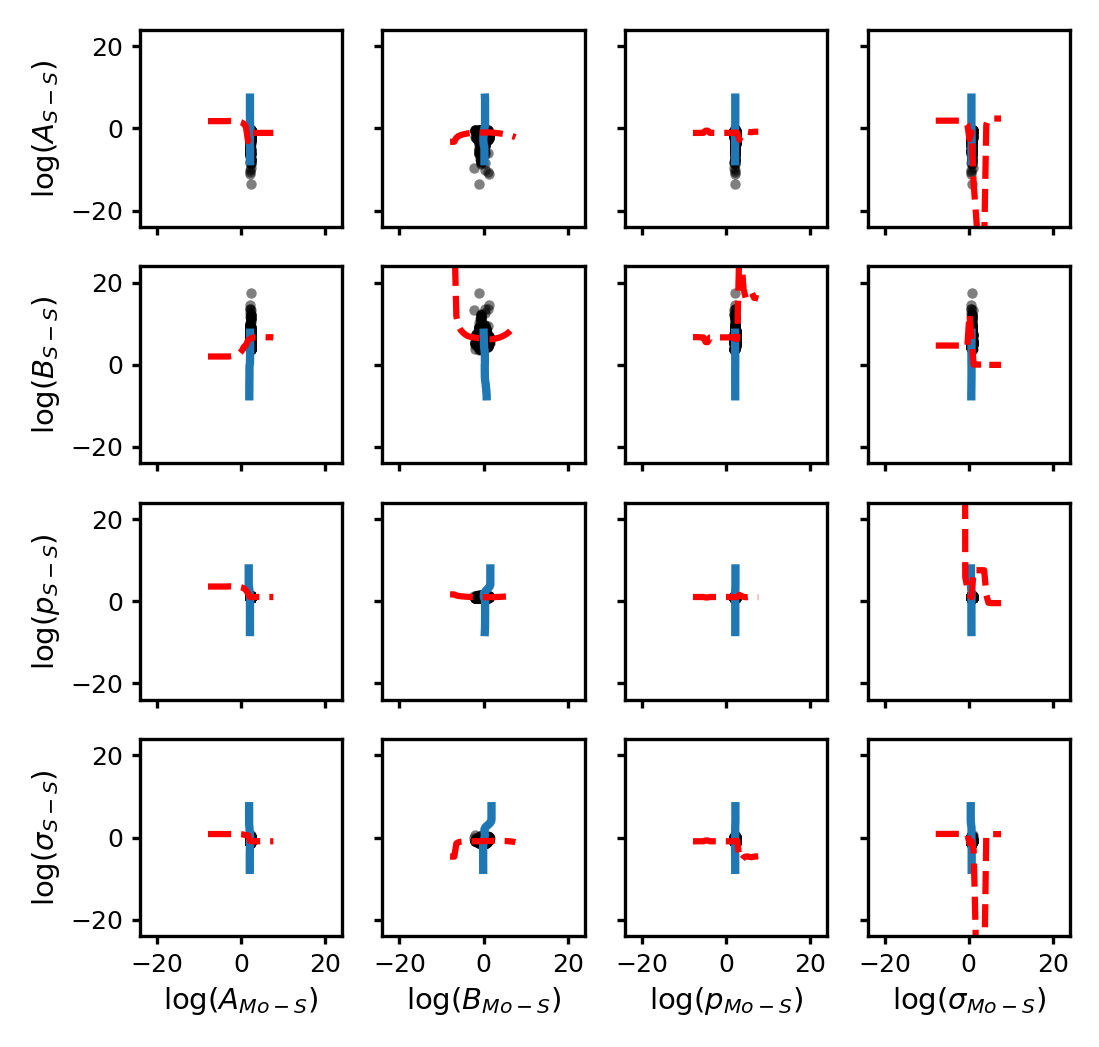}
    \caption[UQ results for SW potential Mo--S and S--S parameters at $T = 1.71\times10^{-2}~T_0$]{
        Profile likelihood and MCMC samples for Mo--S and S--S parameters at sampling temperature $1.71\times10^{-2}~T_0$ for the SW MoS$_2$ potential.
    }
\end{figure*}

\begin{figure*}[!h]
    \centering
    \includegraphics[width=0.6\textwidth]{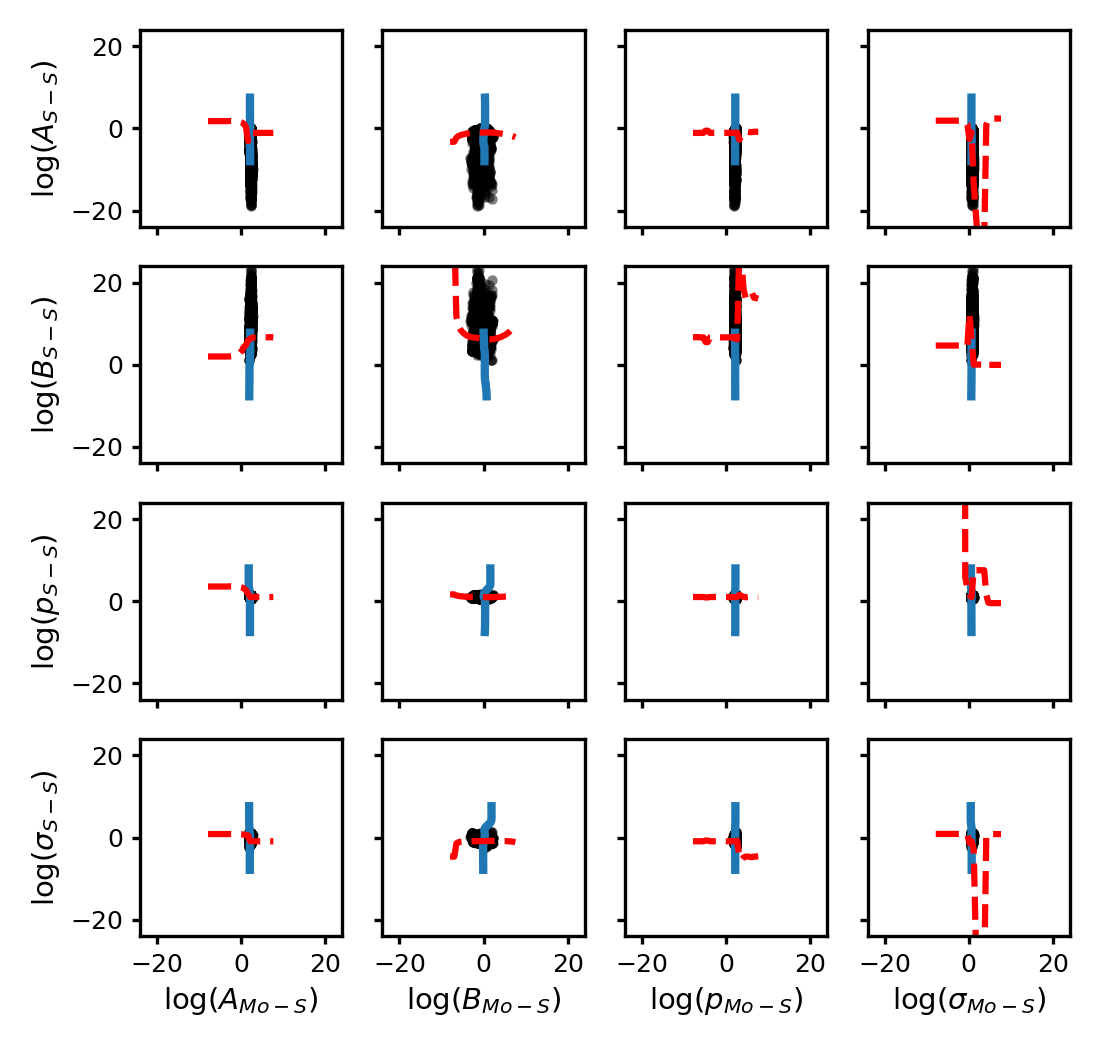}
    \caption[UQ results for SW potential Mo--S and S--S parameters at $T = 5.40\times10^{-2}~T_0$]{
        Profile likelihood and MCMC samples for Mo--S and S--S parameters at sampling temperature $5.40\times10^{-2}~T_0$ for the SW MoS$_2$ potential.
    }
\end{figure*}

\begin{figure*}[!h]
    \centering
    \includegraphics[width=0.6\textwidth]{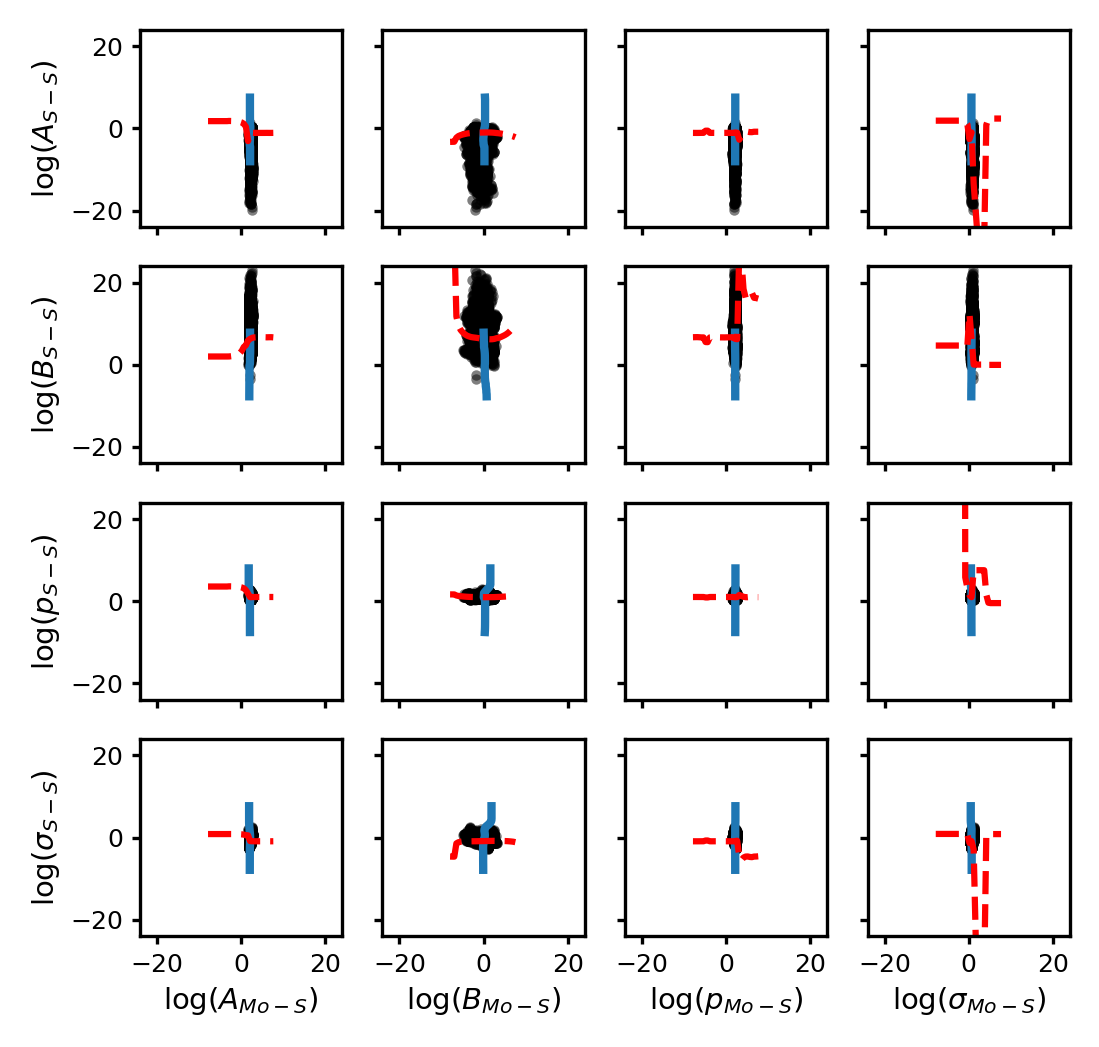}
    \caption[UQ results for SW potential Mo--S and S--S parameters at $T = 1.71\times10^{-1}~T_0$]{
        Profile likelihood and MCMC samples for Mo--S and S--S parameters at sampling temperature $1.71\times10^{-1}~T_0$ for the SW MoS$_2$ potential.
    }
\end{figure*}

\begin{figure*}[!h]
    \centering
    \includegraphics[width=0.6\textwidth]{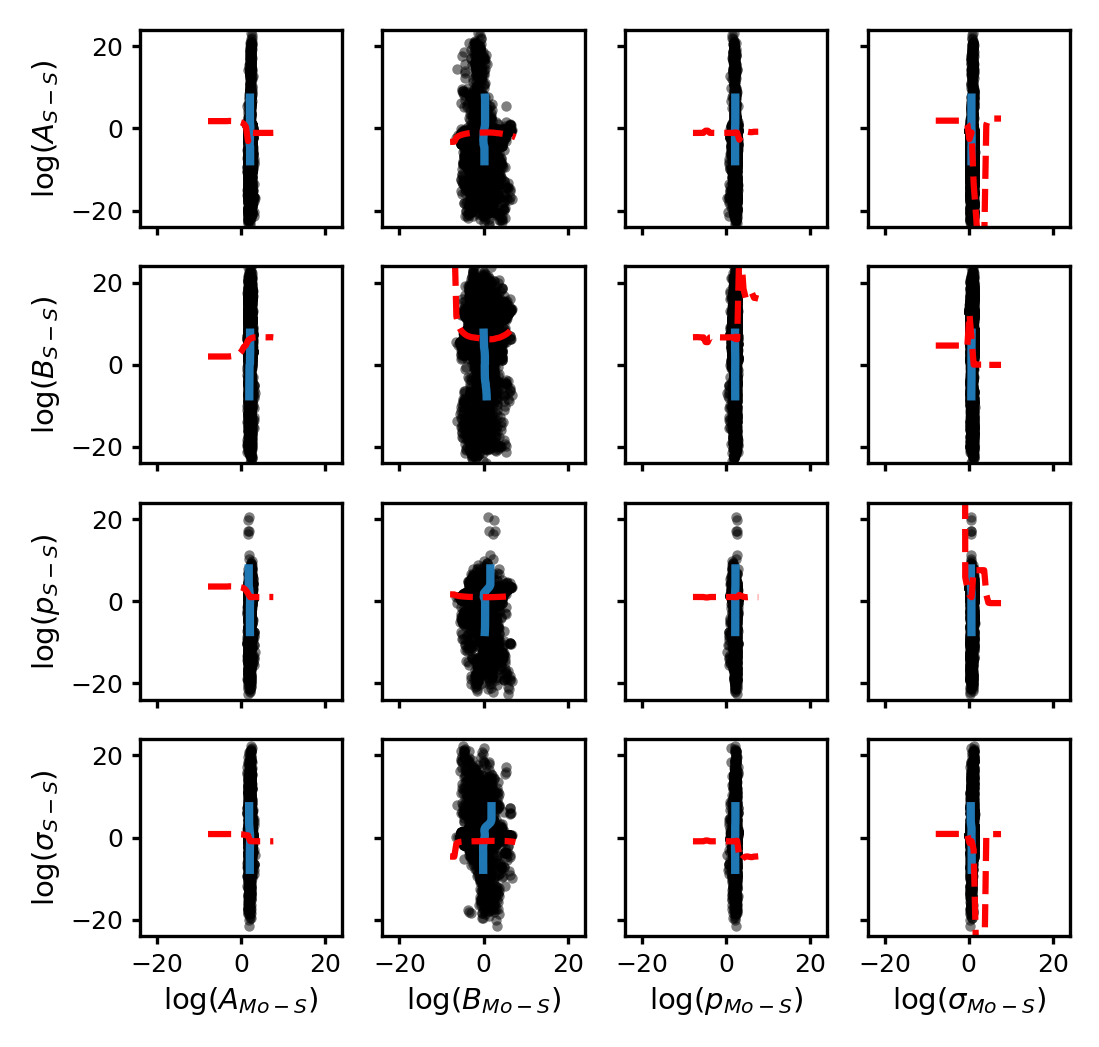}
    \caption[UQ results for SW potential Mo--S and S--S parameters at $T = 5.40\times10^{-1}~T_0$]{
        Profile likelihood and MCMC samples for Mo--S and S--S parameters at sampling temperature $5.40\times10^{-1}~T_0$ for the SW MoS$_2$ potential.
    }
\end{figure*}

\ifincludeTo
    \begin{figure*}[!h]
        \centering
        \includegraphics[width=0.6\textwidth]{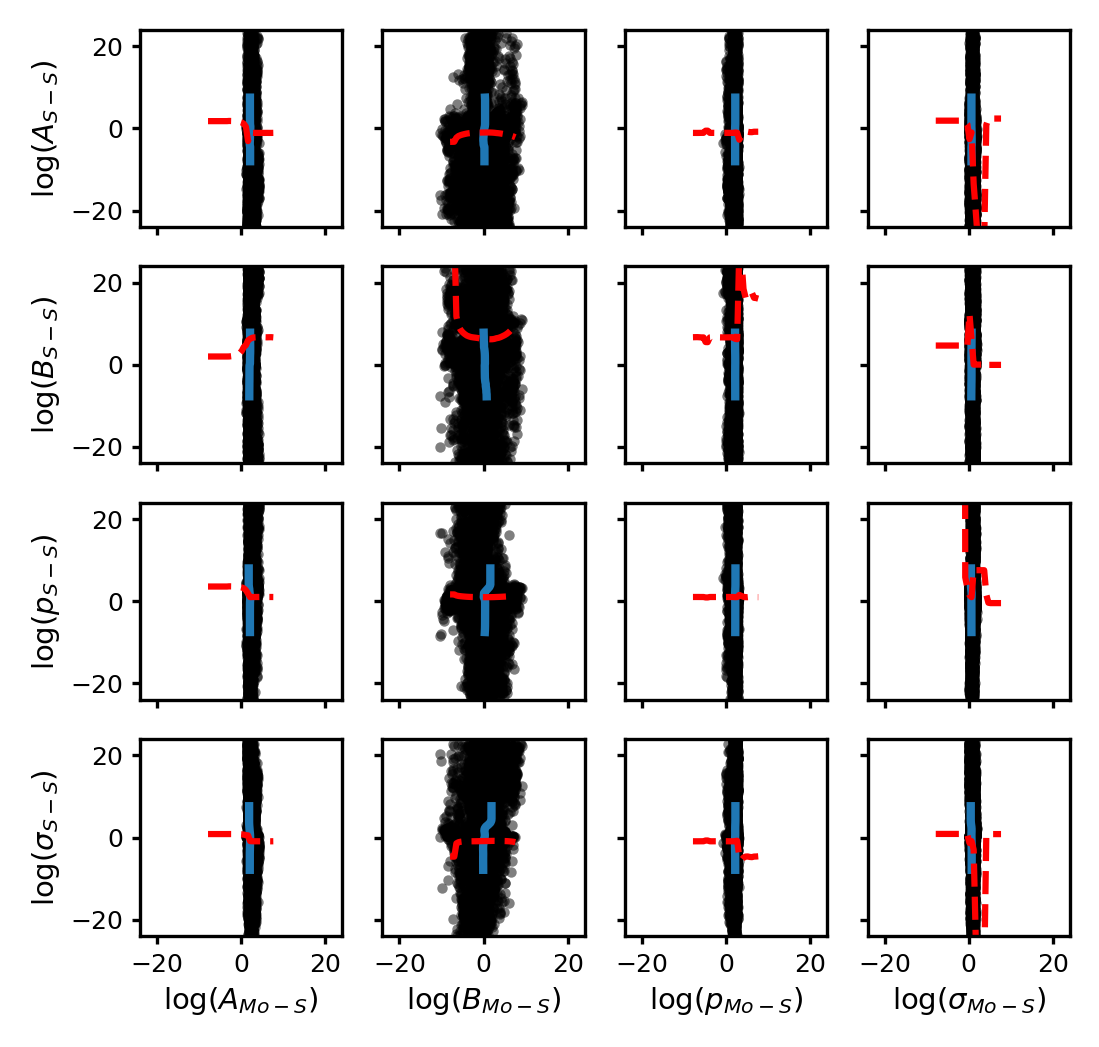}
        \caption[UQ results for SW potential Mo--S and S--S parameters at $T = T_0$]{
            Profile likelihood and MCMC samples for Mo--S and S--S parameters at sampling temperature $T_0$ for the SW MoS$_2$ potential.
        }
    \end{figure*}
\fi

\begin{figure*}[!h]
    \centering
    \includegraphics[width=0.6\textwidth]{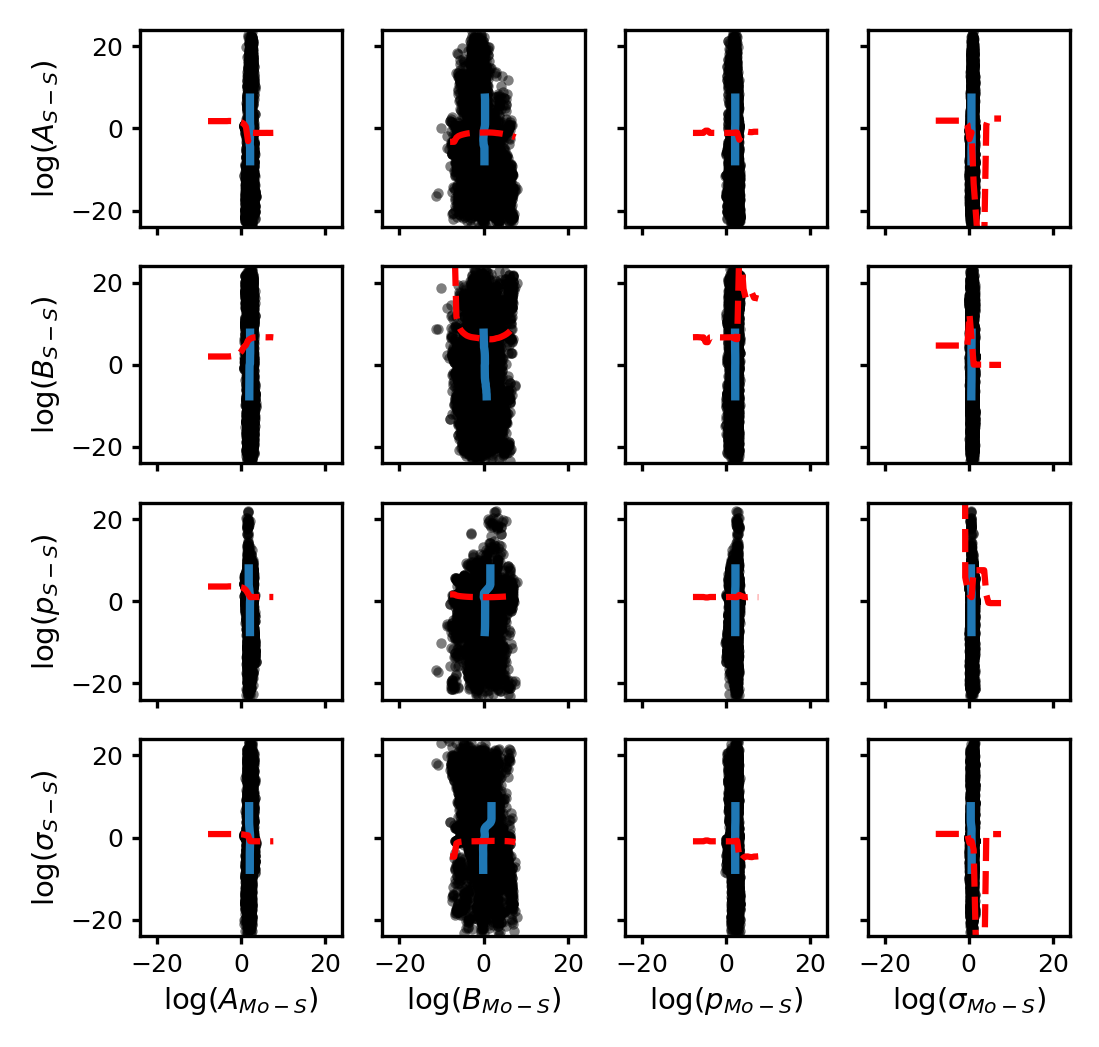}
    \caption[UQ results for SW potential Mo--S and S--S parameters at $T = 1.71~T_0$]{
        Profile likelihood and MCMC samples for Mo--S and S--S parameters at sampling temperature $1.71~T_0$ for the SW MoS$_2$ potential.
    }
\end{figure*}

\begin{figure*}[!h]
    \centering
    \includegraphics[width=0.6\textwidth]{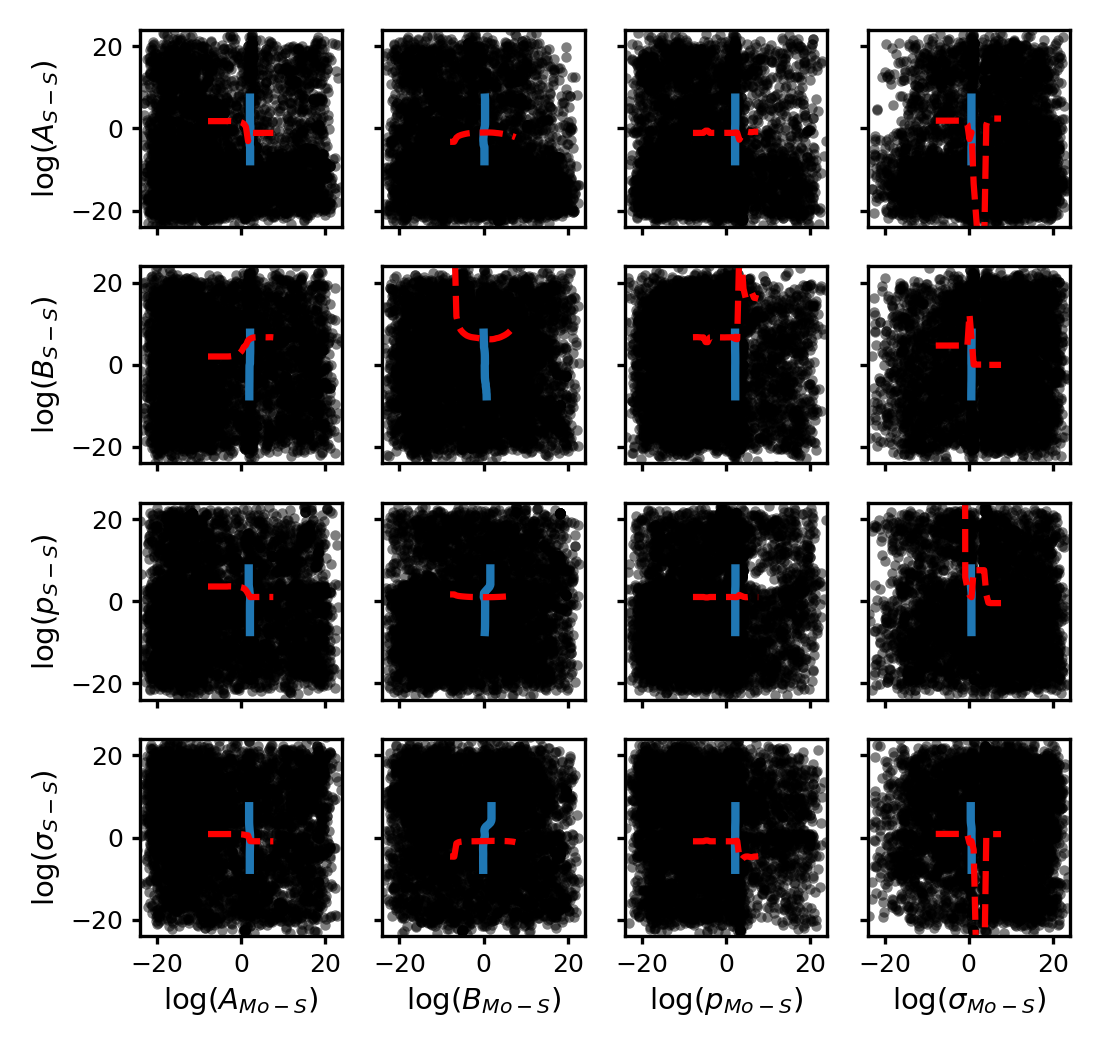}
    \caption[UQ results for SW potential Mo--S and S--S parameters at $T = 5.40~T_0$]{
        Profile likelihood and MCMC samples for Mo--S and S--S parameters at sampling temperature $5.40~T_0$ for the SW MoS$_2$ potential.
    }
\end{figure*}

\cleardoublepage

\subsection{Mo--S and 3-body parameters}
\label{subsec:Mo-S_3-body}
Profile likelihood and MCMC samples between Mo--S and 3-body parameters.
Notice that there is a lack of correlation between between parameters corresponding to different interaction types.

\begin{figure*}[!h]
    \centering
    \includegraphics[width=0.6\textwidth]{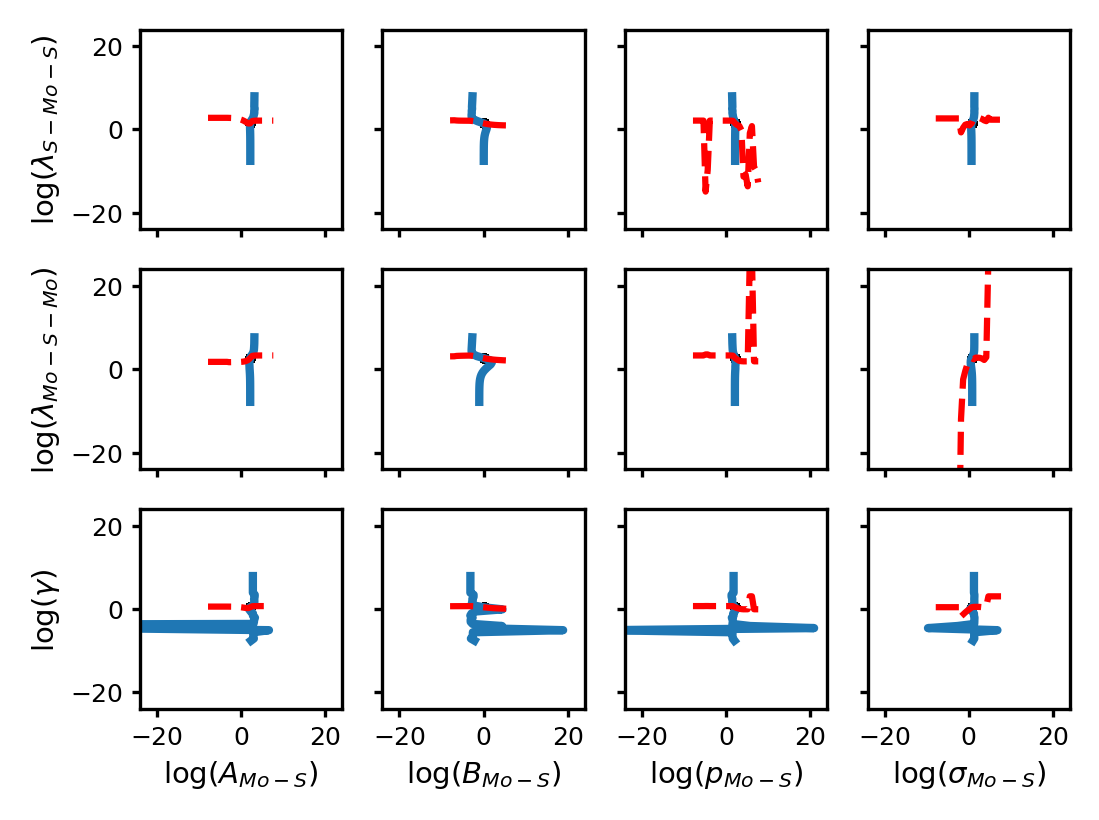}
    \caption[UQ results for SW potential Mo--S and 3-body parameters at $T = 5.40\times10^{-6}~T_0$]{
        Profile likelihood and MCMC samples for Mo--S and 3-body parameters at sampling temperature $5.40\times10^{-6}~T_0$ for the SW MoS$_2$ potential.
    }
\end{figure*}

\begin{figure*}[!h]
    \centering
    \includegraphics[width=0.6\textwidth]{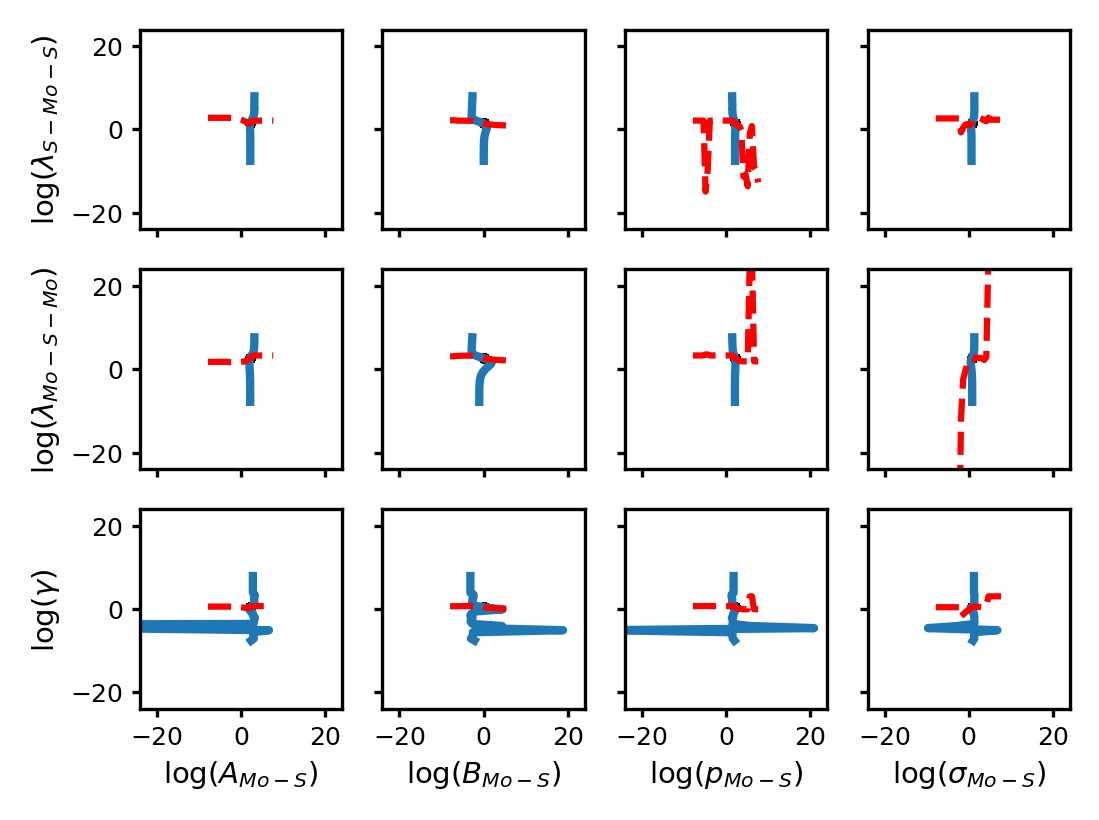}
    \caption[UQ results for SW potential Mo--S and 3-body parameters at $T = 1.71\times10^{-5}~T_0$]{
        Profile likelihood and MCMC samples for Mo--S and 3-body parameters at sampling temperature $1.71\times10^{-5}~T_0$ for the SW MoS$_2$ potential.
    }
\end{figure*}

\begin{figure*}[!h]
    \centering
    \includegraphics[width=0.6\textwidth]{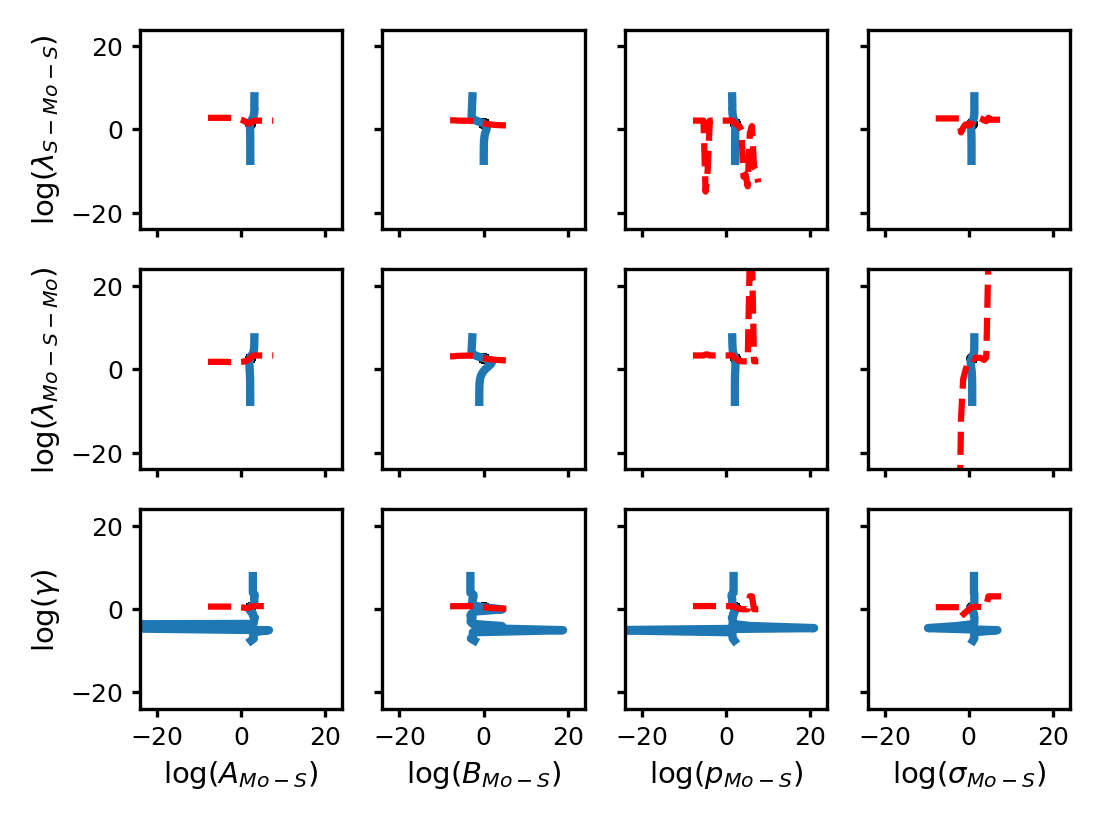}
    \caption[UQ results for SW potential Mo--S and 3-body parameters at $T = 5.40\times10^{-5}~T_0$]{
        Profile likelihood and MCMC samples for Mo--S and 3-body parameters at sampling temperature $5.40\times10^{-5}~T_0$ for the SW MoS$_2$ potential.
    }
\end{figure*}

\begin{figure*}[!h]
    \centering
    \includegraphics[width=0.6\textwidth]{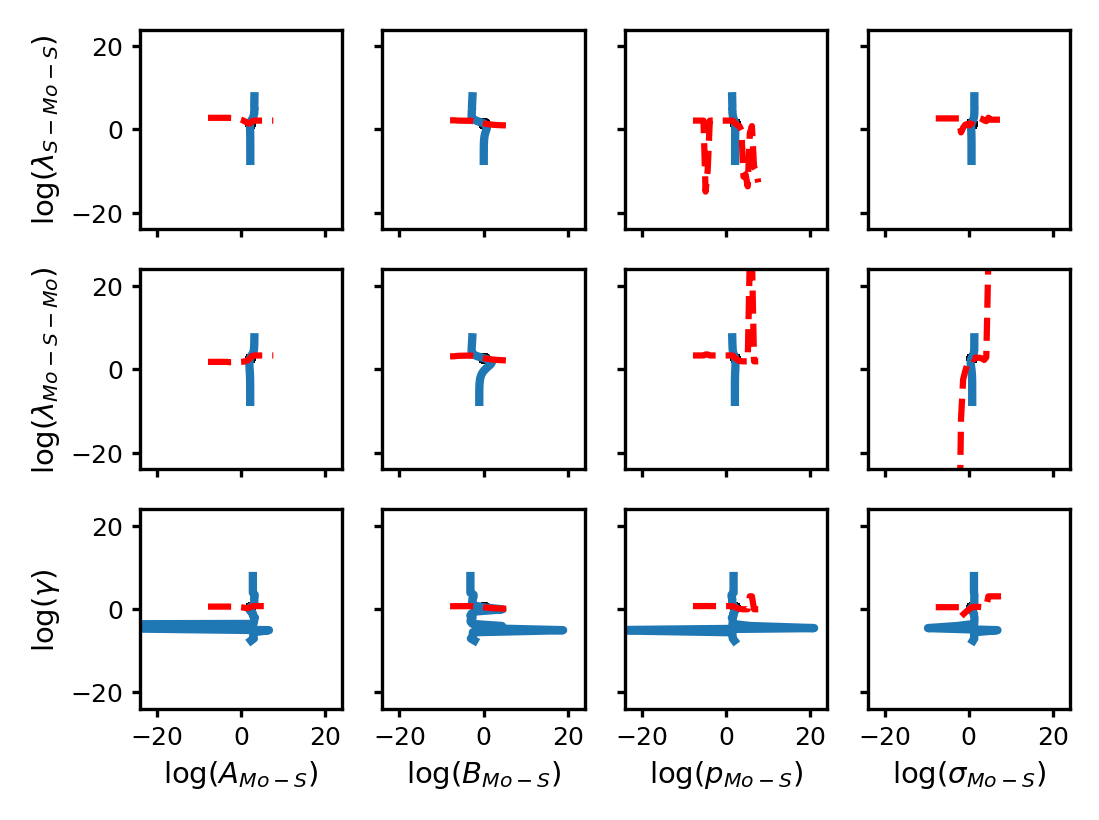}
    \caption[UQ results for SW potential Mo--S and 3-body parameters at $T = 1.71\times10^{-4}~T_0$]{
        Profile likelihood and MCMC samples for Mo--S and 3-body parameters at sampling temperature $1.71\times10^{-4}~T_0$ for the SW MoS$_2$ potential.
    }
\end{figure*}

\begin{figure*}[!h]
    \centering
    \includegraphics[width=0.6\textwidth]{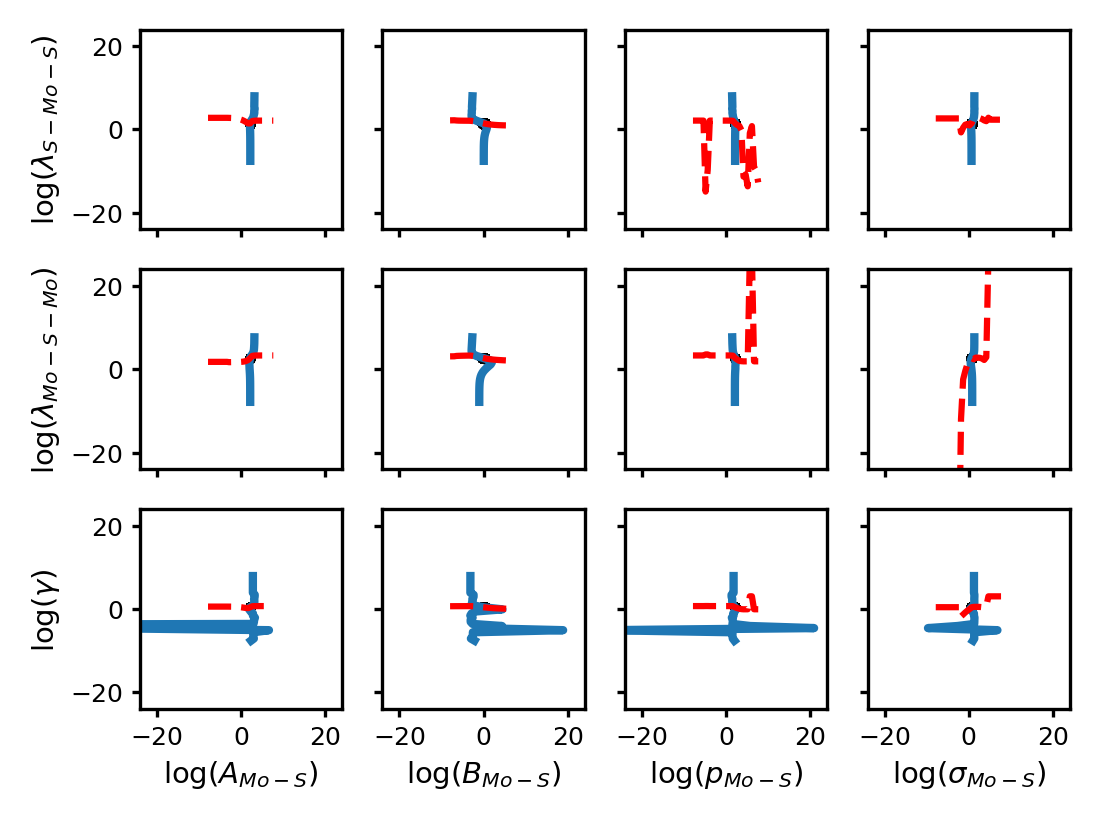}
    \caption[UQ results for SW potential Mo--S and 3-body parameters at $T = 5.40\times10^{-4}~T_0$]{
        Profile likelihood and MCMC samples for Mo--S and 3-body parameters at sampling temperature $5.40\times10^{-4}~T_0$ for the SW MoS$_2$ potential.
    }
\end{figure*}

\begin{figure*}[!h]
    \centering
    \includegraphics[width=0.6\textwidth]{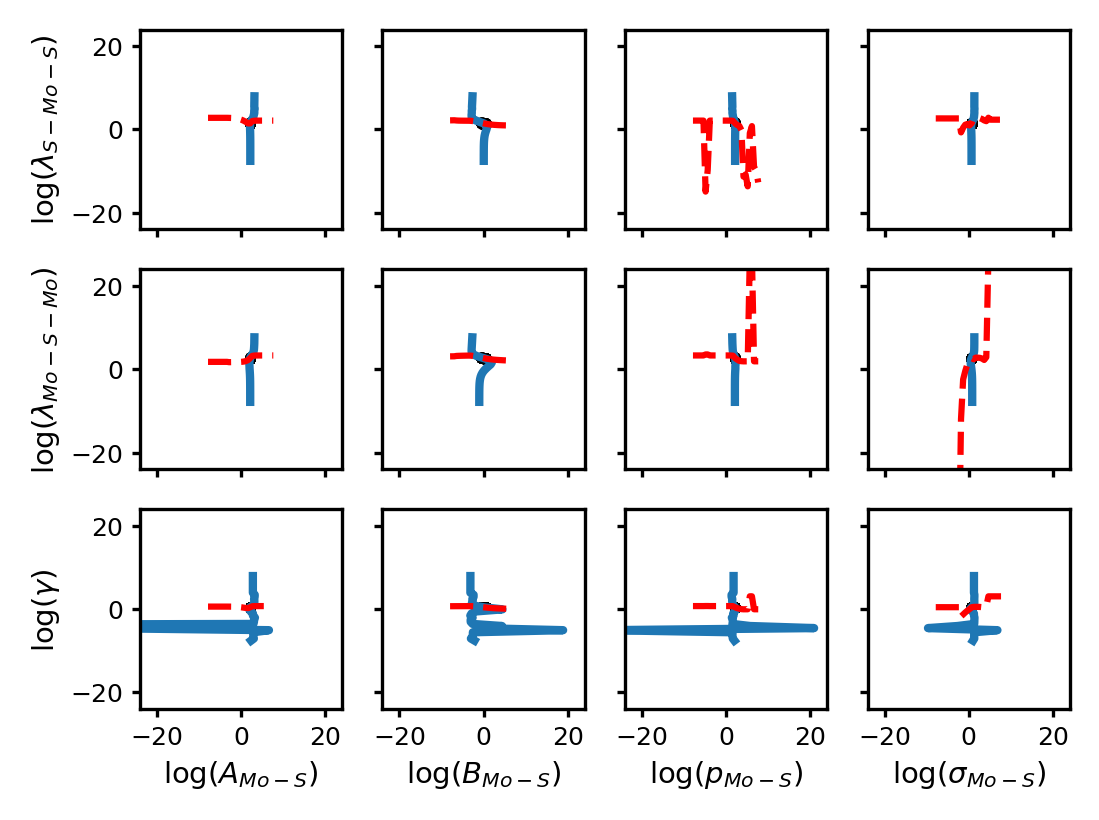}
    \caption[UQ results for SW potential Mo--S and 3-body parameters at $T = 1.71\times10^{-3}~T_0$]{
        Profile likelihood and MCMC samples for Mo--S and 3-body parameters at sampling temperature $1.71\times10^{-3}~T_0$ for the SW MoS$_2$ potential.
    }
\end{figure*}

\begin{figure*}[!h]
    \centering
    \includegraphics[width=0.6\textwidth]{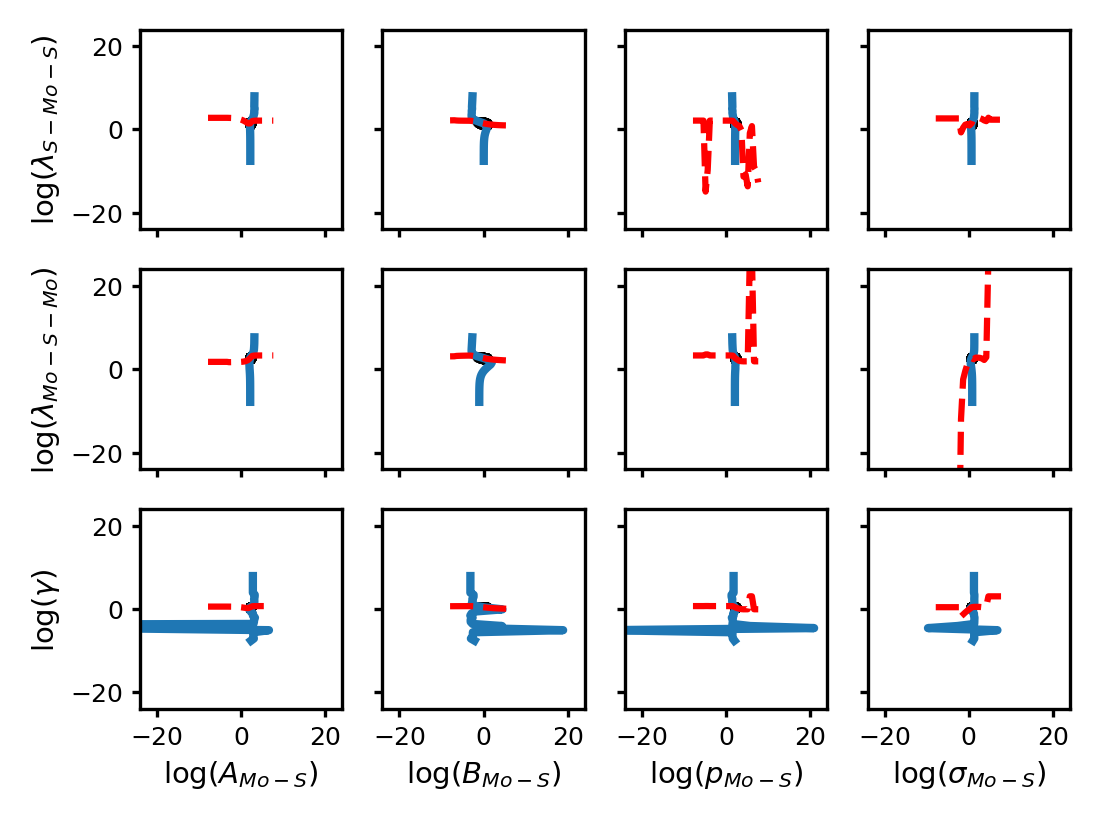}
    \caption[UQ results for SW potential Mo--S and 3-body parameters at $T = 5.40\times10^{-3}~T_0$]{
        Profile likelihood and MCMC samples for Mo--S and 3-body parameters at sampling temperature $5.40\times10^{-3}~T_0$ for the SW MoS$_2$ potential.
    }
\end{figure*}

\begin{figure*}[!h]
    \centering
    \includegraphics[width=0.6\textwidth]{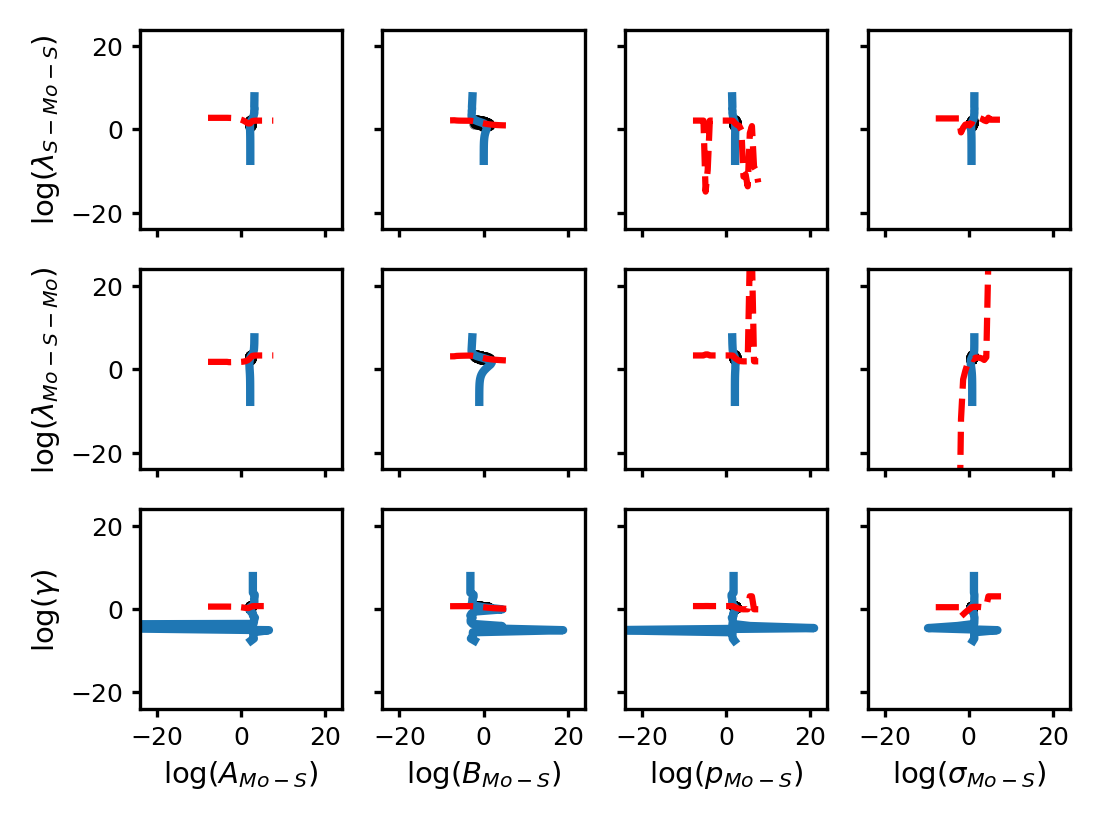}
    \caption[UQ results for SW potential Mo--S and 3-body parameters at $T = 1.71\times10^{-2}~T_0$]{
        Profile likelihood and MCMC samples for Mo--S and 3-body parameters at sampling temperature $1.71\times10^{-2}~T_0$ for the SW MoS$_2$ potential.
    }
\end{figure*}

\begin{figure*}[!h]
    \centering
    \includegraphics[width=0.6\textwidth]{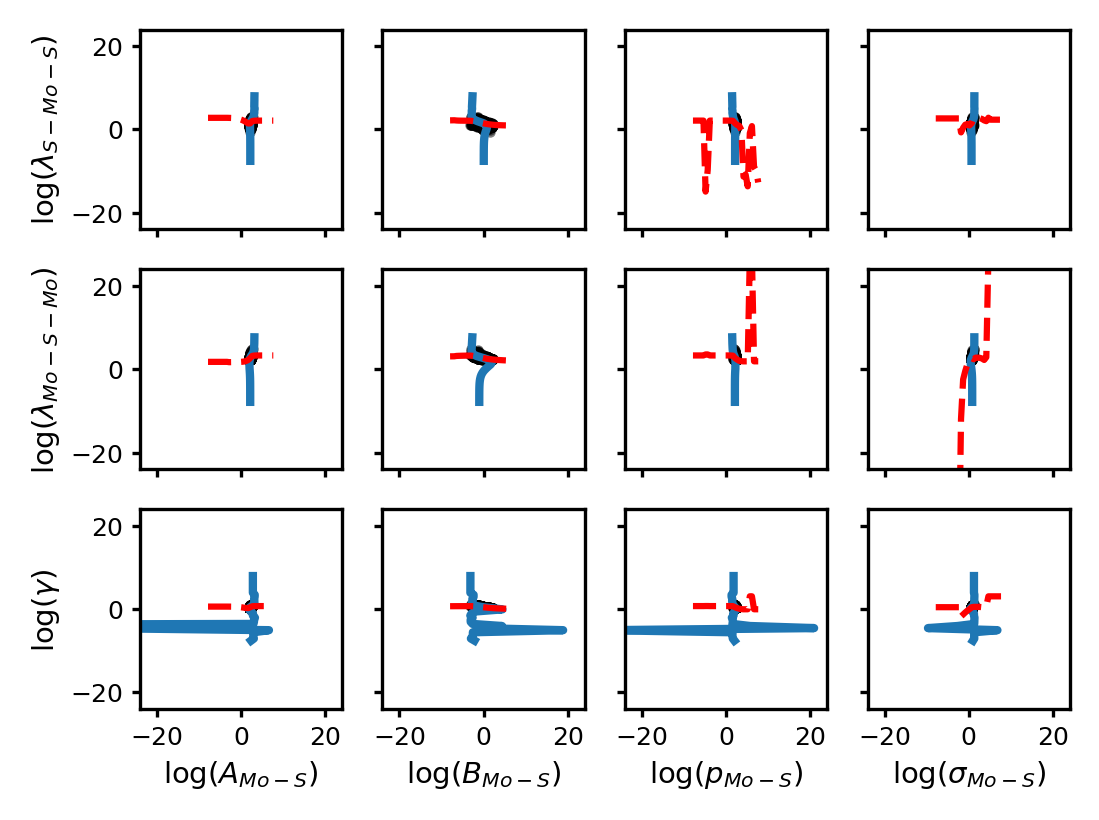}
    \caption[UQ results for SW potential Mo--S and 3-body parameters at $T = 5.40\times10^{-2}~T_0$]{
        Profile likelihood and MCMC samples for Mo--S and 3-body parameters at sampling temperature $5.40\times10^{-2}~T_0$ for the SW MoS$_2$ potential.
    }
\end{figure*}

\begin{figure*}[!h]
    \centering
    \includegraphics[width=0.6\textwidth]{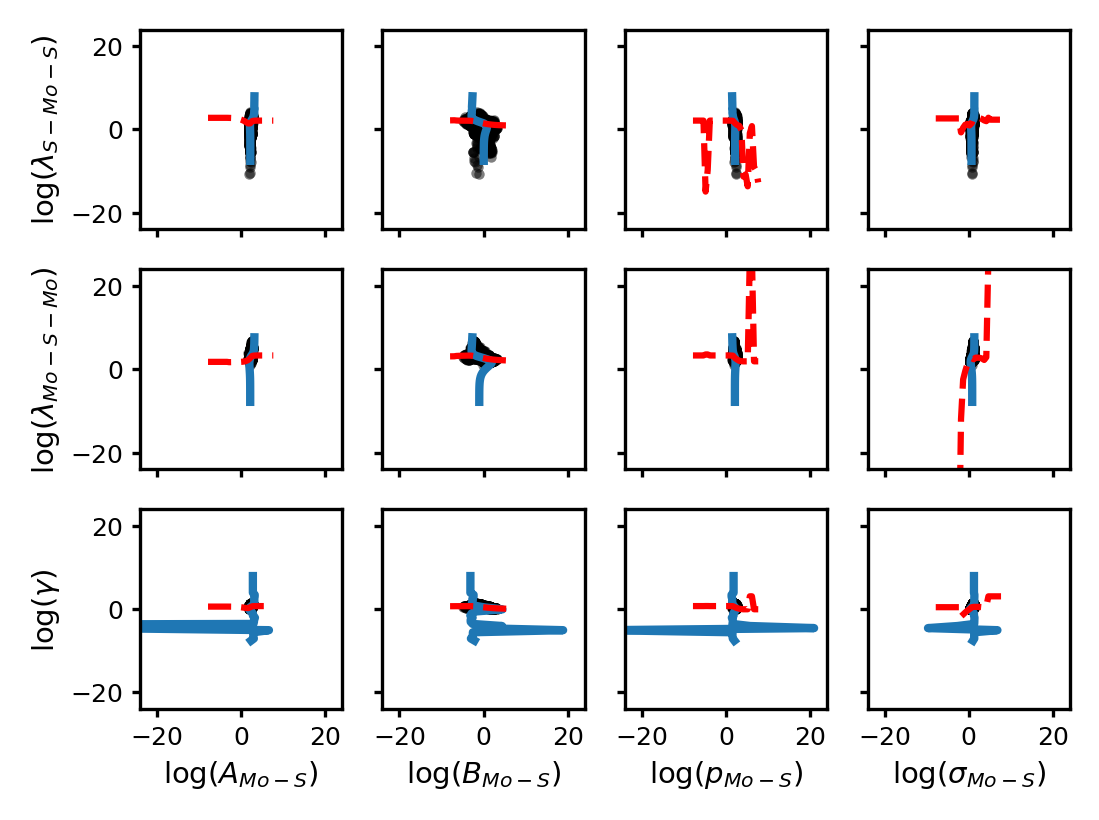}
    \caption[UQ results for SW potential Mo--S and 3-body parameters at $T = 1.71\times10^{-1}~T_0$]{
        Profile likelihood and MCMC samples for Mo--S and 3-body parameters at sampling temperature $1.71\times10^{-1}~T_0$ for the SW MoS$_2$ potential.
    }
\end{figure*}

\begin{figure*}[!h]
    \centering
    \includegraphics[width=0.6\textwidth]{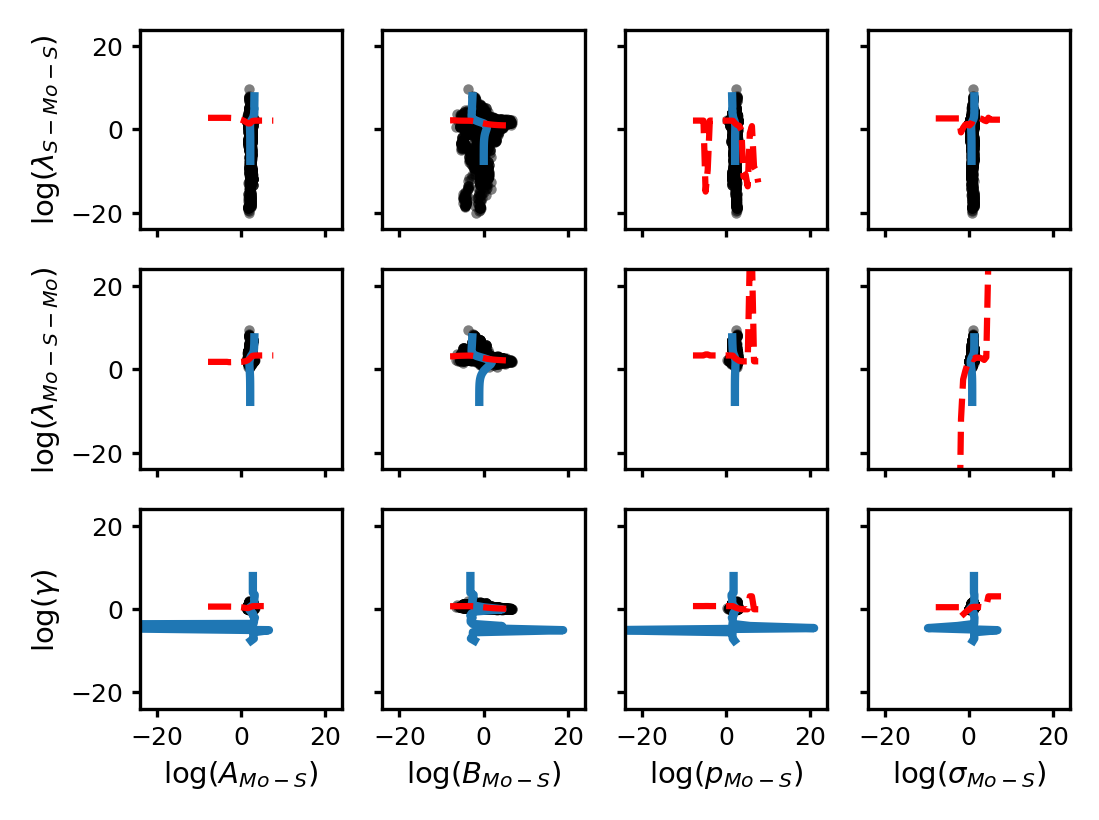}
    \caption[UQ results for SW potential Mo--S and 3-body parameters at $T = 5.40\times10^{-1}~T_0$]{
        Profile likelihood and MCMC samples for Mo--S and 3-body parameters at sampling temperature $5.40\times10^{-1}~T_0$ for the SW MoS$_2$ potential.
    }
\end{figure*}

\ifincludeTo
    \begin{figure*}[!h]
        \centering
        \includegraphics[width=0.6\textwidth]{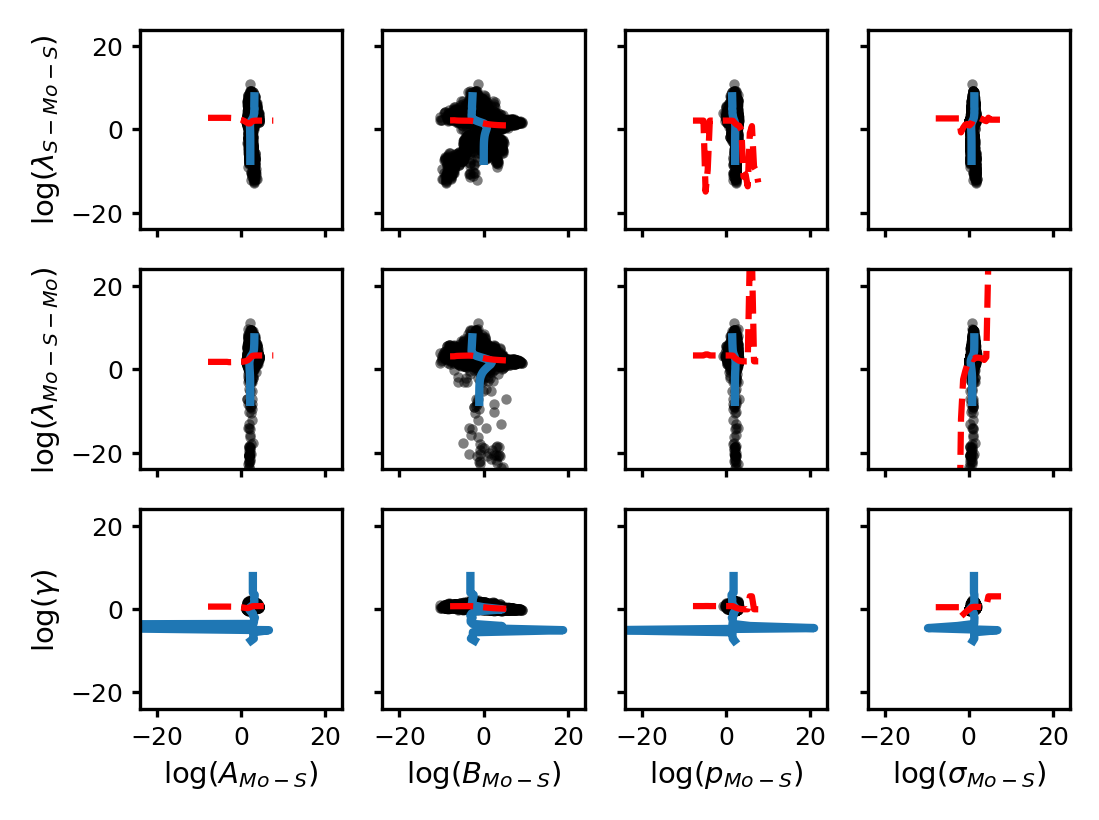}
        \caption[UQ results for SW potential Mo--S and 3-body parameters at $T = T_0$]{
            Profile likelihood and MCMC samples for Mo--S and 3-body parameters at sampling temperature $T_0$ for the SW MoS$_2$ potential.
        }
    \end{figure*}
\fi

\begin{figure*}[!h]
    \centering
    \includegraphics[width=0.6\textwidth]{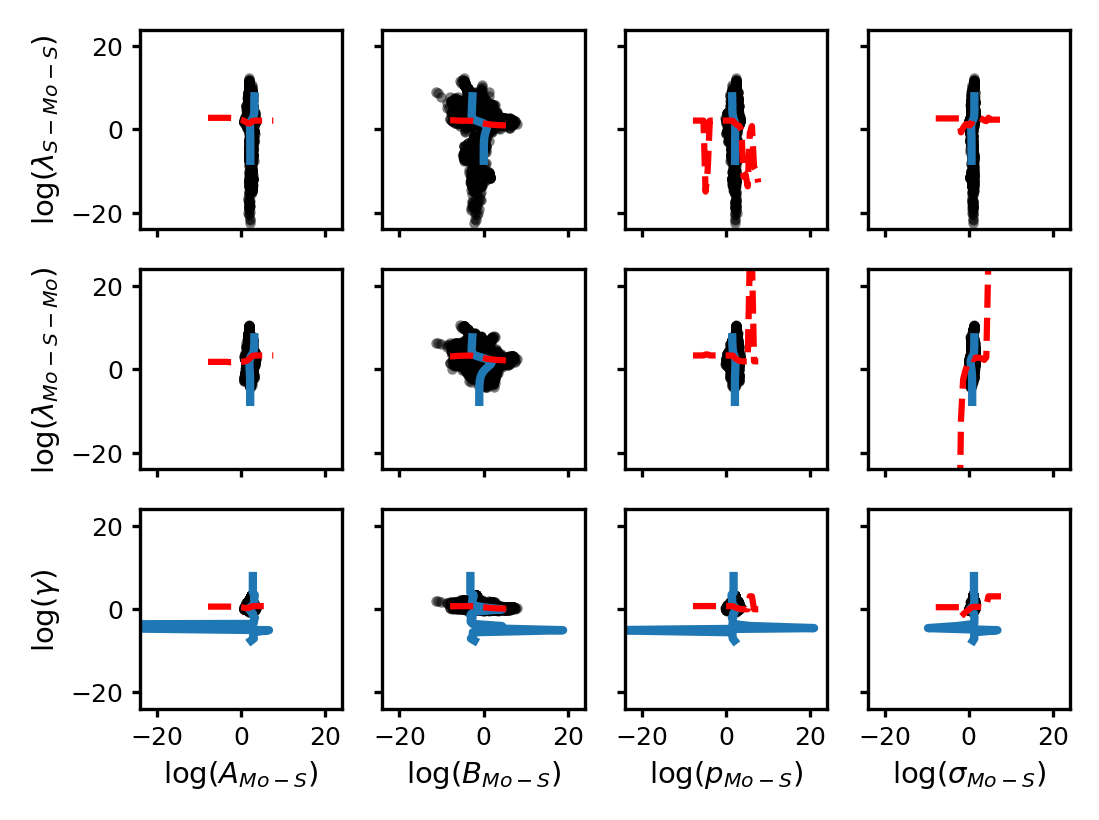}
    \caption[UQ results for SW potential Mo--S and 3-body parameters at $T = 1.71~T_0$]{
        Profile likelihood and MCMC samples for Mo--S and 3-body parameters at sampling temperature $1.71~T_0$ for the SW MoS$_2$ potential.
    }
\end{figure*}

\begin{figure*}[!h]
    \centering
    \includegraphics[width=0.6\textwidth]{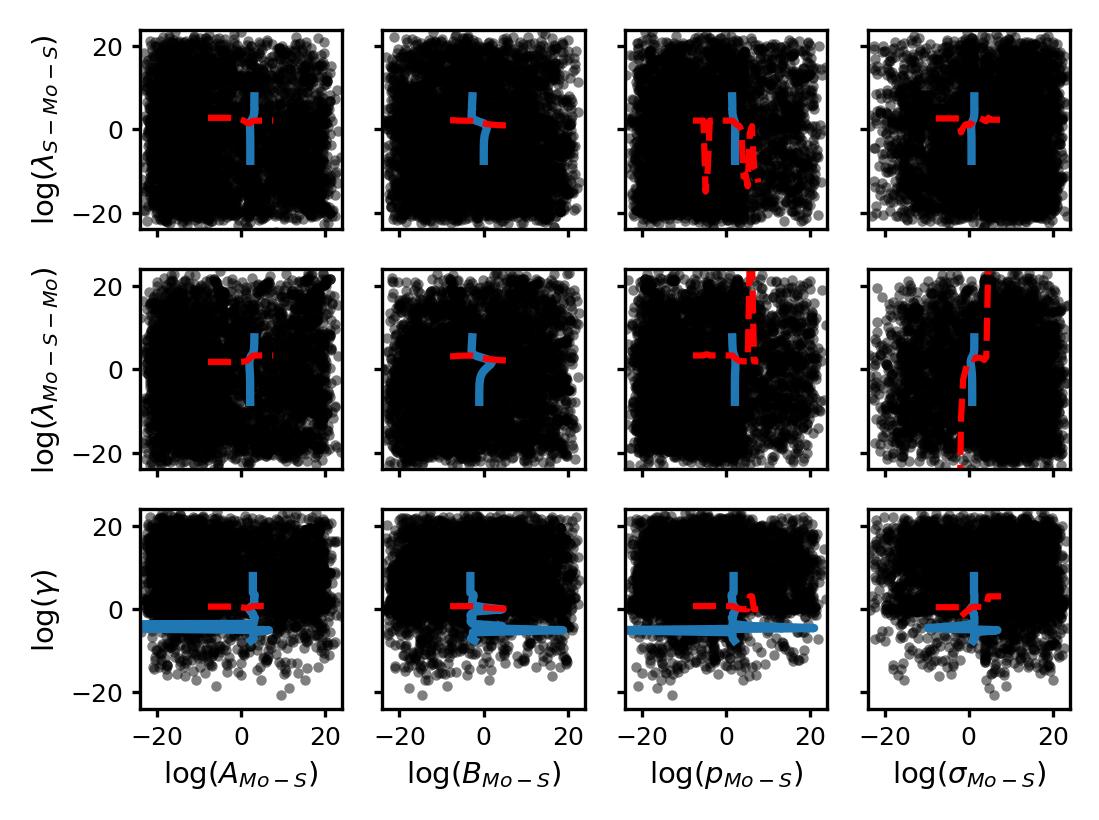}
    \caption[UQ results for SW potential Mo--S and 3-body parameters at $T = 5.40~T_0$]{
        Profile likelihood and MCMC samples for Mo--S and 3-body parameters at sampling temperature $5.40~T_0$ for the SW MoS$_2$ potential.
    }
\end{figure*}

\cleardoublepage

\subsection{S--S and 3-body parameters}
\label{subsec:S-S_3-body}
Profile likelihood and MCMC samples between S--S and 3-body parameters.
Notice that there is a lack of correlation between between parameters corresponding to different interaction types.

\begin{figure*}[!h]
    \centering
    \includegraphics[width=0.6\textwidth]{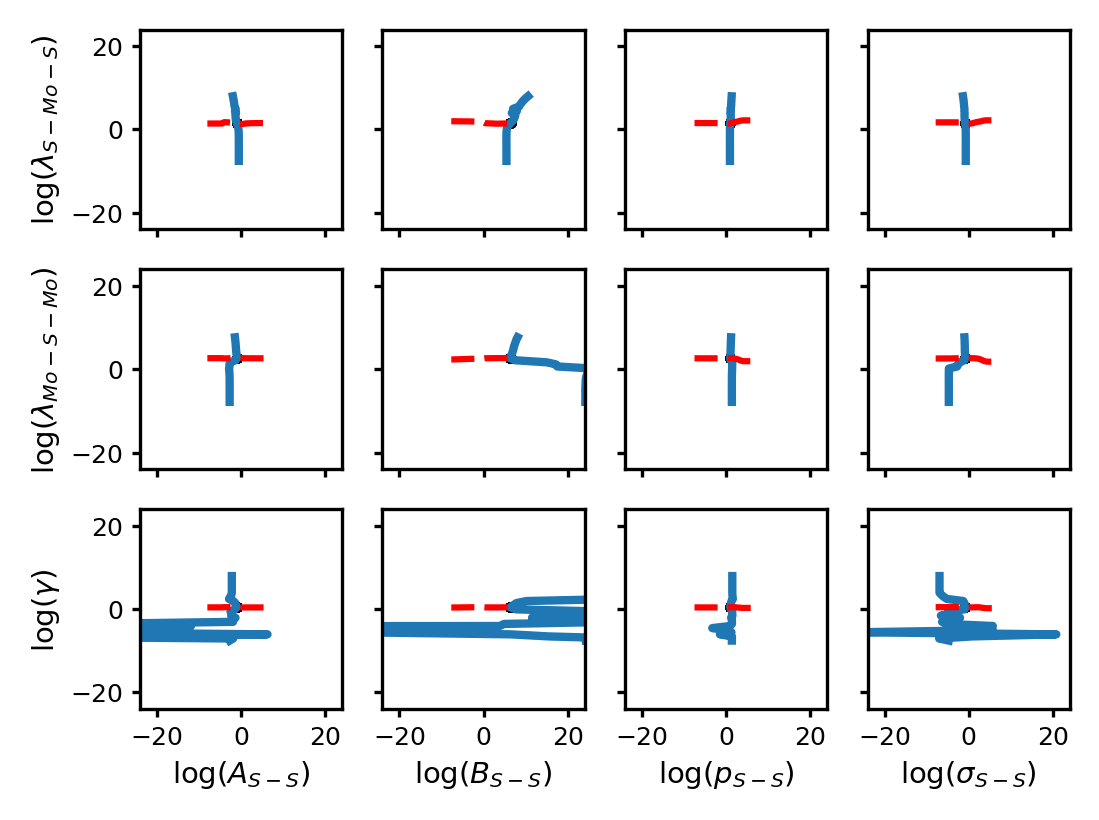}
    \caption[UQ results for SW potential S--S and 3-body parameters at $T = 5.40\times10^{-6}~T_0$]{
        Profile likelihood and MCMC samples for S--S and 3-body parameters at sampling temperature $5.40\times10^{-6}~T_0$ for the SW MoS$_2$ potential.
    }
\end{figure*}

\begin{figure*}[!h]
    \centering
    \includegraphics[width=0.6\textwidth]{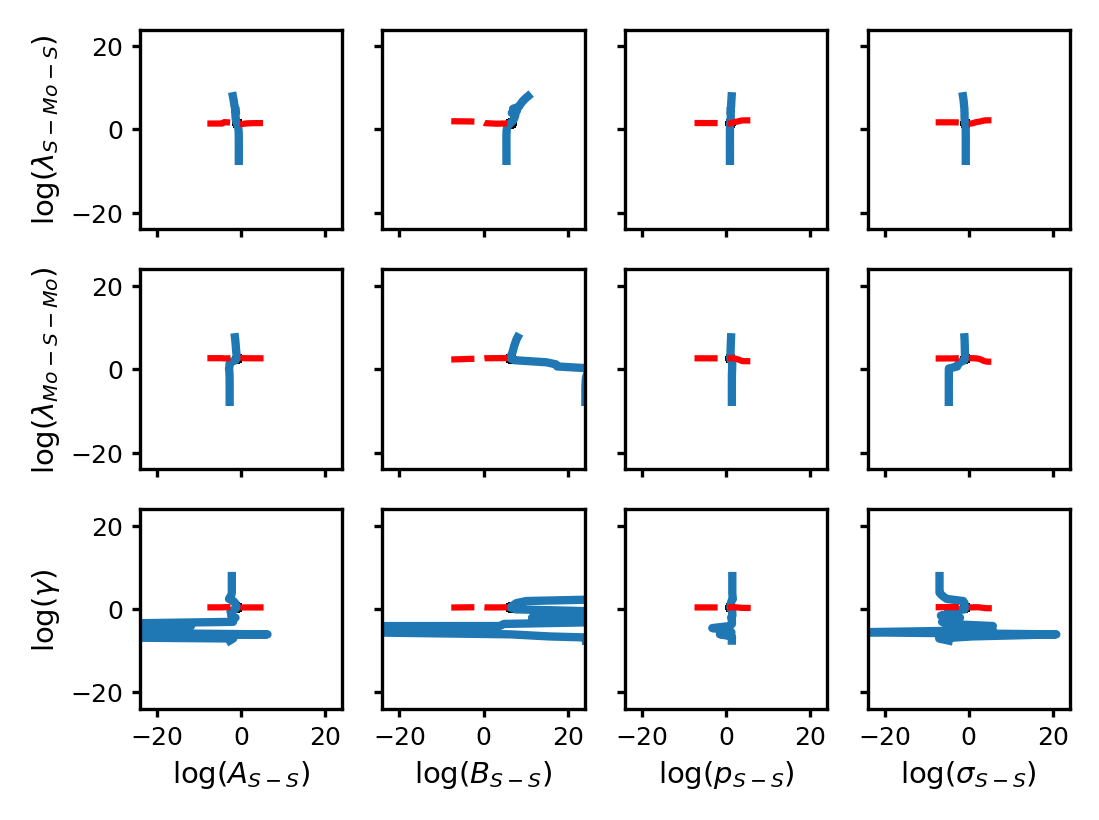}
    \caption[UQ results for SW potential S--S and 3-body parameters at $T = 1.71\times10^{-5}~T_0$]{
        Profile likelihood and MCMC samples for S--S and 3-body parameters at sampling temperature $1.71\times10^{-5}~T_0$ for the SW MoS$_2$ potential.
    }
\end{figure*}

\begin{figure*}[!h]
    \centering
    \includegraphics[width=0.6\textwidth]{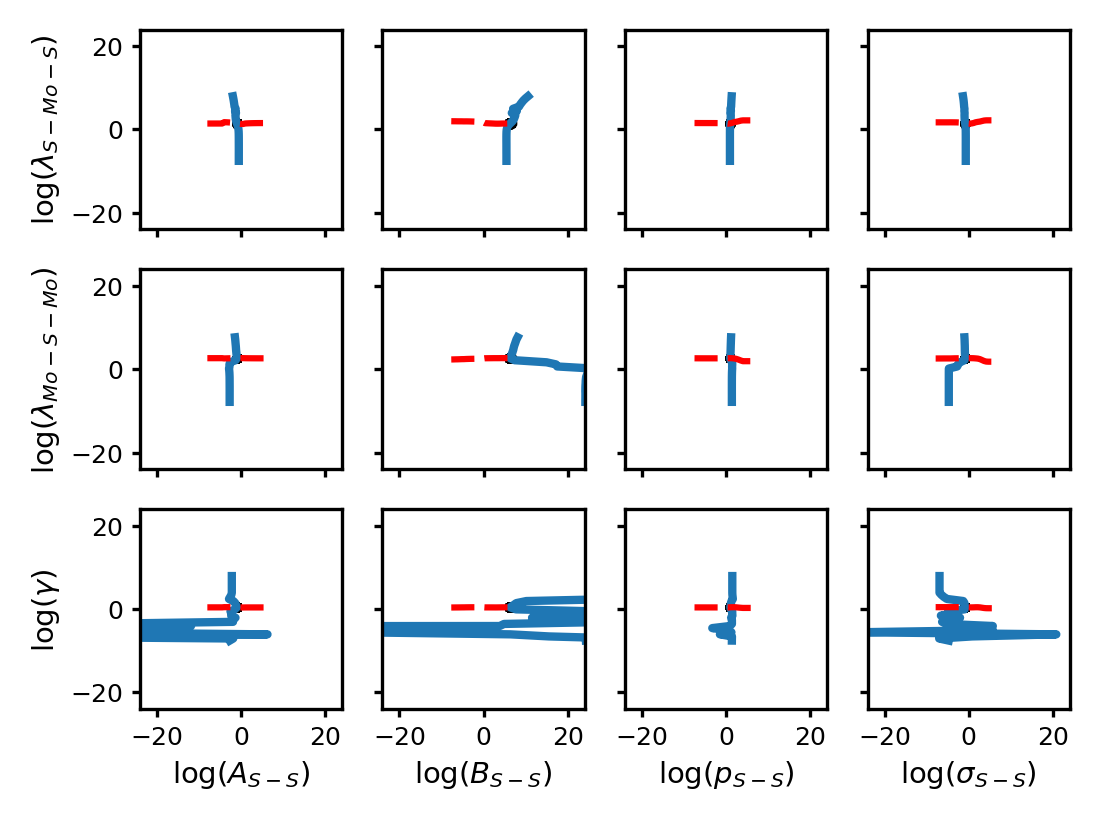}
    \caption[UQ results for SW potential S--S and 3-body parameters at $T = 5.40\times10^{-5}~T_0$]{
        Profile likelihood and MCMC samples for S--S and 3-body parameters at sampling temperature $5.40\times10^{-5}~T_0$ for the SW MoS$_2$ potential.
    }
\end{figure*}

\begin{figure*}[!h]
    \centering
    \includegraphics[width=0.6\textwidth]{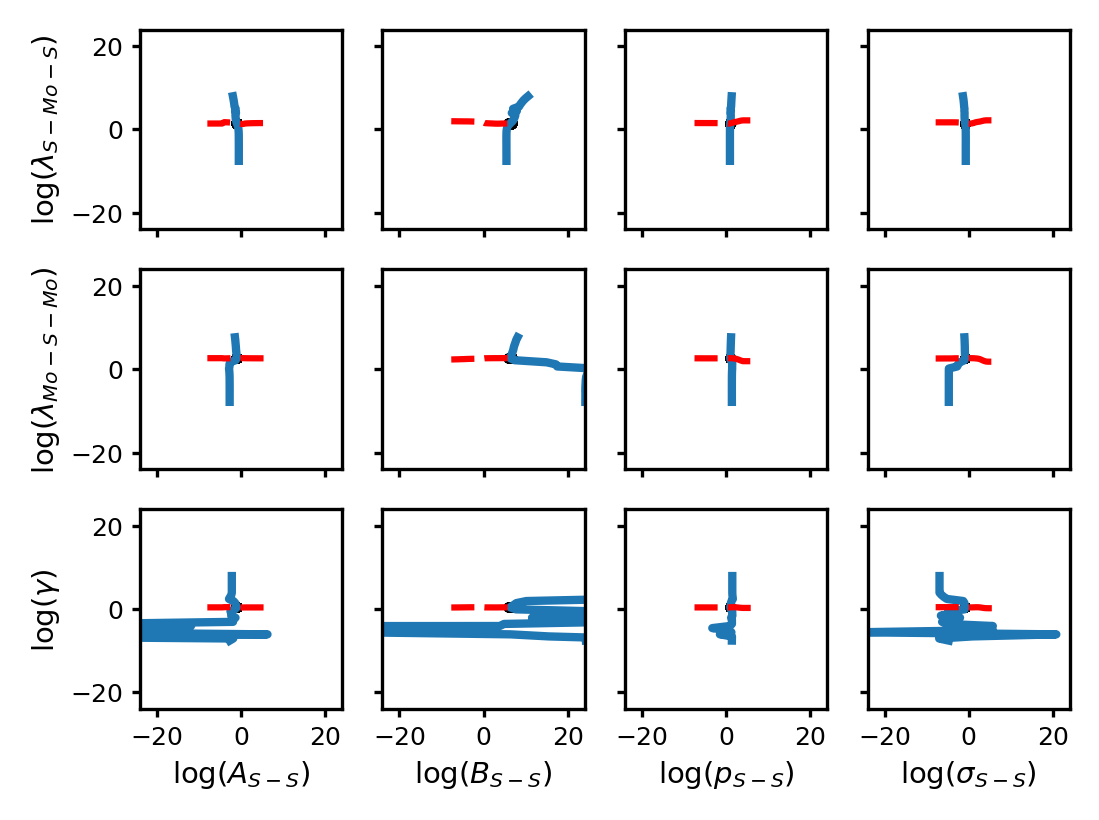}
    \caption[UQ results for SW potential S--S and 3-body parameters at $T = 1.71\times10^{-4}~T_0$]{
        Profile likelihood and MCMC samples for S--S and 3-body parameters at sampling temperature $1.71\times10^{-4}~T_0$ for the SW MoS$_2$ potential.
    }
\end{figure*}

\begin{figure*}[!h]
    \centering
    \includegraphics[width=0.6\textwidth]{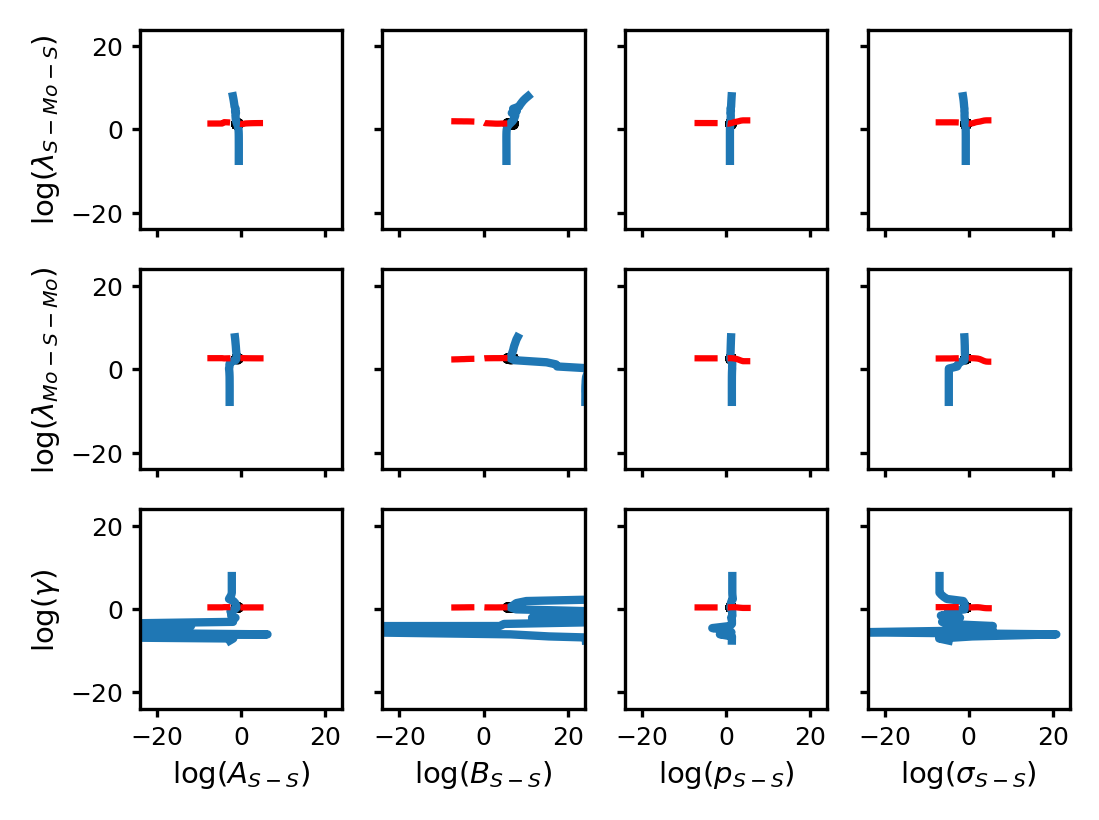}
    \caption[UQ results for SW potential S--S and 3-body parameters at $T = 5.40\times10^{-4}~T_0$]{
        Profile likelihood and MCMC samples for S--S and 3-body parameters at sampling temperature $5.40\times10^{-4}~T_0$ for the SW MoS$_2$ potential.
    }
\end{figure*}

\begin{figure*}[!h]
    \centering
    \includegraphics[width=0.6\textwidth]{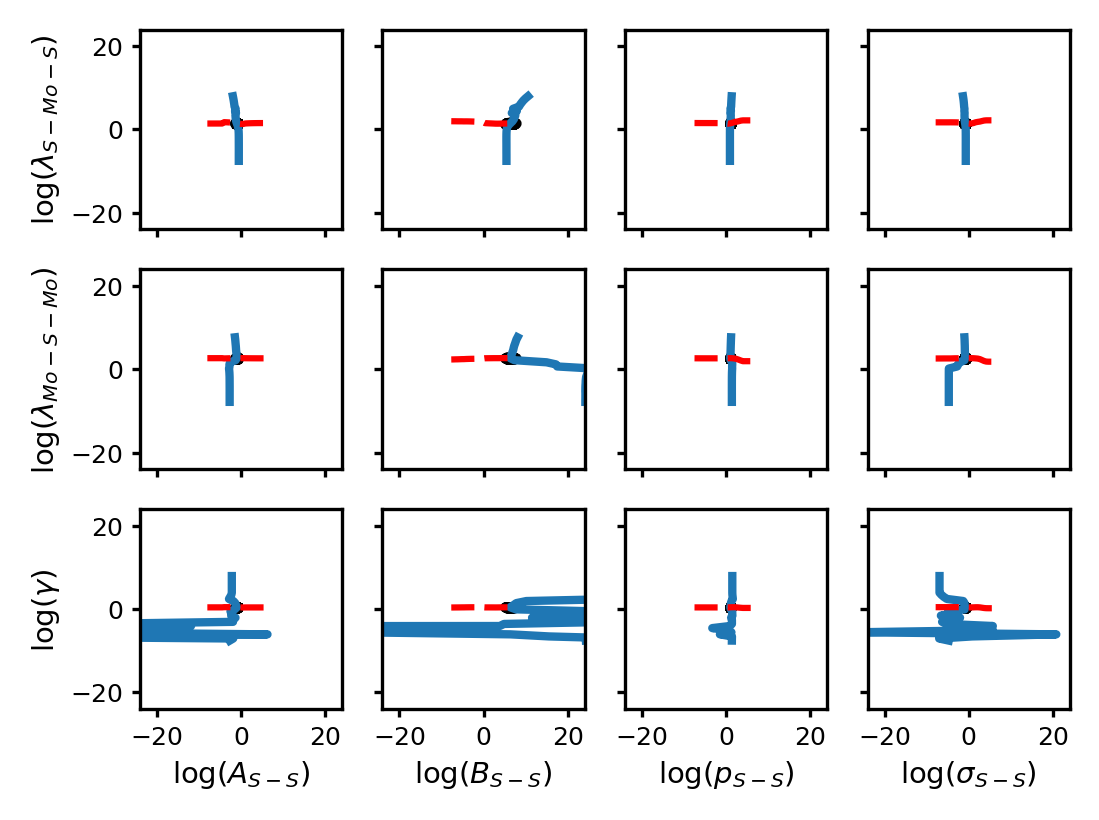}
    \caption[UQ results for SW potential S--S and 3-body parameters at $T = 1.71\times10^{-3}~T_0$]{
        Profile likelihood and MCMC samples for S--S and 3-body parameters at sampling temperature $1.71\times10^{-3}~T_0$ for the SW MoS$_2$ potential.
    }
\end{figure*}

\begin{figure*}[!h]
    \centering
    \includegraphics[width=0.6\textwidth]{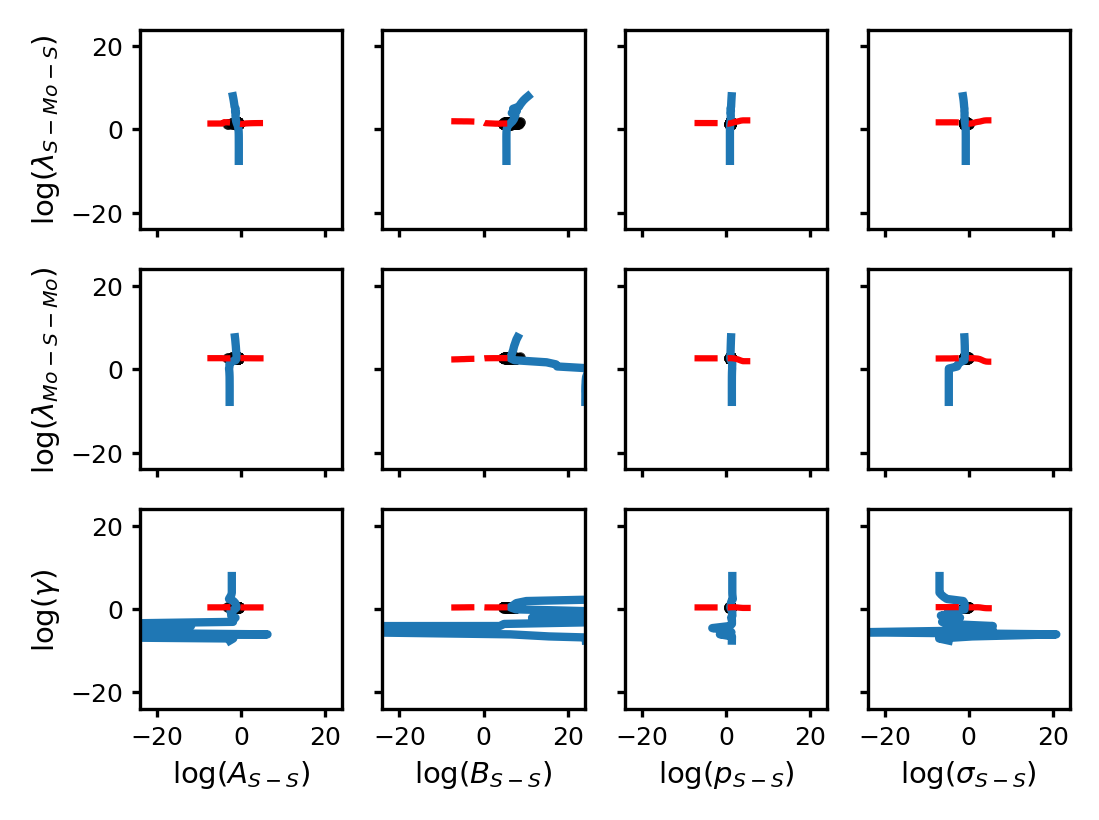}
    \caption[UQ results for SW potential S--S and 3-body parameters at $T = 5.40\times10^{-3}~T_0$]{
        Profile likelihood and MCMC samples for S--S and 3-body parameters at sampling temperature $5.40\times10^{-3}~T_0$ for the SW MoS$_2$ potential.
    }
\end{figure*}

\begin{figure*}[!h]
    \centering
    \includegraphics[width=0.6\textwidth]{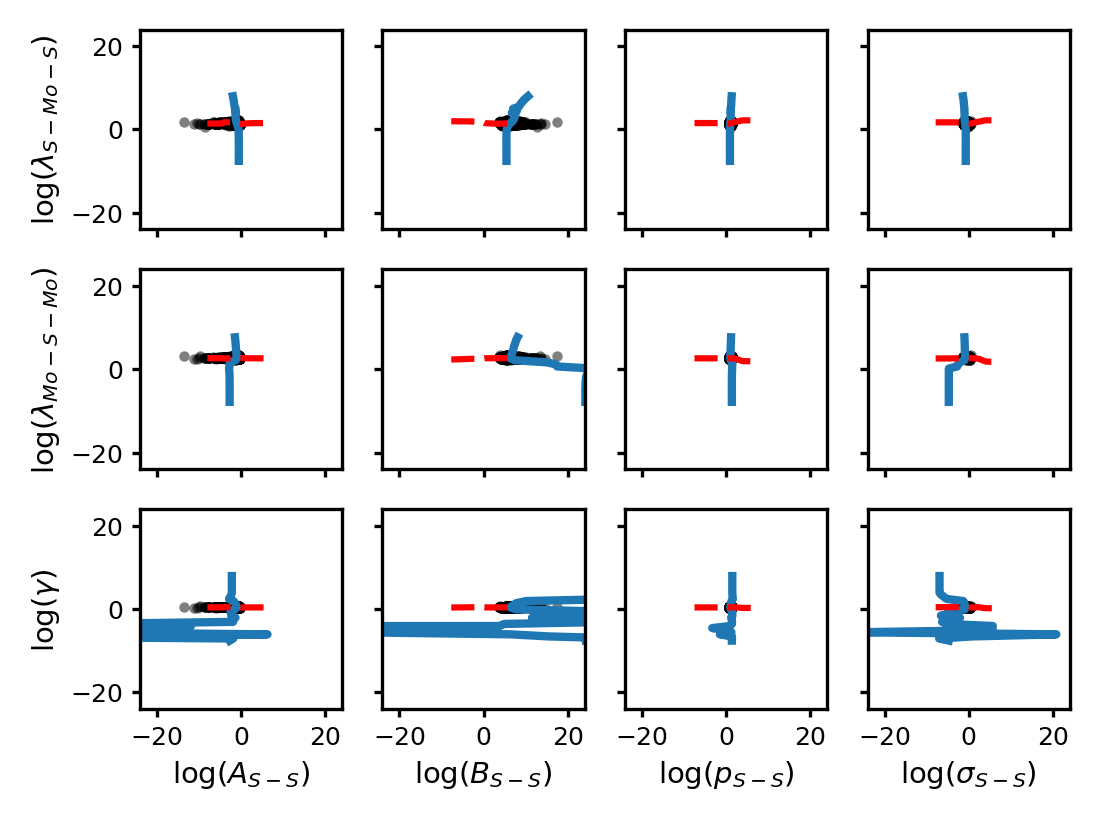}
    \caption[UQ results for SW potential S--S and 3-body parameters at $T = 1.71\times10^{-2}~T_0$]{
        Profile likelihood and MCMC samples for S--S and 3-body parameters at sampling temperature $1.71\times10^{-2}~T_0$ for the SW MoS$_2$ potential.
    }
\end{figure*}

\begin{figure*}[!h]
    \centering
    \includegraphics[width=0.6\textwidth]{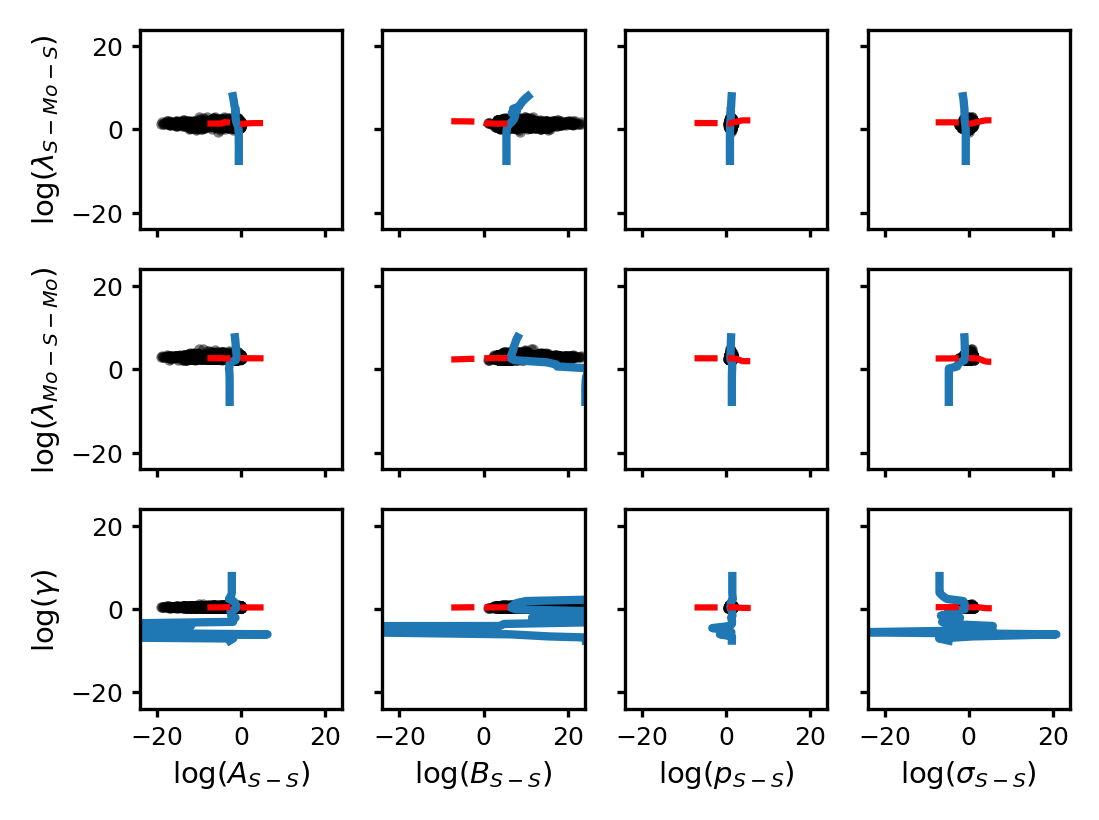}
    \caption[UQ results for SW potential S--S and 3-body parameters at $T = 5.40\times10^{-2}~T_0$]{
        Profile likelihood and MCMC samples for S--S and 3-body parameters at sampling temperature $5.40\times10^{-2}~T_0$ for the SW MoS$_2$ potential.
    }
\end{figure*}

\begin{figure*}[!h]
    \centering
    \includegraphics[width=0.6\textwidth]{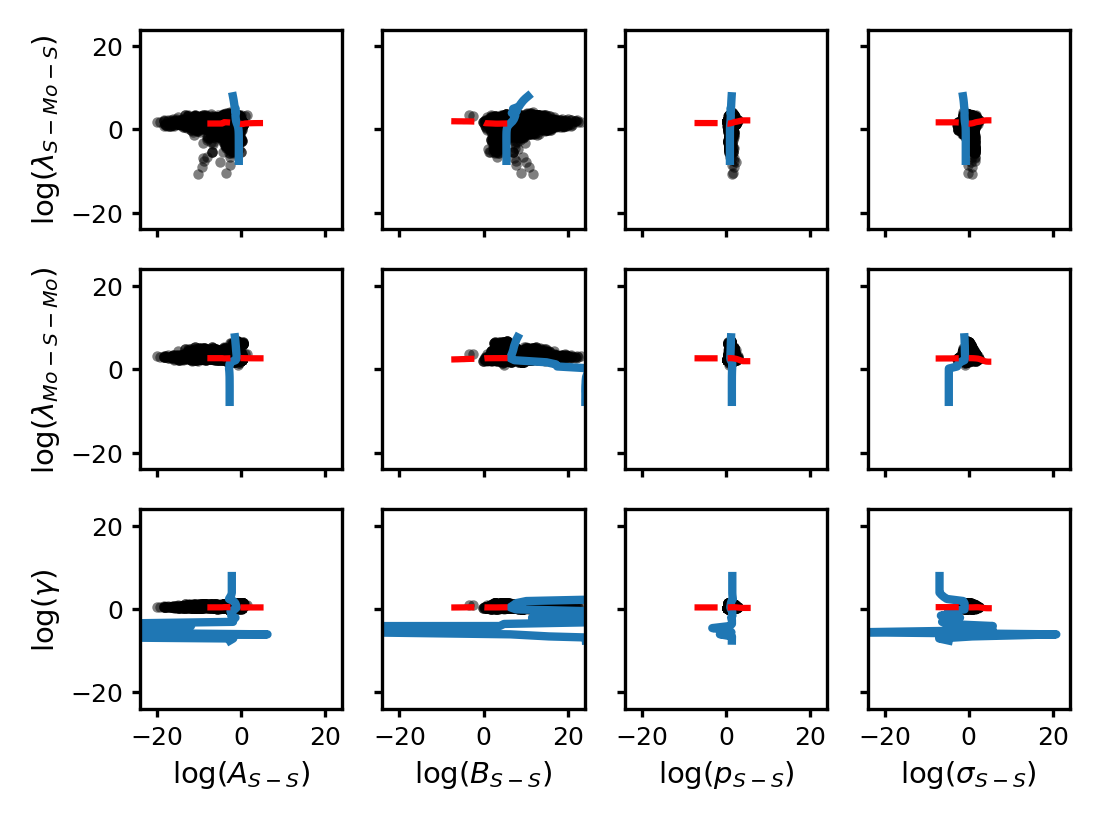}
    \caption[UQ results for SW potential S--S and 3-body parameters at $T = 1.71\times10^{-1}~T_0$]{
        Profile likelihood and MCMC samples for S--S and 3-body parameters at sampling temperature $1.71\times10^{-1}~T_0$ for the SW MoS$_2$ potential.
    }
\end{figure*}

\begin{figure*}[!h]
    \centering
    \includegraphics[width=0.6\textwidth]{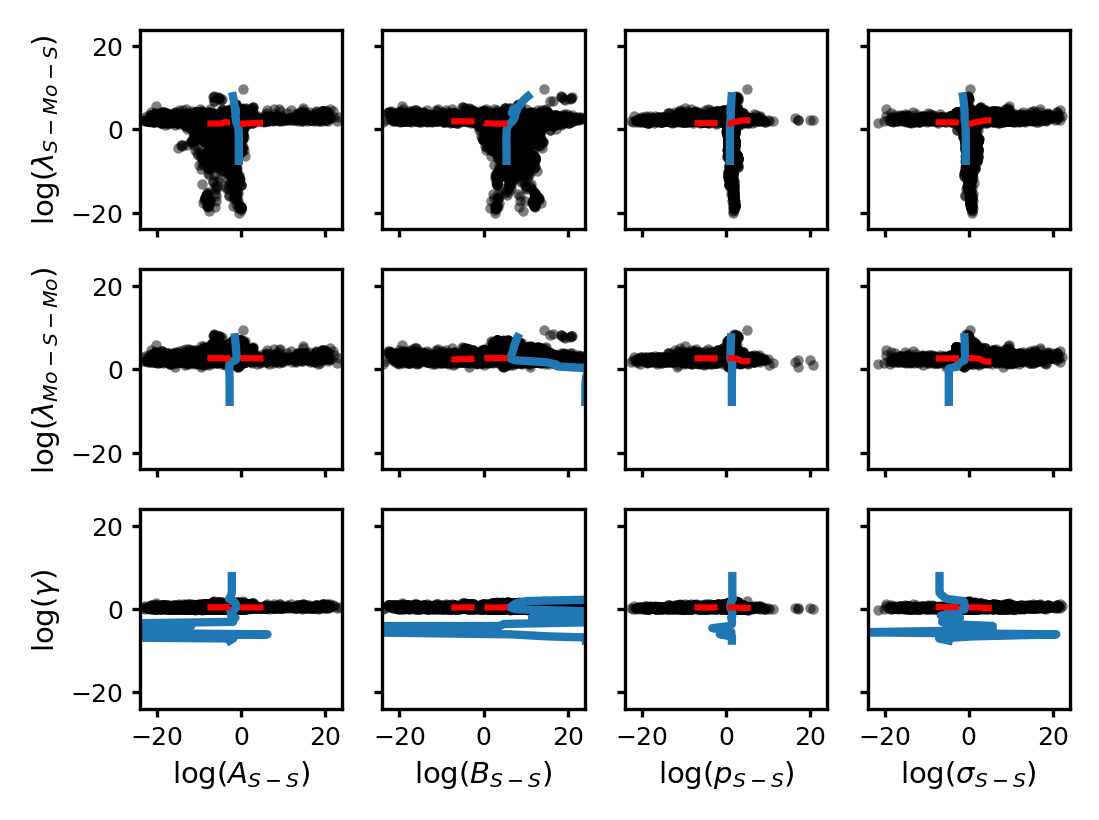}
    \caption[UQ results for SW potential S--S and 3-body parameters at $T = 5.40\times10^{-1}~T_0$]{
        Profile likelihood and MCMC samples for S--S and 3-body parameters at sampling temperature $5.40\times10^{-1}~T_0$ for the SW MoS$_2$ potential.
    }
\end{figure*}

\ifincludeTo
    \begin{figure*}[!h]
        \centering
        \includegraphics[width=0.6\textwidth]{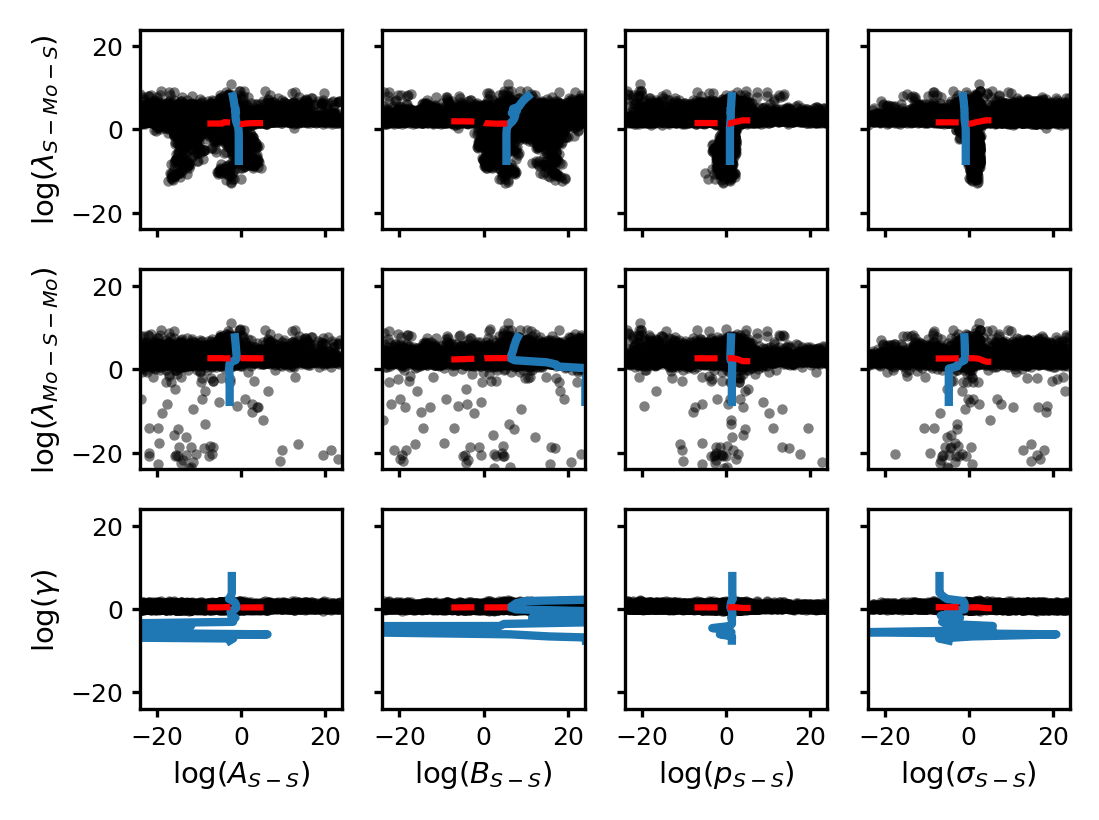}
        \caption[UQ results for SW potential S--S and 3-body parameters at $T = T_0$]{
            Profile likelihood and MCMC samples for S--S and 3-body parameters at sampling temperature $T_0$ for the SW MoS$_2$ potential.
        }
    \end{figure*}
\fi

\begin{figure*}[!h]
    \centering
    \includegraphics[width=0.6\textwidth]{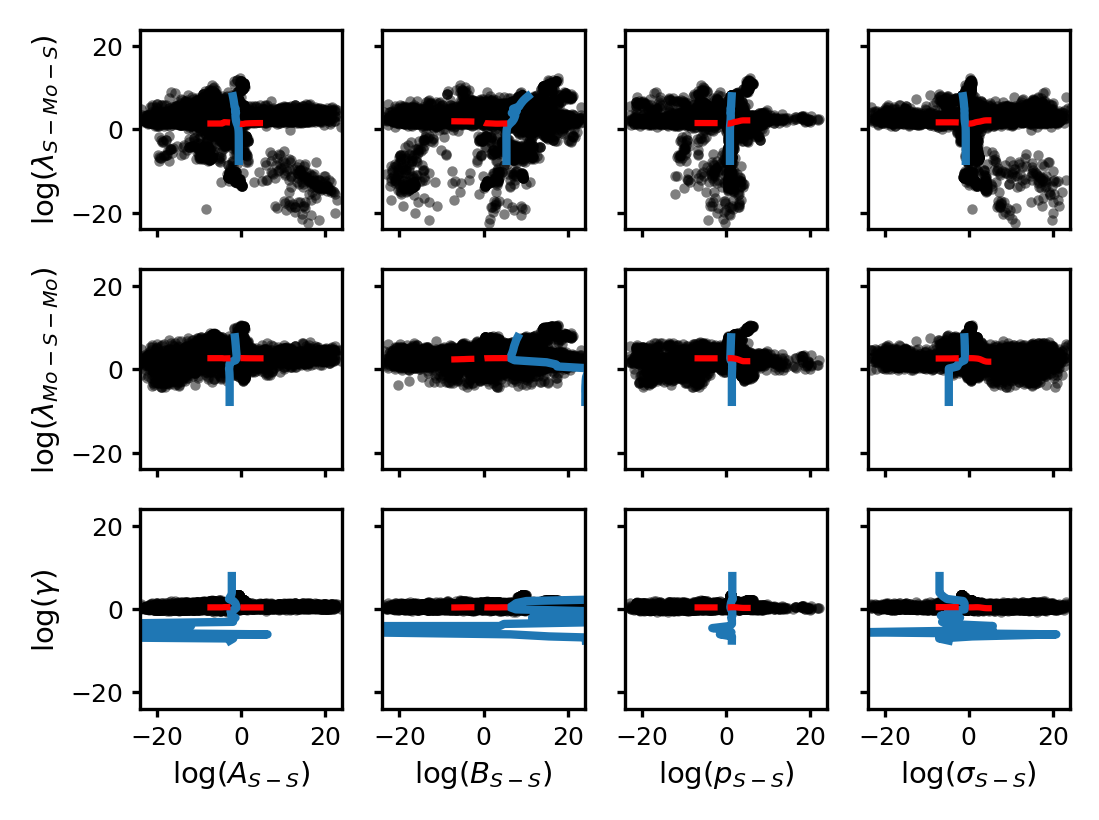}
    \caption[UQ results for SW potential S--S and 3-body parameters at $T = 1.71~T_0$]{
        Profile likelihood and MCMC samples for S--S and 3-body parameters at sampling temperature $1.71~T_0$ for the SW MoS$_2$ potential.
    }
\end{figure*}

\begin{figure*}[!h]
    \centering
    \includegraphics[width=0.6\textwidth]{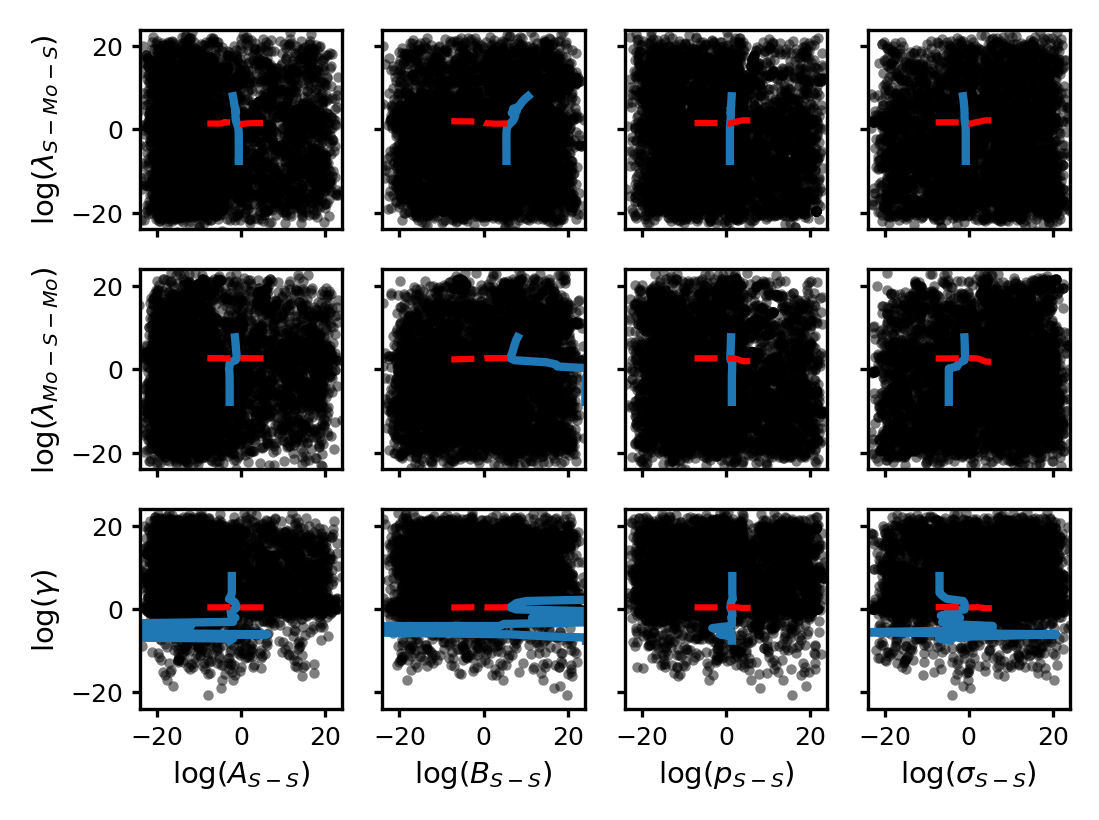}
    \caption[UQ results for SW potential S--S and 3-body parameters at $T = 5.40~T_0$]{
        Profile likelihood and MCMC samples for Mo--S and 3-body parameters at sampling temperature $5.40~T_0$ for the SW MoS$_2$ potential.
    }
\end{figure*}

\section{MCMC samples with uniform prior in linear parameterization}
\label{sec:sup_uq_sw_lin}

We show the MCMC sample of the SW potential for MoS$_2$ system, where the sampling is done using a linear (original) parameterization of the model.
The prior is a uniform distribution in a rectangular region bounded by $0 < \theta_i < 100$ and $0 < B_\text{S--S} < 1000$, where $\theta_i$ are the parameters in the potential (we increase the upper bound of the prior support for $B_\text{S--S}$ to include its best fit value).
The results shown here should be compared to the figures in Sec.~\ref{sec:sup_uq_sw_log}.

\begin{figure*}[!h]
    \centering
    \includegraphics[width=\textwidth]{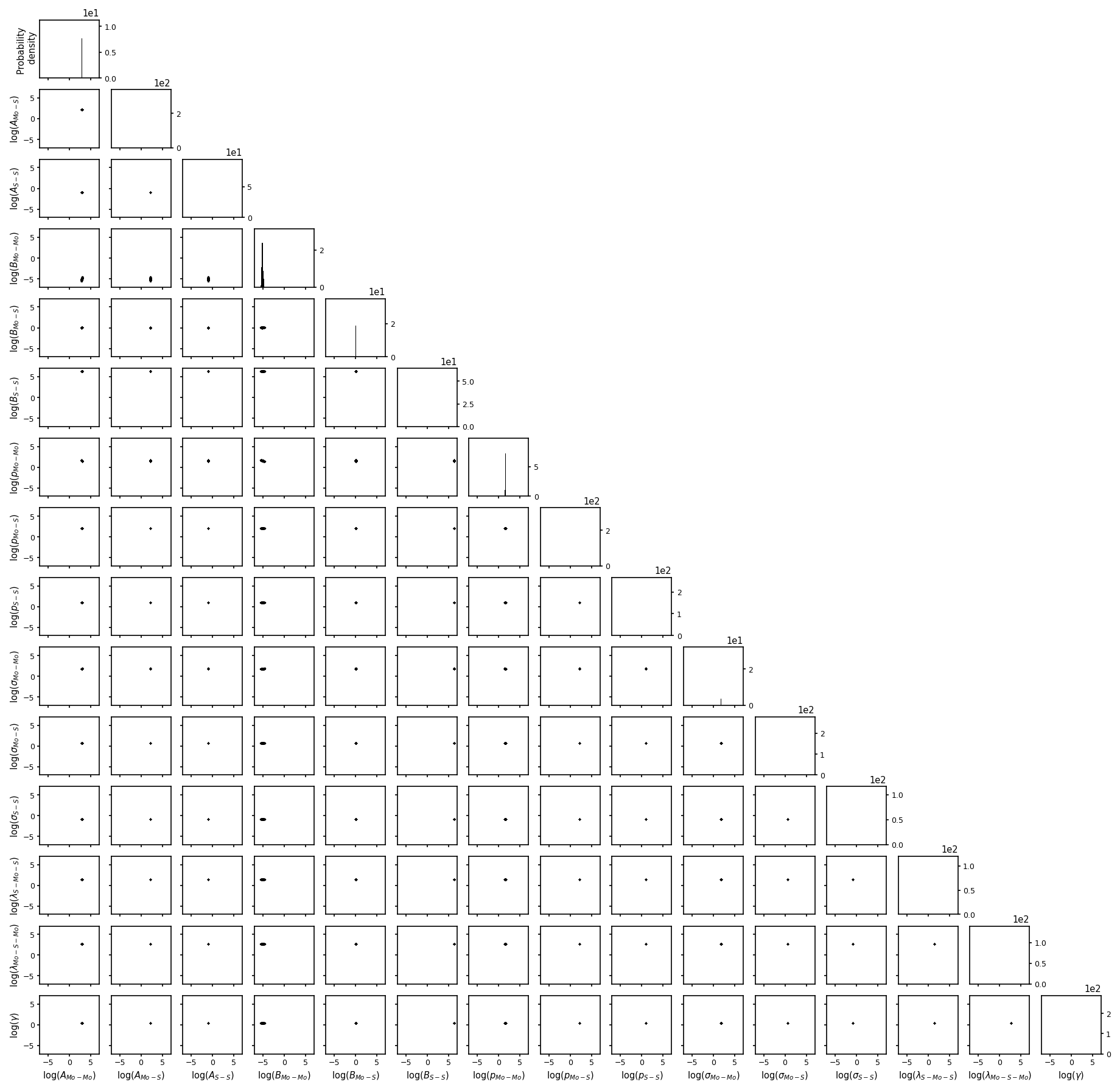}
    \caption[UQ results for SW potential in linear parameterization at $T = 5.40 \times 10^{-6}~T_0$]{
        MCMC samples with uniform prior in linear parameterization at sampling temperature $5.40 \times 10^{-6}~T_0$ for the SW MoS$_2$ potential.
    }
\end{figure*}

\begin{figure*}[!h]
    \centering
    \includegraphics[width=\textwidth]{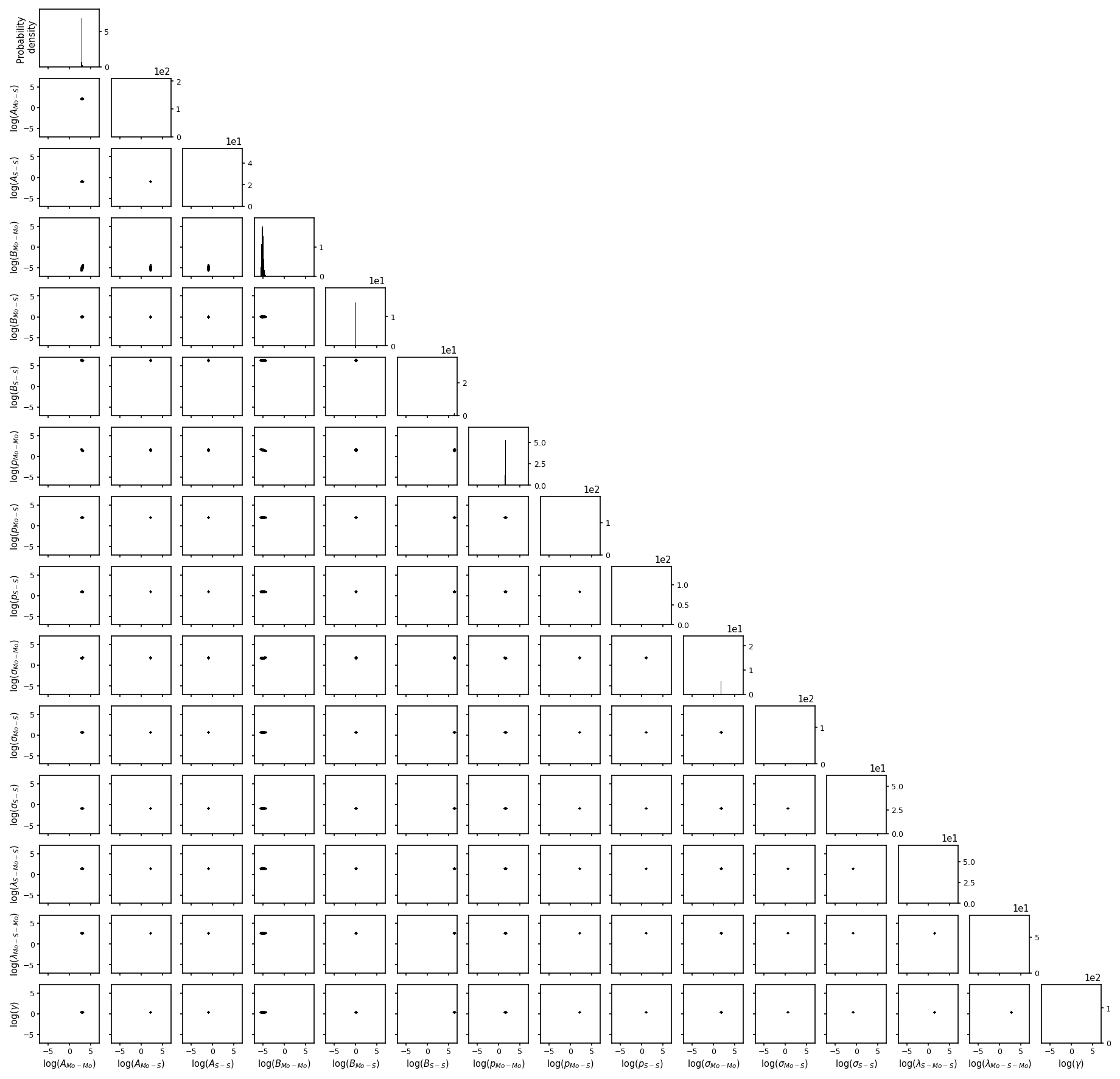}
    \caption[UQ results for SW potential in linear parameterization at $T = 1.71 \times 10^{-5}~T_0$]{
        MCMC samples with uniform prior in linear parameterization at sampling temperature $1.71 \times 10^{-5}~T_0$ for the SW MoS$_2$ potential.
    }
\end{figure*}

\begin{figure*}[!h]
    \centering
    \includegraphics[width=\textwidth]{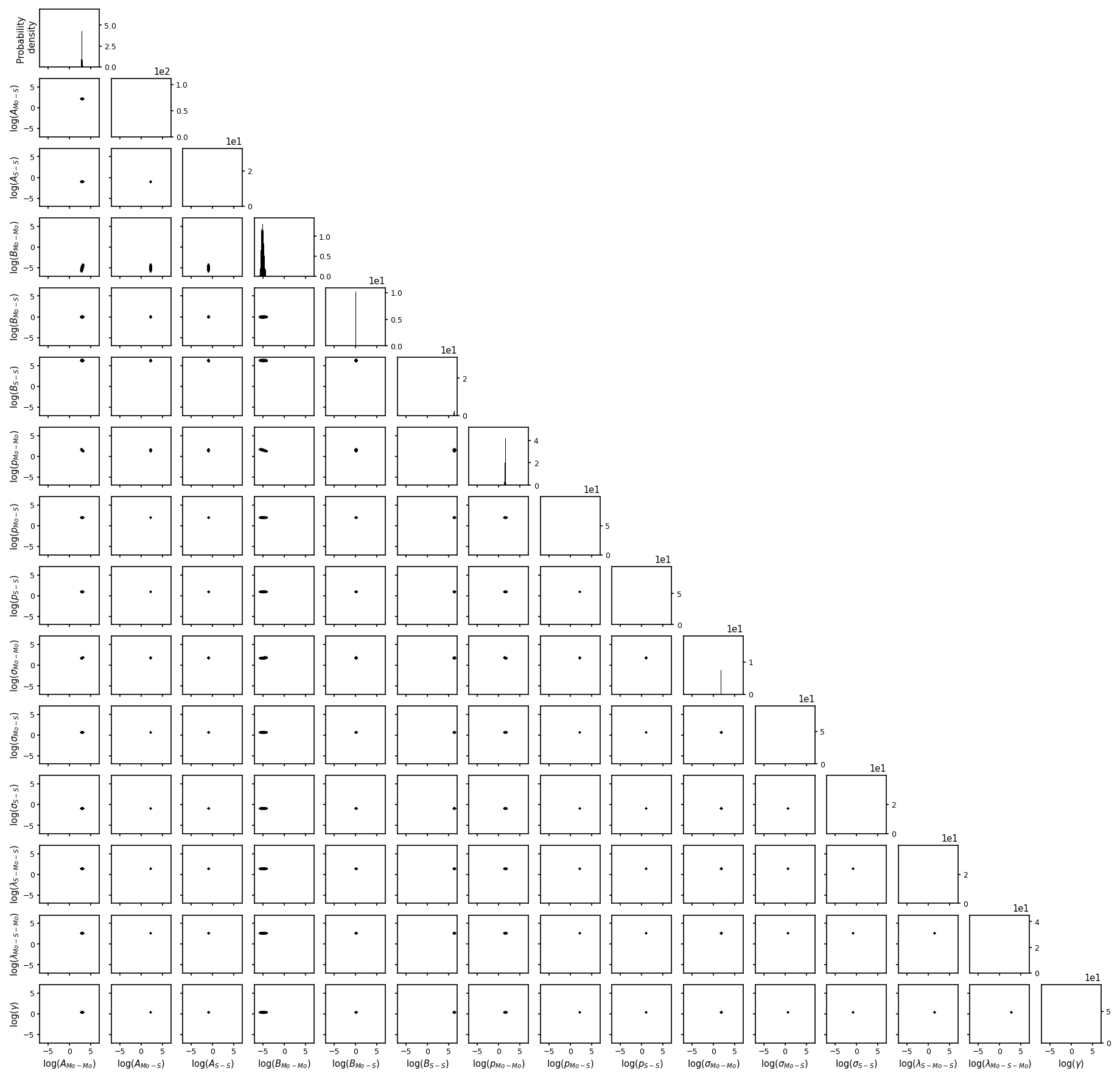}
    \caption[UQ results for SW potential in linear parameterization at $T = 5.40 \times 10^{-5}~T_0$]{
        MCMC samples with uniform prior in linear parameterization at sampling temperature $5.40 \times 10^{-5}~T_0$ for the SW MoS$_2$ potential.
    }
\end{figure*}

\begin{figure*}[!h]
    \centering
    \includegraphics[width=\textwidth]{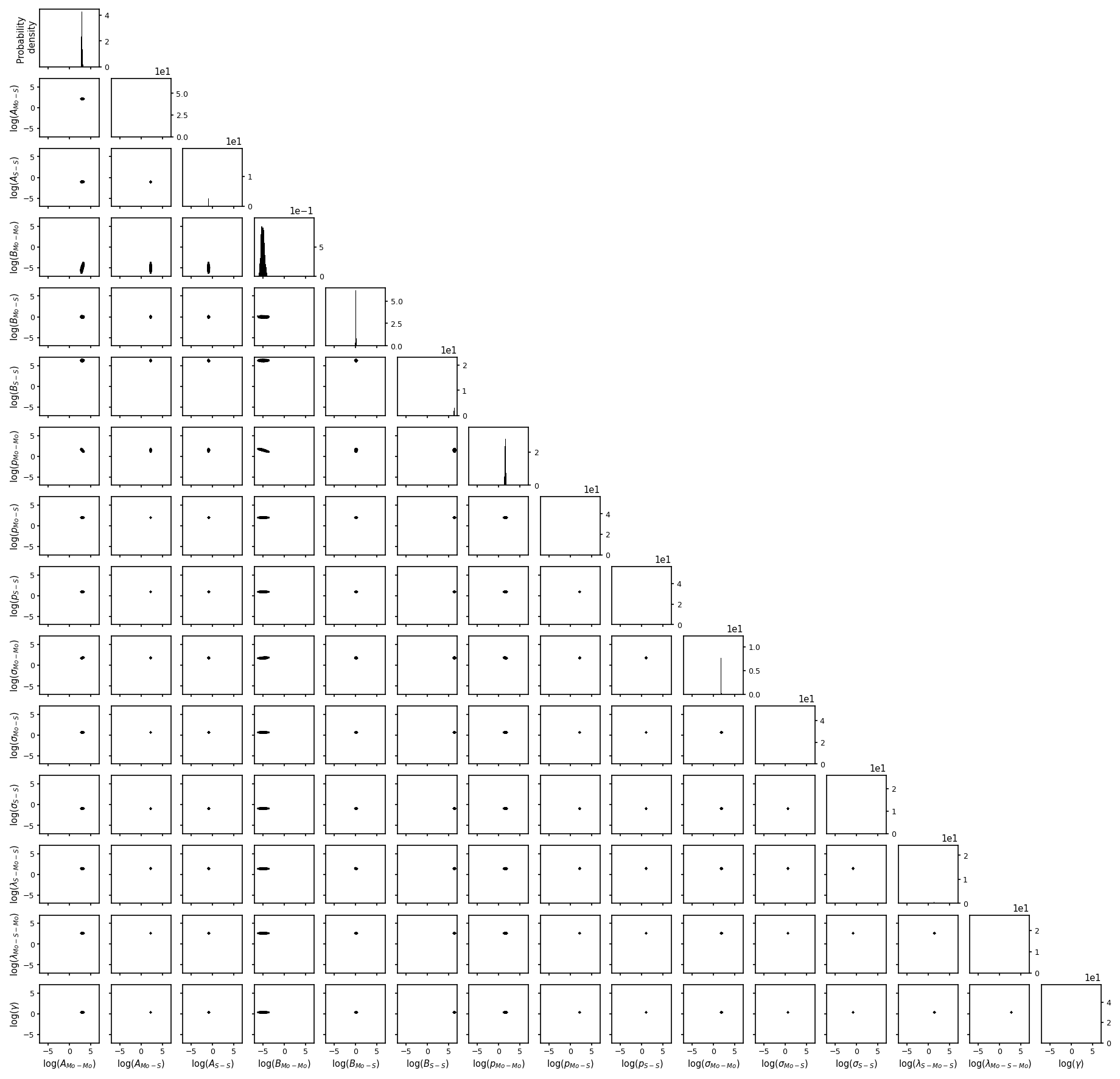}
    \caption[UQ results for SW potential in linear parameterization at $T = 1.71 \times 10^{-4}~T_0$]{
        MCMC samples with uniform prior in linear parameterization at sampling temperature $1.71 \times 10^{-4}~T_0$ for the SW MoS$_2$ potential.
    }
\end{figure*}

\begin{figure*}[!h]
    \centering
    \includegraphics[width=\textwidth]{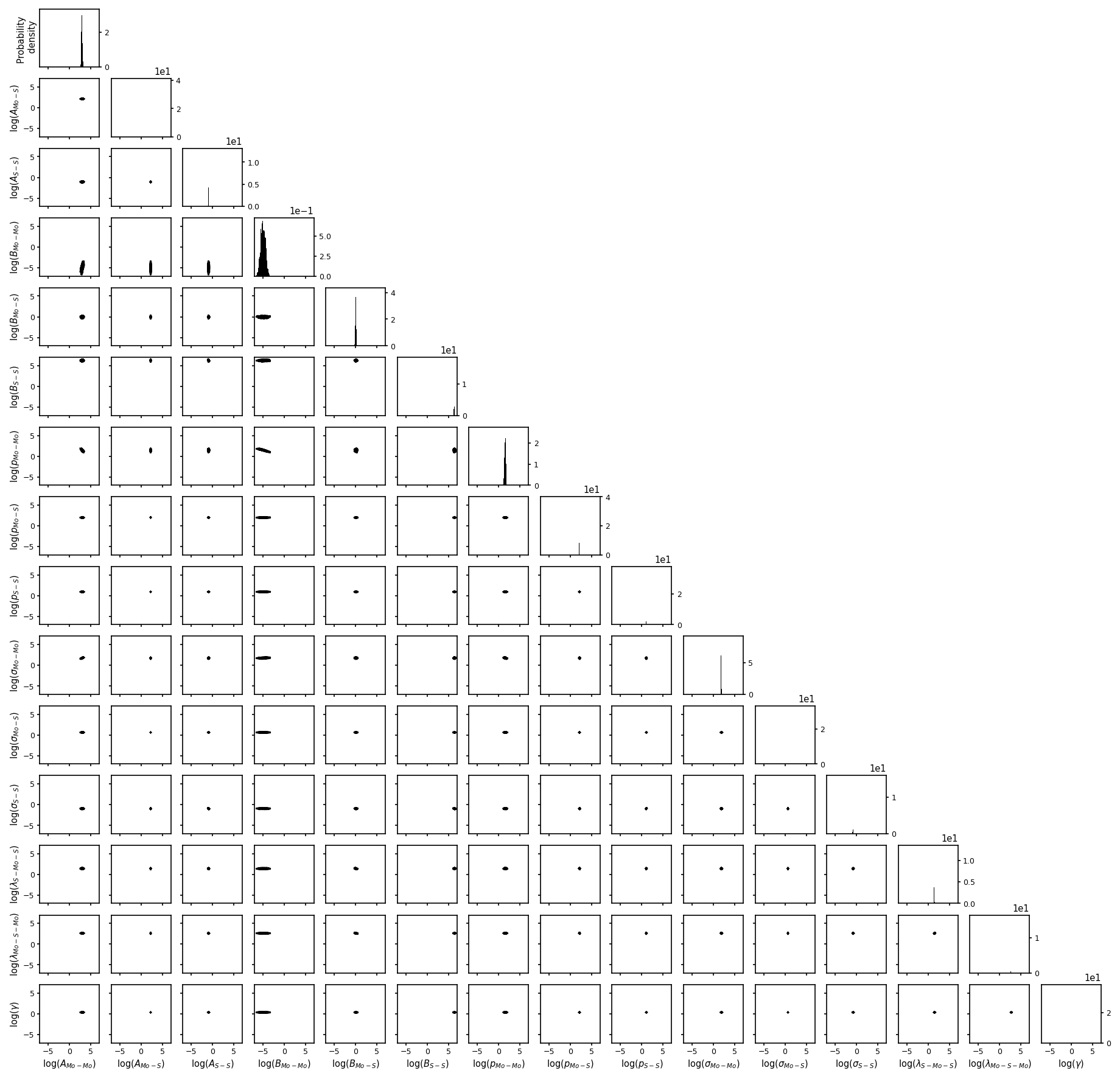}
    \caption[UQ results for SW potential in linear parameterization at $T = 5.40 \times 10^{-4}~T_0$]{
        MCMC samples with uniform prior in linear parameterization at sampling temperature $5.40 \times 10^{-4}~T_0$ for the SW MoS$_2$ potential.
    }
\end{figure*}

\begin{figure*}[!h]
    \centering
    \includegraphics[width=\textwidth]{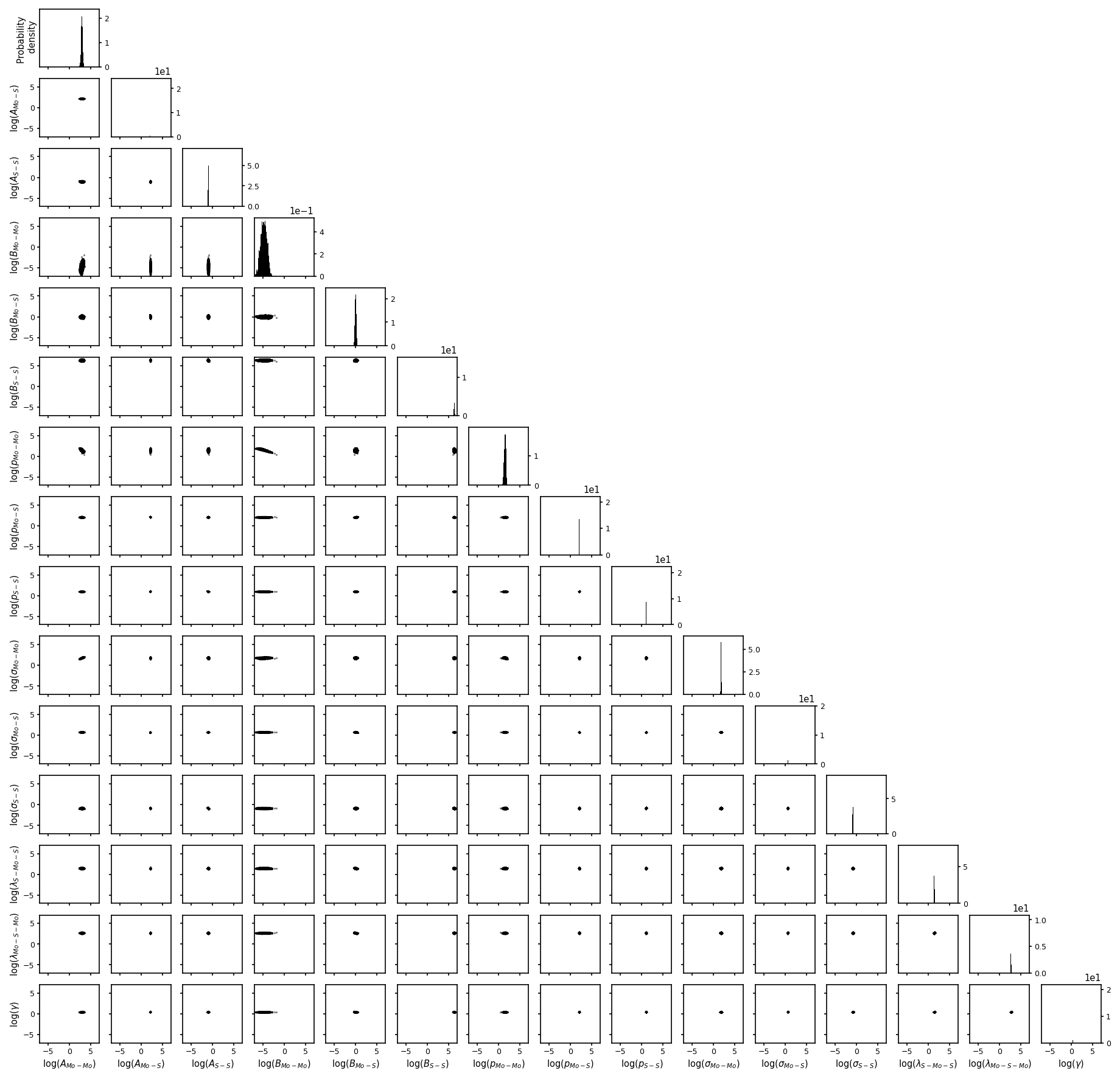}
    \caption[UQ results for SW potential in linear parameterization at $T = 1.71 \times 10^{-3}~T_0$]{
        MCMC samples with uniform prior in linear parameterization at sampling temperature $1.71 \times 10^{-3}~T_0$ for the SW MoS$_2$ potential.
    }
\end{figure*}

\begin{figure*}[!h]
    \centering
    \includegraphics[width=\textwidth]{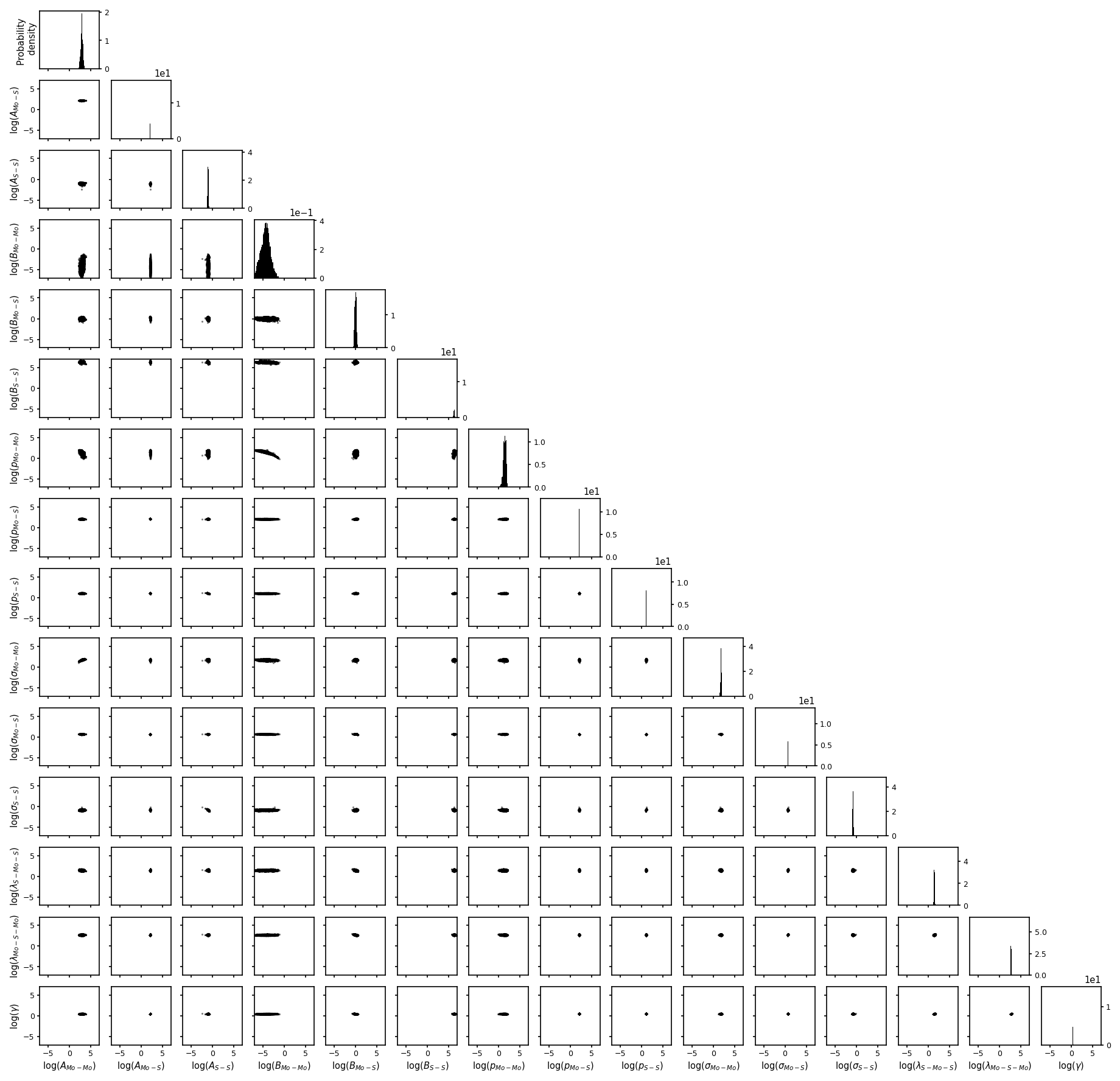}
    \caption[UQ results for SW potential in linear parameterization at $T = 5.40 \times 10^{-3}~T_0$]{
        MCMC samples with uniform prior in linear parameterization at sampling temperature $5.40 \times 10^{-3}~T_0$ for the SW MoS$_2$ potential.
    }
\end{figure*}

\begin{figure*}[!h]
    \centering
    \includegraphics[width=\textwidth]{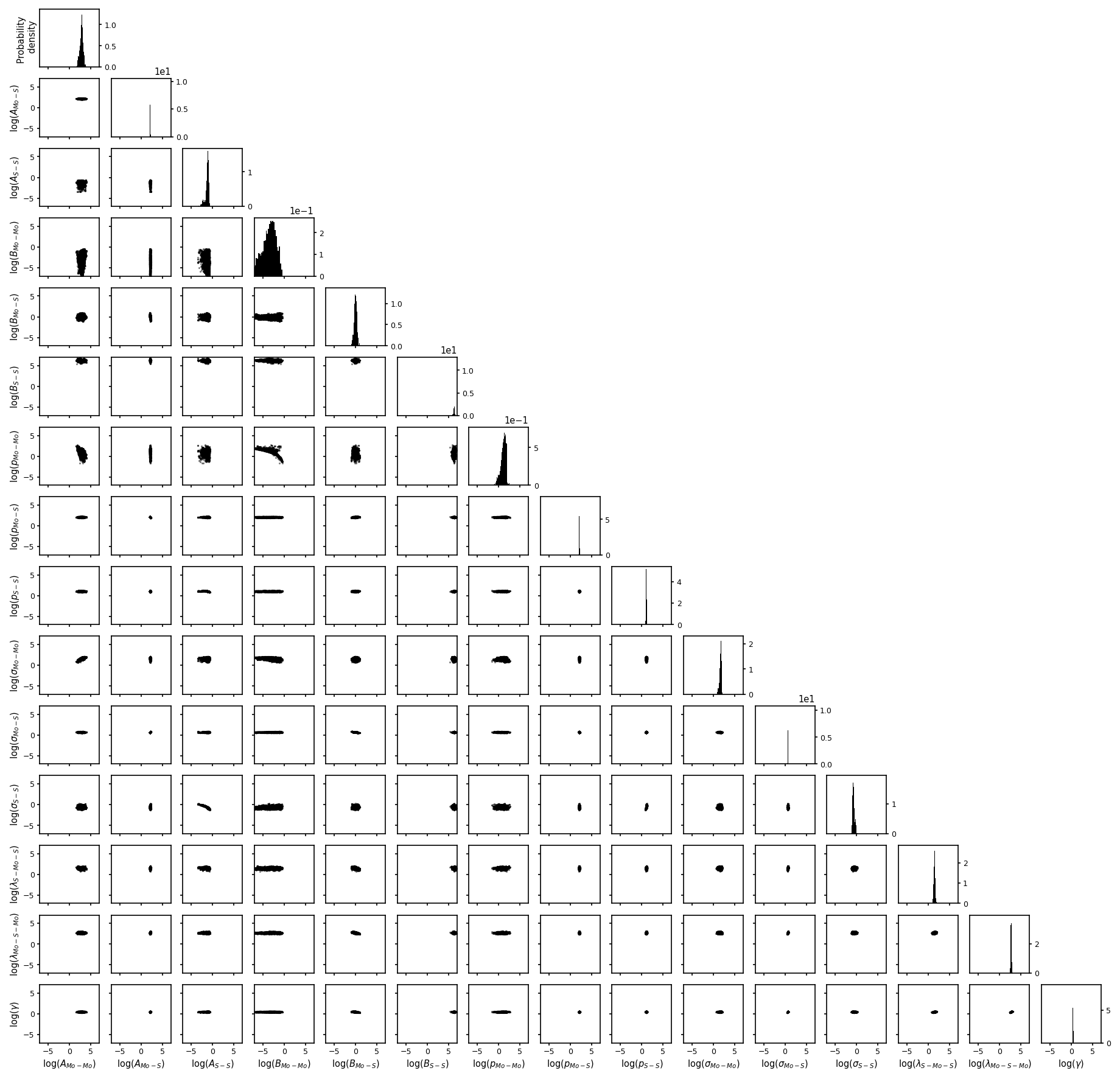}
    \caption[UQ results for SW potential in linear parameterization at $T = 1.71 \times 10^{-2}~T_0$]{
        MCMC samples with uniform prior in linear parameterization at sampling temperature $1.71 \times 10^{-2}~T_0$ for the SW MoS$_2$ potential.
    }
\end{figure*}

\begin{figure*}[!h]
    \centering
    \includegraphics[width=\textwidth]{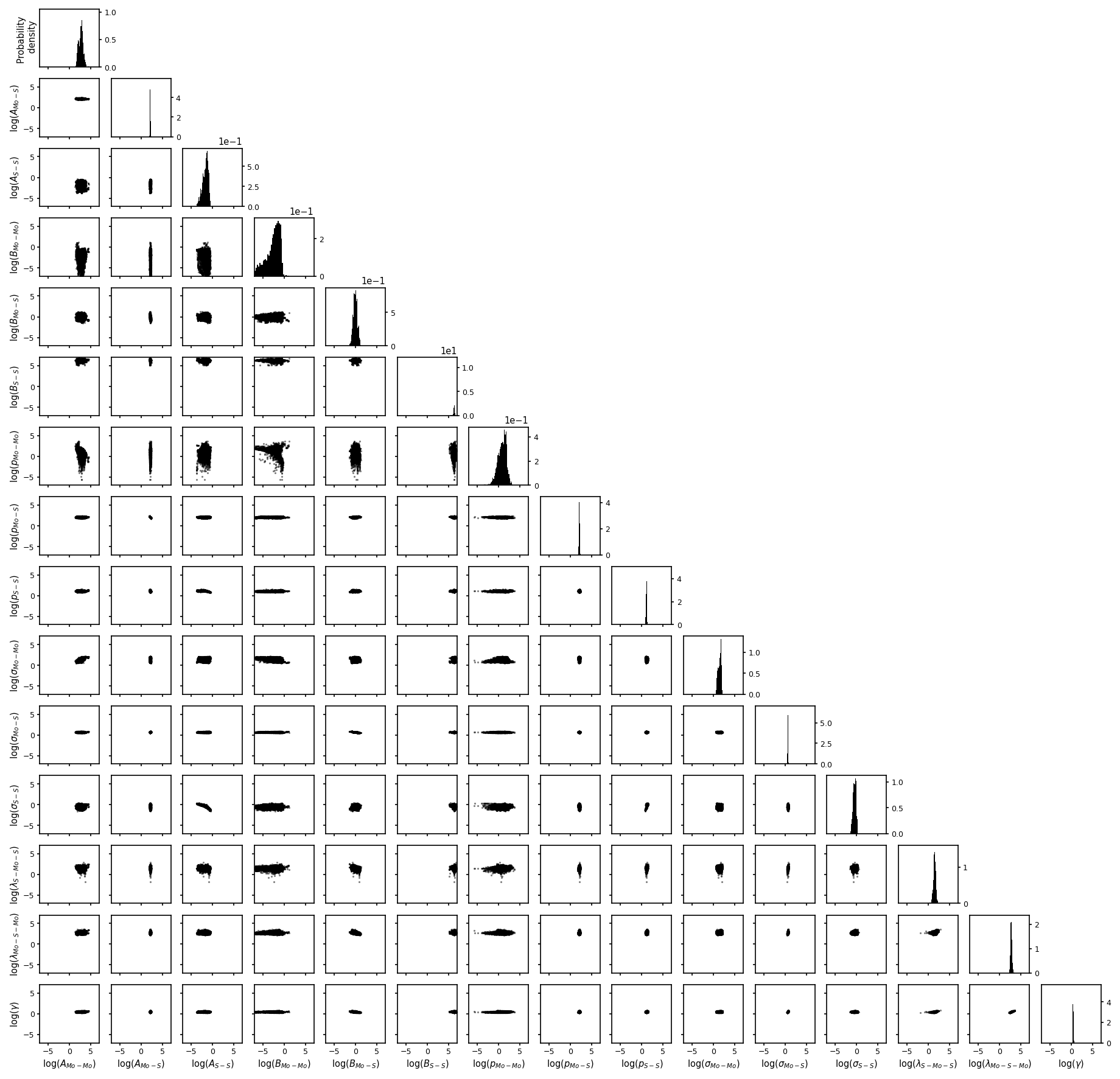}
    \caption[UQ results for SW potential in linear parameterization at $T = 5.40 \times 10^{-2}~T_0$]{
        MCMC samples with uniform prior in linear parameterization at sampling temperature $5.40 \times 10^{-2}~T_0$ for the SW MoS$_2$ potential.
    }
\end{figure*}

\begin{figure*}[!h]
    \centering
    \includegraphics[width=\textwidth]{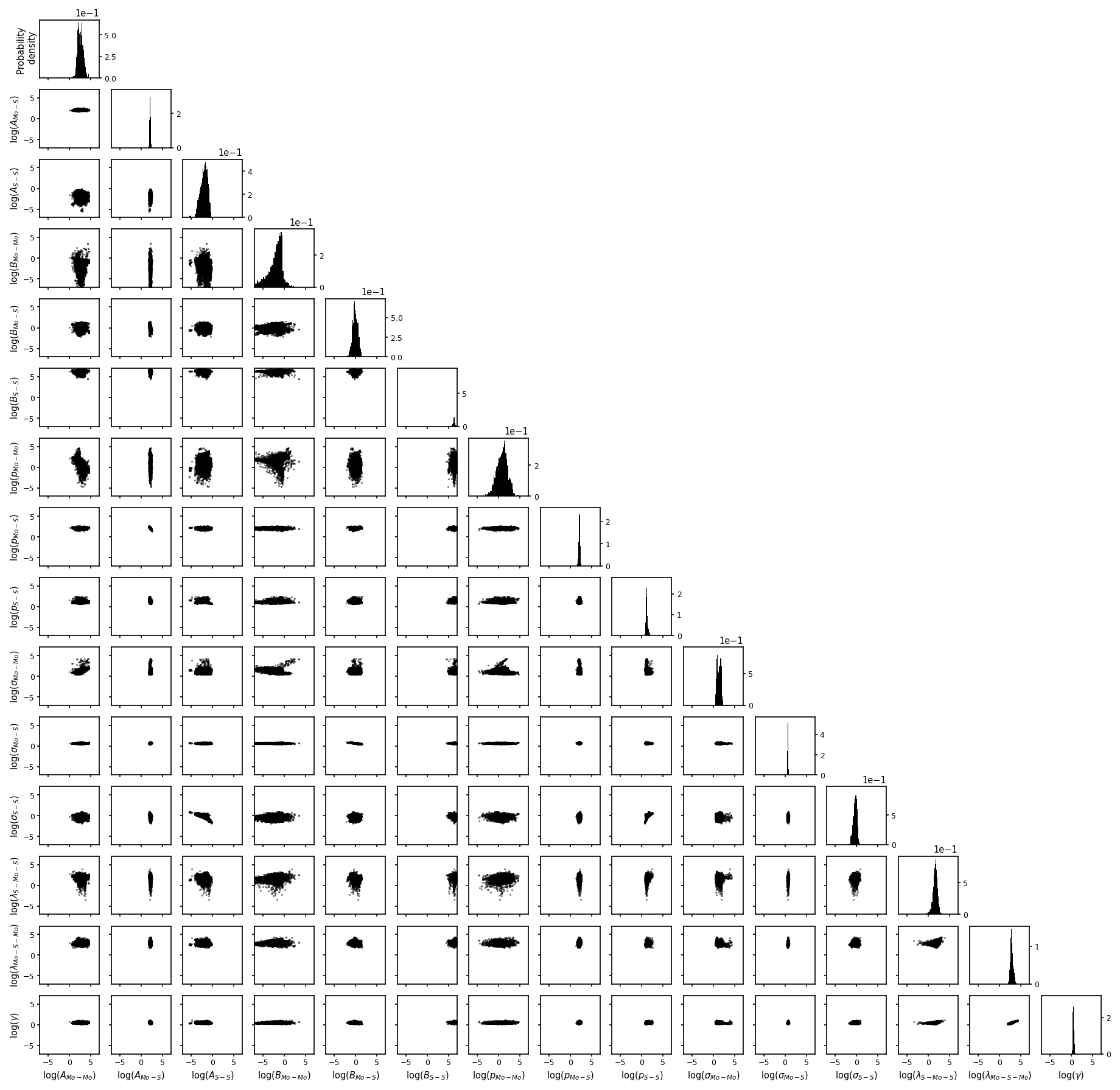}
    \caption[UQ results for SW potential in linear parameterization at $T = 1.71 \times 10^{-1}~T_0$]{
        MCMC samples with uniform prior in linear parameterization at sampling temperature $1.71 \times 10^{-1}~T_0$ for the SW MoS$_2$ potential.
    }
\end{figure*}

\begin{figure*}[!h]
    \centering
    \includegraphics[width=\textwidth]{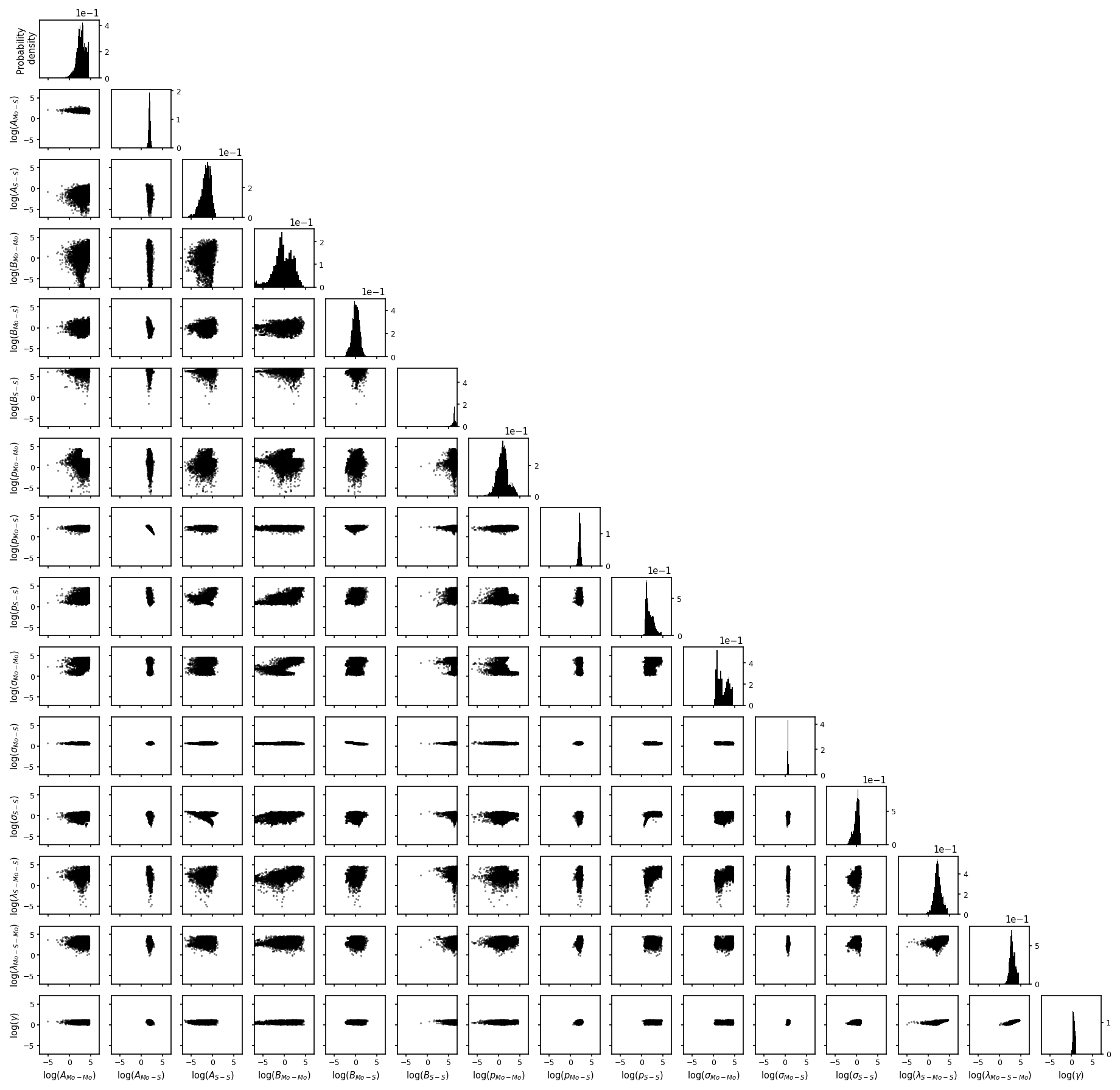}
    \caption[UQ results for SW potential in linear parameterization at $T = 5.40 \times 10^{-1}~T_0$]{
        MCMC samples with uniform prior in linear parameterization at sampling temperature $5.40 \times 10^{-1}~T_0$ for the SW MoS$_2$ potential.
    }
\end{figure*}

\ifincludeTo
    \begin{figure*}[!h]
        \centering
        \includegraphics[width=\textwidth]{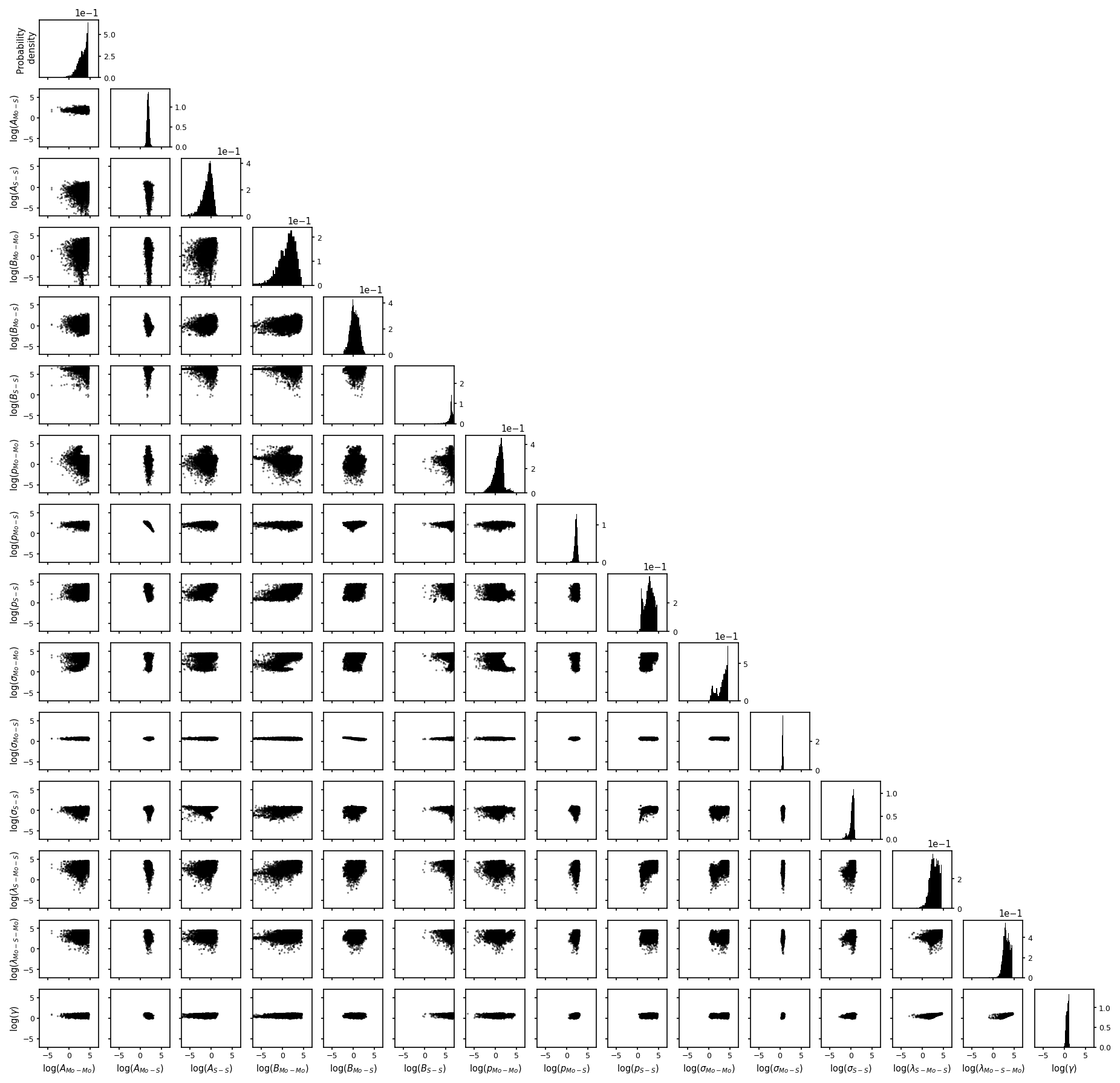}
        \caption[UQ results for SW potential in linear parameterization at $T = T_0$]{
            MCMC samples with uniform prior in linear parameterization at sampling temperature $T_0$ for the SW MoS$_2$ potential.
        }
    \end{figure*}
\fi

\begin{figure*}[!h]
    \centering
    \includegraphics[width=\textwidth]{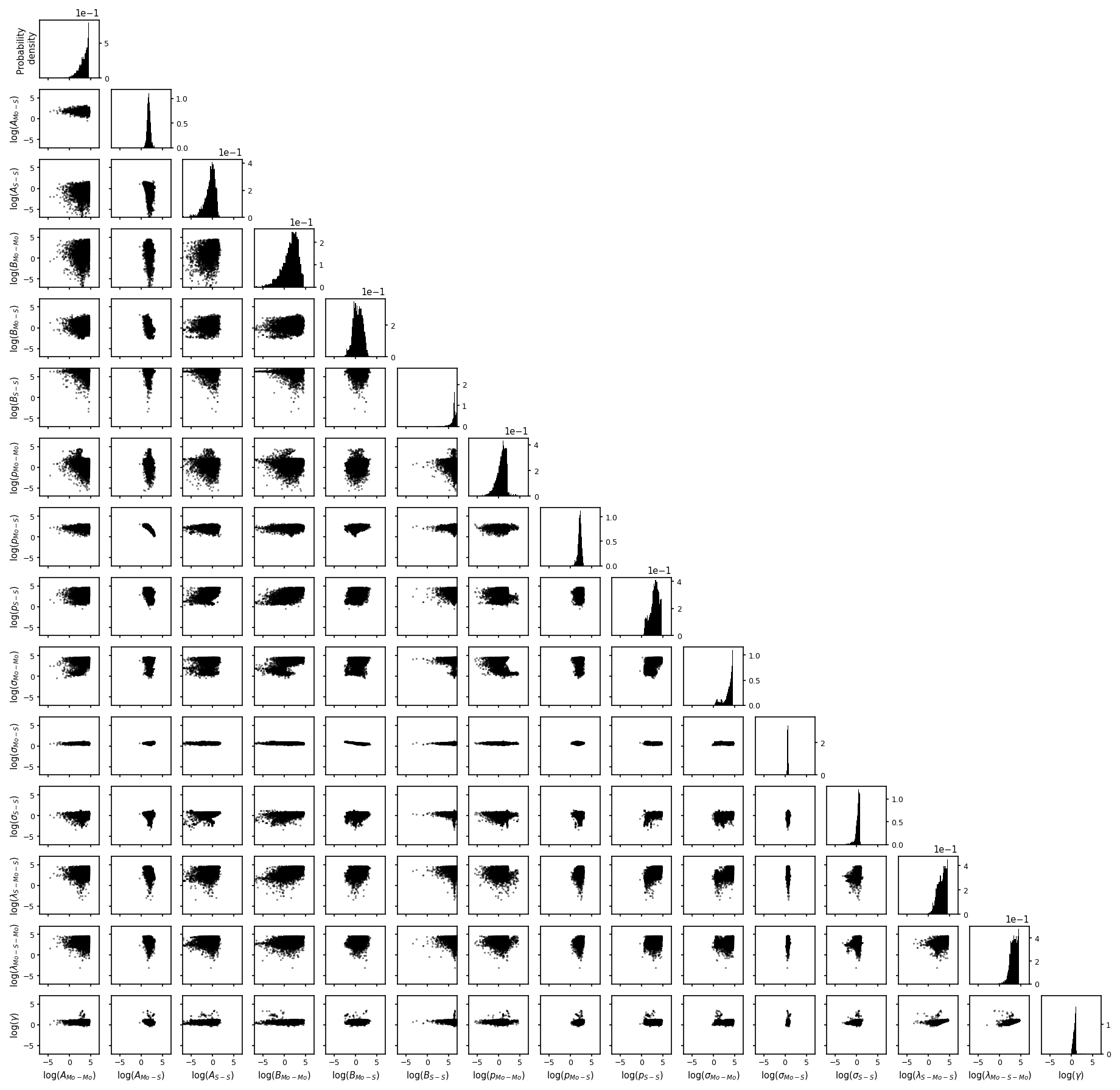}
    \caption[UQ results for SW potential in linear parameterization at $T = 1.71~T_0$]{
        MCMC samples with uniform prior in linear parameterization at sampling temperature $1.71~T_0$ for the SW MoS$_2$ potential.
    }
\end{figure*}

\begin{figure*}[!h]
    \centering
    \includegraphics[width=\textwidth]{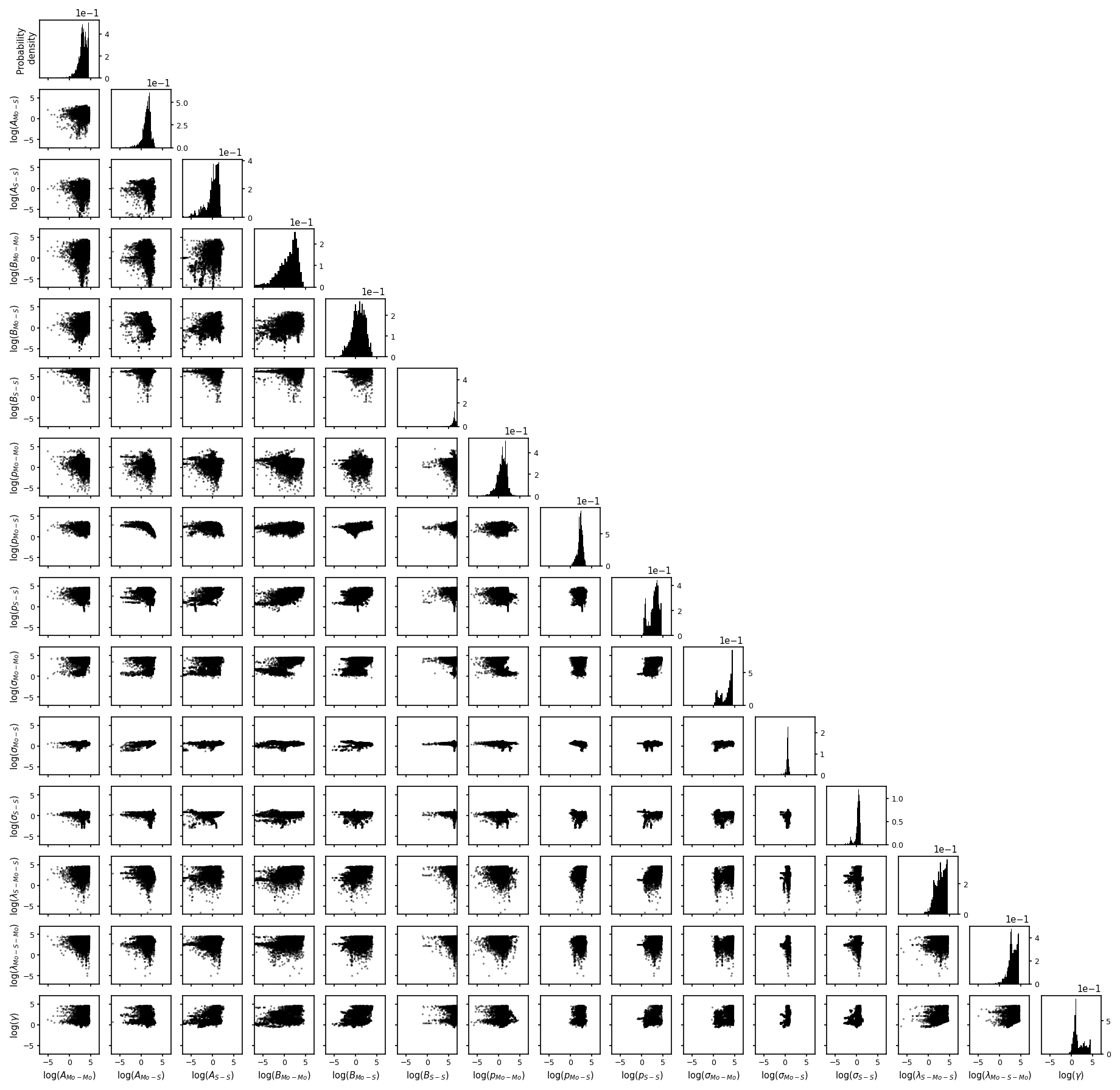}
    \caption[UQ results for SW potential in linear parameterization at $T = 5.40~T_0$]{
        MCMC samples with uniform prior in linear parameterization at sampling temperature $5.40~T_0$ for the SW MoS$_2$ potential.
    }
\end{figure*}